\documentclass[aps,prb,twocolumn,english,balance,superscriptaddress,floats,showpacs,longbibliography,nofootinbib]{revtex4-2}
\usepackage{amsmath}
\usepackage{amssymb}
\usepackage{stmaryrd}
\usepackage{graphicx}
\usepackage{esint}
\usepackage{subfigure}
\usepackage{multirow}
\usepackage{mathtools}
\usepackage{xcolor}
\usepackage{gensymb}
\usepackage{MnSymbol}
\usepackage[normalem]{ulem}
\usepackage[T1]{fontenc}

\makeatletter

\newcommand{\beq}{\begin{equation}}
\newcommand{\eeq}{\end{equation}}
\newcommand{\bea}{\begin{eqnarray}}
\newcommand{\eea}{\end{eqnarray}}
\newcommand{\bwt}{\begin{widetext}}
\newcommand{\ewt}{\end{widetext}}
\@ifundefined{textcolor}{}
{%
 \definecolor{BLACK}{gray}{0}
 \definecolor{WHITE}{gray}{1}
 \definecolor{RED}{rgb}{1,0,0}
 \definecolor{GREEN}{rgb}{0,1,0}
 \definecolor{BLUE}{rgb}{0,0,1}
 \definecolor{CYAN}{cmyk}{1,0,0,0}
 \definecolor{MAGENTA}{cmyk}{0,1,0,0}
 \definecolor{YELLOW}{cmyk}{0,0,1,0}
}

\newcommand{\be}{\mathbf{e}}
\newcommand{\bj}{\mathbf{j}}

\newcommand{\bN}{\mathbf{N}}
\newcommand{\bk}{\mathbf{k}}

\newcommand{\br}{\mathbf{r}}
\newcommand{\bR}{\mathbf{R}}
\newcommand{\bE}{\mathbf{E}}
\newcommand{\bA}{\mathbf{A}}

\newcommand{\bx}{\mathbf{x}}
\newcommand{\by}{\mathbf{y}}
\newcommand{\bz}{\mathbf{z}}
\newcommand{\bB}{\mathbf{B}}
\newcommand{\eps}{\epsilon}

\usepackage{babel}
\makeatother
\usepackage{babel}

\newcommand{\newtext}[1]{\textcolor{black}{#1}}
\renewcommand{\sout}[1]{\unskip}

\begin{document}

\title{Observation of giant nonlinear Hall conductivity in Bernal bilayer graphene}

\author{Dmitry V. Chichinadze$^{\parallel}$}
%\thanks{These authors contributed equally to this work}
\email{chichinadze@magnet.fsu.edu}
\affiliation{National High Magnetic Field Laboratory, Tallahassee, Florida, 32310, USA}
\author{Naiyuan James Zhang$^{\parallel}$}
%\thanks{These authors contributed equally to this work}
\affiliation{Department of Physics, Brown University, Providence, Rhode Island, 02912, USA}
\author{Jiang-Xiazi Lin}
\affiliation{Department of Physics, Brown University, Providence, Rhode Island, 02912, USA}
\author{\newtext{Erin Morissette}}
\affiliation{Department of Physics, Brown University, Providence, Rhode Island, 02912, USA}
\author{Xiaoyu Wang}
\affiliation{National High Magnetic Field Laboratory, Tallahassee, Florida, 32310, USA}
\author{Kenji Watanabe}
\affiliation{Research Center for Functional Materials, National Institute for Materials Science, 1-1 Namiki, Tsukuba, 305-0044, Japan}
\author{Takashi Taniguchi}
\affiliation{International Center for Materials Nanoarchitectonics, National Institute for Materials Science, 1-1 Namiki, Tsukuba, 305-0044, Japan}
\author{Oskar Vafek}
\email{vafek@magnet.fsu.edu}
\affiliation{National High Magnetic Field Laboratory, Tallahassee, Florida, 32310, USA}
\affiliation{Department of Physics,
Florida State University, Tallahassee, Florida, 32306, USA}
\author{J.I.A. Li}
\email{jia\_li@brown.edu}
\affiliation{Department of Physics, Brown University, Providence, Rhode Island, 02912, USA}

\def\thefootnote{$\parallel$}\footnotetext{These authors contributed equally to this work}

\begin{abstract}
 In a system of two-dimensional electrons, a combination of broken symmetry, interactions, and nontrivial topology can conspire to give rise to a nonlinear transport regime, where electric current density scales as the square of electric field. This regime has become a venue for exciting discoveries such as the nonlinear Hall effect and diode-like nonreciprocal transport. However, interpretation of experimental data is challenging in the nonlinear regime as DC transport is described by a rank-3 conductivity tensor with 6 free parameters. 
Here, we resolve this challenge by analytically solving for the nonlinear potential distribution across the disk sample for an arbitrary linear and nonlinear conductivity tensors. This allows us to unambiguously extract all components of the nonlinear tensor from experimental measurement. Using this novel tool, we identify giant nonlinear Hall effect in Bernal bilayer graphene. Our methodology provides the first systematic framework for interpreting nonlinear transport and uncovers a new route towards understanding quasi-2D materials. 
\end{abstract}

\maketitle

\begin{figure*}
\includegraphics[width=0.99\linewidth]{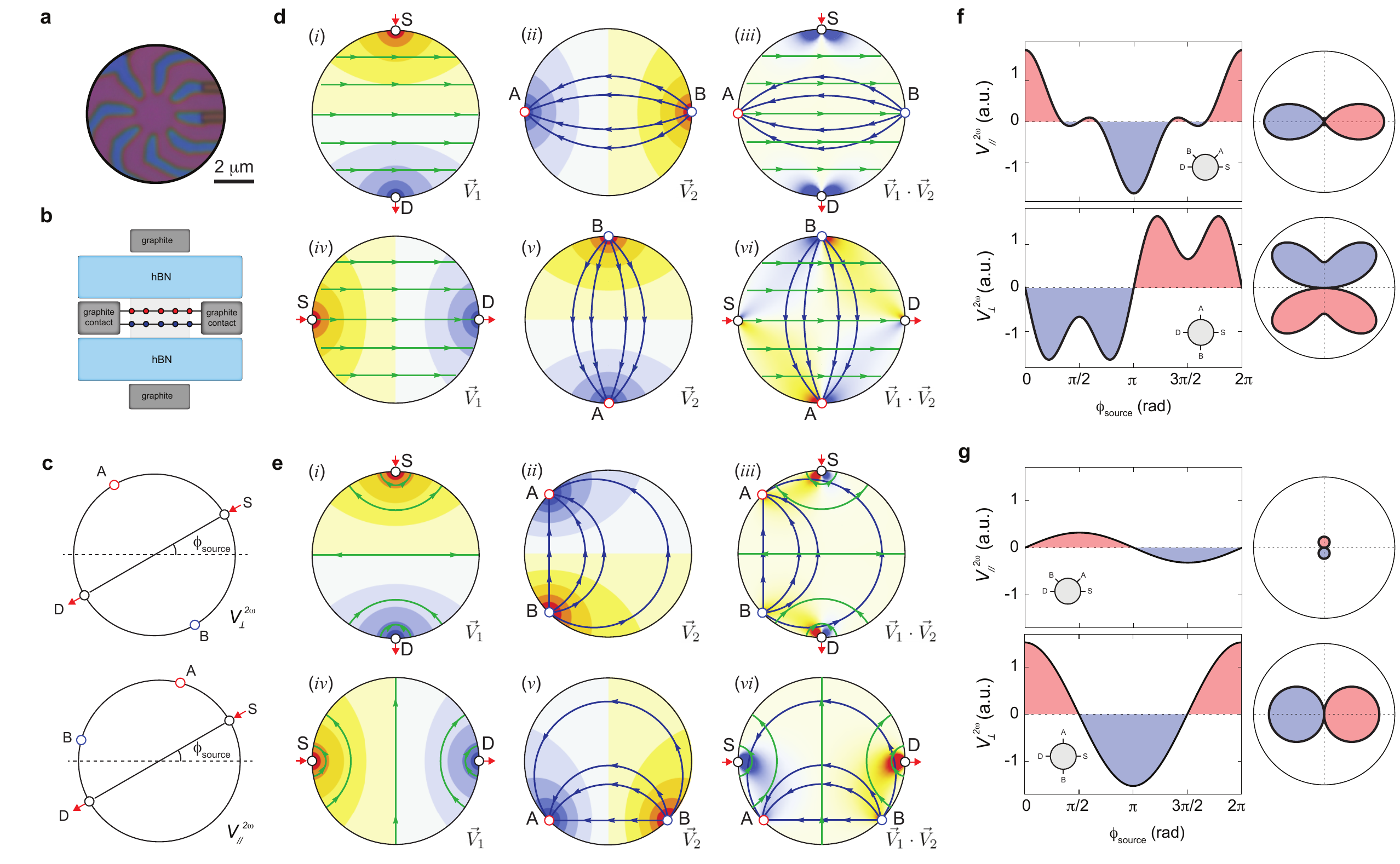}
\caption{ {\bf{Nonlinear voltage response due to different components of the nonlinear conductivity tensor.}} 
(a) An optical image of the sample used in this work. The BBG is shown in purple (sample and contacts), blue is the dielectric substrate underneath the sample. The scale bar denotes 2 $\mu$m. (b) Schematic cross section of the BBG heterostructure. (c) Schematic of the angle-resolved measurement setup for $V_{\perp}$ (top) and  $V_{\parallel}$ (bottom) in a disk-shaped sample.  (d-e) The expected nonlinear voltage response under different measurement configurations for different nonlinear tensors: (d) $V_{\perp}$ of a tensor containing only 
$\tilde{\sigma}_{xxx}$
component, and (e) $V_{\parallel}$ of a tensor containing only the nonlinear Hall component. For illustration we took $\mathcal{C}$ along the $x-$axis. Three columns show a breakdown of Eqn.~\ref{main_eq:marvelous}: the left column shows vector map $\vec{V}_1=\tilde{\mathcal{E}}^{(2)}(\br')$ \newtext{with color scheme depicting distribution of linear electrostatic potential}, the middle column shows $\vec{V}_2=\nabla_{\br'}\left(G_{N}(\br_\bA,\br')-G_{N}(\br_\bB,\br')\right)$ \newtext{with color scheme depicting distribution of fictitious potential $G_{N}(\br_\bA,\br')-G_{N}(\br_\bB,\br')$}, the right column shows their scalar product, $\vec{V}_1\cdot\vec{V}_2$, \newtext{and its magnitude distribution}. Top and bottom rows correspond to current source located at \newtext{$\phi_{\text{Source}}=\pi/2$} and \newtext{$\phi_{\text{Source}}=\pi$}, respectively.  (f) The angle dependence of \newtext{$V_{\parallel}$} (top) and \newtext{$V_{\perp}$}  (bottom) for the scenario described by panel (d) with 
$\tilde{\sigma}_{xxx}$ as the only nonzero tensor component. \sout{(f) and} \newtext{(g)} The angle dependence of $V_{\parallel}$ (top) and   $V_{\perp}$ (bottom) for the scenario described by panel (e) with the nonlinear Hall conductivity as the only nonzero tensor component. }
\label{fig1}
\end{figure*}

\textbf{Introduction}
The Hall effect is one of the most studied  
phenomena in condensed matter physics. In its linear form, it relates the electrical potential difference perpendicular to the electrical current via a direct proportionality. 
Despite being more than a century-old there's still room for new discoveries as exemplified by the recent observation of
the fractional quantum anomalous Hall effect in 2D van der Waals moire heterostructures \cite{Park2023_FCI,Lu2024_FCI,Fan_Xu_PRX_FCI,Zeng2023_FCI}. Beyond the linear response regime, \sout{a recent} theoretical \sout{proposal} \newtext{proposals} pointed to a novel form of Hall effect, where the electric current in the nonlinear regime is always perpendicular to the local electric field ~\cite{GenkinMednis1968,Sodemann_Fu} and microscopically induced by a Berry curvature dipole (BCD). Exploring the transport response in the nonlinear regime, including the nonlinear Hall effect, opens new pathways to investigate exotic electronic orders emerging from the interplay between correlation and topology, which has attracted intense research efforts \cite{Sodemann_Fu,KinFai2019,Pablo2019,TBGnonlinearPRL2022,GaoScience2023,Han2024,JamesLeo2024NatMat,TelluriumPRL2024,TelluriumPRB2025,KoenigDzeroLevchenkoPesin,Du2021NatComm,Du2021,KaplanPRL2024}. While the concept of nonlinear transport holds promise of new scientific discoveries as well as novel technological applications \cite{IsobeSciAdv2020,ZhangTerahertzPNAS2021}, so far the field has many open questions.

Nonlinear electrical transport of interest here is defined as the second-order in electric field response when the Ohm's law acquires quadratic corrections and the current density $\bj$ satisfies 
\begin{equation}
    j_{\alpha} = \sigma_{\alpha\mu} E_{\mu} + \tilde{\sigma}_{\alpha \mu \nu} E_{\mu} E_{\nu}.
    \label{Ohms_law}
\end{equation}
Here $\alpha,\mu,\nu$ are spatial indices, $E_{\mu}$ is the electric field driving the current, while $\sigma_{\alpha \mu}$ and $\tilde{\sigma}_{\alpha\mu \nu}$ are the linear and nonlinear conductivity tensors respectively (repeated indices are summed). We focus on two spatial dimensions where, in general, $\sigma_{\alpha \mu}$ has 4 independent components while in the DC limit nonlinear conductivity tensor $\tilde{\sigma}_{\alpha \mu \nu}$ has 6. 
The linear Hall conductivity $\sigma_H$ corresponds to the antisymmetric component of $\sigma_{\alpha \mu}$,
while the remaining three specify the standard dissipative (Ohmic) response, its anisotropy and the direction of principal axes. The nonlinear Hall current is most conveniently expressed in a coordinate-free form as $\bj^{\text{nl}}_{\text{Hall}}=\hat{\bz}\times\textbf{E} \left( \mathcal{C} \cdot \textbf{E} \right)$. The nonlinear Hall conductivity can therefore be completely specified by a two-component vector $\mathcal{C}$.
Unlike its linear counterpart, nonzero $\tilde{\sigma}_{\alpha \mu \nu}$ requires inversion symmetry to be broken. This is evident from Eq. \eqref{Ohms_law} given that both $\textbf{j}$ and \textbf{E} are odd under inversion. 
As we explain later in the text, it also requires out-of plane rotational symmetry to be completely broken, i.e., it would be prohibited in the presence on an $n$-fold symmetry for any $n>1$.  
Unlike its linear counterpart \cite{Onsager1931}, the nonlinear Hall effect does not require the time-reversal (TR) symmetry to be broken \cite{Sodemann_Fu,MorimotoNagaosa2018SciReps,Du2021,MichishitaNagaosaPRB2022}. 
Since nonlinear response reflects symmetry breaking to which its linear counterpart may be insensitive, the observation of nonlinear transport response highlights new opportunities to investigate the rich interplay between correlations, broken symmetries, and topology underlying strongly correlated 2D electrons.
For example, proposed microscopic mechanisms of nonlinear electrical response range from 
scattering induced by disorder (extrinsic effects) \cite{KoenigDzeroLevchenkoPesin,Du2021,Du2021NatComm,AtenciaNHE2023} or interactions \cite{MorimotoNagaosa2018SciReps}
to topology \cite{GenkinMednis1968,Sodemann_Fu} or quantum geometry \cite{Gao2014PRLNHEmetric,Wang2021PRL,MichishitaNagaosaPRB2022,AtenciaNHE2023,AgarwalQG2023PRB,KaplanPRL2024} (intrinsic effects).    
However, it remains an outstanding challenge to properly analyze nonlinear transport data and to extract the components of $\tilde{\sigma}_{\alpha \mu \nu}$ from an experiment without ambiguity \newtext{and, therefore, to pinpoint its microscopic origin}. 

Presently, experiments designed to obtain the nonlinear transport response in quasi-2D heterostructures \cite{KinFai2019,Pablo2019,GaoScience2023,JamesLeo2024NatMat} involve measurements of nonlinear potential difference between two contacts, which are either aligned with the direction of injected current or perpendicular to it. In previous reports, transverse nonlinear voltage difference was \sout{often} considered as evidence of nonlinear Hall effect ~\cite{Pablo2019,KinFai2019,GaoScience2023}. However, equating the nonlinear Hall effect with the transverse \sout{transport} response may lead to erroneous conclusions. As is well known, a transverse response in the linear regime can result from a reduced rotational symmetry even for $\sigma_H=0$ ~\cite{Wu2017nematic}. In the nonlinear regime, the richer structure of $\tilde{\sigma}_{\alpha \mu \nu}$ unlocks \sout{an even} \newtext{a} more diverse range of possibilities to generate a voltage difference transverse to the input current direction, some of which can occur even at $\mathcal{C}=0$ and are therefore manifestly unrelated to the nonlinear Hall effect (this is illustrated in the Fig. \ref{fig1}). \newtext{One can also see that the Hall bar geometry is insufficient for uniquely identifying the nonlinear Hall effect because, even in an ideal case, it can only access 2 out of 6 components of $\tilde{\sigma}_{\alpha \mu \nu}$. Moreover, as we show in the Methods section, these 2 components contain a combination of dissipative and Hall contributions, which cannot be disentangled even by utilizing symmetry arguments.} A genuine identification of the nonlinear Hall effect therefore requires detailed and quantitative understanding of signatures of each independent component of the nonlinear conductivity tensor $\tilde{\sigma}_{\alpha \mu \nu}$.

In this work we utilize Bernal bilayer graphene (BBG) samples shaped in disk geometry and for the first time unambiguously extract the structure of nonlinear conductivity tensor. This is achieved in two steps. First, by \sout{analytically} \newtext{theoretically} determining the distribution of nonlinear electrical potential across the disk-shaped sample with an arbitrary (uniform) $\sigma_{\alpha \mu}$, $\tilde{\sigma}_{\alpha \mu \nu}$, and
for a current $I$ injected at an arbitrary source (S) and removed at an arbitrary drain (D) at the perimeter of the disk. Second, by employing angle-resolved transport measurements in both $I$ and $I^2$ regimes -- with a large number of source-drain-contact \sout{configurations} \newtext{wiring installations} shown in Figs. \ref{fig1} and \ref{fig3}a -- as an experimental input to which the analytical solution is fitted. 
The new capability allows us to identify the giant nonlinear Hall effect with record values of nonlinear conductivity exceeding previously reported values for giant nonlinearities in graphene-based structures \cite{He2022} and exceeding the values expected from known microscopic mechanisms by 3 to 7 orders of magnitude.
The rest of this paper is organized as follows. 
We first demonstrate the possibility to generate transverse nonlinear response in the absence of nonlinear Hall conductivity i.e. for $\mathcal{C}=0$, illustrating the challenge we aim to address. Second, we present our solution which makes use of three basis functions for the spatial distribution of nonlinear potential (i.e. $\sim I^2$) that we evaluate analytically at the perimeter of the disk. Finally, we describe a new measurement and analysis scheme for angle-resolved transport in the nonlinear regime, which gives rise to full identification of the nonlinear conductivity tensor and helps us uncover the presence of a giant nonlinear Hall effect in Bernal bilayer graphene.

\begin{figure*}
\includegraphics[width=0.99\linewidth]{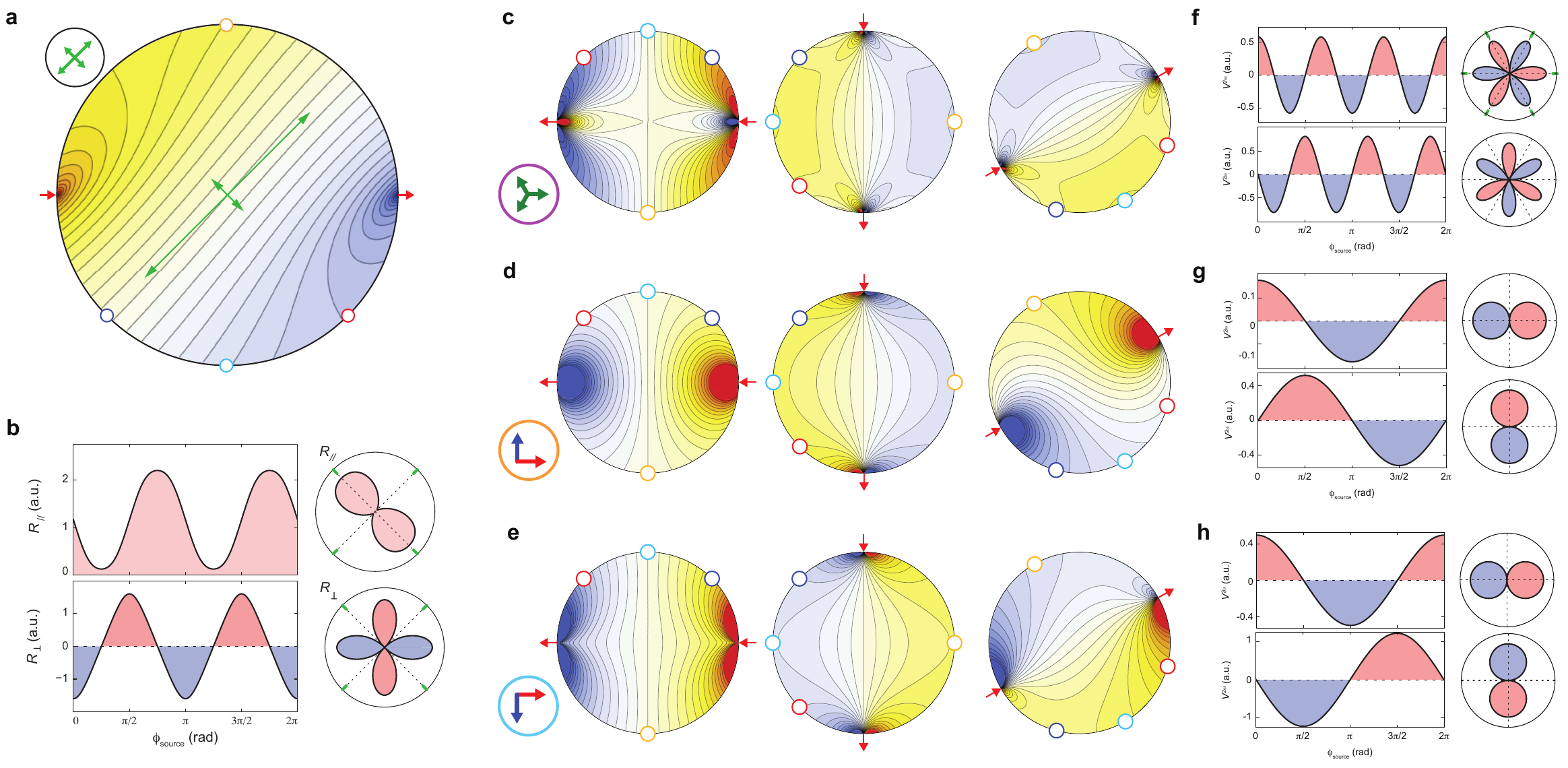}
\caption{{\bf{Basis functions of nonlinear electrical potential. } }
(a) Linear electrical potential from an anisotropic linear conductivity 
tensor. (b) Example of the angle dependence of linear voltage response $V_{\parallel}$ (top) and $V_{\perp}$ (bottom) of an anisotropic linear 
tensor. (c-e) Nonlinear electrical potential from the $(E_x - iE_y)^2$ (c), $ (E_x+iE_y)^2$ (d), and $E_x^2+E_y^2$ (e) basis functions with current source located at \newtext{$\phi_{\text{Source}}=0$} (left column), \newtext{$\phi_{\text{Source}}=\pi/2$} (middle column), and \newtext{$\phi_{\text{Source}}=7\pi/6$}  (right column). (f-g) Angle dependence of the nonlinear voltage response $V_\parallel$ and $V_\perp$ of each basis function: $(E_x - iE_y)^2$ (f), $ (E_x+iE_y)^2$ (g), and $E_x^2+E_y^2$ (h). For all potential maps, current source and drain are indicated by the red arrows, $V_{\parallel}$ A-B contacts are indicated by blue and red circles respectively, and $V_{\perp}$ A-B contacts are indicated by cyan and orange circles respectively. 
}
\label{fig2}
\end{figure*}

\textbf{Conductivity tensors and their properties in 2D}
To better understand the behavior of the nonlinear conductivity tensor under rotations and reflections it is convenient to rewrite Eq. \eqref{Ohms_law} in the complex notation as
\begin{equation}
    \begin{gathered}
        j_x + i j_y = \Xi^{(1)}_+ (E_x + i E_y) +  \Xi^{(1)}_- (E_x - i E_y) + \\ + \Xi^{(2)}_- (E_x - i E_y)^2 + \Xi^{(2)}_+ (E_x + i E_y)^2 + \Xi^{(2)}_0 \left( E_x^2 + E_y^2 \right).
        \label{Ohms_law_complex}
        \end{gathered}
\end{equation}
The complex parameters $\Xi^{(a)}_{b}$ are directly related to the components of the conductivity tensors, see Supplementary Information \newtext{(SI)} for explicit formulas. A rotation of the coordinate system by an angle $\alpha$ results in a simple multiplication of a complex vector by a phase factor:
$
    v_x \pm i v_y \rightarrow e^{\pm i \alpha} ( v_x \pm i v_y ).
$
The $\Xi^{(2)}_-$-term is thus readily seen to be invariant under $\pm 2\pi/3$ rotations about an out-of plane axis. On the other hand, both $\Xi^{(2)}_+$ and $\Xi^{(2)}_0$ terms completely break a rotation symmetry, and therefore must vanish in the presence of such symmetry.
 
Under the mirror reflection about the $xz$ plane, $M_y$, the $y-$ component of a polar vector changes sign, which is equivalent to complex conjugating the complex vector. Therefore, if $M_y$ is a symmetry, $\Xi^{(2)}_b$ must be purely real. We can understand the consequences of a mirror reflection about an arbitrary (vertical) plane by first rotating by an angle $\alpha$ and then applying $M_y$ in the new coordinate system. If such a combined operation is a symmetry, then $e^{-3i\alpha}\Xi^{(2)}_-$,
$e^{i\alpha}\Xi^{(2)}_+$ and $e^{-i\alpha}\Xi^{(2)}_0$ must each be purely real. 
We will return to this result later in the text.
Extracting $\Xi^{(2)}_b$ from the experimental data thus offers valuable information about rotational and mirror symmetries.

Hall currents are non-dissipative.
In order to separate the components of the tensor into dissipative and non-dissipative, we note that the non-dissipative components lead to current without generating any heat. By demanding that the total current density $\textbf{j}^{\text{nd}}$ is perpendicular to $\textbf{E}$ for {\it any} in-plane orientation of $\textbf{E}$, we readily find that the most general form of the non-dissipative current is $\textbf{j}^{\text{nd}} = \sigma_H  \textbf{E}\times \hat{\bz} + \hat{\bz}\times\textbf{E} \left( \mathcal{C} \cdot \textbf{E} \right)$. 
The real vector $\mathcal{C}$ is related to the complex parameters by $i \mathcal{C}_x + \mathcal{C}_y= \Xi^{(2)}_{+}-{\Xi_0^{(2)}}^\ast$. 
Note that because $\mathcal{C}$ is independent of $\Xi^{(2)}_-$, the nonlinear Hall current density $\textbf{j}^{\text{nl}}_{\text{Hall}}$, i.e. the last term in $\textbf{j}^{\text{nd}}$ above, does not contribute to the 3-fold symmetric nonlinear current density and 
is in fact prohibited in the presence of an $n$-fold rotational symmetry for any $n>1$, unlike the linear Hall effect. 
The presence of a mirror plane 
prohibits nonzero $\sigma_H$ and selects a preferred direction of the $\mathcal{C}$-vector.
The dissipative part of $\tilde{\sigma}_{\alpha \mu \nu}$ consists of two contributions: the 3-fold symmetric one and a purely longitudinal. For the latter, $\bE$ and the nonlinear current density it generates, $\bj^{\text{nl}}_\parallel$, are always aligned: $\textbf{j}^{\text{nl}}_{\parallel} = \textbf{E} \left( \mathcal{B} \cdot \textbf{E} \right)$, where
$\mathcal{B}_x + i \mathcal{B}_y={\Xi_+^{(2)}}^\ast+\Xi_0^{(2)}$. 
As we discuss in the later sections, a vertical mirror plane forces the vectors $\mathcal{C}$ and $\mathcal{B}$ to be orthogonal.
Finally, the 3-fold contribution can also be expressed in the coordinate-free form as 
$
\textbf{j}^{\text{nl}}_{\rightY} = \mathcal{A} \left( \left( \mathcal{A}\cdot \textbf{E}\right)^2 - \left( \mathcal{A} \times \textbf{E} \right)^2 \right) + 2 \mathcal{A} \times \left(\mathcal{A}\times \textbf{E}\right) \left(\mathcal{A} \cdot \textbf{E} \right)
$
with $\mathcal{A}_x + i \mathcal{A}_y = \sqrt[3]{\Xi^{(2)}_-}.$ Because there are three complex solutions to taking the cube root differing by $\pm2\pi/3$ phase, the orientation of  $\mathcal{A}$ is defined only up to $120^{\circ}$-rotations. 
The total nonlinear current density can thus be expressed as $\textbf{j}^{\text{nl}}=\textbf{j}^{\text{nl}}_{\rightY}+\textbf{j}^{\text{nl}}_\parallel+\textbf{j}^{\text{nl}}_{\text{Hall}}$.

\textbf{Nonlinear transport in disk geometry: experimental setup and an electrostatics problem}
To study nonlinear transport in BBG we prepare a disk-shaped sample with 8 leads attached to it, where every lead can serve as source/drain for current or measurement point for voltage, see Fig. \ref{fig1}a. 
We employ an experimental setup in which the current $I$ is injected at source (S) and removed at drain (D) while the potential difference between two leads A and B is measured. Inversion symmetry in this setup is explicitly broken by an applied perpendicular displacement field from top and bottom gates, as shown in Fig.~\ref{fig1}b.
In general, current-voltage characteristic of such setup is nonlinear. For small magnitude of injected current $I$, the measured linear and nonlinear potentials are proportional to $I$ and $I^2$ respectively.

To analyze the measurements we make the assumption of local and uniform conductivity tensors (see Fig. \newtext{S3} in the SI) and map the nonlinear transport problem to an electrostatics problem of potential distribution across the disk sample. To this end, we expand the current density and electric field in powers of $I$ as $\bj=\bj^{(1)}+\bj^{(2)}+\ldots$ and $\bE=-\nabla \Phi^{(1)}-\nabla\Phi^{(2)}+\ldots$, and substitute this expansion in the Eq. \ref{Ohms_law}. To linear order in $I$ we recover the Ohm's law $j^{(1)}_\alpha=-\sigma_{\alpha\mu}\nabla_\mu \Phi^{(1)} (\br)$.
The total current density $\bj$ satisfies the continuity equation 
$\nabla \cdot \bj = I (f( \br - \br_S) - f( \br - \br_D)),$
where $f$ is the distribution function describing the finite size source and drain, modeled by a box distribution for arc-shaped leads in the experiment. Within our expansion the above continuity equation is saturated by $\bj^{(1)}$, i.e. $\nabla \cdot \bj^{(1)} = \nabla \cdot \bj$, and therefore $\nabla \cdot \bj^{(2)}=0$.
Taking the divergence of Ohm's law and combining it with the continuity equation for $\bj^{(1)}$ results in a second-order partial differential equation (PDE) that becomes a Poisson PDE along the principal axes after simple coordinate rescaling \cite{Oskar_solo}. We require that the normal components of both $\bj^{(1)}$ and $\bj^{(2)}$ vanish at the disk boundary. The exact solution for $\Phi^{(1)}$ for an arbitrary $\sigma_{\alpha\mu}$ is presented in the Ref. \cite{Oskar_solo}. 
Fitting our experimental data to this solution, we find that $\sigma_H=0$, i.e. there is no anomalous Hall effect. \sout{Focusing on $\sigma_H=0$, the vanishing normal components of $\bj$ at the boundary specify the normal derivative of the potential $\Phi^{(1)}$, leading to Neumann boundary conditions} 
The resulting $\Phi^{(1)} (\br)$ is proportional to $I$ and enters the equation for the order $I^2$ contributions, which is of our primary interest, as $\nabla \cdot \bj^{(2)} = -\nabla_{\alpha} \sigma_{\alpha\mu} \nabla_\mu \Phi^{(2)} (\br)+\nabla\cdot \tilde{\mathcal{E}}^{(2)}(\br) = 0,$
where $\tilde{\mathcal{E}}^{(2)}_{\alpha} = \tilde{\sigma}_{\alpha \mu \nu} \left(-\nabla_\mu \Phi^{(1)}\right) \left(-\nabla_\nu \Phi^{(1)}\right)$. After a simple coordinate rescaling, this \sout{also} has the form of a Poisson equation with the Neumann boundary condition \newtext{\cite{jackson2021classical} for $\Phi^{(2)} (\br)$} and a source term given by $\nabla\cdot \tilde{\mathcal{E}}^{(2)}$.
Our analysis of the experimental data shows that the anisotropy of $\sigma_{\alpha\mu}$ is $<5\%$ (in addition to, as mentioned, $\sigma_{H}=0$). Therefore, in what follows, we focus on $\Phi^{(2)} (\br)$ for isotropic $\sigma_{\alpha\mu}$, but allowing for a  completely general $\tilde{\sigma}_{\alpha\mu\nu}$. The general solution for an arbitrary linear anisotropy \newtext{and the details of the derivation are} presented in the SI. 
Thus, upon application of the Green's theorem, we can express the solution of the Poisson equation as
\begin{eqnarray}
\Phi^{(2)} (\br)=
\frac{1}{\sigma}\int d^2\br' \tilde{\mathcal{E}}^{(2)}(\br') \cdot \left(\nabla_{\br'}G_{N}(\br,\br')\right).
\label{main_eq:marvelous}
\end{eqnarray}
Here $G_{N}(\br,\br')$ is the Neumann Green's function, the integral is restricted to the interior of the disk, and $\sigma=\sigma_{xx}=\sigma_{yy}$.
The $I^2$-potential difference between contacts A and B  measured in the experiment would then correspond to $V^{(2)}_{AB} = \Phi^{(2)} (\br_A) - \Phi^{(2)} (\br_B)$, with implicit dependence on the location of source and drain.

Important insights follow from Eq. \ref{main_eq:marvelous} as illustrated in the Fig. \ref{fig1}. Panels d) and e) show distributions of (i) $\tilde{\mathcal{E}}^{(2)}(\br')$, (ii) the difference of gradients of the Green's function evaluated at measurement points A and B -- which has a simple physical interpretation as the electric field produced by a fictitious unit source and drain at B and A to linear order -- and (iii) the scalar product entering as an integrand in the Eq. \ref{main_eq:marvelous}. Without computing the precise value of the integral it can be seen in the Fig.~\ref{fig1}d that the transverse voltage drop is nonzero even if the only nonzero component \sout{of the nonlinear conductivity tensor} is $\tilde{\sigma}_{xxx}$. Note that this implies vanishing of the nonlinear Hall conductivity which can depend only on components of $\tilde{\sigma}_{\alpha\mu\nu}$ with two nonequal indices. Similarly, as seen in the Fig.~\ref{fig1}e, longitudinal voltage drop is nonzero even for a purely nonlinear Hall conductivity, i.e for $\mathcal{A}=\mathcal{B}=0$ and a nonzero $\mathcal{C}\parallel \hat{\bx}$.  This indicates, that the mere presence or absence of nonlinear signal in  specific measurement configurations cannot serve as a conclusive evidence even for a qualitative structure of $\tilde{\sigma}_{\alpha \mu \nu}$. 

It follows directly from the Eq. \ref{main_eq:marvelous} that
in the $I^2$ regime the nonlinear potential 
can be expressed as a \textit{linear} combination of three independent contributions
$$
\Phi^{\left(2 \right)} (\br)=\Phi^{(2_{-})} (\br)+\Phi^{(2_{+})} (\br)+\Phi^{(2_{0})} (\br).
$$ \newtext{These} form our basis functions, each linearly proportional to one of $\Xi^{(2)}_{-,+,0}$. 
For the case of isotropic linear conductivity, \sout{they} \newtext{the basis functions} take the following form at the boundary of the disk:  \begin{widetext}
\begin{eqnarray}
&&\Phi^{(2_{-})}_{iso}(\br)=
\frac{\Xi^{(2)}_{-} I^2}{\sigma^3\pi^2 a}
\biggr(\frac{1}{2zz^2_S}+\frac{1}{2zz^2_D}
 - \frac{ z^* (z_S-z_D) + z^*_S (z_D-z) - z^*_D (z_S-z)}{(z_S-z)(z_D-z)(z_S-z_D)} \biggr)+c.c. \label{basis1} \\
&&\Phi^{(2_{+})}_{iso}(\br)=\frac{\Xi^{(2)}_{+} I^2}{\sigma^3 \pi^2 a}
\left(\frac{z^2_S}{2(z-z_S)}+\frac{z^2_D}{2(z-z_D)}-\frac{1}{z^*_S-z^*_D} \ln \frac{z_S - z}{z_D -z}  \right)+c.c.\\
&&\Phi^{(2_{0})}_{iso}(\br)=\frac{\Xi^{(2)}_{0} I^2}{\sigma^3 \pi^2 a}
\left( \frac{1}{2(z_S-z)}\left(\mathcal{L}_\lambda + \ln \frac{z_S (z-z_D)}{(z-z_S)(z_S-z_D)} \right) + S\leftrightarrow D \right) +c.c.
\label{basis3}
\end{eqnarray}
\end{widetext}
Here \newtext{$a$ is the radius of the sample that we mapped to a unit disk,} $z = x+iy=e^{i\vartheta}$ is the complex representation of the location of a measurement point on the circumference of \sout{a} \newtext{the} unit disk, $z_{S,D}=e^{i\vartheta_{S,D}}$ \sout{are} \newtext{represent} locations of the source and the drain \def\thefootnote{$1$}\footnote{\textcolor{black}{$\ln(z)$ has a branch cut discontinuity in the complex $z$ plane running from $0$ to $-\infty$.}}, \sout{(also with unit amplitude), $a$ is the radius of the sample,} $c.c.$ is complex conjugate, and 
$\mathcal{L}_\lambda\approx -\frac{3}{2}+\ln\lambda$ with $\lambda$ being the angular width of a source and drain \newtext{in radians}. In Fig. \ref{fig2} we show the angular dependence of basis functions and their potential distribution across the entire sample. Certain measurement configurations experience a cancellation of $\lambda$-dependent contribution. For a general case of finite anisotropy of linear conductivity tensor the solution is also expressed via a linear combination of three basis functions. These basis functions are more complicated and are shown in the SI.

\begin{figure*}
\includegraphics[width=0.99\linewidth]{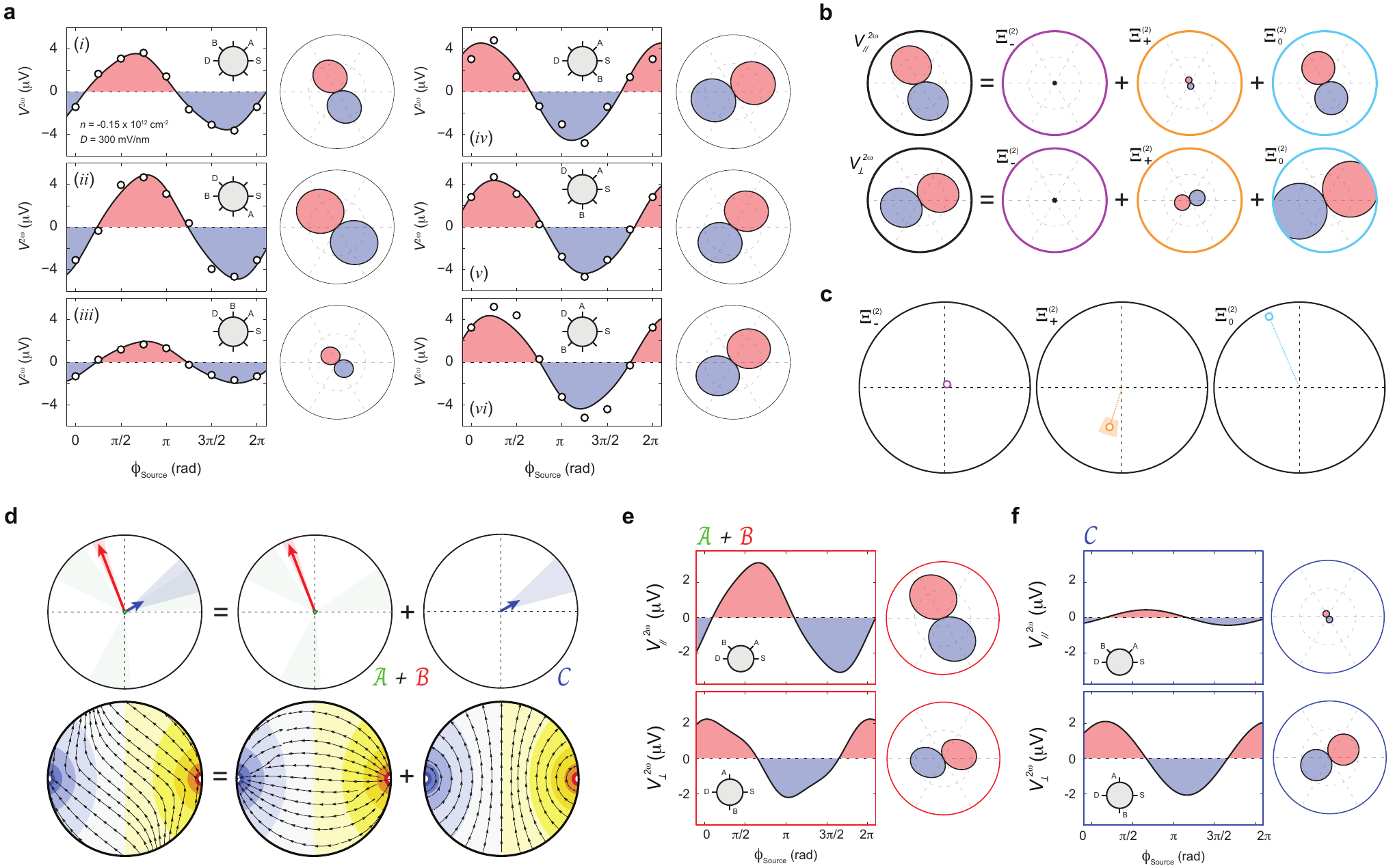}
\caption{{\bf{Extraction of the nonlinear conductivity tensor.}}
(a) Nonlinear voltage measurement (black circles) versus \sout{current source location} \newtext{$\phi_{\text{Source}}$} of 6 different measurement configurations, with 8 different 
\newtext{wiring installations}
\sout{of source} per configuration, \sout{of the same state} and the theoretically expected dependence  
(black lines) plotted using the 
extracted
nonlinear conductivity tensor. \sout{All} Measurements \newtext{in this figure} are performed in a Bernal bilayer graphene sample at $n=-0.15\times 10^{12}$ cm$^{-2}$, $D=300$ mV/nm, $T=20$mK. The measurement configurations are indicated as insets. The remarkable agreement between 48 independent measurements and their theoretical fit from the 
 experimentally extracted tensor demonstrates the uniformity of the nonlinear conductivity. (b) Polar plots of $V_{\parallel}$ (top) and $V_{\perp}$ (bottom) signals decomposed into three basis functions using the extracted nonlinear conductivity tensor. \newtext{Each of $\Xi^{(2)}$ parameters was determined by fitting of the measured signal with the solutions for nonlinear potential, see Eqs. \eqref{basis1}-\eqref{basis3} and Methods.} (c) The fitted $\Xi^{(2)}_-$, $\Xi^{(2)}_{+}$, and $\Xi^{(2)}_{0}$ plotted in polar coordinates. The shaded \sout{cones} \newtext{regions} indicate the fitting uncertainties in angles. The extracted parameters read $\Xi^{(2)}_- = 0.75 e^{i0.31 \pi} \frac{\mu \text{m}}{\Omega \cdot \text{V}}, \Xi^{(2)}_{+} = 7.9 e^{-i0.6 \pi} \frac{\mu \text{m}}{\Omega \cdot \text{V}}, \Xi^{(2)}_{0} = 14.7 e^{i0.63 \pi} \; \frac{\mu \text{m}}{\Omega \cdot \text{V}}.$
 (d) \newtext{Distribution of the $\tilde{\mathcal{E}}^{(2)}$ contribution to the nonlinear current $\bj^{\text{nl}}$}  calculated for  \newtext{the source and the drain 180$^{\circ}$ apart} \sout{configuration} (bottom) using the full extracted tensor decomposed into the dissipative component and non-dissipative component (top). 
 Shaded cones indicate the fitting uncertainties of angles. Plots of the \newtext{calculated} angular dependence of dissipative (e) and non-dissipative (f) component of $V_{\parallel}$ (top) and $V_\perp$ (bottom) \newtext{for the extracted $\Xi^{(2)}_{-,0,+}$}. }
\label{fig3}
\end{figure*}

\begin{figure*}
\includegraphics[width=0.8\linewidth]{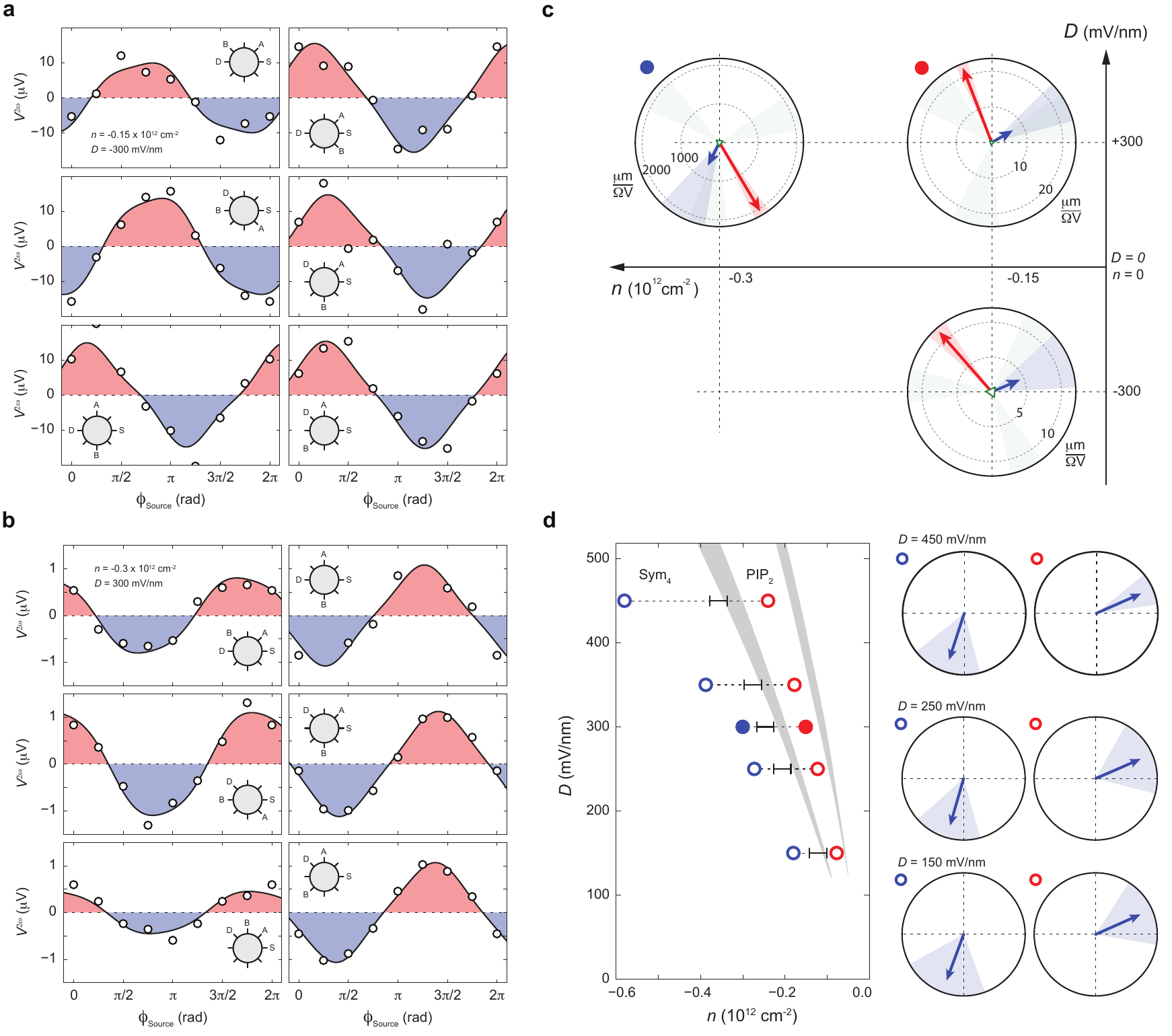}
\caption{{\bf{$n-D$ dependence of the nonlinear conductivity behavior of Bernal bilayer graphene.} }
(a-b) Nonlinear voltage measurement (black circles) versus \sout{current source location} \newtext{$\phi_{\text{Source}}$} of 6 different measurement configurations, with 8 different  
\newtext{wiring installations}
\sout{of source} per configuration, and the theoretical 
curve plotted using the experimentally extracted tensor (black lines) at (a) $n=-0.15\times 10^{12}$ cm$^{-2}$, $D=-300$ mV/nm and (b) $n=-0.3\times 10^{12}$ cm$^{-2}$, $D=300$ mV/nm. All measurements are performed in the same Bernal bilayer graphene sample at $T=20$mK. The measurement configurations  are indicated as insets. (c) The vector representation of the extracted $\tilde{\sigma}_{\alpha \mu \nu}$ for three points in the $n-D$ phase diagram. Length of vectors indicate the magnitude of the corresponding component of $\tilde{\sigma}$.  Shaded cones indicate the fitting uncertainties of angles. (d) The orientation of the non-dissipative $\mathcal{C}$-vector extracted in the PIP$_2$ and Sym$_4$ phases for different values of $n$ and $D$. Left panel shows the $n-D$ phase diagram with gray areas corresponding to phase transitions as reported in \cite{Andrea_BBG_Science2022}. Black bar indicates the region in the $n-D$ plane where the nonlinear signal changes its orientation by 180$^{\circ}$.  Blue and red circles indicate values of $n$ and $D$ within Sym$_4$ and PIP$_2$ phases at which we check orientation of the $\mathcal{C}$-vector. The orientation of the $\mathcal{C}$-vector is robust within the phases as we show in the right panel.
}
\label{D300Transition}
\end{figure*}

\textbf{Extraction and analysis of nonlinear conductivity tensor in Bernal bilayer graphene} 
We apply our newly-developed methodology to study nonlinear electron transport in BBG; \newtext{see Methods for details}. 
The optical image and the schematic sketch of the device are shown in Fig. \ref{fig1}a,b. Top and bottom electrostatic gates allow us to control both electron density, $n$, and the out-of-plane displacement field, $D$. We focus our attention on the region of the phase diagram where the cascade of electronic transitions into correlated metallic states was previously observed in measurements of electron compressibility \cite{Andrea_BBG_Science2022}.
Specifically, we study the region 
that spans partially isospin-polarized phase with 2 large Fermi surfaces (PIP$_2$) and the phase \sout{where} \newtext{with} 4 large Fermi surfaces (FSs) \sout{are present}, where spin-valley symmetry is believed to be restored (Sym$_4$).

First, we analyze linear potential difference 
and extract $\sigma_{\alpha \mu}$ 
following \cite{Oskar_solo,Zhang2022electronic}. We find that linear conductivity is weakly anisotropic with the anisotropy $<5 \%$ and that $\sigma_H \simeq 0$ 
across our area of interest in the $n-D$ plane.
This indicates that within our experimental accuracy  
$\sigma_{\alpha \mu}$ is isotropic
and no anomalous Hall contribution is present.

To analyze nonlinear signal we apply slowly varying ($\sim 10$ Hz) AC current 
and measure potential difference between two leads A, B at the second harmonic frequency. First, we verify that nonlinear potential is \newtext{proportional to} $I^2$.
Then, we perform measurements of nonlinear potential using \sout{measurement} configurations that differ in locations of source, drain, A, and B; \newtext{each configuration is defined as a collection of 8 wiring installations with the same relative positions of source, drain, A, B, and related by a global rotation by $\pi/4$}.
A representative set of measured nonlinear potentials \sout{data} at $D=300$mV/nm and $n=-0.15\times 10^{12}$ cm$^{-2}$ (PIP$_2$ phase)  for 6 different 
configurations is shown in Fig. \ref{fig3}a as a function of source angle \newtext{$\phi_{\text{Source}}$},  defined in Fig. \ref{fig1}c. \sout{We apply our methodology, discussed in SI, to extract 3 complex parameters $\Xi^{(2)}_{-,+,0}$ (6 real parameters), which we graphically represent in Fig. \ref{fig3}b in the complex plane, out of 48 independent configurations. The extracted parameters allow us to plot the expected nonlinear voltage drop for every measurement configuration and for any source angle.} \newtext{
We apply our methodology to these 48 independent wiring installations and extract 3 complex parameters $\Xi^{(2)}_{-,+,0}$, as discussed in SI and Methods. The parameters are graphically represented in the complex plane in Fig. \ref{fig3}c, allowing us to plot $\Phi^{(2)}(\br_A)-\Phi^{(2)}(\br_B)$ for every measurement configuration and for any \newtext{$\phi_{\text{Source}}$} using Eqs. \eqref{basis1}-\eqref{basis3}.
}  The expected voltage drop for every configuration is shown in Fig. \ref{fig3}a and it fits the data very well.

Using the extracted $\Xi^{(2)}_{-,+,0}$ parameters we \sout{find the components of $\tilde{\sigma}_{\alpha \mu \nu}$ and express them via} \newtext{obtain} $\mathcal{A},\mathcal{B},\mathcal{C}$ vectors, \newtext{see Fig. \ref{fig3}d}. Interestingly, we find that while rotational symmetry appears intact in linear response, it is manifestly broken in nonlinear response as \newtext{evidenced by sizable  $\mathcal{B},\mathcal{C}$-vectors. Compared to $\mathcal{B},\mathcal{C}$, the} 3-fold contribution from the $\mathcal{A}$-vector is 
\sout{very} \newtext{negligibly} small \newtext{as marked by the tiny green triangle in Fig. \ref{fig3}d.} \sout{compared to contributions from $\mathcal{B},\mathcal{C}$-vectors}.
\sout{In addition, we observe a} \newtext{The observed}  nonlinear Hall signal 
\sout{which} is consistent with a presence of a single mirror plane since the extracted $\mathcal{B}$ and $\mathcal{C}$ vectors are orthogonal within experimental accuracy as we show in Figs. \ref{fig3}d and \ref{D300Transition}c. 
The magnitudes of the nonlinear conductivity components in the PIP$_2$ phase are $\left|  
\mathcal{A}^3 \right| = 0.75 \pm 0.65 \frac{\mu \text{m}}{\Omega \cdot \text{V}}, \left|  
\mathcal{B} \right| = 22.5 \pm 3.75\frac{\mu \text{m}}{\Omega \cdot \text{V}}, \left|  
\mathcal{C} \right| = 6.88 \pm 2.05\; \frac{\mu \text{m}}{\Omega \cdot \text{V}}$ , which is about $100 \%$ larger than the recently observed giant nonlinear conductivity in moir\'{e} graphene superlattices \cite{He2022}.
In Fig. \ref{fig3}d we also show calculated nonlinear current patterns across the disk sample from dissipative and nondissipative contributions. \sout{As we show in the Figs. \ref{fig3}e,f,} \newtext{We note in passing that} the magnitude of the voltage drop in the perpendicular configuration is comparable for $\mathcal{A}+\mathcal{B}\approx \mathcal{B}$ and $\mathcal{C}$-generated contributions, even though $|\mathcal{C}| \sim  |\mathcal{B}|/3$, \newtext{as we show in the Figs. \ref{fig3}e,f}.

To gain more insights into possible origins of nonlinear conductivity, especially the $\mathcal{B},\mathcal{C}$ vectors, we study nonlinear transport across the transition from PIP$_2$ to Sym$_4$ phase. 
In Fig. \ref{D300Transition}b we show nonlinear signal inside Sym$_4$ phase with $D=300$mV/nm and $n=-0.3\times 10^{12}$ cm$^{-2}$.  \sout{Unexpectedly} \newtext{Surprisingly}, the Sym$_4$ phase \newtext{also} exhibits a \sout{nonzero} \newtext{finite nonlinear} response \sout{in the nonlinear channel}. Like in the PIP$_2$ phase, we utilize the angular fit to extract $\mathcal{A}$, $\mathcal{B}$, and $\mathcal{C}$. We find the $\mathcal{A}$ vector to be very small, whereas \newtext{within our experimental accuracy} both $\mathcal{B}$ and $\mathcal{C}$ rotate  by nearly 180$^{\circ}$ \sout{within our experimental accuracy} from the orientation of the same vectors in PIP$_2$, see Fig. \ref{D300Transition}c.
By performing angle-resolved measurement with varying $n$ over a wide range of $D$, we find that the $180^{\circ}$ rotation in $\mathcal{B}$ and $\mathcal{C}$ is well-aligned with the transition line extracted from the compressibility measurements \cite{Andrea_BBG_Science2022} (Fig. \ref{D300Transition}d). That the  PIP$_2$ to Sym$_4$ transition prominently manifests itself in nonlinear transport offers a new constraint for understanding the nature of the states on each side of the transition. 
We also \sout{check} \newtext{observe} that the structure of $\tilde{\sigma}_{\alpha \mu \nu}$ remains nearly the same and $\mathcal{C}$ vector orientation is quite robust  within each of the two phases, see Fig. \ref{D300Transition}d and SI \newtext{Fig. S5} \sout{for all extracted components of the tensor}. Interestingly, the magnitude of nonlinear conductivity \sout{tensor} extracted in Sym$_4$ phase is almost 100 \newtext{times} larger than that in PIP$_2$ phase with $\left|  
\mathcal{A}^3 \right| = 100 \pm 82.9\frac{\mu \text{m}}{\Omega \cdot \text{V}}, \left|  
\mathcal{B} \right| = 2072 \pm 873\frac{\mu \text{m}}{\Omega \cdot \text{V}}, \left|  
\mathcal{C} \right| = 668 \pm 339 \; \frac{\mu \text{m}}{\Omega \cdot \text{V}}$. To our knowledge, these are the record \sout{numbers} \newtext{values} reported for nonlinear conductivity \sout{so far}.  The larger uncertainty in magnitudes of $\mathcal{A}$, $\mathcal{B}$, and $\mathcal{C}$ \newtext{in the Sym$_4$ phase compared to the PIP$_2$} is caused by the larger uncertainty in the value of linear conductivity due to proximity of the Sym$_4$ phase to the ballistic regime.

\sout{Dissipative contribution to nonlinear current has four possible origins:} \newtext{Known dissipative contributions to nonlinear current include} (i) disorder, e.g., skew scattering ~\cite{KoenigDzeroLevchenkoPesin,IsobeSciAdv2020,Du2021NatComm,AtenciaNHE2023}; (ii) interactions ~\cite{MorimotoNagaosa2018SciReps}; (iii) quantum geometry 
~\cite{MichishitaNagaosaPRB2022,AgarwalQG2023PRB,KaplanPRL2024}; (iv) second-order Drude contributions to $\tilde{\sigma}_{\alpha\mu \nu}$ \cite{GenkinMednis1968,Wang2021PRL,KaplanPRL2024}. Both (iii) and (iv) require TR breaking \newtext{and could also generate non-dissipative contribution}. On the other hand, the BCD can \sout{naturally account for} \newtext{in principle generate} the non-dissipative component, the nonlinear Hall effect, in the absence of TR breaking. 
If the BCD alone is responsible for the nonlinear Hall effect, reversing \newtext{the sign of} $D$ is expected to \sout{influence} \newtext{reverse} the orientation of the $\mathcal{C}$-vector  ~\cite{Sodemann_Fu}, \newtext{provided that the source of the rotational symmetry breaking is unaffected by the reversal of $D$. This is because the sign of the Berry curvature in a given valley reverses under $D \rightarrow -D$.} Alternatively, BCD-induced $\mathcal{C}$-vector could preserve its orientation \newtext{provided that the source of the rotational symmetry breaking changes sign under $D \rightarrow -D$. However, in this case $\mathcal{A}$ and $\mathcal{B}$ reverse sign under $D \rightarrow -D$ assuming they originate from quantum geometric or second-order Drude contributions, see SI.} \sout{while $\mathcal{A}$ and $\mathcal{B}$, given by quantum geometric or second-order Drude contributions, reverse sign under $D \rightarrow -D$, given that the order parameter associated with the nematic perturbation is linked to the direction of $D$.} An example of this scenario is the \newtext{TR-breaking} charge $E_u$ order parameter. \sout{, which breaks TR symmetry.} 
We \sout{examine} \newtext{test} these scenarios by extracting $\mathcal{A},\mathcal{B},\mathcal{C}$ vectors at $D= \pm 300$mV/nm. Fig. \ref{D300Transition}c shows the angular dependence of nonlinear response taken at $D=-300$mV/nm and $n=-0.15\times 10^{12}$ cm$^{-2}$. The magnitudes are given by $\left|  
\mathcal{A}^3 \right| = 0.92 \pm 0.58\frac{\mu \text{m}}{\Omega \cdot \text{V}}, \left|
\mathcal{B} \right| = 11.5 \pm 2.14\frac{\mu \text{m}}{\Omega \cdot \text{V}}, \left|  
\mathcal{C} \right| = 4.46 \pm 1.69 \; \frac{\mu \text{m}}{\Omega \cdot \text{V}}.$ The orientations of both the $\mathcal{C}$-vector and $\mathcal{B}$-vector remain the same as in Fig. \ref{fig3}d. 
Therefore, the invariance of \newtext{both} $\mathcal{B}$-\sout{vector} and $\mathcal{C}$-vector  \sout{combined,} indicates that nonlinear Hall effect is not induced by the BCD mechanism. \sout{as the order parameter associated with the nematic perturbation appears invariant under a sign change in $D$.}
Moreover, \newtext{using scattering time extracted from linear conductivity,} the typical magnitude of nonlinear Hall conductivity associated with BCD \sout{is $\sim 10^{-5} \div 10^{-7} \; \frac{\mu \text{m}}{\Omega \cdot \text{V}}$} \newtext{ranges from $\sim 10^{-5}$ to $10^{-7} \; \frac{\mu \text{m}}{\Omega \cdot \text{V}}$} \cite{KinFai2019,Pablo2019,He2022} -- several orders of magnitude smaller than $|\mathcal{C}|$ in both PIP$_2$ and Sym$_4$ phases.

Combined, our observations 
pose a peculiar conundrum: 
linear conductivity appears to be nearly isotropic while
nonlinear conductivity manifestly breaks 
rotational symmetry, seemingly down to a single mirror plane. The apparent decoupling between the symmetry of linear and nonlinear signal is also reported in a previous observation \cite{JamesLeo2024NatMat}. Notably, rotational symmetry is broken for both the dissipative and nondissipative contributions to nonlinear current, which is evidenced by the angular dependence in Fig.~\ref{fig3}e and f, and for both PIP$_2$ and Sym$_4$ phases. We also observe that the magnitude of measured nonlinear signal increases with $D$ (see Fig. \newtext{S7} in SI) ~\cite{Lin2023Arxiv}.

According to 
Fig.~\ref{D300Transition}c,
our observations in the nonlinear regime exhibit vanishingly small 3-fold contribution. This rules out skew scattering and side jumps as the origin of nonlinearity, since both produce a large 3-fold contribution to nonlinear current even in the presence of rotational symmetry breaking with magnitude $\sim 1$meV, see SI.  
A typical value of quantum geometry-induced nonlinear conductivity \sout{is $\sim 10^{-4} \div 10^{-5} \; \frac{\mu \text{m}}{\Omega \cdot \text{V}}$} \newtext{ranges from $\sim 10^{-4}$ to $10^{-5} \; \frac{\mu \text{m}}{\Omega \cdot \text{V}}$} \cite{Han2024}, which is orders of magnitude smaller than the extracted values in Fig. \ref{D300Transition}. On a qualitative level, one can account for the observed phenomena within the second-order Drude contribution to nonlinear conductivity by assuming the presence of a weak spin-orbit coupling and of a weak order parameter that breaks time reversal and rotational symmetries, \newtext{see SI Secs. XV and XVII}. In such case one can correctly reproduce the hierarchy of $\left| \mathcal{A}^3 \right|, \left|  
\mathcal{B} \right|, \left|  
\mathcal{C} \right|$, the presence of a mirror axis, and the reorientation transition of $\mathcal{B}, \mathcal{C}$-vectors between PIP$_2$ and Sym$_4$ phases. \newtext{Namely, for such a weak order parameter Onsager relation restricts linear conductivity anisotropy to be very small, while smallness of the 3-fold nonlinear term is guaranteed by the properties of the second-order Drude contribution to nonlinear conductivity, see SI Section XV A. } However, as we show in the SI, none of the known microscopic mechanisms
can account for the magnitude of the signal as they fall short orders of magnitude, e.g., the skew scattering which was estimated to induce $\tilde{\sigma} \lesssim 0.01 \; \frac{\mu \text{m}}{\Omega \cdot \text{V}}$ \cite{IsobeSciAdv2020,He2022} in graphene-based systems. Overall, the observed structures and magnitudes of linear and nonlinear conductivity tensors in PIP$_2$ and Sym$_4$ phases, the existence of the reorientation transition of nonlinear signal between the two phases, and the mean free path restriction (less than the diameter of the disk in the diffusive regime) put severe constraints which cannot be satisfied by any known microscopic mechanism of nonlinear transport.

\textbf{Discussion and outlook}
Here we presented a systematic way of analyzing nonlinear transport measurements and unambiguously extracting nonlinear conductivity tensor from experimental data, 
which sets up the foundation for all future experimental works in the vibrant field of nonlinear  
electron transport. We used our approach to determine nonlinear conductivity tensor in Bernal bilayer graphene and show the presence of giant nonlinear Hall effect. We also found that nonlinear signal strongly breaks rotational symmetry whereas in the linear regime this symmetry appears intact. The new way of extracting nonlinear conductivity tensor allowed us to conclude that 
none of the known microscopic mechanisms for nonlinear transport can account for the structure and/or magnitude of the measured nonlinear conductivity tensor. 
The methodology presented in this work allows to systematically study nonlinear transport of quasi-2D materials \cite{Cao2018,Yankowitz_Science_2019,Wong2020,Zondiner2020,Choi2021,Park2021,Kim2022,Liu2022,Zhou2021,Andrea_BBG_Science2022,Zhang2023,Zeng2023_FCI,Park2023_FCI,Lu2024_FCI,Fan_Xu_PRX_FCI,TelluriumPRL2024} and paves the way to future discoveries.

\textbf{Acknowledgments}
We thank \newtext{Fernando de Juan,} Cyprian Lewandowski, Kin Fai Mak, Di Xiao, Binghai Yan, and Andrea Young for fruitful discussions.
D.V.C. acknowledges financial support from the
National High Magnetic Field Laboratory through a Dirac
Fellowship, which is funded by the National Science Foundation (Grant No. DMR-2128556) and the State of Florida. O.V.
was funded by the Gordon and Betty Moore Foundation’s
EPiQS Initiative Grant GBMF11070. The experimental portion of this material is based on the work supported by the Air Force Office of Scientific Research under award no. FA9550-23-1-0482. N.J.Z. and J.I.A.L. acknowledge support from the Air Force Office of Scientific Research. J.I.A.L. acknowledges partial support from the Alfred P. Sloan Foundation through the Sloan Research Fellowship (FG-2023-19817). \newtext{Part of this work was enabled by the use of pyscan (github.com/sandialabs/pyscan), scientific measurement software made available by the Center for Integrated Nanotechnologies, an Office of Science User Facility operated for the U.S. Department of Energy.}

\bibliography{biblio}

\newtext{\section*{Methods}}

\newtext{\subsection*{Limitations of Hall bar geometry}}

\newtext{
Here we discuss the case of Hall bar and demonstrate why this geometry is insufficient to unambiguously determine nonlinear Hall effect. 
Specifically, we consider a rectangular sample with the uniform current density source that coincides with the entire left edge and drain that coincides with the entire right edge of the rectangle. For simplicity we also restrict our discussion to the ideal case of isotropic linear conductivity tensor with zero Hall component, a best-case scenario. Then, the electric field $\bE^{(1)} = - \nabla \Phi^{(1)}$ generated at linear order in injected current $I$ is strictly along the $x-$axis, see Fig. \ref{ext_fig1}. Hence, only $\tilde{\sigma}_{xxx}$ and $\tilde{\sigma}_{yxx}$ can be experimentally accessed as we verify using Finite Element Method, see Fig. \ref{ext_fig1}. This is because no electric field in $y-$direction is generated at the linear order in $I$, and because $\nabla \cdot \bj^{(2)} = 0$ implies $\sigma \nabla^2 \Phi^{(2)} = \nabla_{\alpha} \left[ \tilde{\sigma}_{\alpha \mu \nu} \left(-\nabla_\mu \Phi^{(1)}\right) \left(-\nabla_\nu \Phi^{(1)}\right) \right]$. Therefore, only $\tilde{\sigma}_{xxx}$ and $\tilde{\sigma}_{yxx}$ enter the right-hand-side of the Poisson PDE.  However, as we show in the Supplementary Information, 
\begin{equation}
    \begin{aligned}
        &\tilde{\sigma}_{xxx} = \mathcal{A}_x^3 - 3 \mathcal{A}_x \mathcal{A}_y^2 + \mathcal{B}_x, \\
        &\tilde{\sigma}_{yxx} = 3 \mathcal{A}_x^2 \mathcal{A}_y - \mathcal{A}_y^3 + \mathcal{C}_x.
    \end{aligned}
    \label{conductivities_relations}
\end{equation}
Hall bar measurements determine $\tilde{\sigma}_{xxx}$ and $\tilde{\sigma}_{yxx}$, and give no information about the remaining components of $\tilde{\sigma}$. But the two equations above contain four unknown variables, making them underdetermined. Therefore, regardless of the measured values of $\tilde{\sigma}_{xxx}$ and $\tilde{\sigma}_{yxx}$, it is impossible to guarantee that $\mathcal{C}_x$ is nonzero, which is necessary to prove the very existence of the nonlinear Hall effect.
In the case of finite anisotropy of linear conductivity tensor  
it is also impossible to guarantee nonzero $\mathcal{C}$,
as electric field might acquire $y-$component due to linear conductivity principal axes being misaligned with the direction along which the Hall bar was etched. 
}

\begin{figure*}
\includegraphics[width=0.99\linewidth]{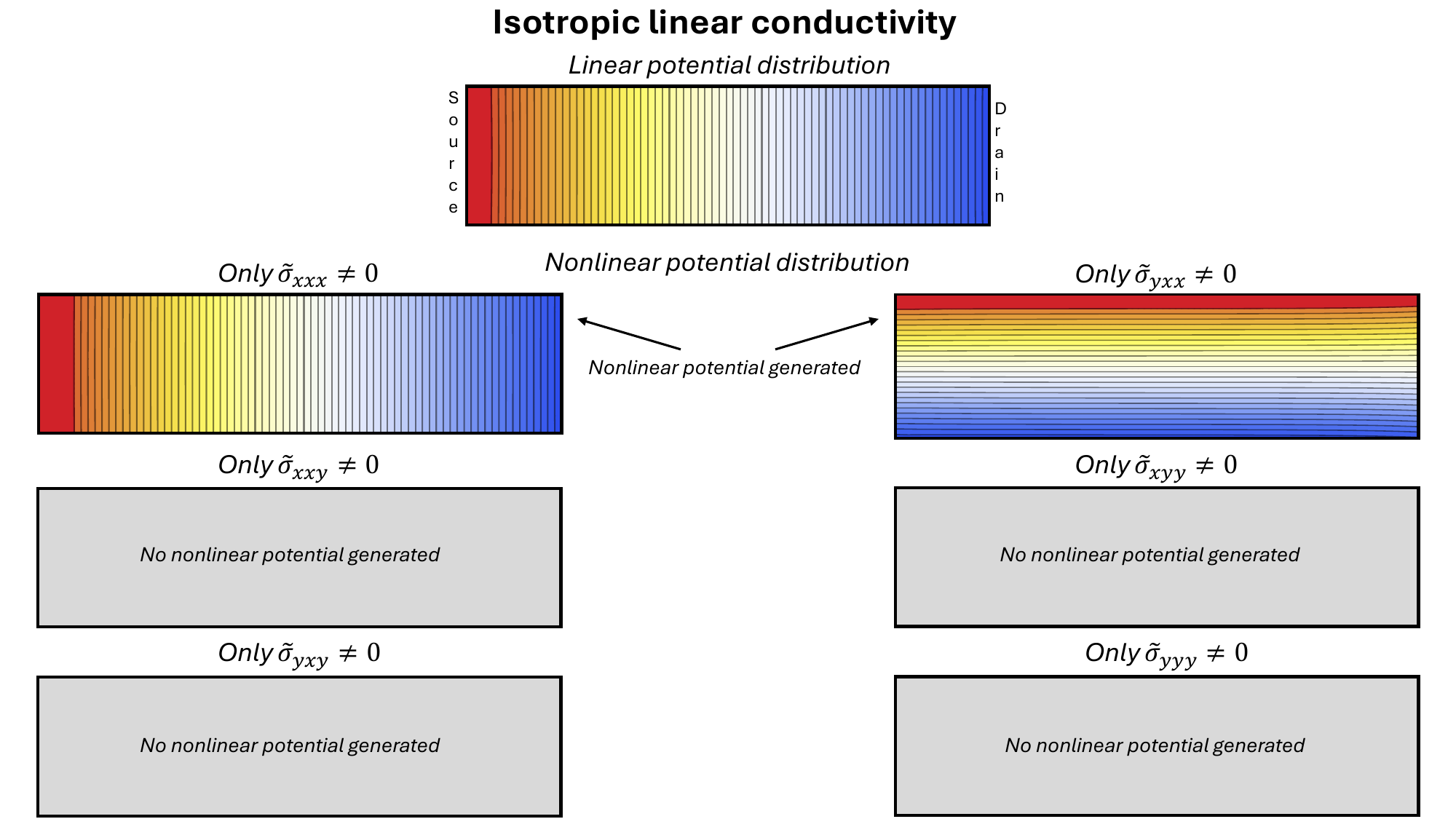}
\caption{ \newtext{{\bf{Nonlinear voltage response due to different components of the nonlinear conductivity tensor in Hall bar geometry.}} 
A Finite Element Method (FEM) simulation of linear (top panel) and nonlinear (6 bottom panels) electrostatic potentials in Hall bar geometry. Current is injected from the left side of the bar and drained from the right side. For simplicity, we assume isotropic linear conductivity with no linear Hall contribution. In such case, electric field $\bE^{(1)} = - \nabla \Phi^{(1)}$ at linear order in injected current has no component along the $y-$axis. Therefore, the only components of nonlinear conductivity tensor that can generate any nonlinear potential $\Phi^{(2)}$ are $\tilde{\sigma}_{xxx}$ and $\tilde{\sigma}_{yxx}$; the rest generate no nonlinear potential, as we verify using FEM.}    }
\label{ext_fig1}
\end{figure*}

\newtext{
In the presence of a mirror axis $M$, there are additional constraints on the structure of nonlinear conductivity tensor. Specifically for
$M_x: x \rightarrow -x, \; y \rightarrow y$, and $\tilde{\sigma}_{xxx}, \tilde{\sigma}_{xyy}, \tilde{\sigma}_{yxy}$ must all vanish:
\begin{equation}
    \begin{aligned}
        &\tilde{\sigma}_{xxx} = \mathcal{A}_x^3 - 3 \mathcal{A}_x \mathcal{A}_y^2 + \mathcal{B}_x = 0, \\
        &\tilde{\sigma}_{yxx} = 3 \mathcal{A}_x^2 \mathcal{A}_y - \mathcal{A}_y^3 + \mathcal{C}_x, \\
        &\tilde{\sigma}_{xyy} = - \left( \mathcal{A}_x^3 - 3 \mathcal{A}_x \mathcal{A}_y^2 \right) - \mathcal{C}_y = 0, \\
        &\tilde{\sigma}_{yxy} = - \left( \mathcal{A}_x^3 - 3 \mathcal{A}_x \mathcal{A}_y^2 \right) + \frac{\mathcal{B}_x + \mathcal{C}_y}{2} = 0.
    \end{aligned}
    \label{conductivities_relations_mirrorY}
\end{equation}
The vanishing of $\tilde{\sigma}_{xxx}$ in the presence of $M_x$ can be seen in experimental data for the Hall bar geometry in the Fig. 2d of Ref. \cite{Pablo2019}.
For $M_y: x \rightarrow x, \; y \rightarrow -y $ we show in the SI that $\tilde{\sigma}_{yxx}, \tilde{\sigma}_{xxy}, \tilde{\sigma}_{yyy}$ must all vanish, i.e.,
\begin{equation}
    \begin{aligned}
        &\tilde{\sigma}_{xxx} = \mathcal{A}_x^3 - 3 \mathcal{A}_x \mathcal{A}_y^2 + \mathcal{B}_x, \\
        &\tilde{\sigma}_{yxx} = 3 \mathcal{A}_x^2 \mathcal{A}_y - \mathcal{A}_y^3 + \mathcal{C}_x = 0, \\
        &\tilde{\sigma}_{xxy} = 3 \mathcal{A}_x^2 \mathcal{A}_y - \mathcal{A}_y^3 + \frac{\mathcal{B}_y - \mathcal{C}_x}{2} = 0,\\
        &\tilde{\sigma}_{yyy} = -\left( 3 \mathcal{A}_x^2 \mathcal{A}_y - \mathcal{A}_y^3 \right) + \mathcal{B}_y = 0.
    \end{aligned}
    \label{conductivities_relations_mirrorX}
\end{equation}
Regardless, each system of equations remains underdetermined. 
To see this explicitly when $M_x$ is a symmetry, we note that Eq. \eqref{conductivities_relations_mirrorY} gives $\mathcal{B}_x=\mathcal{C}_y=\mathrm{Re}\left(\mathcal{A}_y+i\mathcal{A}_y\right)^3=0$, and $\tilde{\sigma}_{yxx} = \mathrm{Im} \left(\mathcal{A}_y+i\mathcal{A}_y\right)^3 + \mathcal{C}_x$. But the latter does not determine $\mathcal{C}_x$. Therefore, a nonzero $\tilde{\sigma}_{yxx}$ cannot guarantee that $\mathcal{C}_x\neq 0$.
Thus, the Hall bar geometry is incapable of disentangling the nonlinear Hall component from the 3-fold component even if perfectly aligned with the crystalline mirror axis.
} 

\begin{figure*}
\includegraphics[width=0.8\linewidth]{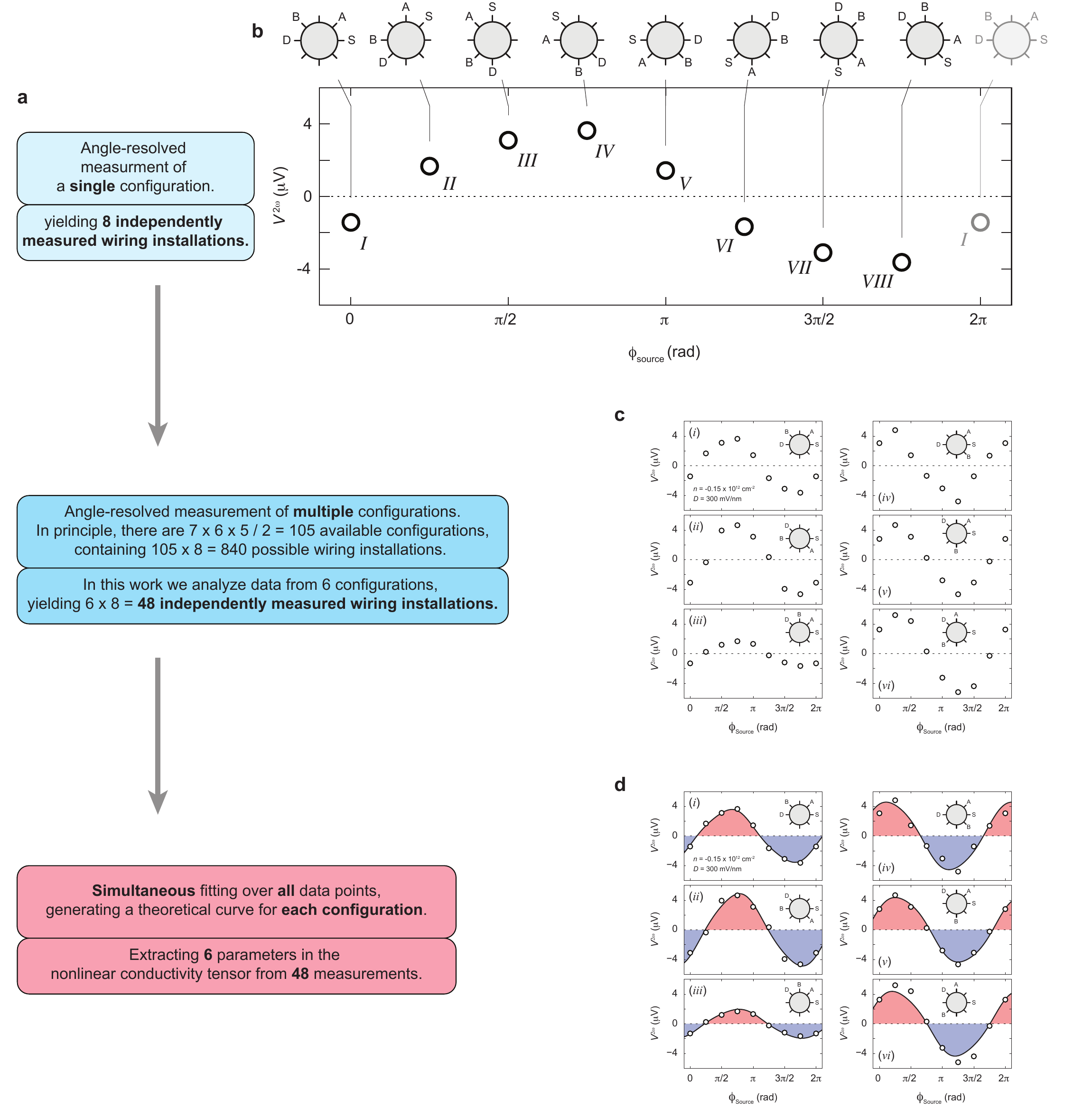}
\caption{ \newtext{ {\bf{Measurement procedure for conductivity fitting} } 
(a) A flow chart summarizing the procedure of nonlinear conductivity extraction. (b) Angle-resolved transport measurement of nonlinear potential difference for a \textit{single configuration}, with \textit{8 independent wiring installations} illustrated on top. Notice how every next wiring installation is obtained from a previous one by counterclockwise rotation. In other words, a single configuration has 8 rotationally equivalent wiring installations which measure voltage drop for current injected along different directions. (c) Angle-resolved transport measurement of nonlinear potential difference for \textit{multiple configurations}, each with 8 rotationally equivalent wiring installations. In principle, there are 105 different configurations allowed; in the main text we show data from 6 configurations, with a total of $6\times 8= 48$ \textit{independently measured wiring installations} for Sample 1 (in the SI we show data for Sample 2 with 72 independently measured wiring installations). (d) All $48$ measured data points are then \textit{simultaneously fitted} with one nonlinear conductivity tensor, i.e. 6\textit{ free parameters}. Using the extracted tensor, we can generate an expected voltage drop for any given measurement configuration and a continuously varying $\phi_{\text{Source}}$, shown as the black curves in panel (d), illustrating the good agreement between the theory and the data. For this figure we used the same data as in Fig.~\ref{fig3}a.}
}
\label{fig_measure}
\end{figure*}

\newtext{\subsection*{Measurement procedure for conductivity extraction}}

\newtext{Here we present an overview of the measurement procedure preparing the data set for the extraction of conductivity in the disk geometry. We use nonlinear conductivity extraction as an example. }

\newtext{ Fig.~\ref{fig_measure}a summarizes the procedure in a flow chart.
As we explained in the Main Text, a measurement configuration is defined as a collection of 8 wiring installations with the same relative positions of source, drain, and two contacts A, B, and related by a global rotation by $\pi/4$. For a disk-shaped sample with eight contacts, we have $7\times6\times5\div2=105$ distinct configurations, each with $8$ possible wiring installations, in total allowing us for $840$ independent measurements. In this work, we perform experiments with 6 different configurations, yielding a total of $48$ independent measurements at each point in the phase diagram (see Fig.~\ref{fig_measure}c, also see Fig.~\ref{fig3}a and Fig.~\ref{D300Transition}a-b). This places us well within the overdetermined regime for the fit, since a general nonlinear conductivity tensor in 2D has only 6 independent components in the DC limit.
}

\newtext{After the measurements, all $48$ data points are fitted \textit{simultaneously} to the same nonlinear conductivity tensor. As such, we extract 6 parameters for the nonlinear conductivity tensor using $48$ independent measurements, demonstrating a well-constrained model with a favorable data-to-parameter ratio that suppresses overfitting and highlights the reliability of the tensor extraction. Using the extracted parameters of the nonlinear conductivity tensor we plot the expected voltage drop $\Phi^{(2)}(\br_A)-\Phi^{(2)}(\br_B)$ for every measurement configuration and for any \newtext{$\phi_{\text{Source}}$} using Eqs. \eqref{basis1}-\eqref{basis3}, as demonstrated in Fig.~\ref{fig_measure}d. }

\newtext{For linear conductivity extraction, the measurement procedure is essentially the same. The data-to-parameter ratio is enhanced because there are only 4 parameters defining the linear conductivity tensor in 2D. Specific details of linear and nonlinear conductivity extraction are discussed below.}

\newtext{\subsection*{Details of the linear conductivity extraction}}

\newtext{Following the previous section, suppose we have $n$ independent wiring installations. For $i$-th wiring installation we measure linear in current potential difference (voltage drop $V^{\textrm{EXP}}_i$) between two leads A$_i$ and B$_i$ with current injected at source S$_i$ and drained at D$_i$. These measurements are done for fixed electron density, displacement field, and injected current $I$. 
Ref.~\cite{Oskar_solo} provides an expression for the linear electrostatic potential given arbitrary $\sigma_{\mu \nu}$ and $\text{A}_i, \text{B}_i$, and $\text{S}_i, \text{D}_i$ located anywhere on the circumference of the disk. Note that Eqn. 2 in Ref.~\cite{Oskar_solo} involves a nonlinear function and, therefore, does not have a straightforward inverse that allows for a direct computation of $\sigma_{\mu \nu}$ based on $V^\textrm{EXP}_i$. Thus, we approach this extraction numerically: by plugging locations of $\text{S}_i$, $\text{D}_i$, $\text{A}_i$, $\text{B}_i$, and the value of $I$ into Eqn. 2 of Ref.~\cite{Oskar_solo}, we obtain an array of theoretically predicted voltage drops $V^\textrm{THE}_i (\sigma_{\mu \nu})$, that is dependent only on the linear conductivity matrix $\sigma_{\mu \nu}$. We define the relative root mean squared error (RRMSE) as the loss function, as in Ref.~\cite{JamesLeo2024NatMat}: 
\begin{equation}
    \textrm{RRMSE}=\frac{\sqrt{\frac{1}{n}\sum_i(V_i^\textrm{EXP}-V_i^\textrm{THE})^2}}{\sqrt{\sum_i(V_i^\textrm{EXP})^2}},
\end{equation}
which is also dependent only on the linear conductivity tensor $\sigma_{\mu \nu}$. Using Nelder–Mead algorithm, we minimize RRMSE and thus complete the fitting for $\sigma_{\mu \nu}$.} 

\newtext{Generally, we consider a RRMSE around or less than $10\%$ as a good fit, as shown in SI Fig. S4. The sharp dip in the RRMSE when sweeping one fitting parameter of $\sigma_{\mu \nu}$ shows the uniqueness of the parameter choice. For example, a definitive identification of $\bar{\sigma}$ (the isotropic part of linear conductivity tensor). On the other hand, a flat RRMSE with respect to sweeping the parameter shows that the fit is not sensitive to that parameter, for example, the lack of response in RRMSE with respect to the principal axes orientation $\alpha$ illustrates the weakness of anisotropy in these states. }

\newtext{\subsection*{Details of the nonlinear conductivity extraction}}

\newtext{
The goal of fitting is to extract 6 independent components of nonlinear conductivity tensor from the data vector $V^{\textrm{NL-EXP}}_i$ ($i=1,2,...,n$). Mathematically, one can pose this problem as a system of linear equations:
\begin{equation}
    \vec{V}^{\textrm{NL-EXP}} = \hat{A} \cdot \vec{c},
\end{equation}
where $\vec{V}^{\textrm{NL-EXP}}$ is the vector of experimental data containing $n$ data points (each point is a nonlinear potential difference between two leads A and B for fixed source and drain), $\vec{c}$ is the unknown 6-component vector which depends on the components $\Xi^{(2)}_{-,+,0}$ of nonlinear conductivity tensor, and $\hat{A}$ is the $n \times 6$ matrix ($i$-th column of matrix $\hat{A}$ contains 6 numbers). The goal of fitting is to find $\vec{c}$. Here we use the following definition of $\vec{c}$:
\begin{equation}
    \vec{c} = \left( \mathrm{Re} \Xi^{(2)}_{-}, \mathrm{Im} \Xi^{(2)}_{-}, \mathrm{Re} \Xi^{(2)}_{+}, \mathrm{Im} \Xi^{(2)}_{+}, \mathrm{Re} \Xi^{(2)}_{0}, \mathrm{Im} \Xi^{(2)}_{0} \right).
\end{equation}
Matrix $\hat{A}$ contains theoretical values of nonlinear electrostatic potential difference $\Phi^{\left(2 \right)} (\br_A) - \Phi^{\left(2 \right)} (\br_B)$ for a given ($i$-th) wiring installation; they are found using Eqs. \eqref{basis1}-\eqref{basis3} for known angular width of source/drain. The vector $\vec{c}$ can be straightforwardly found using
\begin{equation}
    \vec{c} = \left( \hat{A}^{T} \hat{A} \right)^{-1} \hat{A}^{T} \cdot \vec{V}^{\textrm{NL-EXP}}.
    \label{fitting_eq}
\end{equation}
In the case of perfect data (no noise) this system is overcomplete, as we use $n$ points to extract $6$ parameters. In the realistic case of noisy data, Eq. \eqref{fitting_eq} is an exact analog of the linear least squares fit. The variation of vector of fitted parameters $\vec{c}$ can be estimated using the standard formula for linear regression, see SI. 
}

\newtext{
An explicit implementation of matrix $\hat{A}$ depends on the results of a linear fit. For non-vanishing linear anisotropy one first needs to make sure that the experimental data is expressed in the reference frame aligned with principal axis ($x-$axis is along the higher resistance direction). After that one generates matrix $\hat{A}$ using the anisotropic basis functions with $\Delta \sigma / \bar{\sigma}$ extracted from the linear fit. In a simpler case of vanishing (small) linear anisotropy one doesn't need to adjust the reference frame of data vector. 
}

\clearpage
\onecolumngrid

\begin{center}
\textbf{\large Supplementary information for: Observation of giant nonlinear Hall conductivity in Bernal bilayer graphene}
\end{center}

\setcounter{figure}{0}

\renewcommand{\thefigure}{S\arabic{figure}}

\maketitle

\tableofcontents

\section{Experimental details}

\subsection{Second harmonic and quadratic regime}

\begin{figure*}[b]
\includegraphics[width=0.5\linewidth]{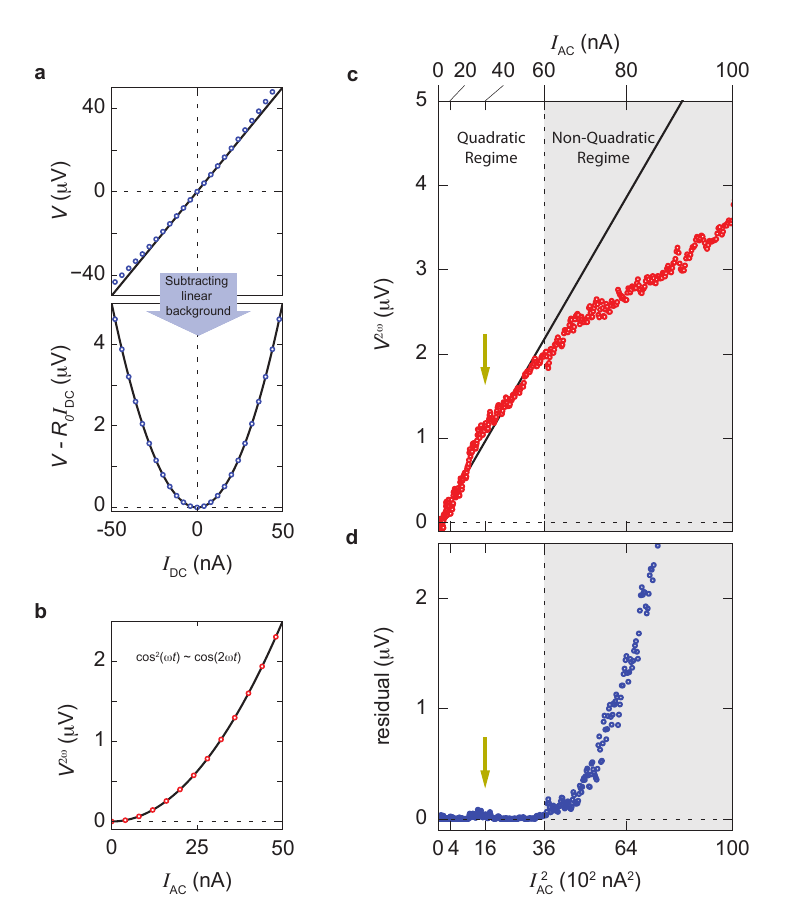}
\caption{{\bf{Second harmonic measurement and quadratic regime} }
(a) An exmaple $I-V$ relation, illustrating how second order voltage response is extracted using DC measurement. (b) Assuming the same $I-V$ relation, using second harmonic AC measurement to extract second order voltage response. 
(c) Experimental data of $V^{2\omega}$ as a function of the square of AC current amplitude $I_{AC}^2$, showing how quadratic regime is determined. }
\label{figSI_IV}
\end{figure*}

Isolating the second order response from the dominating first order part using a DC current source can be experimentally tedious. One would need to first measure the total voltage response as a function of current, then remove the first order component by performing a linear fit at $I_{\textrm{DC}}=0$ (See Fig.~\ref{figSI_IV}a). 
Experimentally we measure the second harmonic response of the injected AC current (while keeping the frequency around $13$Hz to be at DC limit) to extract the second order current response. This can be understood easily by plugging $I_{\textrm{AC}}=I_0 \cos(\omega t)$ into $V=aI + bI^2$. With some trigonometry, one can see that the second order term gives raise to a signal with frequency $2\omega$, which can be easily isolated out with the second harmonic function on a lock-in amplifier. As shown Fig.~\ref{figSI_IV}b, the isolated second harmonic signal follows a quadratic relation with respect to the amplitude of the AC current and is a factor of 2 smaller than DC measurement from the trigonometry relation. 

Just like the pure DC measurement, the second harmonic measurement can pickup contributions from higher order terms. To ensure that we are working within the current limit where the quadratic term is the only non-vanishing nonlinear term (the quadratic regime), we measure the second harmonic signal as a function of $I_0$, the amplitude of the injected AC bias (See Fig.~\ref{figSI_IV}c top panel). Then we find the largest $I_0$ below which the relation $V^{2\omega} \propto I_0^2$ still holds (black straight line when plotting in power scale). Such cutoff ($60$nA in this example), is indicated by vanishing residual below $I_0$ and a sharp onset above $I_0$ (Fig.~\ref{figSI_IV}c bottom panel). Working at a current below the cutoff ($40$nA vs $60$nA in this work) ensures the quadratic regime requirement is fulfilled. 

\begin{figure*}[t]
\includegraphics[width=1\linewidth]{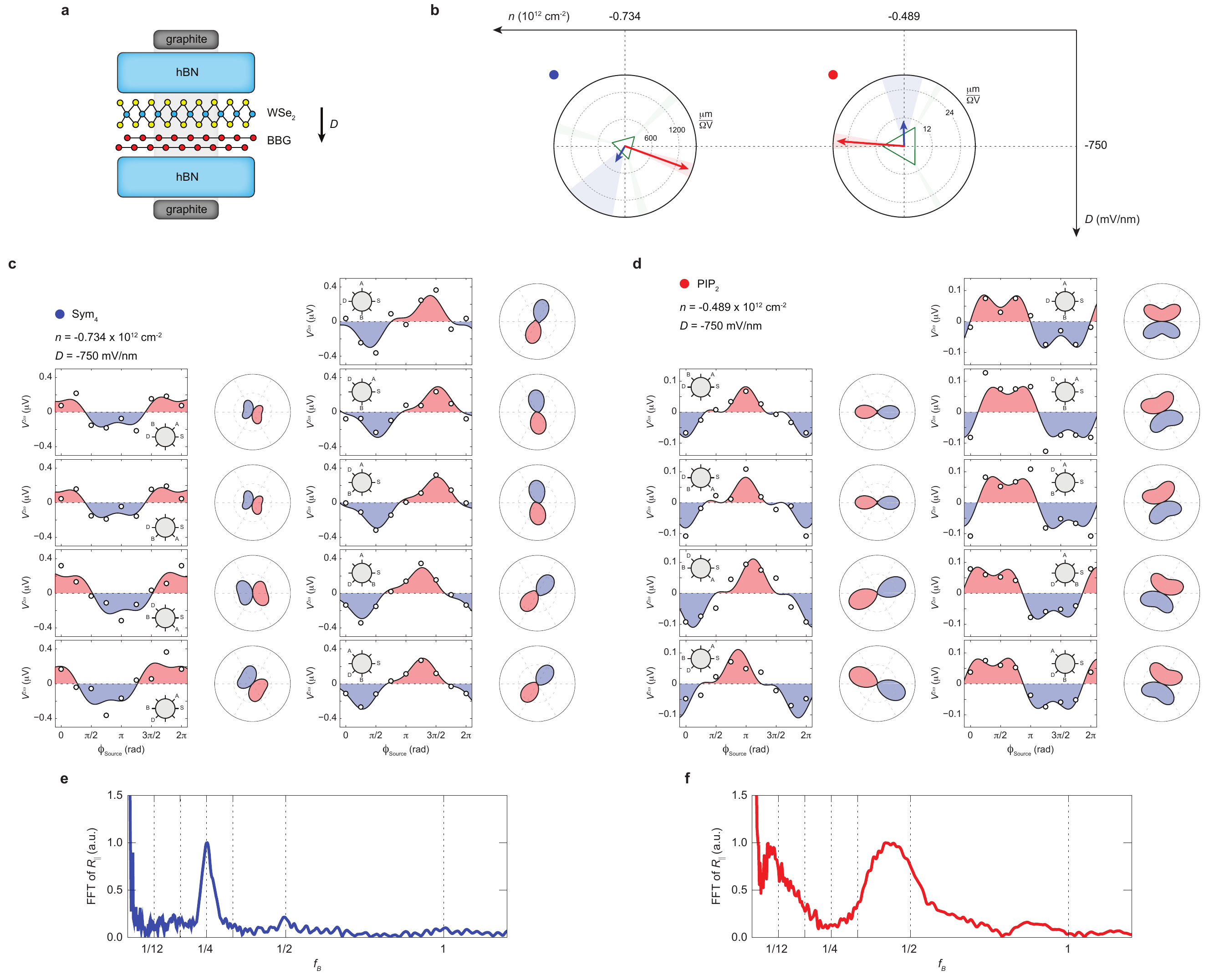}
\caption{ \newtext{{\bf{$n - D$ dependence of nonlinear conductivity from Sample 2} }
(a) Schematic of cross section of the Sample 2 structure. Negative $D$ corresponds to electrons far away from the WSe$_2$, minimizing the SOC from the proximity effect. (b) The vector representation of the extracted $\tilde\sigma_{\alpha\mu\nu}$ in the $n-D$ phase diagram of Sample 2. The sizes of green triangles, red and blue arrows denote $|\mathcal{A}^3|$, $|\mathcal{B}|$ and $|\mathcal{C}|$ vectors, respectively. The shades denote the fitting error in the angles of the vectors. (c-d) Angle resolved nonlinear voltage measurement (black circles) of 9 measurement configurations, with 8 different wiring installations per configuration, in total of 72 independent measurements at (c) $n=-0.734 \times 10^{12}$ cm$^{-2}$, $D=-750$ mV/nm (Sym$_4$ phase), and (d) $n=-0.489 \times 10^{12}$ cm$^{-2}$, $D=-750$ mV/nm (PIP$_2$ phase). Black curves denote the theoretically calculated expected nonlinear voltage drop using the extracted tensors. (e-f) FFT of longitudinal resistance, $R_\parallel$, as a function of frequency in $B$, measured at (e) $n=-0.734 \times 10^{12}$ cm$^{-2}$, $D=-750$ mV/nm, and (f) $n=-0.489 \times 10^{12}$ cm$^{-2}$, $D=-750$ mV/nm. Quantum oscillations peaking at $1/4$ and slightly below $1/2$ confirm our identification of Sym$_4$ and PIP$_2$ phases, respectively. All data measured in Sample 2 at $T=20$ mK. $B=0$ for data in panels c) and d).  } } 
\label{figSI_sample2}
\end{figure*}

\begin{figure*}[b]
\includegraphics[width=0.5\linewidth]{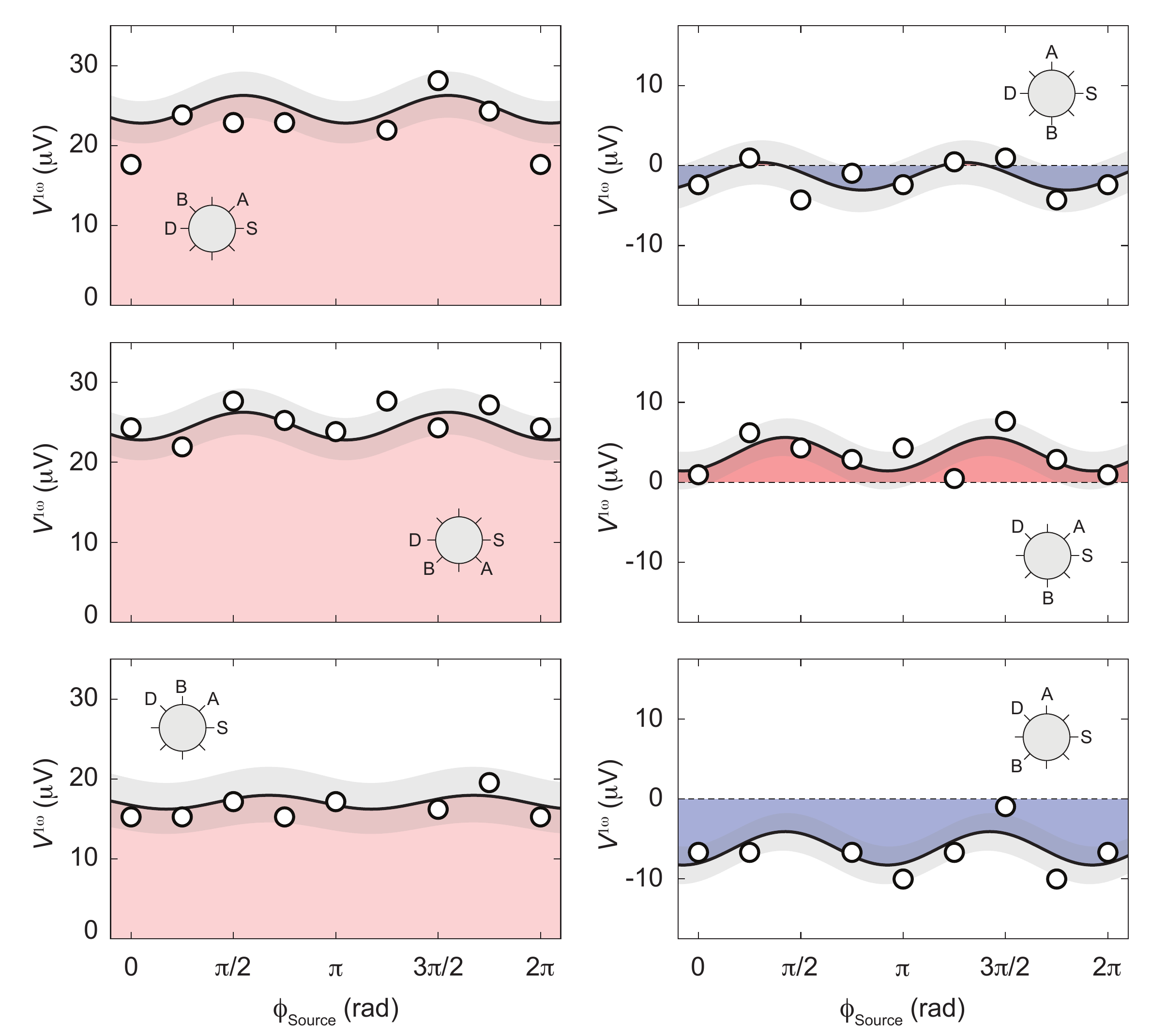}
\caption{{\bf{Angle-resolved transport in the linear regime.} } Transport response in the linear regime was measured using six different configurations, with eight distinct orientations \newtext{(wiring installations)} of current flow per configuration. Solid black lines represent the expected transport response derived from a single conductivity matrix in the linear regime, while the gray shaded stripes denote the anticipated measurement error due to the finite width of the graphene leads. Notably, certain configurations probe the transport response near the corners of the disk-shaped sample, such as the configuration depicted in the bottom left panel. The consistent fit of all measurements to the same conductivity matrix strongly supports the conclusion that the transport response is both local and uniform across the sample.
}
\label{full_linear}
\end{figure*}

\subsection{Sample and measurement}
The sample consists of a Bernal bilayer graphene dual-encapsulated by top and bottom graphite gates, separated with hexagonal BN. The stack is picked up with a standard poly(bisphenol A carbonate) (PC)/polydimethylsiloxane (PDMS) stamp on a glass slide. The stack is then released on to a doped Si/SiO${}_2$ substrate and patterned using electron beam lithography with a minimum resolution of about $100$nm, yielding a contact with angular width of $18^\circ$ for a disk of diameter $2\mu$m. Metal contacts are then made with Cr/Au (2nm/100nm respectively). 

The dual-gate structure allows separate control over charge carrier density and displacement field. Transport measurements are performed with Standford Research 860 and 830 Lock-in amplifiers at about 13Hz, in a BlueFors LD400 dilution refrigerator with a base temperature of ~20mK. The fast configuration switching is allowed by Keysight 34980A data acquisition system and a custom designed breakout box.

\begin{figure*}
\includegraphics[width=1\linewidth]{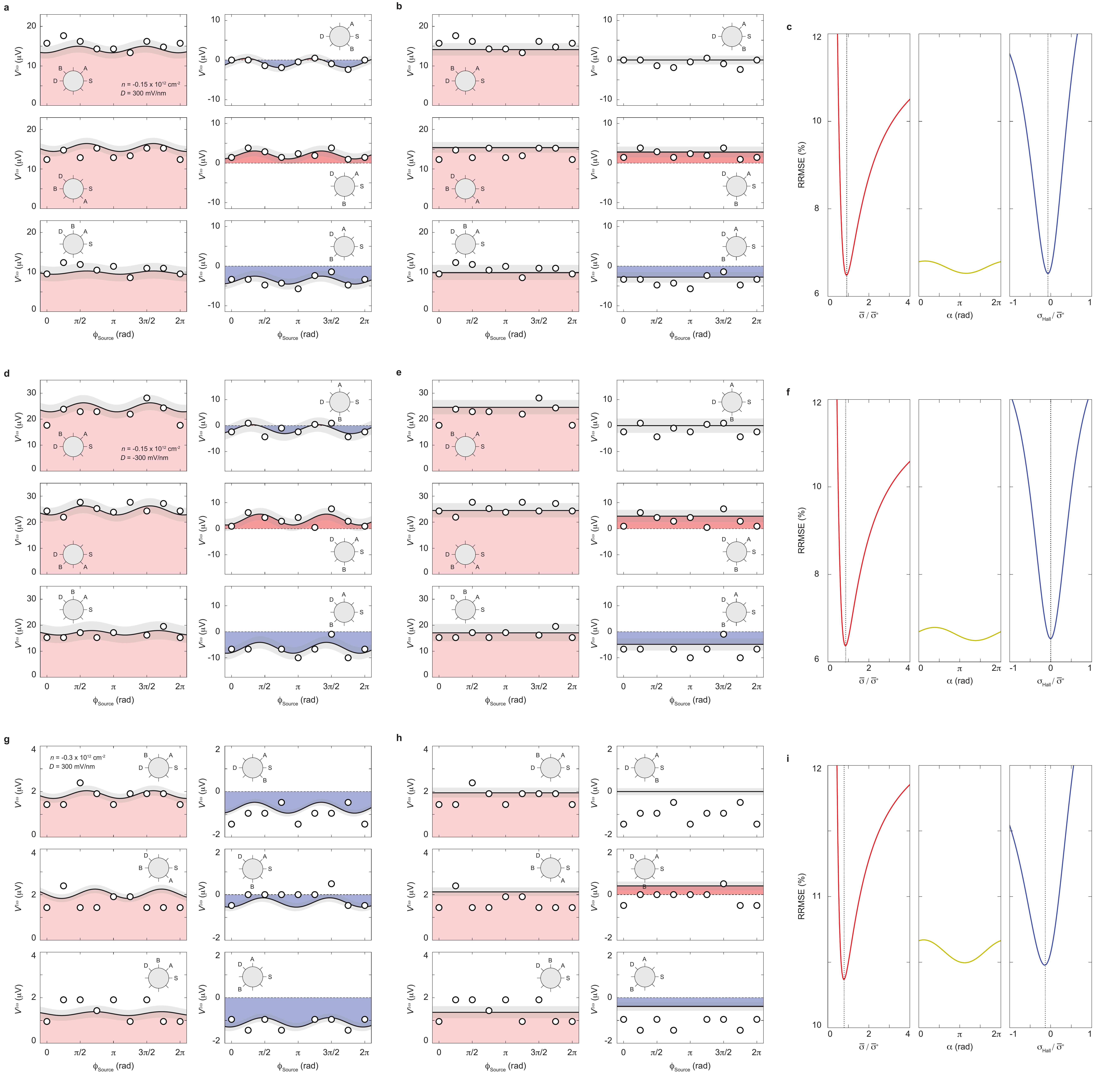}
\caption{{\bf{Linear full fit} }
Full fit for extracting the linear conductivity matrix for (a-c) $n=-0.15 \times 10^{12}$cm${}^{-2}$ and $D=300$ mV/nm, (d-f) $n=-0.15 \times 10^{12}$cm${}^{-2}$ and $D=-300$ mV/nm, and (g-i) $n=-0.3 \times 10^{12}$cm${}^{-2}$ and $D=300$ mV/nm. Panel (a,d,g) show the fitting result (black curves) compared against measured data (black open circles) over 48 different wiring installations, allowing anisotropy and Hall conductance. All six lines in each panel come from the same linear conductivity matrix. Grey shades denote the error bar of the fit from the finite width of the sample contacts. In each case $\Delta\sigma / \bar{\sigma} \approx 3\%$, and $\sigma_{H} \simeq 0$. In PIP$_2$ phase linear conductivity is  $\bar{\sigma} \simeq 41 \frac{e^2}{h} $ while in Sym$_4$ phase $\bar{\sigma} \simeq 298 \frac{e^2}{h} $.  Panel (b,e,h) show the fitting result (black curves) compared against measured data (black open circles) over 48 different \newtext{wiring installations} \sout{configurations}, forcing $\Delta\sigma / \bar{\sigma} = 0$, and $\sigma_{H} = 0$. Contrasting (a,d,g) with (g,e,h), we demonstrate the anisotropy and Hall conductance in each state is weak. (d,f,i) shows how \newtext{Relative Root Mean Square Error (RRMSE)} \cite{JamesLeo2024NatMat} changes as we sweep different parameters of the linear conductivity matrix. Left panels shows a sharp dip at $\bar{\sigma}$, indicating an precise extraction of $\bar{\sigma}$. Middle panels shows a flat response as we sweep the principle axis angle. The insensitivity confirms that the anisotropy is weak (see section Fitting Protocol and Error Estimation). Right panel shows a sharp dip near 0, confirming a small $\sigma_H$ compared to $\bar{\sigma}$. \newtext{Mean free path is $\sim 0.3 \mu$m for PIP$_2$ phase and $\sim 1\mu$m for Sym$_4$ phase. } }
\label{figSI_linear}
\end{figure*}

\begin{figure*}
\includegraphics[width=0.5\linewidth]{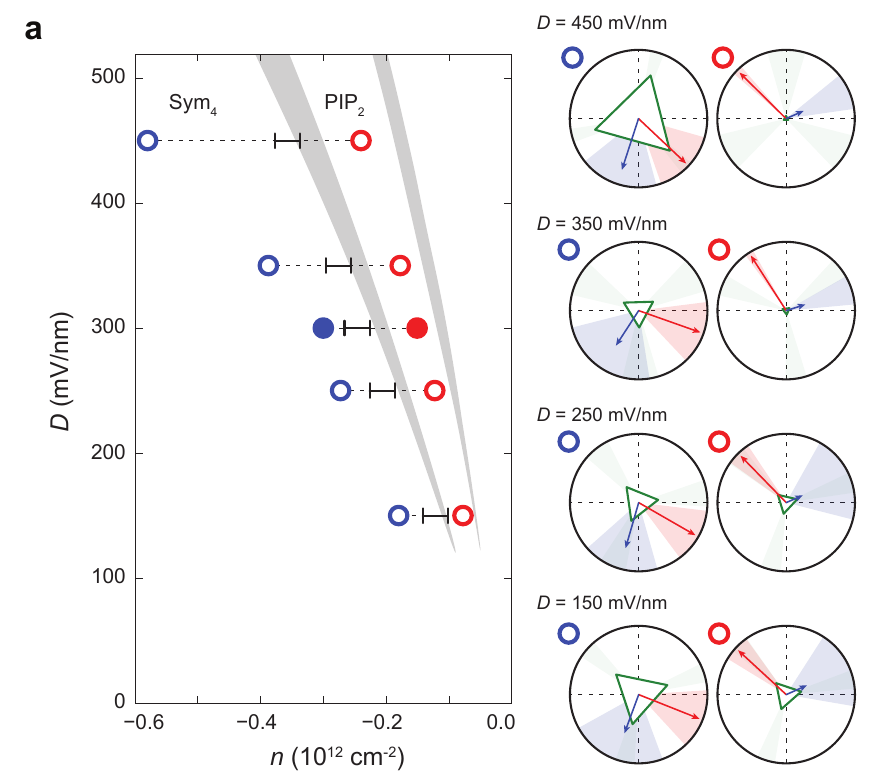}
\caption{{\bf{Full fit in $n-D$ phase space. } }
(a) Same $n-D$ phase diagram as shown in Main Fig. 4d of the main text, with all three fitted vectors plotted on the right hand side. The green triangle depicts $\mathcal{A}^3$, red arrow shows $\mathcal{B}$, and blue arrow indicates $\mathcal{C}$. The size of the triangle and the arrows indicates relative magnitude of different components of the tensor, normalized to each corresponding $\mathcal{B}$ vector. Shaded cones indicate the fitting uncertainties of angles. The larger 3-fold component in the Sym$_4$ phase likely is caused by measurement uncertainty due to lower signal-to-noise ratio in the Sym$_4$ phase.  }
\label{figSI_nD3vec}
\end{figure*}

\begin{figure*}
\includegraphics[width=0.7\linewidth]{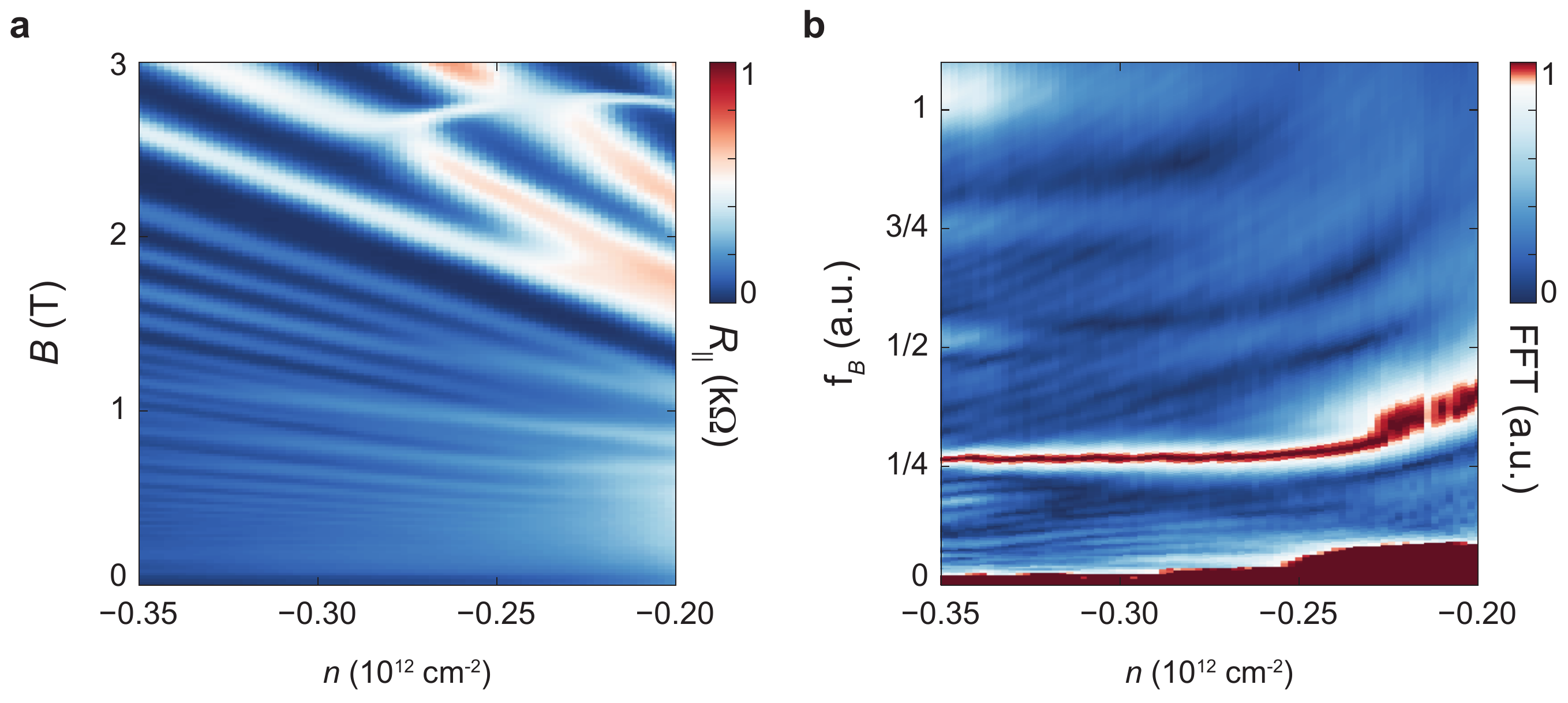}
\caption{ {\bf{Quantum oscillations data for Sym${}_\mathbf{4}$ state.} }
(a) $n-B$ map at $D=300$mV/nm. (b) FFT of panel (a). The peak near $1/4$ at $n=-0.3\times 10^{12}$ cm$^{-2}$ indicates four nearly degenerate Fermi pockets. }
\label{figSI_QO}
\end{figure*}

\begin{figure*}
\includegraphics[width=0.5\linewidth]{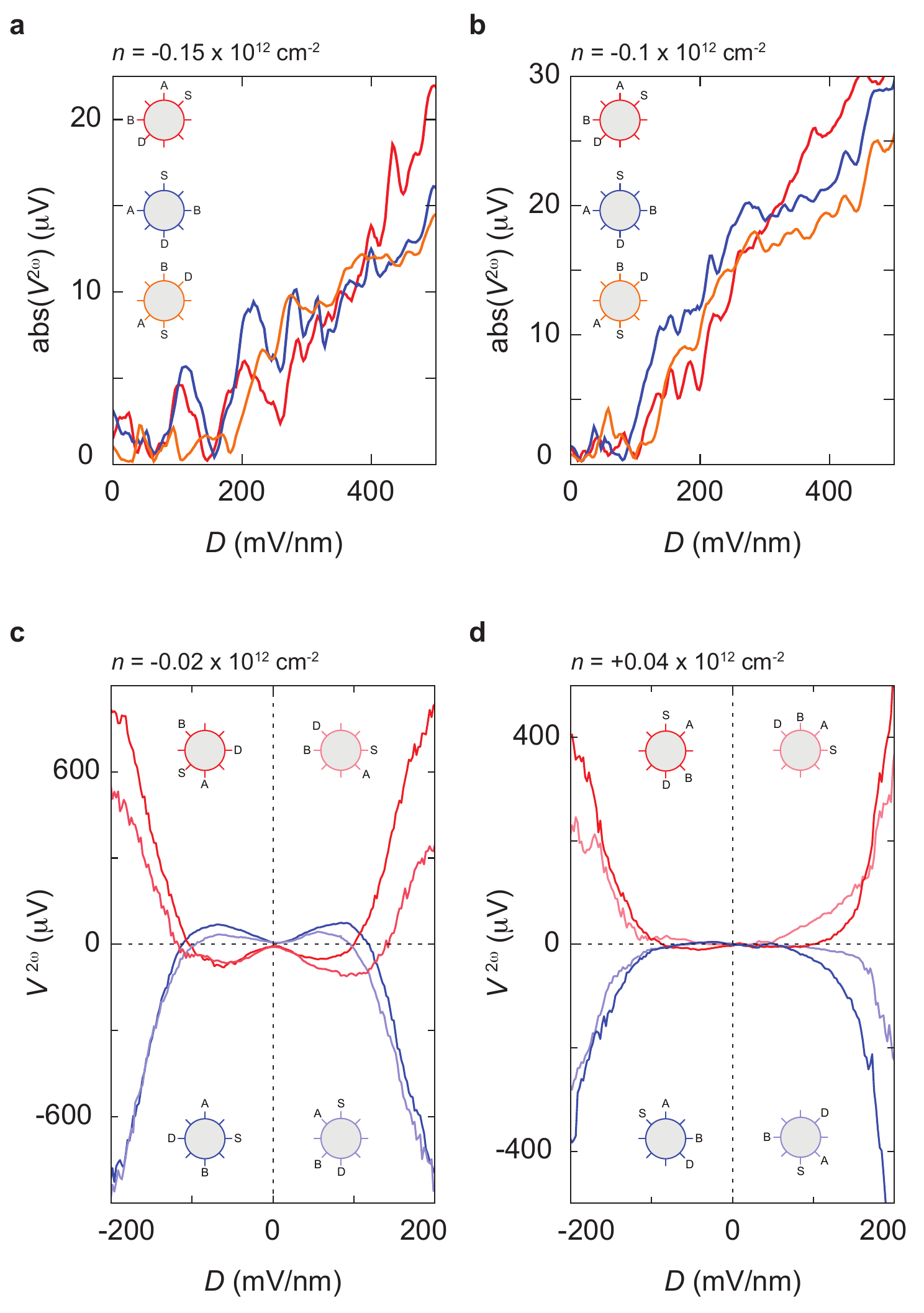}
\caption{{\bf{Displacement field dependence of the second harmonic response. } }
(a-b) Displacement field dependence of the second harmonic response amplitude measured for different configurations, at (a) $n=-0.15 \times 10^{12}$ cm$^{-2}$ and (b) $n=-0.1 \times 10^{12}$ cm$^{-2}$. (c-d) Displacement field dependence of the second harmonic response measured for different configurations, at (c) $n=-0.02 \times 10^{12}$ cm$^{-2}$ and (d) $n=0.04 \times 10^{12}$ cm$^{-2}$. }
\label{figSI_Displacement}
\end{figure*}

\begin{figure*}
\includegraphics[width=0.7\linewidth]{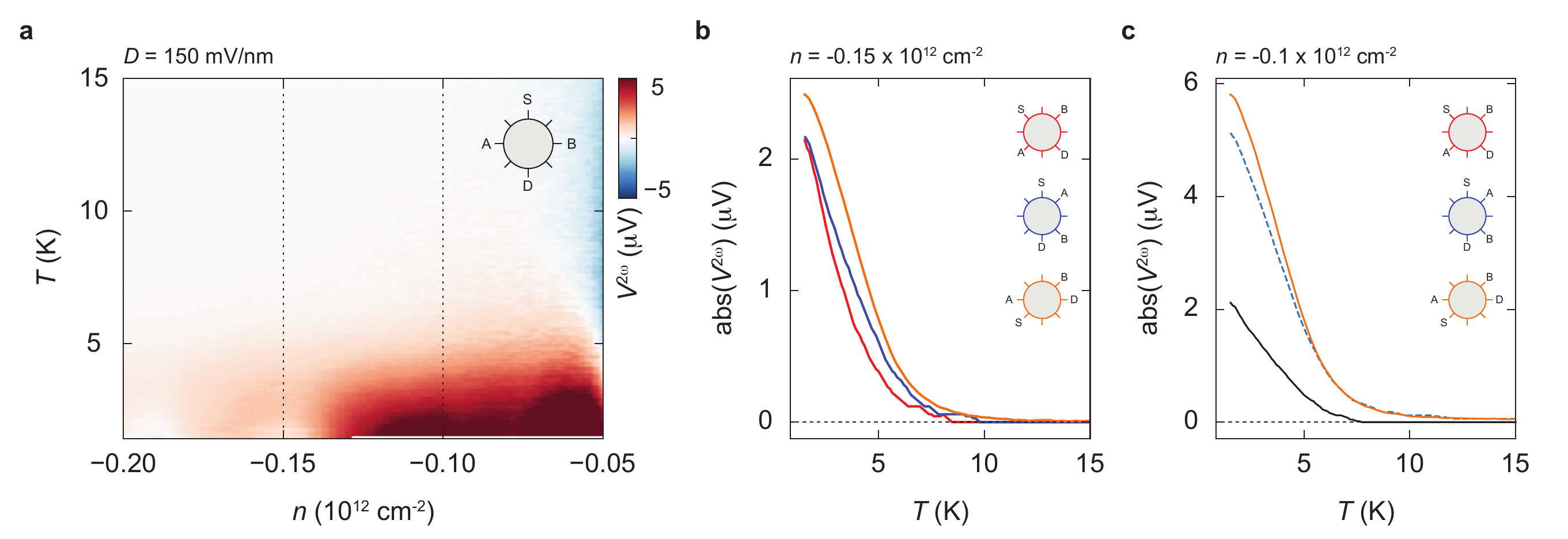}
\caption{{\bf{Temperature dependence of the second harmonic response. } }
(a) $n-T$ map at $D=150$mV/nm. (b-c) Temperature dependence of the second harmonic response measured for different wiring installations, at (b) $n=-0.15 \times 10^{12}$ cm$^{-2}$ and (c) $n=-0.1 \times 10^{12}$ cm$^{-2}$.}
\label{figSI_temperature}
\end{figure*}

\begin{figure*}
\includegraphics[width=0.5\linewidth]{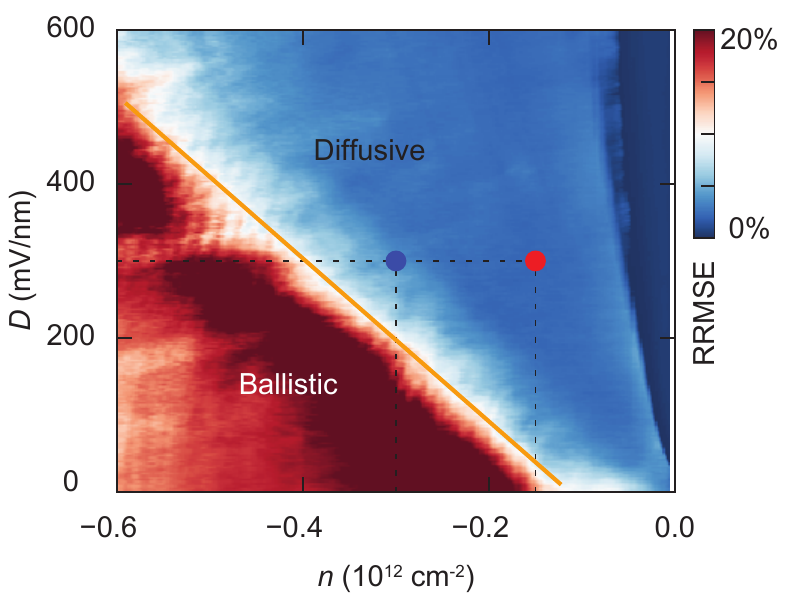}
\caption{\newtext{{\bf{Determining ballistic regime using linear full fit. } }
RRMSE of linear full fit in the $n-D$ phase diagram. We perform angle resolved transport measurements and fit voltage drops using our framework for extraction of the linear conductivity tensor. The quality of the fit is characterized by the Relative Root Mean Square Error (RRMSE). As shown in the color map, there is a sharp onset of RRMSE across the orange solid line. To the right of this line, RRMSE is around or less than $10\%$, indicating that the fit is good and the sample is in diffusive regime. To the left, RRMSE quickly increases beyond $20\%$. The failure of the fit indicates that the sample is becoming less diffusive, approaching the ballistic regime. Red and blue dot indicate the locations in the $n-D$ phase diagram where data used in the Main Fig. 4 was collected. }}
\label{figSI_Ballistic}
\end{figure*}

\newtext{\subsection{Results for Sample 2}}

\newtext{
Here we report the reproducibility of the extraction of nonlinear conductivity, the observation of giant nonlinear Hall effect, and the reorientation of the nonlinear Hall effect across the PIP$_2$ - Sym$_4$ transition in a second sample, denoted as Sample 2. Sample 2 is disk-shaped BBG heterostructure with a few-layer WSe$_2$ at proximity. The sample fabrication and measurement setup is similar to the main sample reported above. A schematic of the Sample 2 structure is shown in Fig.~\ref{figSI_sample2}a. We show nonlinear potential measurement data from two points in the $n-D$ phase diagram for Sample 2: $n=-0.734 \times 10^{12}$ cm$^{-2}$, $D=-750$ mV/nm (Sym$_4$ phase), and $n=-0.489 \times 10^{12}$ cm$^{-2}$, $D=-750$ mV/nm (PIP$_2$ phase). Here, we highlight three observations:}

\newtext{
1. In both regimes, the excellent agreement between the experimental data and the theoretical curves illustrates the successful extraction of the nonlinear conductivity tensor, demonstrating the universality of our framework. Shown in Fig.~\ref{figSI_sample2}c-d, in each regime, we perform angle resolved nonlinear transport measurement using 9 measurement configurations, each with 8 wiring installations, with a total of $9\times 8=72$ independent measurements. All 72 data points are then fitted together, yielding one nonlinear conductivity tensor $\tilde\sigma_{\alpha\mu\nu}$ with 6 free parameters. The extracted 6 components of nonlinear conductivity tensor can then be used to predict the nonlinear voltage drop for any accessible  configuration. 
}

\newtext{
2. The extracted nonlinear conductivity tensors in PIP$_2$ and Sym$_4$ phases have the same structure and similar magnitude for both Sample 1, reported in Main Fig. 3 and 4, and Sample 2. As shown in Fig.~\ref{figSI_sample2}b, in both phases  $\mathcal{A}+\mathcal{B}\approx \mathcal{B}$ and $|\mathcal{C}| \sim  |\mathcal{B}|/3$. In PIP$_2$ phase, the magnitudes of the extracted nonlinear tensor components read $|\mathcal{A}^3|=9.2 \pm 1.3\textrm{ }\frac{\mu\textrm{m}}{\Omega\cdot\textrm{V}}$, $|\mathcal{B}|=28.8 \pm 3.3\textrm{ }\frac{\mu\textrm{m}}{\Omega\cdot\textrm{V}}$, and $|\mathcal{C}|=10.9 \pm 3.2\textrm{ }\frac{\mu\textrm{m}}{\Omega\cdot\textrm{V}}$. In Sym$_4$ phase, we again observe very large magnitude of nonlinear conductivity, namely,  $|\mathcal{A}^3|=296 \pm 64\textrm{ }\frac{\mu\textrm{m}}{\Omega\cdot\textrm{V}}$, $|\mathcal{B}|=1504 \pm 178\textrm{ }\frac{\mu\textrm{m}}{\Omega\cdot\textrm{V}}$, and $|\mathcal{C}|=424 \pm 156\textrm{ }\frac{\mu\textrm{m}}{\Omega\cdot\textrm{V}}$. } 

\newtext{
3. We, in accordance with our report in main text, also observe the nearly-180$^{\circ}$  reorientation of the nonlinear conductivity tensor upon crossing from the PIP$_2$ to Sym$_4$ phase. 
}

\newtext{
The results obtained from Sample 2 highlight the robustness of our framework regarding extracting conductivity tensors and the universality of giant nonlinear conductivity in BBG samples. 
It is important to note that the dual gated structure of Sample 2 allows us to use large negative displacement field to push charge carriers away from the WSe$_2$, therefore minimizing the effect from proximity-induced spin-orbit-coupling (SOC). We identify the corresponding isospin phases in the negative $D$ range using quantum oscillation, as shown in Fig.~\ref{figSI_sample2}e-f. The fact that the observations from Sample 2 are in agreement with the ones from Sample 1 illustrates that at $D=-750$ mV/nm, SOC from WSe$_2$ is diminishing. In the case of positive displacement field, where charge carriers are proximal with WSe$_2$, both linear and nonlinear conductivity behave very differently from what is reported here. A more detailed analysis contrasting positive and negative $D$ in Sample 2, discussing the effect of proximity-induced SOC on nonlinear conductivity would be an important follow-up, but is beyond the scope of this work. 
}

\section{Ohm's law and the symmetries of linear and nonlinear conductivity tensors}

\subsection{Coordinate representation}

Ohm's law relates current $j$ generated in media to the applied electric field $E$. In the standard linear case $j$ depends linearly on $E$. In the nonlinear case, which we consider in this work, $j$ has an additional dependence on $E^2$ and the nonlinear Ohm's law reads
\begin{equation}
    j_{\alpha} = \sigma_{\alpha\mu} E_{\mu} + \tilde{\sigma}_{\alpha \mu \nu} E_{\mu} E_{\nu},
    \label{Ohm_in_coordinate}
\end{equation}
where $E_{\mu}$ -- is the applied electric field, $\sigma_{\alpha \mu}$ -- is the linear conductivity tensor, and $\tilde{\sigma}_{\alpha \mu \nu}$ -- is the nonlinear conductivity tensor and with $\alpha,\mu,\nu$ are spatial indices. In 2 spatial dimensions linear conductivity tensor $\sigma_{\alpha \mu}$ is a $2 \times 2$ matrix, which in the absence of any symmetries has 4 independent components. This statement is valid for both AC and DC currents. The structure of nonlinear conductivity tensor does depend on whether the applied electric field is time-dependent or not. In the case of AC field and in the absence of any spatial symmetries, $\tilde{\sigma}_{\alpha \mu \nu}$ can have up to 8 independent components in 2 spatial dimensions. In the DC case, $\tilde{\sigma}_{\alpha \mu \nu}$ has to be symmetric with respect to $\mu \leftrightarrow \nu$, therefore, the maximal number of independent components is reduced to 6. We restrict our studies solely to the DC case in 2 spatial dimensions. 

In coordinate representation linear and nonlinear tensors can be written in terms of Pauli matrices
\begin{equation}
    \begin{gathered}
        \sigma_{\alpha \mu} = \mathfrak{k}_0 \tau_0 + \mathfrak{k}_1 \tau_1 + i \mathfrak{k}_2 \tau_2 +  \mathfrak{k}_3 \tau_3\\
        \tilde{\sigma}_{\alpha \mu \nu} = \mathfrak{c}_0^{\alpha} \tau_0 + \mathfrak{c}_1^{\alpha} \tau_1 + \mathfrak{c}_3^{\alpha} \tau_3,
    \end{gathered}
\end{equation}
where $\mathfrak{c}_2^{\alpha} \tau_2$ component is absent in the DC limit. 

Spatial (point-group) symmetries set constraints on the structure of conductivity tensors. For example, $C_3$ invariance of the system requires that upon $C_3$ rotations $\sigma_{\alpha \mu}$ and  $\tilde{\sigma}_{\alpha \mu \nu}$ remain invariant, i.e.,
\begin{equation}
\begin{gathered}
R_{\alpha' \alpha} \sigma_{\alpha \mu} (R^{-1})_{\mu \mu'} = \sigma_{\alpha' \mu'} = \sigma_{\alpha \mu}, \\
R_{\alpha' \alpha} \tilde{\sigma}_{\alpha \mu \nu} (R^{-1})_{\mu \mu'} (R^{-1})_{\nu \nu'} = \tilde{\sigma}_{\alpha' \mu' \nu'} = \tilde{\sigma}_{\alpha \mu \nu},
\end{gathered}
\label{C3_tensor_transformation}
\end{equation}
where $R$ -- is the matrix of rotation by $2 \pi /3$. By solving Eq. \eqref{C3_tensor_transformation} for coefficients $\mathfrak{k}_j$ and  $\mathfrak{c}_j^{\alpha}$ with requirement that all coefficients are real we arrive at the constraint for linear tensor
\begin{equation}
    \mathfrak{k}_1 = \mathfrak{k}_3 =0
\end{equation}
and for nonlinear tensor
\begin{equation}
    \begin{gathered}
    \mathfrak{c}_1^2 = -\mathfrak{c}_3^1, \; \; \; \mathfrak{c}_3^2 = \mathfrak{c}_1^1,
    \end{gathered}
\end{equation}
and all other coefficients are zeros.

\subsection{Complex representation}

Equation \eqref{Ohm_in_coordinate} can also be conveniently written in the complex form, where 2D vectors are treated as complex numbers. For the purely linear case Ohm's law in the complex form reads
\begin{equation}
    j_x^{l} + i j_y^{l} = \sigma^{(1)}_{+} e^{i \chi^{(1)}_{+}} \left( E_x + i E_y \right) + \sigma^{(1)}_{-} e^{i \chi^{(1)}_{-}} \left( E_x - i E_y \right).
\end{equation}
The four real parameters $\sigma^{(1)}_{+}, \sigma^{(1)}_{-}, \chi^{(1)}_{+}, \chi^{(1)}_{-}$ in complex representation can be easily related to the four real parameters $\mathfrak{k}_{0}, \mathfrak{k}_1, \mathfrak{k}_2, \mathfrak{k}_3$ of real representation:
\begin{eqnarray}
    &&\mathfrak{k}_0 = \sigma^{(1)}_{+} \cos \chi^{(1)}_{+}, \\
    &&\mathfrak{k}_3 = \sigma^{(1)}_{-} \cos \chi^{(1)}_{-}, \\
    &&\mathfrak{k}_1 = \sigma^{(1)}_{-} \sin \chi^{(1)}_{-}, \\
    &&\mathfrak{k}_2 = - \sigma^{(1)}_{+} \sin \chi^{(1)}_{+}.
\end{eqnarray}

Let us now focus on the nonlinear contribution to the electric current in Eq. \eqref{Ohm_in_coordinate}. There are two ways to derive the complex form of the nonlinear contribution. One is to explicitly use the coordinate representation and match coefficients in front of equivalent terms. The other one involves an analogy with the multiplication of two-dimensional irreducible representations (irreps) of point groups. 
Let us use the more elegant method of multiplying irreps.
The right-hand side of the equation can be thought of as a product of two $E$ irreducible representations (because electric field is a 2-component vector). To use this analogy let us recall the irrep multiplication rule $E \times E = A_1 + A_2 + E$. Imagine that we want to calculate a Kronecker product of two different two-component irreps $E$ with basis vectors $(P_x \pm i P_y)$ and $(Q_x \pm i Q_y)$. Then the $A_1$ combination is $P_x Q_x + P_y Q_y$ (which comes from $(P_x + i P_y)(Q_x - i Q_y) + c.c.$); two $E$ combinations are $P_x Q_x - P_y Q_y$ and $P_x Q_y + P_y Q_x$ (both come from linear combinations of $(P_x + i P_y)(Q_x + i Q_y)$ and $(P_x - i P_y)(Q_x - i Q_y)$); and the $A_2$ combination is $P_x Q_y - P_y Q_x$ (this one comes from
$(P_x + i P_y)(Q_x - i Q_y) - c.c.$). We now use the fact that in our case both $P-$ and $Q-$ vectors are the time-independent electric field $E$, therefore, the $A_2$ combination identically vanishes and there are no two additional independent parameters that would correspond to the $A_2$ contribution. This manifests the symmetry of the nonlinear tensor upon the interchange of $E_x$ and $E_y$ for DC current.
Explicitly substituting $P_x = Q_x = E_x$ and $P_y = Q_y = E_y$ we obtain three possible terms
\begin{eqnarray}
    &&(E_x-iE_y)^2, \\
    &&(E_x+iE_y)^2, \\
    &&E_x^2+E_y^2,
\end{eqnarray}
a linear combination of which will give the nonlinear contribution to current density. Therefore, the full Eq. \eqref{Ohm_in_coordinate} in complex form reads
\begin{equation}
    j_x + i j_y = \sigma^{(1)}_{+} e^{i \chi^{(1)}_{+}} \left( E_x + i E_y \right) + \sigma^{(1)}_{-} e^{i \chi^{(1)}_{-}} \left( E_x - i E_y \right) + \sigma^{(2)}_{-} e^{i\chi^{(2)}_{-}}(E_x-iE_y)^2 + \sigma^{(2)}_{+} e^{i \chi^{(2)}_{+}} (E_x+iE_y)^2 + \sigma^{(2)}_{0} e^{i \chi^{(2)}_{0}} \left( E_x^2+E_y^2 \right).
    \label{Ohm_in_complex}
\end{equation}
In the main text we employed the shorthand notation 
\begin{equation}
    \begin{gathered}
        \Xi^{(1)}_{+,-} = \sigma^{(1)}_{+,-} e^{i \chi^{(1)}_{+,-}}, \\
        \Xi^{(2)}_{+,0,-} = \sigma^{(2)}_{+,0,-} e^{i \chi^{(2)}_{+,0,-}}.
    \end{gathered}
\end{equation}
 for brevity. 

Some symmetry constraints on the nonlinear tensor, e.g., transformation under rotations, automatically follow from the complex representation. In this work we are specifically interested in hexagonal lattice systems, where $C_3$ rotational symmetry is fundamental. In complex form rotation of vectors is extremely simple and is performed by multiplication by $e^{i \phi}$, where $\phi$ - is the rotation angle. It is evident then that the $C_3$-invariant contribution to the nonlinear tensor is given by
\begin{equation}
    \sigma^{(2)}_{-} e^{i\chi^{(2)}_{-}}(E_x-iE_y)^2
\end{equation}
because under $C_3$ a vector $x+iy\rightarrow -\frac{1}{2}x+\frac{\sqrt{3}}{{2}}y+i\left(-\frac{1}{2}y-\frac{\sqrt{3}}{{2}}x\right)=
\left(-\frac{1}{2}-i\frac{\sqrt{3}}{{2}}\right)(x+iy)=e^{-i2\pi/3}(x+iy)$.

\subsection{Relations between components of nonlinear conductivity tensor in coordinate and complex representations}

Consider the nonlinear contribution to the current in coordinate
\begin{equation}
    \begin{gathered}
        E_x^2 \left( \mathfrak{c}_0^1 + \mathfrak{c}_3^1 + i \mathfrak{c}_0^2 + i \mathfrak{c}_3^2 \right) + 2 E_x E_y \left( \mathfrak{c}_1^1 + i \mathfrak{c}_1^2 \right) + E_y^2 \left( \mathfrak{c}_0^1 - \mathfrak{c}_3^1 + i \mathfrak{c}_0^2 - i \mathfrak{c}_3^2  \right)  
    \end{gathered}
\end{equation}
and in complex 
\begin{equation}
    \sigma^{(2)}_{-} e^{i\chi^{(2)}_{-}}(E_x-iE_y)^2 + \sigma^{(2)}_{+} e^{i \chi^{(2)}_{+}} (E_x+iE_y)^2 + \sigma^{(2)}_{0} e^{i \chi^{(2)}_{0}} \left( E_x^2+E_y^2 \right)
    \end{equation}
forms. To understand the relation between the components of the tensor in coordinate representation and coefficients $\sigma^{(2)}_{-}, \sigma^{(2)}_{+}, \sigma^{(2)}_{0}, \chi^{(2)}_{-}, \chi^{(2)}_{+}, \chi^{(2)}_{0}$ of the complex equation it is instructive to consider the simplified case with $\chi^{(2)}_{-} = \chi^{(2)}_{+} = \chi^{(2)}_{0} =0$ first. Then 
\begin{equation}
    \begin{gathered}
        E_x^2 \left( \mathfrak{c}_0^1 + \mathfrak{c}_3^1 + i \mathfrak{c}_0^2 + i \mathfrak{c}_3^2 \right) + 2 E_x E_y \left( \mathfrak{c}_1^1 + i \mathfrak{c}_1^2 \right) + E_y^2 \left( \mathfrak{c}_0^1 - \mathfrak{c}_3^1 + i \mathfrak{c}_0^2 - i \mathfrak{c}_3^2  \right) = \\ =  E_x^2 \left( \sigma^{(2)}_{-} + \sigma^{(2)}_{+} + \sigma^{(2)}_{0} \right) - 2 i E_x E_y \left( \sigma^{(2)}_{-} - \sigma^{(2)}_{+} \right) - E_y^2 \left( \sigma^{(2)}_{-} + \sigma^{(2)}_{+} - \sigma^{(2)}_{0} \right).     
    \end{gathered}
\end{equation}
Since $\sigma^{(2)}_{-}, \sigma^{(2)}_{+}, \sigma^{(2)}_{0}$ are purely real by definition, $\sigma^{(2)}_{-}+ \sigma^{(2)}_{+}+ \sigma^{(2)}_{0}$ and $\sigma^{(2)}_{-}+ \sigma^{(2)}_{+}- \sigma^{(2)}_{0}$ are also purely real. This means that $\mathfrak{c}_0^2 = \mathfrak{c}_3^2 = 0$. At the same time $\mathfrak{c}_1^1 + i \mathfrak{c}_1^2$ has to be purely imaginary, thus $\mathfrak{c}_1^1 = 0$. Therefore, we are left with a simple system of linear equations that fixes relations between the components of the tensor in coordinate and complex forms:
\begin{eqnarray}
    \sigma^{(2)}_{-} + \sigma^{(2)}_{+} + \sigma^{(2)}_{0} = \mathfrak{c}_0^1 + \mathfrak{c}_3^1 \\
    \sigma^{(2)}_{-} + \sigma^{(2)}_{+} - \sigma^{(2)}_{0} = \mathfrak{c}_3^1 - \mathfrak{c}_0^1 \\ 
    \sigma^{(2)}_{-} - \sigma^{(2)}_{+} = -\mathfrak{c}_1^2
\end{eqnarray}
Solving for $\sigma^{(2)}$-coefficients we obtain
\begin{equation}
    \begin{gathered}
        \sigma^{(2)}_{-} = \frac{\mathfrak{c}_3^1 - \mathfrak{c}_1^2}{2}, \\
        \sigma^{(2)}_{+} = \frac{\mathfrak{c}_3^1 + \mathfrak{c}_1^2}{2}, \\
        \sigma^{(2)}_{0} = \mathfrak{c}_0^1.
    \end{gathered}
    \label{Relations_for_c2}
\end{equation}
This relates components of the nonlinear tensor to coefficients $\sigma^{(2)}_{-}, \sigma^{(2)}_{+}, \sigma^{(2)}_{0}$ for $\chi^{(2)}_{-} = \chi^{(2)}_{+} = \chi^{(2)}_{0} =0$.
Let's perform a simple sanity check: consider the $C_3$-symmetric case. There, $\mathfrak{c}_0^1 = 0$, $\mathfrak{c}_1^2 = - \mathfrak{c}_3^1$, hence $\sigma^{(2)}_{+} = \sigma^{(2)}_{0}=0$ and $\sigma^{(2)}_{-} = \mathfrak{c}_3^1$. 

We can also generalize those relations for the case of $\chi^{(2)}_{-}, \chi^{(2)}_{+}, \chi^{(2)}_{0}$ being non-zero by matching the terms between coordinate and complex representations:
\begin{eqnarray}
    \sigma^{(2)}_{-} e^{i \chi^{(2)}_{-}} + \sigma^{(2)}_{+} e^{i \chi^{(2)}_{+}} + \sigma^{(2)}_{0} e^{i \chi^{(2)}_{0}} = \mathfrak{c}_0^1 + \mathfrak{c}_3^1 + i \mathfrak{c}_0^2 + i \mathfrak{c}_3^2 \\
    \sigma^{(2)}_{-} e^{i \chi^{(2)}_{-}} + \sigma^{(2)}_{+} e^{i \chi^{(2)}_{+}} - \sigma^{(2)}_{0} e^{i \chi^{(2)}_{0}} = \mathfrak{c}_3^1 - \mathfrak{c}_0^1 + i \mathfrak{c}_3^2 - i \mathfrak{c}_0^2 \\ 
    \sigma^{(2)}_{-} e^{i \chi^{(2)}_{-}} - \sigma^{(2)}_{+} e^{i \chi^{(2)}_{+}} = -\mathfrak{c}_1^2 + i \mathfrak{c}_1^1.
\end{eqnarray}
We solve for $\sigma^{(2)}$'s and $\chi^{(2)}$'s to obtain the most general relations between the coefficients
\begin{equation}
    \begin{gathered}
        \sigma^{(2)}_{-} = \frac{\sqrt{ \left( \mathfrak{c}_3^1 - \mathfrak{c}_1^2 \right)^2 + \left( \mathfrak{c}_3^2 + \mathfrak{c}_1^1 \right)^2 }}{2}, \;\;\;
        \tan \chi^{(2)}_{-} = \frac{\mathfrak{c}_3^2 + \mathfrak{c}_1^1}{\mathfrak{c}_3^1 - \mathfrak{c}_1^2},        \\
        \sigma^{(2)}_{+} = \frac{ \sqrt{\left( \mathfrak{c}_3^1 + \mathfrak{c}_1^2 \right)^2 + \left( \mathfrak{c}_3^2 - \mathfrak{c}_1^1 \right)^2 }}{2}, \;\;\;
        \tan \chi^{(2)}_{+} = \frac{\mathfrak{c}_3^2 - \mathfrak{c}_1^1}{\mathfrak{c}_3^1 + \mathfrak{c}_1^2}, \\
        \sigma^{(2)}_{0} = \sqrt{ \left( \mathfrak{c}_0^1 \right)^2 + \left( \mathfrak{c}_0^2 \right)^2}, \; \; \;
        \tan \chi^{(2)}_{0} = \frac{\mathfrak{c}_0^2}{\mathfrak{c}_0^1}.
    \end{gathered}
    \label{General_relations_for_c2}
\end{equation}
In the limit of $\chi^{(2)}_{-} = \chi^{(2)}_{+} = \chi^{(2)}_{0}=0$ these equations match Eqs. \eqref{Relations_for_c2}. 

\subsection{Dissipative vs non-dissipative components of conductivity tensor}

In linear response theory Onsager relations uniquely relate time-reversal symmetry breaking (non-reciprocity) and dissipation. It is well known that Hall component of linear conductivity tensor is non-dissipative and requires broken time-reversal symmetry to be present. The rest of the tensor is dissipative. This can be inferred from Joule-Lenz law applied to current density in linear regime. First, we split the overall current into two parts, which we label dissipative and non-dissipative
$$
j = j^{d} + j^{nd}. 
$$
This separation is performed for linear and nonlinear contributions to current separately.
Non-dissipative components of the tensor are those that lead to current not generating any heat \underline{for any orientation of the electric field}, i.e., such that 
\begin{equation}
    j^{nd} \cdot E = 0.
    \label{Non-dissipative_criterion}
\end{equation}
To find the non-dissipative component we calculate the dot product of the full current $j=j^{d} + j^{nd}$ with the electric field and look for the contribution that satisfies Eq. \eqref{Non-dissipative_criterion}.
In linear regime, the only component satisfying this condition is $\mathfrak{k}_2$, which corresponds to Hall conductivity:
\begin{equation}
    \begin{gathered}
        j^{l} \cdot E = j_{\alpha}^{l} E_{\alpha} = \sigma_{\alpha \mu} E_{\mu} E_{\alpha} = E_{\alpha} (\mathfrak{k}_0 \tau_0 + \mathfrak{k}_1 \tau_1 + i \mathfrak{k}_2 \tau_2 +  \mathfrak{k}_3 \tau_3)_{\alpha \mu} E_{\mu} = \\ 
        =\mathfrak{k}_0 \left( E_x^2 + E_y^2 \right) + \mathfrak{k}_3 \left( E_x^2 - E_y^2 \right) + 2 \mathfrak{k}_1 E_x E_y.         
    \end{gathered}
\end{equation}
The same conclusion can be reached using the complex representation with $E_x + i E_y = |E| e^{i \lambda}$:
\begin{equation}
    \begin{gathered}
        j^{l} \cdot E = \mathrm{Re} \left[ (j_x^{l} + i j_y^{l}) (E_x - i E_y) \right] = \mathrm{Re} \left[ \sigma^{(1)}_{+} e^{i \chi^{(1)}_{+}} \left( E_x^2 + E_y^2 \right) + \sigma^{(1)}_{-} e^{i \chi^{(1)}_{-}} \left( E_x - i E_y \right)^2 \right] = \\ 
        = |E|^2 \mathrm{Re} \left[ \sigma^{(1)}_{+} e^{i \chi^{(1)}_{+}} +  \sigma^{(1)}_{-} e^{i \chi^{(1)}_{-}} e^{-2 i \lambda} \right]=0,
    \end{gathered}
\end{equation}
where we request the equality to hold for \underline{any} orientation of the electric field, i.e., for any $\lambda$. Therefore, the only component of the current that satisfies this condition is non-dissipative component
\begin{equation}
    \left(j^{nd}_x + i j^{nd}_y\right)^l = \frac{ \sigma^{(1)}_{+} e^{i \chi^{(1)}_{+}} - \sigma^{(1)}_{+} e^{-i \chi^{(1)}_{+}} }{2} \left( E_x + i E_y \right)
\end{equation}
which is exactly the linear Hall component, and the rest provides the linear dissipative part: 
\begin{equation}
    \left( j^{d}_x + i j^{d}_y \right)^l = \frac{ \sigma^{(1)}_{+} e^{i \chi^{(1)}_{+}} + \sigma^{(1)}_{+} e^{-i \chi^{(1)}_{+}} }{2} \left( E_x + i E_y \right) + \sigma^{(1)}_{-} e^{i \chi^{(1)}_{-}} \left( E_x - i E_y \right).
\end{equation}

For nonlinear component of the current the situation is more complicated. There is no consensus on Onsager relations for nonlinear conductivity tensor; this is a topic of an extensive ongoing research. Therefore, we adopt the Joule-Lenz criterion \eqref{Non-dissipative_criterion} for microscopic current and apply it directly to the nonlinear contribution to current density in complex representation with $E_x + i E_y = |E| e^{i \lambda}$:
\begin{equation}
    \begin{gathered}
        j^{nl} \cdot E = \mathrm{Re} \left[ (j_x^{nl} + i j_y^{nl}) (E_x - i E_y) \right] = \\ = \mathrm{Re} \left[ \sigma^{(2)}_{-} e^{i\chi^{(2)}_{-}}(E_x-iE_y)^3 + \sigma^{(2)}_{+} e^{i \chi^{(2)}_{+}} (E_x+iE_y)(E_x^2+E_y^2) + \sigma^{(2)}_{0} e^{i \chi^{(2)}_{0}} (E_x-iE_y) \left( E_x^2+E_y^2 \right) \right] = \\ = 
        |E|^3 \mathrm{Re} \left[ \sigma^{(2)}_{-} e^{i\chi^{(2)}_{-}} e^{-3 i \lambda} + \sigma^{(2)}_{+} e^{i \chi^{(2)}_{+}} e^{i \lambda} + \sigma^{(2)}_{0} e^{i \chi^{(2)}_{0}} e^{-i \lambda} \right] = 0.
    \end{gathered}
\end{equation}
For the non-dissipative contribution, this condition has to be satisfied for any $\lambda$, hence
\begin{eqnarray}
    \sigma^{(2)}_{-} \cos \left( \chi^{(2)}_{-} - 3 \lambda  \right) + \sigma^{(2)}_{+} \cos \left( \chi^{(2)}_{+} + \lambda \right) + \sigma^{(2)}_{0} \cos \left( \chi^{(2)}_{0} - \lambda \right) = 0.
\end{eqnarray}
Since $\sin n \lambda, \cos m \lambda$ are linearly independent, we obtain that
non-dissipative component of the nonlinear tensor requires
\begin{eqnarray}
    &&\mathrm{Re} \left( \sigma^{(2)}_{-} e^{i \chi^{(2)}_{-}} \right) = \mathrm{Im} \left( \sigma^{(2)}_{-} e^{i \chi^{(2)}_{-}} \right) = 0, \\
    &&\mathrm{Re} \left( \sigma^{(2)}_{+} e^{i \chi^{(2)}_{+}} \right) + \mathrm{Re} \left( \sigma^{(2)}_{0} e^{i \chi^{(2)}_{0}} \right) = 0, \\
    &&\mathrm{Im} \left( \sigma^{(2)}_{+} e^{i \chi^{(2)}_{+}} \right) - \mathrm{Im} \left( \sigma^{(2)}_{0} e^{i \chi^{(2)}_{0}} \right) = 0.
    \label{nondis_conds}
\end{eqnarray}
Thus the full nonlinear current reads
$$
j^{nl} = j^{nl}_d + j^{nl}_{nd},
$$
where the non-dissipative component of the current satisfies
\begin{equation}
    \begin{gathered}
        \left(j_x^{nd} + i j_y^{nd} \right)^{nl} = \frac{\sigma^{(2)}_{+} e^{i \chi^{(2)}_{+}} - \sigma^{(2)}_{0} e^{-i \chi^{(2)}_{0}}}{2} (E_x + i E_y)^2 - \frac{\sigma^{(2)}_{+} e^{-i \chi^{(2)}_{+}} - \sigma^{(2)}_{0} e^{i \chi^{(2)}_{0}}}{2} \left( E_x^2 + E_y^2 \right).
    \end{gathered}
    \label{nonlin_nondiss}
\end{equation}
and the rest of the nonlinear tensor corresponds to the dissipative component of the nonlinear current and is given by
\begin{equation}
    \begin{gathered}
        \left(j^{d}_{x} + i j^{d}_{y}\right)^{nl} = \sigma^{(2)}_{-} e^{i \chi^{(2)}_{-}} (E_x - i E_y)^2 + \frac{\sigma^{(2)}_{+} e^{i \chi^{(2)}_{+}} + \sigma^{(2)}_{0} e^{-i \chi^{(2)}_{0}}}{2} (E_x + i E_y)^2 + \frac{\sigma^{(2)}_{+} e^{-i \chi^{(2)}_{+}} + \sigma^{(2)}_{0} e^{i \chi^{(2)}_{0}}}{2} \left( E_x^2 + E_y^2 \right).
    \end{gathered}
\end{equation}
Note that Eq. \eqref{nonlin_nondiss} can be written as 
\begin{equation}
    \left( \hat{x} + i \hat{y} \right) \cdot \textbf{j}^{nl}_{nd} = \left( \hat{x} + i \hat{y} \right) \cdot \left(\hat{z} \times \textbf{E} \right) \left( \mathcal{C} \cdot \textbf{E} \right),\
    \label{nonlin_nondiss_Liang_form}
\end{equation}
where $\textbf{E}$ -- is the electric field, $\hat{x},\hat{y},\hat{z}$ -- are unit vectors in $x,y$ and $z$ (out-of-plane) directions correspondingly, and 
 a 2-component vector $\mathcal{C}$ is given by 
 \begin{equation}
    \begin{gathered}
    \mathcal{C}_x = \mathrm{Im} \left[ \sigma^{(2)}_{+} e^{i \chi^{(2)}_{+}} - \sigma^{(2)}_{0} e^{-i \chi^{(2)}_{0}}\right], \\
    \mathcal{C}_y = \mathrm{Re} \left[ \sigma^{(2)}_{+} e^{i \chi^{(2)}_{+}} - \sigma^{(2)}_{0} e^{-i \chi^{(2)}_{0}} \right].
    \end{gathered}
\end{equation}
Note that Eq.  \eqref{nonlin_nondiss_Liang_form} becomes exactly the Berry curvature dipole-induced nonlinear Hall effect if $\mathcal{C} = \mathcal{D}$ -- is the Berry curvature dipole vector \cite{Sodemann_Fu}. In our work we do not immediately associate vector $\mathcal{C}$ with Berry curvature dipole, as nonlinear Hall effect in general can be caused by various mechanisms. Nonetheless, since our expression for non-dissipative nonlinear current has the same form as expressions for nonlinear Hall effect, we will use the terms "nonlinear non-dissipative" and "nonlinear Hall" interchangeably.

It is important to mention that 3-fold contribution $\sigma^{(2)}_{-} e^{i \chi^{(2)}_{-}} (E_x - i E_y)^2$ does not contribute to the nonlinear Hall and is different from the 1-fold dissipative contribution: it always contains both transverse and longitudinal contributions and neither of them can be made vanishing throughout the entire sample, see next section. 

\subsection{Parallel component of nonlinear current}

The other condition that can be easily obtained from the complex representation of Ohm's law is the condition on nonlinear component of the current to always be longitudinal to the applied electric field (no transverse component). For microscopic current density and electric field it is given by
\begin{equation}
    j^{nl} \times E = 0,
\end{equation}
which for nonlinear contribution to current in complex representation translates to 
\begin{equation}
    \begin{gathered}
        j^{nl} \times E = -\mathrm{Im} \left[ (j_x^{nl} + i j_y^{nl}) (E_x - i E_y) \right] = -|E|^3 \mathrm{Im} \left[ \sigma^{(2)}_{-} e^{i\chi^{(2)}_{-}} e^{-3 i \lambda} + \sigma^{(2)}_{+} e^{i \chi^{(2)}_{+}} e^{i \lambda} + \sigma^{(2)}_{0} e^{i \chi^{(2)}_{0}} e^{-i \lambda} \right] = 0.
    \end{gathered}
\end{equation}
Then for any $\lambda$ 
\begin{eqnarray}
    \sigma^{(2)}_{-} \sin \left( \chi^{(2)}_{-} - 3 \lambda  \right) + \sigma^{(2)}_{+} \sin \left( \chi^{(2)}_{+} + \lambda \right) + \sigma^{(2)}_{0} \sin \left( \chi^{(2)}_{0} - \lambda \right) = 0.
\end{eqnarray}
Again, $\sin n \lambda, \cos m \lambda$ are linearly independent, therefore,
the condition is satisfied for any $\lambda$ if
\begin{eqnarray}
    &&\mathrm{Re} \left( \sigma^{(2)}_{-} e^{i \chi^{(2)}_{-}} \right) = \mathrm{Im} \left( \sigma^{(2)}_{-} e^{i \chi^{(2)}_{-}} \right) = 0, \\
    &&\mathrm{Re} \left( \sigma^{(2)}_{+} e^{i \chi^{(2)}_{+}} \right) - \mathrm{Re} \left( \sigma^{(2)}_{0} e^{i \chi^{(2)}_{0}} \right) = 0, \\
    &&\mathrm{Im} \left( \sigma^{(2)}_{+} e^{i \chi^{(2)}_{+}} \right) + \mathrm{Im} \left( \sigma^{(2)}_{0} e^{i \chi^{(2)}_{0}} \right) = 0.
\end{eqnarray}
As we see from those conditions, the 3-fold component cannot be made purely longitudinal as it always will have transverse components in addition to longitudinal. 

The 1-fold (longitudinal) component of dissipative nonlinear current can also be cast into the coordinate-free way similar to Eq.  \eqref{nonlin_nondiss_Liang_form} so that the full non-dissipative nonlinear current satisfies
\begin{equation}
    \left( \hat{x} + i \hat{y} \right) \cdot \textbf{j}^{nl}_{d} = \sigma^{(2)}_{-} e^{i \chi^{(2)}_{-}} (E_x - i E_y)^2 +  \left( \hat{x} + i \hat{y} \right) \cdot \textbf{E} \left( \mathcal{B} \cdot \textbf{E} \right)
\end{equation}
upon the introduction of an auxiliary vector $\mathcal{B}$, components of which are given by
\begin{equation}
    \begin{gathered}
        \mathcal{B}_x = \mathrm{Re} \left[ \sigma^{(2)}_{+} e^{-i \chi^{(2)}_{+}} + \sigma^{(2)}_{0} e^{i \chi^{(2)}_{0}} \right], \\
        \mathcal{B}_y = \mathrm{Im} \left[ \sigma^{(2)}_{+} e^{-i \chi^{(2)}_{+}} + \sigma^{(2)}_{0} e^{i \chi^{(2)}_{0}} \right].
    \end{gathered}
\end{equation}
The relative orientation of $\mathcal{C}$ and $\mathcal{B}$ vectors can be inferred from their dot product
\begin{equation}
    \mathcal{B} \cdot \mathcal{C} = 2 \sigma^{(2)}_{+} \sigma^{(2)}_{0} \sin \left( \chi^{(2)}_{+} + \chi^{(2)}_{0} \right).
\end{equation}
In the presence of mirror axis $\chi^{(2)}_{+} = - \chi^{(2)}_{0}$ the dot product vanishes indicating that the two vectors are orthogonal. In the absence of mirror axis the two vectors can have arbitrary orientation.

\subsection{Relations between the $\mathcal{A}, \mathcal{B}, \mathcal{C}$ vectors and components of the $\tilde{\sigma}$ in the coordinate representation}

Using Eqs. \eqref{Ohm_in_coordinate} and \eqref{Ohm_in_complex} we can directly relate components of nonlinear conductivity tensor in coordinate representation to $\Xi^{(2)}_{-,0,+}$:
\begin{equation}
    \begin{aligned}
        &\tilde{\sigma}_{xxx} = \mathrm{Re} \Xi^{(2)}_{-} + \mathrm{Re} \Xi^{(2)}_{+} + \mathrm{Re} \Xi^{(2)}_{0} = \sigma^{(2)}_{-} \cos \chi^{(2)}_{-}  + \sigma^{(2)}_{+} \cos \chi^{(2)}_{+} + \sigma^{(2)}_{0} \cos \chi^{(2)}_{0}, \\
        &\tilde{\sigma}_{xxy} = \mathrm{Im} \Xi^{(2)}_{-} - \mathrm{Im} \Xi^{(2)}_{+} = \sigma^{(2)}_{-} \sin \chi^{(2)}_{-}  - \sigma^{(2)}_{+} \sin \chi^{(2)}_{+} ,\\
        &\tilde{\sigma}_{xyy} = -\mathrm{Re} \Xi^{(2)}_{-} - \mathrm{Re} \Xi^{(2)}_{+} + \mathrm{Re} \Xi^{(2)}_{0} = -\sigma^{(2)}_{-} \cos \chi^{(2)}_{-}  - \sigma^{(2)}_{+} \cos \chi^{(2)}_{+} + \sigma^{(2)}_{0} \cos \chi^{(2)}_{0}, \\
        &\tilde{\sigma}_{yxx} = \mathrm{Im} \Xi^{(2)}_{-} + \mathrm{Im} \Xi^{(2)}_{+} + \mathrm{Im} \Xi^{(2)}_{0} = \sigma^{(2)}_{-} \sin \chi^{(2)}_{-}  + \sigma^{(2)}_{+} \sin \chi^{(2)}_{+} + \sigma^{(2)}_{0} \sin \chi^{(2)}_{0}, \\
        &\tilde{\sigma}_{yxy} = -\mathrm{Re} \Xi^{(2)}_{-} + \mathrm{Re} \Xi^{(2)}_{+} = -\sigma^{(2)}_{-} \cos \chi^{(2)}_{-}  + \sigma^{(2)}_{+} \cos \chi^{(2)}_{+} ,\\
        &\tilde{\sigma}_{yyy} = -\mathrm{Im} \Xi^{(2)}_{-} - \mathrm{Im} \Xi^{(2)}_{+} + \mathrm{Im} \Xi^{(2)}_{0} = -\sigma^{(2)}_{-} \sin \chi^{(2)}_{-}  - \sigma^{(2)}_{+} \sin \chi^{(2)}_{+} + \sigma^{(2)}_{0} \sin \chi^{(2)}_{0}.
    \end{aligned}
    \label{Complex_Carthesian_1}
\end{equation}
Introducing vector representation for the $\Xi^{(2)}_-$ term as  $$\mathcal{A}_x + i \mathcal{A}_y = \sqrt[3]{\Xi^{(2)}_-}$$ the 3-fold component of the current
$$
(j_x + i j_y)_{\rightY} = \sigma^{(2)}_{-} e^{i\chi^{(2)}_{-}}(E_x-iE_y)^2
$$
can be cast into the coordinate-free form
\begin{equation}
    \mathbf{j}^{\text{nl}}_{\rightY} = \mathcal{A} \left( \left( \mathcal{A}\cdot \mathbf{E}\right)^2 - \left( \mathcal{A} \times \mathbf{E} \right)^2 \right) + 2 \mathcal{A} \times \left(\mathcal{A}\times \mathbf{E}\right) \left(\mathcal{A} \cdot \mathbf{E} \right).
\end{equation}
Carthesian and vector representations can be linked using the definitions of $\mathcal{A}, \mathcal{B}, \mathcal{C}$ and Eq. \eqref{Complex_Carthesian_1}:
\begin{equation}
    \begin{aligned}
        &\tilde{\sigma}_{xxx} = \mathcal{A}_x^3 - 3 \mathcal{A}_x \mathcal{A}_y^2 + \mathcal{B}_x, \\
        &\tilde{\sigma}_{xxy} = 3 \mathcal{A}_x^2 \mathcal{A}_y - \mathcal{A}_y^3 + \frac{\mathcal{B}_y - \mathcal{C}_x}{2},\\
        &\tilde{\sigma}_{xyy} = - \left( \mathcal{A}_x^3 - 3 \mathcal{A}_x \mathcal{A}_y^2 \right) - \mathcal{C}_y, \\
        &\tilde{\sigma}_{yxx} = 3 \mathcal{A}_x^2 \mathcal{A}_y - \mathcal{A}_y^3 + \mathcal{C}_x, \\
        &\tilde{\sigma}_{yxy} = - \left( \mathcal{A}_x^3 - 3 \mathcal{A}_x \mathcal{A}_y^2 \right) + \frac{\mathcal{B}_x + \mathcal{C}_y}{2},\\
        &\tilde{\sigma}_{yyy} = -\left( 3 \mathcal{A}_x^2 \mathcal{A}_y - \mathcal{A}_y^3 \right) + \mathcal{B}_y.
    \end{aligned}
\end{equation}

\subsection{Symmetry properties of nonlinear conductivity tensor}

Nonlinear conductivity tensor has a higher rank than  linear conductivity tensor, therefore, its transformation properties are less obvious. In this section we discuss symmetry constraints on the components of nonlinear conductivity tensor.

The most fundamental symmetry constraint on the nonlinear conductivity tensor is the absence of inversion symmetry. The absence of inversion is evident from Eq. \eqref{Ohm_in_coordinate}: in order to have nonzero linear response the nonlinear tensor $\tilde{\sigma}_{\alpha \mu \nu}$ has to be odd under inversion to be nonzero because both $j$ and $E$ vectors are odd under inversion.
In this regard nonlinear conductivity tensor is very similar to the optical second-order nonlinear susceptibility, which is responsible for generation of second optical harmonic in noncentrosymmetric materials or from the surface of centrosymmetric materials.

We start the discussion of point-group symmetries by considering the effects of rotation. We focus solely on nonlinear contribution to current density and, for convenience, resort to the complex representation \eqref{Ohm_in_complex}. In complex representation rotation of vectors by angle $\alpha$ is performed by a simple multiplication by a phase factor:
\begin{equation}
    v_x \pm i v_y \rightarrow e^{\pm i \alpha} ( v_x \pm i v_y ).
\end{equation}
Then under rotation nonlinear component of Eq. \eqref{Ohm_in_complex} transforms as
\begin{equation}
    e^{i \alpha} ( j_x + i j_y )^{nl} = \sigma^{(2)}_{-} e^{i\chi^{(2)}_{-}} e^{-2 i \alpha} (E_x-iE_y)^2 + \sigma^{(2)}_{+} e^{2 i \alpha} e^{i \chi^{(2)}_{+}} (E_x+iE_y)^2 + \sigma^{(2)}_{0} e^{i \chi^{(2)}_{0}} \left( E_x^2+E_y^2 \right).
\end{equation}
Dividing both sides of the equation by $e^{i \alpha}$ we arrive at 
\begin{equation}
     j_x^{nl} + i j_y^{nl}  = \sigma^{(2)}_{-} e^{i\chi^{(2)}_{-}} e^{-3 i \alpha} (E_x-iE_y)^2 + \sigma^{(2)}_{+} e^{i \chi^{(2)}_{+}} e^{i \alpha} (E_x+iE_y)^2 + \sigma^{(2)}_{0} e^{i \chi^{(2)}_{0}} e^{-i \alpha} \left( E_x^2+E_y^2 \right).
\end{equation}
Rotational symmetry requires the prefactors containing $\alpha$ to vanish. The first term ($\sigma^{(2)}_{-}$-term), therefore, is invariant upon $C_3$ rotation, as $e^{- 3 i \alpha} = 1$ for $\alpha = \pm \frac{2\pi}{3}$. The other two terms, $\sigma^{(2)}_{+}$ and $\sigma^{(2)}_{0}$ are only invariant upon the full $2 \pi$ rotation. All three terms break $C_{2z}$ (in-plane inversion) symmetry as they change sign for $\alpha = \pi$. 

Another important point-group symmetry is mirror. In 2D there are two generic mirror symmetries, $M_x$ and $M_y$, that act as 
\newtext{
\begin{equation}
    \begin{gathered}
        M_x:  x \rightarrow -x, \; \; \; y \rightarrow  y, \\
        M_y: x \rightarrow x, \; \; \; y \rightarrow - y.
    \end{gathered}
\end{equation}
}
Under \newtext{$M_y$} all complex vectors in Eq. \eqref{Ohm_in_complex} experience complex conjugation:
\begin{equation}
     j_x^{nl} - i j_y^{nl}  = \sigma^{(2)}_{-} e^{i\chi^{(2)}_{-}} (E_x + iE_y)^2 + \sigma^{(2)}_{+} e^{i \chi^{(2)}_{+}} (E_x - iE_y)^2 + \sigma^{(2)}_{0} e^{i \chi^{(2)}_{0}} \left( E_x^2+E_y^2 \right).
\end{equation}
Applying complex conjugation to the whole expression we obtain
\begin{equation}
     j_x^{nl} + i j_y^{nl}  = \sigma^{(2)}_{-} e^{-i\chi^{(2)}_{-}} (E_x - iE_y)^2 + \sigma^{(2)}_{+} e^{-i \chi^{(2)}_{+}} (E_x + iE_y)^2 + \sigma^{(2)}_{0} e^{-i \chi^{(2)}_{0}} \left( E_x^2+E_y^2 \right).
\end{equation}
Therefore, the presence of \newtext{$M_y$} symmetry requires all phase factors $\chi^{(2)}_{-}, \chi^{(2)}_{+}, \chi^{(2)}_{0}$ to be $0$ or $\pi$. Under \newtext{$M_x$} Eq. \eqref{Ohm_in_complex} becomes 
\begin{equation}
     - j_x^{nl} + i j_y^{nl}  = \sigma^{(2)}_{-} e^{i\chi^{(2)}_{-}} (E_x + iE_y)^2 + \sigma^{(2)}_{+} e^{i \chi^{(2)}_{+}} (E_x - iE_y)^2 + \sigma^{(2)}_{0} e^{i \chi^{(2)}_{0}} \left( E_x^2+E_y^2 \right).
\end{equation}
Applying complex conjugation to the whole equation and multiplying both sides by -1 we arrive at
\begin{equation}
     j_x^{nl} + i j_y^{nl}  = -\sigma^{(2)}_{-} e^{-i\chi^{(2)}_{-}} (E_x - iE_y)^2 - \sigma^{(2)}_{+} e^{-i \chi^{(2)}_{+}} (E_x + iE_y)^2 - \sigma^{(2)}_{0} e^{-i \chi^{(2)}_{0}} \left( E_x^2+E_y^2 \right).
\end{equation}
Therefore, to preserve invariance one has to satisfy $-e^{-i\chi^{(2)}_{-}} =e^{i\chi^{(2)}_{-}};  -e^{-i\chi^{(2)}_{+}} = e^{i\chi^{(2)}_{+}}; -e^{-i\chi^{(2)}_{0}} = e^{i\chi^{(2)}_{0}}$, which leads to all phase factors $\chi^{(2)}_{-}, \chi^{(2)}_{+}, \chi^{(2)}_{0}$ being $\pm \pi/2$, which is expected as $x$ and $y$ axes are orthogonal.

For linear current, enforcing mirror symmetry will make Hall component vanishing. In the nonlinear case, there are two independent parameters that correspond to Hall response. Having a mirror axis reduces the number of independent parameters from 2 to 1, hence, nonlinear Hall effect is not prohibited by mirror symmetry.

\section{Nonlinear transport in disk geometry as an electrostatics problem}

In this section we set up the electrostatics problem which describes the current flow inside the sample in disk geometry. 

We consider a sample made of conducting material in the shape of a disk with source and drain attached to two arbitrary points on the boundary of the disk. We further assume that both linear and nonlinear conductivity are uniform and that the sample is in diffusive regime. These are the only two fundamental assumptions of our approach. We consider the case of source and drain being point-like. Generalization to contacts being of finite size is done by integrating over a distribution of point-like sources. We start by considering the nonlinear Ohm's law Eq. \eqref{Ohm_in_complex} 
\begin{equation}
    j_x + i j_y = \sigma^{(1)}_{+} e^{i \chi^{(1)}_{+}} \left( E_x + i E_y \right) + \sigma^{(1)}_{-} e^{i \chi^{(1)}_{-}} \left( E_x - i E_y \right) + \sigma^{(2)}_{-} e^{i\chi^{(2)}_{-}}(E_x-iE_y)^2 + \sigma^{(2)}_{+} e^{i \chi^{(2)}_{+}} (E_x+iE_y)^2 + \sigma^{(2)}_{0} e^{i \chi^{(2)}_{0}} \left( E_x^2+E_y^2 \right),
\end{equation}
and the continuity equation for electrical current:
\begin{equation}
    \nabla \cdot j = I \left( \delta (\br - \br_S) - \delta (\br - \br_D) \right).\,
    \label{Continuity_equation}
\end{equation}
where $\br_{S,D}$ -- are positions of point-like source and drain correspondingly, and $I$ -- is the integrated current density injected into the sample. We are looking for a potential solution such that the electric field is given by a gradient of the electrostatic potential
\begin{equation}
    E (\br)=-\nabla_{\br} \Phi(\br).
\end{equation}
For arbitrary magnitude of the injected current this nonlinear problem is intractable due to severe nonlinearities appearing in the nonlinear Ohm's law equation. For small magnitudes of injected current, however, this problem becomes treatable perturbatively. Consider the power-law expansion of current density in powers of the injected current magnitude $I$:
\begin{equation}
    j_{\alpha} = \sum_{n=1}^{\infty} \newtext{j^{(n)}_{\alpha}},
\end{equation}
where $\newtext{j^{(n)}_{\alpha}} \propto I^n$. We can then look for solutions to the nonlinear problem order by order in $I$. At the first order in $I$ the Ohm's law becomes linear
\begin{equation}
     j_x^{\newtext{(1)}} + i j_y^{\newtext{(1)}} = \sigma^{(1)}_{+} e^{i \chi^{(1)}_{+}} \left( E_x^{\newtext{(1)}} + i E_y^{\newtext{(1)}} \right) + \sigma^{(1)}_{-} e^{i \chi^{(1)}_{-}} \left( E_x^{\newtext{(1)}} - i E_y^{\newtext{(1)}} \right)
\end{equation}
and the continuity equation contains only $j^{\newtext{(1)}}$:
\begin{equation}
    \nabla \cdot j^{\newtext{(1)}} = I \left( \delta (\br - \br_S) - \delta (\br - \br_D) \right).
\end{equation}
We make use of the linear electrostatic potential definition
\begin{equation}
    E^{\newtext{(1)}} (\br)=-\nabla_{\br} \Phi^{\newtext{(1)}}(\br),
\end{equation} 
and substitute it into the linear Ohm's law to express the 0th-order current density via the gradient of the 0th-order electrostatic field
\begin{equation}
     j_x^{\newtext{(1)}} + i j_y^{\newtext{(1)}} = -\sigma^{(1)}_{+} e^{i \chi^{(1)}_{+}} \left( \nabla_{x} \Phi^{\newtext{(1)}}(\br) + i \nabla_{y} \Phi^{\newtext{(1)}}(\br) \right) - \sigma^{(1)}_{-} e^{i \chi^{(1)}_{-}} \left( \nabla_{x} \Phi^{\newtext{(1)}}(\br) - i \nabla_{y} \Phi^{\newtext{(1)}}(\br) \right).
\end{equation}
We now apply divergence operator to both sides of the expression above to combine its right-hand-side with the right-hand-side of the continuity equation. The result of such operation is a second-order partial differential equation
\begin{equation}
    -\sigma^{(1)}_{+} e^{i \chi^{(1)}_{+}} \left( \nabla^2_{x} \Phi^{\newtext{(1)}}(\br) + i \nabla^2_{y} \Phi^{\newtext{(1)}}(\br) \right) - \sigma^{(1)}_{-} e^{i \chi^{(1)}_{-}} \left( \nabla^2_{x} \Phi^{\newtext{(1)}}(\br) - i \nabla^2_{y} \Phi^{\newtext{(1)}}(\br) \right) = 
    I \left( \delta (\br - \br_S) - \delta (\br - \br_D) \right),
\end{equation}
which in the absence of linear Hall conductivity and for principal axes being aligned with the reference frame can be cast into the convenient form \cite{Oskar_solo}
\begin{equation}
   - \left( \bar{\sigma} + \Delta \sigma \right)  \nabla^2_{x} \Phi^{\newtext{(1)}}(\br) - \left( \bar{\sigma} - \Delta \sigma \right) \nabla^2_{y} \Phi^{\newtext{(1)}}(\br) = I \left( \delta (\br - \br_S) - \delta (\br - \br_D) \right),
\end{equation}
where 
$$
\bar{\sigma} + \Delta \sigma = \sigma^{(1)}_{+} + \sigma^{(1)}_{-}, \;\;\;
\bar{\sigma} - \Delta \sigma = \sigma^{(1)}_{+} - \sigma^{(1)}_{-}.
$$
The boundary condition accompanying this differential equation is based on the constraint that current cannot leave the sample anywhere but at the drain. This requirement forces the boundary condition to be Neumann. We present the explicit form of boundary conditions in the following subsections. 

We now are seeking $j^{\newtext{(2)}}$, the first correction to the current density due to nonlinear dependence on electric field. This correction stems from having nonlinearity in \eqref{Ohm_in_coordinate} (or in \eqref{Ohm_in_complex}) due to the nonvanishing nonlinear conductivity tensor. To find $j^{\newtext{(2)}}$  we need to consider terms of the order $I^2$ in Ohm's law \eqref{Ohm_in_coordinate}. At this order Eq. \eqref{Ohm_in_coordinate} reads
\begin{equation}
    j^{\newtext{(2)}}_{\alpha} = \sigma_{\alpha \mu} E_{\mu}^{\newtext{(2)}} + \tilde{\sigma}_{\alpha \mu \nu} E_{\mu}^{\newtext{(1)}} E_{\nu}^{\newtext{(1)}},
\end{equation}
where $E_{\mu}^{\newtext{(1)}}$ -- is the electric field generated by the electrostatic potential in the linear case (0th-order term, linear in $I$) and $E_{\mu}^{\newtext{(2)}}$ -- is the electric field generated by the correction of the order $I^2$ to the electrostatic potential due to nonlinearity. The correction to current density $j^{\newtext{(2)}}$ satisfies a trivial continuity equation
\begin{equation}
    \nabla \cdot j^{\newtext{(2)}} = 0
\end{equation}
as there are no sources at the boundary which are quadratic in $I$. From the physical perspective this corresponds to the fact that nonlinear current is generated within the system itself due to the nonlinearity and not by external sources. In other words, in the perturbative regime the source for corrections to the linear regime at order $n$ stems from the solution at order $n-1$. Because the divergence of $j^{\newtext{(2)}}$ vanishes and because inside the disk the curl of $E^{\newtext{(2)}}$ must vanish (due to path independence of the work done by a unit charge between any two points inside), we can express 
$$
E^{\newtext{(2)}} = - \nabla_{\br} \Phi^{\newtext{(2)}} (\br).
$$
Combining the last two equations together with the definition of nonlinear correction to electrostatic potential $E^{\newtext{(2)}}$
we arrive at the second-order partial differential equation for $\Phi^{\newtext{(2)}}$:
\begin{eqnarray}
&&j^{\newtext{(2)}}=\sigma\cdot E^{\newtext{(2)}}+\mathcal{E}^{(2)}=-\sigma\cdot \nabla\Phi^{\newtext{(2)}}+\mathcal{E}^{(2)}, \\
&&\nabla\cdot j^{\newtext{(2)}}=0\;\Rightarrow\; \nabla\cdot\sigma\cdot \nabla\Phi^{\newtext{(2)}}=\nabla\cdot\mathcal{E}^{(2)} \Rightarrow\ \\
&&\bar{\sigma}\left(\left(1+\frac{\Delta\sigma}{\bar{\sigma}}\right)\frac{\partial^2}{\partial x^2}+
\left(1-\frac{\Delta\sigma}{\bar{\sigma}}\right)\frac{\partial^2}{\partial y^2}
\right)\Phi^{\newtext{(2)}}(x,y)=\frac{\partial \mathcal{E}^{(2)}_x(x,y)}{\partial x}+\frac{\partial \mathcal{E}^{(2)}_y(x,y)}{\partial y},
\label{eq:Lap2}
\end{eqnarray}
where 
\begin{equation}
    \mathcal{E}^{(2)} (\br) = \tilde{\sigma}_{\alpha \mu \nu} E_{\mu}^{\newtext{(1)}} (\br)  E_{\nu}^{\newtext{(1)}} (\br)
\end{equation}
-- is the source field for nonlinear potential stemming from the electric fields generated in the linear regime.
Note that for $\Delta \sigma=0$ Eq. \eqref{eq:Lap2} becomes the Poisson equation. For $\Delta \sigma\neq0$ one can perform a transormation of variables to cast \eqref{eq:Lap2} into the Poisson form. The boundary condition is again of Neumann type.

Below we separately discuss the problem setup for the case of isotropic linear conductivity and anisotropic linear conductivity. In both cases we consider linear Hall component of the tensor to be negligible. 

\subsection{The case of isotropic linear conductivity tensor}

In the case of isotropic linear conductivity tensor 
$$
\sigma^{(1)}_{+} = \sigma, \; \; \; \sigma^{(1)}_{-} = \chi^{(1)}_{+} = \chi^{(1)}_{-} = 0,
$$
For notational simplicity, we define $\mathcal{E}^{(2)}$ by writing
\begin{eqnarray}
&&j^{\newtext{(2)}}=\sigma E^{\newtext{(2)}}+\mathcal{E}^{(2)}\\
\text{where}\;\;\;&&\mathcal{E}_x^{(2)}+i\mathcal{E}_y^{(2)}=\sigma^{(2)}_{-} e^{i\chi^{(2)}_{-}}\left( E_x^{\newtext{(1)}}-iE_y^{\newtext{(1)}} \right)^2 + \sigma^{(2)}_{+} e^{i \chi^{(2)}_{+}} \left( E_x^{\newtext{(1)}}+iE_y^{\newtext{(1)}} \right)^2 + \sigma^{(2)}_{0} e^{i \chi^{(2)}_{0}} \left[ \left(E_x^{\newtext{(1)}}\right)^2 + \left(E_y^{\newtext{(1)}}\right)^2 \right].
\end{eqnarray}
Therefore, the differential equations for both $\Phi^{\newtext{(1)}} (\br)$ and $\Phi^{\newtext{(2)}} (\br)$ are of Poisson type:
\begin{eqnarray}
    &&- \nabla^2_{\br} \Phi^{\newtext{(1)}} (\br) = \frac{I}{\sigma} \left( \delta (\br - \br_S) - \delta (\br - \br_D) \right), \\
    &&\nabla^2_{\br} \Phi^{\newtext{(2)}} (\br) = \frac{1}{\sigma} \nabla_{\br} \cdot \mathcal{E}^{(2)} (\br),
    \label{eqn:poisson}
\end{eqnarray}
with current being tangential to the boundary
\begin{eqnarray}
&&\hat{n}\cdot j^{\newtext{(1)}}=0\\
\Rightarrow&& \hat{n}\cdot \nabla\Phi^{\newtext{(1)}}|_{boundary}=0; \\
&&\hat{n}\cdot j^{\newtext{(2)}}=\sigma \hat{n}\cdot E^{\newtext{(2)}}+\hat{n}\cdot\mathcal{E}^{(2)} = 0\\
\Rightarrow&& \hat{n}\cdot \nabla\Phi^{\newtext{(2)}}|_{boundary}=\frac{1}{\sigma}\hat{n}\cdot\mathcal{E}^{(2)}|_{boundary},
\label{eqn:boundary condition}\end{eqnarray}
which translates to Neumann boundary conditions.

\subsection{The case of anisotropic linear conductivity tensor}

Consider the first order nonlinear correction to the current density:
\begin{eqnarray}
j^{\newtext{(2)}}=\sigma\cdot E^{\newtext{(2)}}+\mathcal{E}^{(2)}=-\sigma\cdot \nabla_{\br}\Phi^{\newtext{(2)}}(\br)+\mathcal{E}^{(2)},
\end{eqnarray}
where $\sigma$ is in general anisotropic linear conductivity tensor with vanishing Hall component.
Then, since all the current at the source and drain ($I$) has been accounted for in the 
continuity equation for
the linear term $j^{\newtext{(1)}}$, we must have
\begin{eqnarray}
\nabla\cdot j^{\newtext{(2)}}=0\;\Rightarrow\; \nabla\cdot\sigma\cdot \nabla\Phi^{\newtext{(2)}} (\br)=\nabla\cdot\mathcal{E}^{(2)}\label{eq:Lap}.
\end{eqnarray}
In addition, we must have no current along the direction normal to the circle at its boundary (because current cannot escape the sample):
\begin{eqnarray}
\left.\hat{n}\cdot j^{\newtext{(2)}}\right|_{\text{circ.}}=0\;\Rightarrow\; \left.\hat{n}\cdot\sigma\cdot \nabla\Phi^{\newtext{(2)}} (\br)
\right|_{\text{circ.}}=\left.\hat{n}\cdot\mathcal{E}^{(2)}\right|_{\text{circ.}}\label{eq:normal}.
\end{eqnarray}

Now, let us assume that our coordinate system has been set up along the principal axes of the conductivity tensor $\sigma$, i.e., without loss of generality
$$
\sigma^{(1)}_{+} = \bar{\sigma}, \; \; \; \sigma^{(1)}_{-} = \Delta \sigma,
$$ 
and we consider $\Delta \sigma <0$. 
Then, Eq. (\ref{eq:Lap}) reads:
\begin{eqnarray}
\bar{\sigma}\left(\left(1+\frac{\Delta\sigma}{\bar{\sigma}}\right)\frac{\partial^2}{\partial x^2}+
\left(1-\frac{\Delta\sigma}{\bar{\sigma}}\right)\frac{\partial^2}{\partial y^2}
\right)\Phi^{\newtext{(2)}} (x,y)=\frac{\partial \mathcal{E}^{(2)}_x(x,y)}{\partial x}+\frac{\partial \mathcal{E}^{(2)}_y(x,y)}{\partial y}.
\label{eq:Lap2an}
\end{eqnarray}
Now, let us define
\begin{eqnarray}
X&=&\frac{x}{\sqrt{1+\frac{\Delta\sigma}{\bar{\sigma}}}}\\
Y&=&\frac{y}{\sqrt{1-\frac{\Delta\sigma}{\bar{\sigma}}}}\\
\phi^{\newtext{(2)}} (X,Y)&=&\Phi^{\newtext{(2)}} (x,y)\\
\tilde{\mathcal{E}}^{(2)}_X(X,Y)&=&\frac{1}{\sqrt{1+\frac{\Delta\sigma}{\bar{\sigma}}}}\mathcal{E}^{(2)}_x(x,y)\\
\tilde{\mathcal{E}}^{(2)}_Y(X,Y)&=&\frac{1}{\sqrt{1-\frac{\Delta\sigma}{\bar{\sigma}}}}\mathcal{E}^{(2)}_y(x,y)
\end{eqnarray}
Then, Eq. (\ref{eq:Lap2an}) becomes
\begin{eqnarray}
\bar{\sigma}\left(\frac{\partial^2}{\partial X^2}+
\frac{\partial^2}{\partial Y^2}
\right)\phi^{\newtext{(2)}} (X,Y)=\frac{\partial \tilde{\mathcal{E}}^{(2)}_X(X,Y)}{\partial X}+\frac{\partial \tilde{\mathcal{E}}^{(2)}_Y(X,Y)}{\partial Y}.
\end{eqnarray}
So in the $X,Y$ coordinates, $\bR$, the above has the form of Poisson equation, with a Laplacian on the left hand side and a divergence of a vector
field on the right hand side:
\begin{eqnarray}
\nabla^2_{\bR}\phi^{\newtext{(2)}} (\bR) =\frac{1}{\bar{\sigma}}\nabla_{\bR}\cdot \tilde{\mathcal{E}}^{(2)}.
\end{eqnarray}
The right hand side is assumed known through the \newtext{1st}-order solution, when it is found. Notice, that in new coordinates $\bR$ the initial area (disk) becomes an ellipse.

Now, the boundary condition in Eq. (\ref{eq:normal}) can be written as
\begin{eqnarray}
&&\left.\bar{\sigma}\left(\left(1+\frac{\Delta\sigma}{\bar{\sigma}}\right)x\frac{\partial \Phi^{\newtext{(2)}} (x,y)}{\partial x}+
\left(1-\frac{\Delta\sigma}{\bar{\sigma}}\right)y\frac{\partial \Phi^{\newtext{(2)}} (x,y)}{\partial
y}\right)\right|_{x^2+y^2=1}=\left.\left(x\mathcal{E}^{(2)}_x(x,y)+y\mathcal{E}^{(2)}_y(x,y)\right)\right|_{x^2+y^2=1}\\
&&\left.\bar{\sigma}\left(\left(1+\frac{\Delta\sigma}{\bar{\sigma}}\right)X\frac{\partial \phi^{\newtext{(2)}} (X,Y)}{\partial X}+
\left(1-\frac{\Delta\sigma}{\bar{\sigma}}\right)Y\frac{\partial \phi^{\newtext{(2)}} (X,Y)}{\partial
Y}\right)\right|_{\left(1+\frac{\Delta\sigma}{\bar{\sigma}}\right)X^2+\left(1-\frac{\Delta\sigma}{\bar{\sigma}}\right)Y^2=1}=\\
&&\left.\left(\left(1+\frac{\Delta\sigma}{\bar{\sigma}}\right)X\tilde{\mathcal{E}}^{(2)}_X(X,Y)+
\left(1-\frac{\Delta\sigma}{\bar{\sigma}}\right)Y\tilde{\mathcal{E}}^{(2)}_Y(X,Y)\right)
\right|_{\left(1+\frac{\Delta\sigma}{\bar{\sigma}}\right)X^2+\left(1-\frac{\Delta\sigma}{\bar{\sigma}}\right)Y^2=1}
\end{eqnarray}
The equation above has the form of
\begin{eqnarray}
\bN\cdot \nabla_\bR\phi^{\newtext{(2)}}=\bN\cdot\frac{1}{\bar{\sigma}}\tilde{\mathcal{E}}^{(2)}\;\;\text{evaluated at}\;\;
\left(1+\frac{\Delta\sigma}{\bar{\sigma}}\right)X^2+\left(1-\frac{\Delta\sigma}{\bar{\sigma}}\right)Y^2=1,
\end{eqnarray}
where the vector $\bN$ can be expressed as
\begin{eqnarray}
(N_X,N_Y)=\left(\left(1+\frac{\Delta\sigma}{\bar{\sigma}}\right)X, \left(1-\frac{\Delta\sigma}{\bar{\sigma}}\right)Y \right)
\;\;\text{evaluated at}\;\; \left(1+\frac{\Delta\sigma}{\bar{\sigma}}\right)X^2+\left(1-\frac{\Delta\sigma}{\bar{\sigma}}\right)Y^2=1.
\end{eqnarray}
This vector is parallel to the normal of the elliptical boundary in the $X,Y$ plane. To see this, consider contour lines of the function
$f(X,Y)=\left(1+\frac{\Delta\sigma}{\bar{\sigma}}\right)X^2+\left(1-\frac{\Delta\sigma}{\bar{\sigma}}\right)Y^2$.
Then, $\nabla_\bR f(X,Y)$ evaluated at $f(X,Y)=1$ must be normal to its level curve at $1$, which is our ellipse.
But, $\left.\nabla_\bR
f(X,Y)\right|_{f=1}=\left.2\left(\left(1+\frac{\Delta\sigma}{\bar{\sigma}}\right)X,\left(\left(1-\frac{\Delta\sigma}{\bar{\sigma}}\right)Y\right)\right)\right|_{f=1}$
which is what we have above up to an irrelevant scaling factor of $2$.
Therefore, scaling both sides by the inverse length of $\bN$, we can write
\begin{eqnarray}
\hat{\bN}\cdot \nabla_\bR\phi^{\newtext{(2)}} (\bR)=\hat{\bN}\cdot\frac{1}{\bar{\sigma}}\tilde{\mathcal{E}}^{(2)}\;\;\text{evaluated at}\;\;
\left(1+\frac{\Delta\sigma}{\bar{\sigma}}\right)X^2+\left(1-\frac{\Delta\sigma}{\bar{\sigma}}\right)Y^2=1.
\end{eqnarray}
This is the Neumann boundary condition of the Poisson equation
\begin{eqnarray}
\nabla^2_{\bR}\phi^{\newtext{(2)}} (\bR)=\frac{1}{\bar{\sigma}}\nabla_{\bR}\cdot \tilde{\mathcal{E}}^{(2)}.
\end{eqnarray}

\section{Green's function approach}
The divergence theorem in 2D states that for a vector field $\bA$
\begin{eqnarray}
\int_{\Omega} d^2\br \nabla\cdot\bA &=& \int_C dl\; \hat{n}\cdot\bA 
\end{eqnarray}
where the first integral encloses a 2D region $\Omega$ and the second encircles its boundary $C$ in a positive sense (moving along the boundary, the region is to the left).

Let $\bA=\Phi \nabla G - G \nabla \Phi$. Then, the divergence theorem guarantees that
\begin{eqnarray}
\int_{\Omega} d^2\br \nabla\cdot \left(\Phi \nabla G - G \nabla \Phi\right) &=& \int_C dl\; \hat{n}\cdot \left(\Phi \nabla G - G \nabla \Phi\right)\\
\Rightarrow \int_{\Omega} d^2\br \left(\Phi \nabla^2 G - G\nabla^2\Phi \right) &=& \int_C dl\; \hat{n}\cdot \left(\Phi \nabla G - G \nabla \Phi\right)
\label{eqn:divTheoremG}\end{eqnarray}

Now, let's say that we are able to construct a 2-variable function $G(\br,\br')$ such that inside the region $\Omega$
\begin{eqnarray}\label{Eqn:Greens function definition}
\nabla^2_{\br'}G(\br,\br')=-\delta(\br-\br') ;\;\;\; \text{for}\;\br,\br'\in\Omega.
\end{eqnarray}
Then, performing the integral in (\ref{eqn:divTheoremG}) over $\br'$ we have
\begin{eqnarray}
-\Phi(\br)-\int_{\Omega} d^2\br' G(\br,\br')\nabla_{\br'}^2 \Phi(\br') &=& \int_C dl'\;  \left(\Phi(\br') \hat{n}\cdot\nabla_{\br'} G(\br,\br') - G(\br,\br') \hat{n}\cdot\nabla \Phi(\br')\right).
\label{eqn:PhiGeneralG}
\end{eqnarray}
In our case, the value of $\Phi (\br)$ on the boundary is unknown. Naively, this poses a problem because, even if one knows $G$, one still doesn't know how to handle the first term on the right hand side; the second term on the right hand side is known for known $G$ because of the tangential current boundary condition (\ref{eqn:boundary condition}).
However, note that a solution of (\ref{Eqn:Greens function definition}) is not unique. This is the case because we can add to it a solution of the Laplace equation $\nabla^2 u =0$ and (\ref{Eqn:Greens function definition}) will still hold. Therefore, Eq. (\ref{eqn:PhiGeneralG}) holds as well. We can use this freedom to our advantage.

Imagine we manage to find such $G(\br,\br')=G_N(\br,\br')$ that 
\begin{eqnarray}
\hat{n}\cdot\nabla_{\br'} G_N(\br,\br')|_{\br'\in boundary}&=&\text{constant of }\br \;\text{and}\; \br'.
\label{eqn:GN general def}
\end{eqnarray}
Then,
\begin{eqnarray}
\int_C dl'\;  \Phi(\br') \hat{n}\cdot\nabla_{\br'} G_N(\br,\br') &=& \hat{n}\cdot\nabla_{\br'} G_N(\br,\br')\int_C dl'\; \Phi(\br')=\text{constant}. 
\end{eqnarray}
Since only potential differences are physical, a constant change to the potential function $\Phi (\br)$ doesn't carry physical information and doesn't pose a problem.
Then, substituting (\ref{eqn:poisson}),(\ref{eqn:boundary condition}), and (\ref{eqn:GN general def}) into (\ref{eqn:PhiGeneralG}) we have
\begin{eqnarray}
\Phi(\br)&=&-\int_{\Omega} d^2\br' G_N(\br,\br') \frac{1}{\sigma}\nabla_{\br'}\cdot\mathcal{E}^{(2)}(\br') +\int_C dl'\;  G_N(\br,\br') \frac{1}{\sigma}\hat{n}\cdot\mathcal{E}^{(2)}+\text{constant}.
\label{Disk_solution_via_Gn}
\end{eqnarray}
So, if we can construct $G_N$ we will be able to obtain the potential $\Phi (\br)$ everywhere inside the region of interest.

\section{Marvelous identity Eq.(\ref{eq:marvelous})}

Consider the generic case of anisotropic linear conductivity tensor.
We solve for $\phi^{\newtext{(2)}} (\bR)$ using Neumann Green's function
\begin{eqnarray}
&&\nabla^2_{\bR'}G_N(\bR,\bR')=-\delta\left(\bR-\bR'\right)\;\; \text{for}\; \bR,\bR'\in \text{ellipse}\;(\Omega)\\
&&\left.\hat{\bN}\cdot\nabla_{\bR'}G_N(\bR,\bR')\right|_{\bR'\in \text{boundary of the ellipse}\; (C)}=\text{constant of}\;\bR\;\text{and}\;\bR'.
\end{eqnarray}
and, up to an overall constant,
\begin{eqnarray}
&&\phi^{\newtext{(2)}}(\bR)=-\int_{\Omega}d^2\bR' G_{N}(\bR,\bR')\nabla^2_{\bR'}\phi^{\newtext{(2)}}(\bR')+\int_C d\ell' G_{N}(\bR,\bR')\hat{\bN}\cdot\nabla_{\bR'}\phi^{\newtext{(2)}}(\bR')\\
&&=-\frac{1}{\bar{\sigma}}\int_{\Omega}d^2\bR' G_{N}(\bR,\bR')\nabla_{\bR'}\cdot \tilde{\mathcal{E}}^{(2)}(\bR')+\frac{1}{\bar{\sigma}}\int_C d\ell'
G_{N}(\bR,\bR')\hat{\bN}\cdot\tilde{\mathcal{E}}^{(2)}(\bR')\\
&&=-\frac{1}{\bar{\sigma}}\int_{\Omega}d^2\bR'\nabla_{\bR'}\cdot\left( G_{N}(\bR,\bR') \tilde{\mathcal{E}}^{(2)}(\bR')\right)+
\frac{1}{\bar{\sigma}}\int_{\Omega}d^2\bR' \left(\nabla_{\bR'}G_{N}(\bR,\bR')\right) \cdot\tilde{\mathcal{E}}^{(2)}(\bR')
+\frac{1}{\bar{\sigma}}\int_C d\ell' G_{N}(\bR,\bR')\hat{\bN}\cdot\tilde{\mathcal{E}}^{(2)}(\bR')\nonumber\\
\label{eq:cancellation}
\end{eqnarray}
But the divergence theorem in 2D for a vector field $\bA$ states that
\begin{eqnarray}
\int_\Omega d^2\br \nabla\cdot\bA=\int_C d\ell\; \hat{n}\cdot\bA.
\end{eqnarray}
Therefore, the first and the last terms in the last line of (\ref{eq:cancellation}) cancel and we have
\begin{eqnarray}
\phi^{\newtext{(2)}} (\bR)=
\frac{1}{\bar{\sigma}}\int_{\Omega}d^2\bR' \left(\nabla_{\bR'}G_{N}(\bR,\bR')\right) \cdot\tilde{\mathcal{E}}^{(2)}(\bR').
\label{eq:marvelous}
\end{eqnarray}
This expression is valid for any anisotropy $\Delta \sigma$ of the linear tensor in the range $0 \leq \left| \frac{\Delta \sigma}{\bar{\sigma}} \right| < 1$. The Eq. \eqref{eq:marvelous} has a physical interpretation -- the nonlinear potential $\phi^{\newtext{(2)}} (\bR)$ is given by a dot product of two electric fields: one is the field $\nabla_{\bR'}G_{N}(\bR,\bR')$ emanating from a $\delta-$source located at the "observation" point; the other one -- is the electric field generated by the nonlinearity $\propto \tilde{\mathcal{E}}^{(2)}$.

\section{Symmetries of the potential upon interchange of source and drain}

It turns out that the Ohm's law equation enjoys nontrivial symmetries upon the interchange of source and drain.

\subsection{180$^{\circ}$-degrees configuration}

Let us provide an example for 180$^{\circ}$ source and drain configuration. We shall consider the $C_3$-symmetric case first and then generalize our results. Consider the Ohm's law equation for the linear component in the first order in $I$ with the continuity equation. Combining the two together we get
\begin{equation}
    - \sigma^{(1)}_{+} \nabla^2 \Phi^{\newtext{(1)}} (\mathbf{r}, \mathbf{r}_S) = I \left( \delta(\mathbf{r} - \mathbf{r}_S) - \delta(\mathbf{r} + \mathbf{r}_S) \right),
    \label{Linear_potential_symmetry}
\end{equation}
where we used that $\mathbf{r}_D = -\mathbf{r}_S$ in the 180$^{\circ}$ configuration and  explicitly substituted $E^{\newtext{(1)}} = - \nabla \Phi^{\newtext{(1)}}$. Let us swap source and drain in the right-hand-side of \eqref{Linear_potential_symmetry}:
$$
I \left( \delta(\mathbf{r} + \mathbf{r}_S) - \delta(\mathbf{r} - \mathbf{r}_S) \right).
$$
The right-hand-side changes sign, therefore, $\Phi^{\newtext{(1)}}$ also has to change sign under such transformation, i.e.,
\begin{equation}
    \Phi^{\newtext{(1)}} (\mathbf{r}, -\mathbf{r}_S) = - \Phi^{\newtext{(1)}} (\mathbf{r}, \mathbf{r}_S).
\end{equation}
Now let us change $\mathbf{r}$ to $-\mathbf{r}$ without swapping source and drain. The right-hand-side of Eq. \eqref{Linear_potential_symmetry} then yields
$$
I \left( \delta(-\mathbf{r} - \mathbf{r}_S) - \delta(-\mathbf{r} + \mathbf{r}_S) \right) = I \left( \delta(\mathbf{r} + \mathbf{r}_S) - \delta(\mathbf{r} - \mathbf{r}_S) \right),
$$
i.e., remains invariant under such transformation since $\delta-$function is even. But the left-hand-side now experiences a sign change, hence, the potential $\Phi^{\newtext{(1)}}$ has to satisfy
\begin{equation}
    \Phi^{\newtext{(1)}} (-\mathbf{r}, \mathbf{r}_S) = - \Phi^{\newtext{(1)}} (\mathbf{r}, \mathbf{r}_S).
\end{equation}

We just established the symmetry property for the linear potential $\Phi^{\newtext{(1)}} (\mathbf{r}, \mathbf{r}_S)$. Let's see what consequences these symmetries have for the nonlinear potential $\Phi^{\newtext{(2)}} (\mathbf{r}, \mathbf{r}_S)$. The nonlinear potential is defined by the equation 
\begin{equation}
\begin{gathered}
    \nabla^2 \Phi^{\newtext{(2)}} (\mathbf{r}, \mathbf{r}_S) = \frac{1}{\sigma^{(1)}_{+}} \nabla \cdot \mathcal{E}^{(2)} (\mathbf{r}, \mathbf{r}_S), \\
    \hat{n}\cdot \nabla\Phi^{\newtext{(2)}} |_{boundary}=\frac{1}{\sigma^{(1)}_{+}}\hat{n}\cdot\mathcal{E}^{(2)}|_{boundary}
    \end{gathered}
    \label{Nonlinear_potential_eq_C3}
\end{equation}
$\mathcal{E}^{(2)} (\mathbf{r}, \mathbf{r}_S) 
\propto (E^{\newtext{(1)}})^2$ by construction, therefore, it is even under $\mathbf{r} \to -\mathbf{r}, \mathbf{r}_S \to -\mathbf{r}_S$. Differential operator $\nabla$ doesn't change under $\mathbf{r}_S \to -\mathbf{r}_S$ but it flips sign under $\mathbf{r} \to -\mathbf{r}$. Thus, we obtain the following symmetry properties of $\Phi^{\newtext{(2)}}$:
\begin{eqnarray}
    \Phi^{\newtext{(2)}} (\mathbf{r}, \mathbf{r}_S) = \Phi^{\newtext{(2)}} (\mathbf{r}, -\mathbf{r}_S) \\
    \Phi^{\newtext{(2)}} (-\mathbf{r}, \mathbf{r}_S) = - \Phi^{\newtext{(2)}} (\mathbf{r}, \mathbf{r}_S).
\end{eqnarray}

We now generalize our results to the general non-$C_3$-symmetric case in the absence of linear Hall component of the tensor.
The linear current component now acquires an additional term and reads
\begin{equation}
    j_x^{\newtext{(1)}} + i j_y^{\newtext{(1)}} = \sigma^{(1)}_{+} \left( E_x^{\newtext{(1)}} + i E_y^{\newtext{(1)}} \right) + \sigma^{(1)}_{-} \left( E_x^{\newtext{(1)}} - i E_y^{\newtext{(1)}} \right).
\end{equation}
If we take divergence of the left-hand-side we will get 
$$
-(\sigma^{(1)}_{+} \nabla_x^2 + \sigma^{(1)}_{-} \nabla_y^2  ) \Phi^{\newtext{(1)}} = I \left( \delta(\mathbf{r} - \mathbf{r}_S) - \delta(\mathbf{r} + \mathbf{r}_S) \right).
$$
Clearly, this is not a Poisson equation. To cure this we now perform change of variables like in \cite{Oskar_solo}
\begin{eqnarray}
    X = \frac{x}{\sqrt{1+\frac{\sigma^{(1)}_{+} - \sigma^{(1)}_{-}}{\sigma^{(1)}_{+} + \sigma^{(1)}_{-}}}}, \\
    Y = \frac{y}{\sqrt{1-\frac{\sigma^{(1)}_{+} - \sigma^{(1)}_{-}}{\sigma^{(1)}_{+} + \sigma^{(1)}_{-}}}},
\end{eqnarray}
so that we recover the Poisson equation for $\Phi^{\newtext{(1)}}$ like in Eq. \eqref{Linear_potential_symmetry} but in new rescaled variables:
\begin{equation}
    - \nabla^2_{\mathbf{R}} \Phi^{\newtext{(1)}} (\mathbf{R}, \mathbf{R}_S) = \frac{I}{\sqrt{\sigma^{(1)}_{+} \sigma^{(1)}_{-}}} \left( \delta(\mathbf{R} - \mathbf{R}_S) - \delta(\mathbf{R} + \mathbf{R}_S) \right)
    \label{Linear_potential_symmetry_rescaled} 
\end{equation}
with $\mathbf{R} = (X, Y)$. Clearly Eq. \eqref{Linear_potential_symmetry_rescaled} has the same symmetry properties as Eq. \eqref{Linear_potential_symmetry}:
\begin{eqnarray}
    \Phi^{\newtext{(1)}} (\mathbf{R}, -\mathbf{R}_S) = - \Phi^{\newtext{(1)}} (\mathbf{R}, \mathbf{R}_S), \\
    \Phi^{\newtext{(1)}} (-\mathbf{R}, \mathbf{R}_S) = - \Phi^{\newtext{(1)}} (\mathbf{R}, \mathbf{R}_S).
\end{eqnarray}
Now we proceed to the nonlinear potential $\Phi^{\newtext{(2)}}$.
Here the source term has two additional terms and reads
\begin{equation}
 \mathcal{E}^{(2)} (\mathbf{R}, \mathbf{R}_S) = \sigma^{(2)}_{-} (E^{\newtext{(1)}}_x-iE^{\newtext{(1)}}_y)^2 + \sigma^{(2)}_{+}  (E^{\newtext{(1)}}_x+iE^{\newtext{(1)}}_y)^2 + \sigma^{(2)}_{0}  [(E^{\newtext{(1)}}_x)^2+(E^{\newtext{(1)}}_y)^2],  
\end{equation}
where the electric field $E^{\newtext{(1)}}$ depends on $\mathbf{R, R}_S$.
As in the $C_3$-symmetric case, all terms in the right-hand-side are quadratic in the electric field $E^{\newtext{(1)}}$ and, hence, the right-hand-side is even under $\mathbf{R} \to -\mathbf{R}, \mathbf{R}_S \to -\mathbf{R}_S$. 
As in the $C_3$-symmetric case the nonlinear potential is given by
\begin{equation}
    \nabla^2_{\mathbf{R}} \Phi^{\newtext{(2)}} (\mathbf{R}, \mathbf{R}_S) = \nabla_{\mathbf{R}} \cdot \mathcal{E}^{(2)} (\mathbf{R}, \mathbf{R}_S).
\end{equation}
This equation is mathematically identical to Eq. \eqref{Nonlinear_potential_eq_C3}, therefore, the symmetry properties of the non-$C_3$-symmetric case are the same as of the $C_3$-symmetric case.

\section{Neumann Green's function $G_N(\br,\br')$ of a circle}
It can be verified by a direct calculation that for a unit circle
\begin{eqnarray}
G_N(\br,\br')&=&-\frac{1}{4\pi}\ln\left(\left(z-z'\right)\left(z^*-z'^*\right)\right)
-\frac{1}{4\pi}\ln\left(\left(z'-\frac{1}{z^*}\right)\left(z'^*-\frac{1}{z}\right)\right)\\
&=&-\frac{1}{4\pi}\ln\left(\left(z-z'\right)\left(z^*-z'^*\right)\right)
-\frac{1}{4\pi}\ln\left(\left(\frac{1}{z^*}-z'\right)\left(\frac{1}{z}-z'^*\right)\right)
\label{Greens_function_disk}
\end{eqnarray}
satisfies Eq.(\ref{Eqn:Greens function definition}) and Eq.(\ref{eqn:GN general def}). For the verification, is useful to remember that
\begin{eqnarray}
\partial_x&=&\partial_z+\partial_{z^*},\\
\partial_y&=&i\left(\partial_z-\partial_{z^*}\right),\\
\partial_{z^*}\frac{1}{z}&=&\pi\delta(\br).
\end{eqnarray}
To verify the above claim, note that
\begin{eqnarray}
&&\hat{n}\cdot\nabla_{\br'} G_N(\br,\br')=
\frac{z'^*}{2|z'|}\left(\partial_{x'}G_N(\br,\br')+i\partial_{y'}G_N(\br,\br')\right)+
\frac{z'}{2|z'|}\left(\partial_{x'}G_N(\br,\br')-i\partial_{y'}G_N(\br,\br')\right)\\
&&=
\frac{z'^*}{|z'|}\partial_{z'^*}G_N(\br,\br')+
\frac{z'}{|z'|}\partial_{z'}G_N(\br,\br')=\frac{z'^*}{|z'|}\frac{1}{4\pi}\left(\frac{1}{z^*-z'^*}+\frac{1}{\frac{1}{z}-z'^*}\right)+
\frac{z'}{|z'|}\frac{1}{4\pi}\left(\frac{1}{z-z'}+\frac{1}{\frac{1}{z^*}-z'}\right)
\end{eqnarray}
Now, we place $z'$ on the unit circle, i.e. $z'=e^{i\phi}$.
Then,
\begin{eqnarray}
\hat{n}\cdot\nabla_{\br'} G_N(\br,\br')|_{\br'\in boundary}&=&
\frac{1}{4\pi}\left(\frac{e^{-i\phi}}{z^*-e^{-i\phi}}+\frac{e^{-i\phi}}{\frac{1}{z}-e^{-i\phi}}+\frac{e^{i\phi}}{z-e^{i\phi}}+\frac{e^{i\phi}}{\frac{1}{z^*}-e^{i\phi}}\right)\\
&=&\frac{1}{4\pi}\left(\frac{e^{-i\phi}}{z^*-e^{-i\phi}}+\frac{z}{e^{i\phi}-z}+\frac{e^{i\phi}}{z-e^{i\phi}}
+\frac{z^*}{e^{-i\phi}-z^* }\right)\\
&=&-\frac{1}{2\pi}\rightarrow -\frac{1}{2\pi a} 
\end{eqnarray}
where in the last line we restored the proper dimensions.

The Green's function $G_N (\br, \br')$ is valid for any points inside the circle. Therefore, in our subsequent treatment of the problem we will first need to put source and drain inside the circle and then continuously move them to the boundary.

\section{Analytical solution for the isotropic linear conductivity tensor case}

In this section we provide a full analytical solution for the nonlinear potential $\Phi^{\newtext{(2)}} (\br)$ in the case of fully isotropic linear conductivity tensor. Strictly speaking, having a perfectly isotropic linear tensor puts severe restrictions on the nonlinear conductivity tensor. Here we assume that no restrictions are imposed on the nonlinear tensor and consider the most general case: all of its components to be nonzero. In practice our result from this section can be applied to the case of small anisotropy of linear conductivity tensor, when the anisotropy of the linear tensor can be safely disregarded. 

To find the solution we employ the general formula for the potential
\begin{eqnarray}
\phi^{\newtext{(2)}} (\bR)=
\frac{1}{\bar{\sigma}}\int_{\Omega}d^2\bR' \left(\nabla_{\bR'}G_{N}(\bR,\bR')\right) \cdot\tilde{\mathcal{E}}^{(2)}(\bR').
\end{eqnarray}
In the isotropic limit, it becomes
\begin{eqnarray}
&&\Phi^{\newtext{(2)}}_{iso}(\br)=
\frac{1}{\sigma}\int_{\Omega}d^2\br' \left(\nabla_{\br'}G_{N}(\br,\br')\right) \cdot \mathcal{E}^{(2)}(\br')\\
&&=
\frac{1}{\sigma}\int_{\Omega}d^2\br' \left[\left(\frac{\partial}{\partial x'}G_{N}(\br,\br')\right)\mathcal{E}_x^{(2)}(\br')+
\left(\frac{\partial}{\partial y'}G_{N}(\br,\br')\right)\mathcal{E}_y^{(2)}(\br')\right]\\
&&=
\frac{1}{\sigma}\int_{\Omega}d^2\br' \left[\left(\frac{\partial}{\partial
z'}G_{N}(\br,\br')\right)\left(\mathcal{E}_x^{(2)}(\br')+i\mathcal{E}_y^{(2)}(\br')\right)+
\left(\frac{\partial}{\partial z^{'*}}G_{N}(\br,\br')\right)\left(\mathcal{E}_x^{(2)}(\br')-i\mathcal{E}_y^{(2)}(\br')\right)\right] \\
&&=\frac{1}{\sigma}\int_{\Omega}d^2\br' \left(\frac{\partial}{\partial
z'}G_{N}(\br,\br')\right)\left(\mathcal{E}_x^{(2)}(\br')+i\mathcal{E}_y^{(2)}(\br')\right) + c.c.
\label{isotropic_marvel}
\end{eqnarray}
where $\Omega$ -- is the disk of radius 1 and
\begin{eqnarray}
\mathcal{E}_x^{(2)}(\br')+i\mathcal{E}_y^{(2)}(\br')&=&\sigma^{(2)}_{-} e^{i\chi^{(2)}_{-}}\left(E^{\newtext{(1)}}_x-iE^{\newtext{(1)}}_y\right)^2+
\sigma^{(2)}_{+} e^{i\chi^{(2)}_{+}}\left(E^{\newtext{(1)}}_x+iE^{\newtext{(1)}}_y\right)^2+\sigma^{(2)}_{0} e^{i\chi^{(2)}_{0}}\left({E^{\newtext{(1)}}_x}^2+{E^{\newtext{(1)}}_y}^2\right).
\label{E_expansion_in_basis_functions}
\end{eqnarray}
The solution for the \newtext{1st}-order (linear) potential reads
\begin{eqnarray}
\Phi^{\newtext{(1)}}_{iso}(\br)=&&-\frac{I}{4\pi\sigma}\ln\left(\left(z-z_S\right)\left(z^{*}-z^{*}_S\right)\right)
-\frac{I}{4\pi\sigma}\ln\left(\left(\frac{1}{z^{*}}-z_S\right)\left(\frac{1}{z}-z^{*}_S\right)\right)\nonumber\\
&&+\frac{I}{4\pi\sigma}\ln\left(\left(z-z_D\right)\left(z^{*}-z^{*}_D\right)\right)
+\frac{I}{4\pi\sigma}\ln\left(\left(\frac{1}{z^{*}}-z_D\right)\left(\frac{1}{z}-z^{*}_D\right)\right)
\end{eqnarray}
and gives rise to the \newtext{1st}-order electric field $E^{\newtext{(1)}}$
\begin{eqnarray}
&&E^{\newtext{(1)}}_x+iE^{\newtext{(1)}}_y=-\left(\partial_x+i\partial_y\right)\Phi^{\newtext{(1)}}_{iso}(\br)=-2\partial_{z^{*}}\Phi^{\newtext{(1)}}_{iso}(\br)=\nonumber\\
&&=(-2)\left(-\frac{I}{4\pi\sigma}\right)\left(\frac{1}{z^{*}-z^{*}_S}-\frac{1}{z^{*}-z^{*}_D}
+\frac{1}{\frac{1}{z^{*}}-z_S}\frac{-1}{(z^{*})^2}-
\frac{1}{\frac{1}{z^{*}}-z_D}\frac{-1}{(z^{*})^2}\right)\\
&&=\frac{I}{2\pi\sigma}\left(\frac{1}{z^{*}-z^{*}_S}
+\frac{1}{z^{*}-\frac{1}{z_S}}-\frac{1}{z^{*}-z^{*}_D}-\frac{1}{z^{*}-\frac{1}{z_D}}
\right),
\end{eqnarray}
where $z_{S,D} = x_{S,D} + i y_{S,D}$ -- are positions of source and drain on the complex plane and the potential is given at $\br$ specified by $z = x + i y$. Later we will find those additional expressions useful:
\begin{eqnarray}
&&E^{\newtext{(1)}}_x-iE^{\newtext{(1)}}_y
=\frac{I}{2\pi\sigma}\left(\frac{1}{z-z_S}
+\frac{1}{z-\frac{1}{z^{*}_S}}-\frac{1}{z-z_D}-\frac{1}{z-\frac{1}{z^{*}_D}}
\right)
\end{eqnarray}
and
\begin{eqnarray}
&&\frac{\partial}{\partial z'}G_N(\br,\br')=-\frac{1}{4\pi}\left(\frac{1}{z'-z}+\frac{1}{z'-\frac{1}{z^{*}}}\right).
\end{eqnarray}

\subsection{The basis function concept}

Eq. \eqref{E_expansion_in_basis_functions} allows for a very convenient interpretation and analysis of the solution for the potential $\Phi^{\newtext{(2)}}_{iso} (\br) $. In the perturbative approach the source field for the nonlinear response comes from the \underline{linear} combination of three terms, which are quadratic in the electric field that is generated in the linear case. Therefore, the nonlinear potential can be decomposed into three parts, each of which is generated solely by one of the three terms: only $\sigma^{(2)}_{-}$, only $\sigma^{(2)}_{+}$, or only $\sigma^{(2)}_{0}$. Each of these three terms in \eqref{E_expansion_in_basis_functions} carries information about 2 out of 6 parameters that define the nonlinear conductivity tensor. 

In the case of vanishing anisotropy of the linear conductivity tensor one can explicitly write
\begin{eqnarray}
\Phi^{\newtext{\left(2 \right)}}_{iso}(\br)=\Phi^{\newtext{(2_{-})}}_{iso}(\br)+\Phi^{\newtext{(2_{+})}}_{iso}(\br)+\Phi^{\newtext{(2_{0})}}_{iso}(\br),
\end{eqnarray}
where
\begin{eqnarray}
&&\Phi^{\newtext{(2_{-})}}_{iso}(\br)=\frac{1}{\sigma}\int_{\Omega}d^2\br' \left(\frac{\partial}{\partial
z'}G_{N}(\br,\br')\right)\left( \sigma^{(2)}_{-} e^{i\chi^{(2)}_{-}}\left(E^{\newtext{(1)}}_x (\br')-iE^{\newtext{(1)}}_y (\br') \right)^2 \right) + c.c. \\
&&\Phi^{\newtext{(2_{+})}}_{iso}(\br)=\frac{1}{\sigma}\int_{\Omega}d^2\br' \left(\frac{\partial}{\partial
z'}G_{N}(\br,\br')\right)\left( \sigma^{(2)}_{+} e^{i\chi^{(2)}_{+}}\left(E^{\newtext{(1)}}_x (\br') + iE^{\newtext{(1)}}_y (\br') \right)^2 \right) + c.c. \\
&&\Phi^{\newtext{(2_{0})}}_{iso}(\br)=\frac{1}{\sigma}\int_{\Omega}d^2\br' \left(\frac{\partial}{\partial
z'}G_{N}(\br,\br')\right)\left( \sigma^{(2)}_{0} e^{i\chi^{(2)}_{0}}\left({E^{\newtext{(1)}}_x}^2 (\br') + {E^{\newtext{(1)}}_y}^2 (\br') \right) \right) + c.c. 
\end{eqnarray}
-- are the basis functions, since their linear combination provides the full nonlinear potential generated in the system in the $I^2$ regime. Similar decomposition is applied to the anisotropic linear conductivity case. There, the basis functions have to be redefined because of the mixing between the original ones. We discuss anisotropic basis functions later in the text. We also look for basis functions' values only at the boundary of the disk sample, as in realistic experiments the voltage drop is measured between two contacts located at the boundary of the sample.

\begin{figure}[h]
\includegraphics[width=0.5\linewidth]{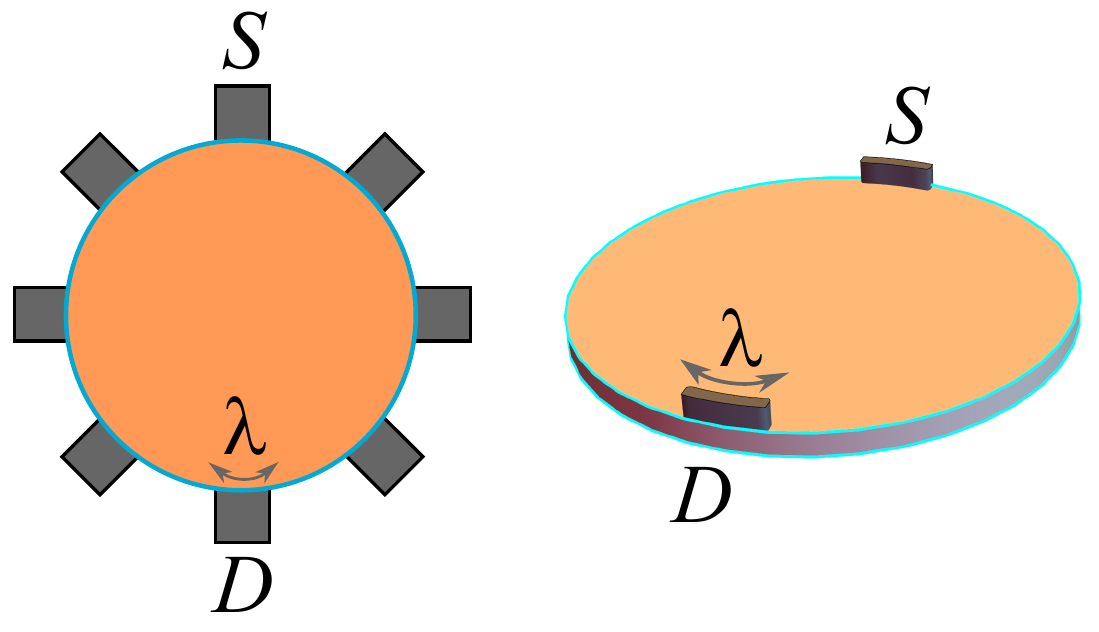}
\centering{}\caption{ Left: sketch illustrating the geometry of the leads (gray rectangles) and the sample (orange disk). Right: illustration of the box distribution regularization of point-like sources for source (S) and drain (D), where $\lambda$ defines the angular size of the arc-shaped contact. The gray rectangle here depicts the distribution function describing the physical finite-size lead.
} 
\label{distr_sketch}
\end{figure}

It is important to mention that the basis functions depend on the shape of the source and drain. In this work we consider only the arc-like contacts, i.e., contacts that on the boundary of the sample are small arcs of the circle, see sketch in Fig. \ref{distr_sketch}. 

\subsection{A simple proof that in the linear isotropic case the nonlinear signal consists of single-harmonic contributions}

In the case of vanishing anisotropy of the linear conductivity tensor one can explicitly prove (without solving for the potential) that the basis functions for every term consist of a single harmonic as a function of the rotation angle $\phi_R$. Below we provide such a simple proof.

\begin{figure}[h]
\includegraphics[width=0.65\linewidth]{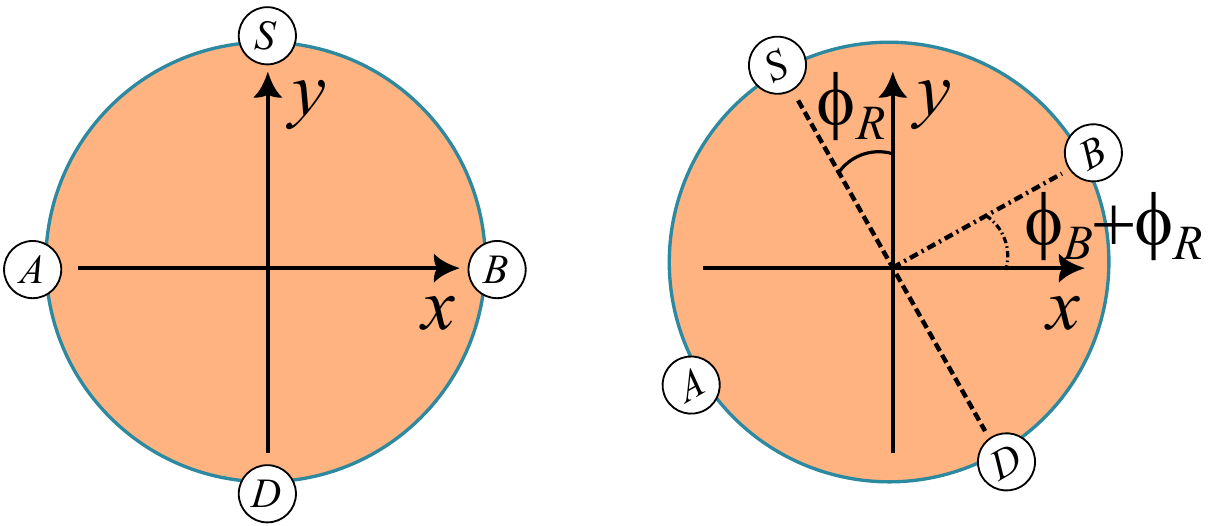}
\centering{}\caption{ Sketch illustrating rotation of the disk sample in the fixed reference frame and the definition of the angle $\phi_R$. The leads ($S, D$ -- source, drain and $A, B$ -- the two measurement points) are depicted in the perpendicular measurement configuration for illustrative purposes.
} 
\label{rot_sketch}
\end{figure}

\subsubsection{The $\sigma^{(2)}_{-}$ term}

As we know from our previous symmetry analysis, this term respects $C_3$ symmetry. Let us consider a generic source and drain configuration. The measurement is performed by rigidly rotating the position of all contacts around the sample, therefore, 
$$
z_S = e^{i (\phi_S + \phi_R)}, \; \; \; z_D = e^{i (\phi_D + \phi_R)}, \; \; \; z = e^{i (\phi + \phi_R)},
$$
where $\phi_R$ is the rotation angle with respect to the original reference frame and $\phi$ is the angle specifying a point on the disc in which the potential is measured, see Fig. \ref{rot_sketch}. According to \eqref{E_expansion_in_basis_functions}, the $\sigma^{(2)}_{-}$ term can be written as 
\begin{equation}
    \sigma^{(2)}_{-} e^{i \chi^{(2)}_{-}} (E^{\newtext{(1)}}_x-iE^{\newtext{(1)}}_y)^2 = \sigma^{(2)}_{-} e^{i \chi^{(2)}_{-}} \frac{I^2}{\pi^2 \sigma^2} \left(\frac{1}{z-z_D}-\frac{1}{z-z_S}\right)^2 = \mathcal{E}_x^{(2)}+i\mathcal{E}_y^{(2)}.
\end{equation}
The integrand in the \eqref{isotropic_marvel} upon substitution of $z, z_S, z_D$ becomes 
\begin{eqnarray}
     &&\left(\frac{\partial}{\partial
z'}G_{N}(\br,\br')\right)\left(\mathcal{E}_x^{(2)}(\br')+i\mathcal{E}_y^{(2)}(\br')\right) + c.c. = - \frac{\sigma^{(2)}_{-} e^{i \chi^{(2)}_{-}} I^2}{2 \pi^3 \sigma^2} \left(\frac{1}{z'-z}\right) \left(\frac{1}{z'-z_D}-\frac{1}{z'-z_S}\right)^2 + c.c. = \nonumber \\
&&= - \frac{\sigma^{(2)}_{-} e^{i \chi^{(2)}_{-}} I^2}{2 \pi^3 \sigma^2} \left(\frac{1}{z'-e^{i (\phi + \phi_R)}}\right) \left(\frac{1}{z'-e^{i (\phi_D + \phi_R)}}-\frac{1}{z'-e^{i (\phi_S + \phi_R)}}\right)^2 + c.c. = \nonumber \\ 
&&=- \frac{\sigma^{(2)}_{-} e^{i \chi^{(2)}_{-}} I^2}{2 \pi^3 \sigma^2} e^{-3 i \phi_R} \left(\frac{1}{z'-e^{i \phi }}\right) \left(\frac{1}{z'-e^{i \phi_D }}-\frac{1}{z'-e^{i \phi_S }}\right)^2 + c.c.,
\end{eqnarray}
where we took advantage of $z'$ being a dummy variable by redefining $z' = z' e^{- i \phi_R}$. We see that the expression above is manifestly single-harmonic and three-fold symmetric in terms of $\phi_R$ dependence.

\subsubsection{The $\sigma^{(2)}_{+}$ term}

Let us now focus on the $\sigma^{(2)}_{+}$ term:
\begin{equation}
    \sigma^{(2)}_{+} e^{i \chi^{(2)}_{+}} (E^{\newtext{(1)}}_x+iE^{\newtext{(1)}}_y)^2 = \sigma^{(2)}_{+} e^{i \chi^{(2)}_{+}} \frac{I^2}{\pi^2 \sigma^2} \left(\frac{1}{z^*-z_D^*}-\frac{1}{z^*-z_S^*}\right)^2.
\end{equation}
Just like in the $\sigma^{(2)}_{-}$ term case we plug in the definitions of $z$'s into the expression for the integrand and obtain
\begin{eqnarray}
     &&\left(\frac{\partial}{\partial
z'}G_{N}(\br,\br')\right)\left(\mathcal{E}_x^{(2)}(\br')+i\mathcal{E}_y^{(2)}(\br')\right) + c.c. = - \frac{\sigma^{(2)}_{+} e^{i \chi^{(2)}_{+}} I^2}{2 \pi^3 \sigma^2} \left(\frac{1}{z'-z}\right) \left(\frac{1}{z'^*-z_D^*}-\frac{1}{z'^*-z_S^*}\right)^2 + c.c. = \nonumber \\
&&= - \frac{\sigma^{(2)}_{+} e^{i \chi^{(2)}_{+}} I^2}{2 \pi^3 \sigma^2} \left(\frac{1}{z'-e^{i (\phi + \phi_R)}}\right) \left(\frac{1}{z'^*-e^{-i (\phi_D + \phi_R)}}-\frac{1}{z'^*-e^{-i (\phi_S + \phi_R)}}\right)^2 + c.c. = \nonumber \\ 
&&=- \frac{\sigma^{(2)}_{+} e^{i \chi^{(2)}_{+}} I^2}{2 \pi^3 \sigma^2} e^{i \phi_R} \left(\frac{1}{z'-e^{i \phi }}\right) \left(\frac{1}{z'^*-e^{-i \phi_D }}-\frac{1}{z'^*-e^{-i \phi_S }}\right)^2 + c.c.,
\end{eqnarray}
where we again took advantage of $z'$ being a dummy variable. The expression above now is manifestly single-harmonic and one-fold symmetric in terms of $\phi_R$ dependence.

\subsubsection{The $\sigma^{(2)}_{0}$ term}

We finally proceed to the $\sigma^{(2)}_{0}$ term
\begin{equation}
    \begin{gathered}
    \sigma^{(2)}_{0} e^{i \chi^{(2)}_{0}}  [(E^{\newtext{(1)}}_x)^2+(E^{\newtext{(1)}}_y)^2] = \sigma^{(2)}_{0} e^{i \chi^{(2)}_{0}} (E^{\newtext{(1)}}_x-iE^{\newtext{(1)}}_y) (E^{\newtext{(1)}}_x+iE^{\newtext{(1)}}_y) = 
    \sigma^{(2)}_{0} e^{i \chi^{(2)}_{0}} \frac{I^2}{\pi^2 \sigma^2} \left(\frac{1}{z-z_D}-\frac{1}{z-z_S}\right) \left(\frac{1}{z^*-z_D^*}-\frac{1}{z^*-z_S^*}\right),
    \end{gathered}
\end{equation}
for which
\begin{eqnarray}
     &&\left(\frac{\partial}{\partial
z'}G_{N}(\br,\br')\right)\left(\mathcal{E}_x^{(2)}(\br')+i\mathcal{E}_y^{(2)}(\br')\right) + c.c. = \nonumber \\
&&= - \frac{\sigma^{(2)}_{0} e^{i \chi^{(2)}_{0}} I^2}{2 \pi^3 \sigma^2} \left(\frac{1}{z'-z}\right) \left(\frac{1}{z'-z_D}-\frac{1}{z'-z_S}\right) \left(\frac{1}{z'^*-z_D^*}-\frac{1}{z'^*-z_S^*}\right) + c.c. = \nonumber \\
&&= - \frac{\sigma^{(2)}_{0} e^{i \chi^{(2)}_{0}} I^2}{2 \pi^3 \sigma^2} \left(\frac{1}{z'-e^{i (\phi + \phi_R)}}\right) \left(\frac{1}{z'-e^{i (\phi_D + \phi_R)}}-\frac{1}{z'-e^{i (\phi_S + \phi_R)}}\right) \left(\frac{1}{z'^*-e^{-i (\phi_D + \phi_R)}}-\frac{1}{z'^*-e^{-i (\phi_S + \phi_R)}}\right) + c.c. \nonumber \\
&&= - \frac{\sigma^{(2)}_{0} e^{i \chi^{(2)}_{0}} I^2}{2 \pi^3 \sigma^2} e^{- i \phi_R} \left(\frac{1}{z'-e^{i \phi}}\right) \left(\frac{1}{z'-e^{i \phi_D }}-\frac{1}{z'-e^{i \phi_S }}\right) \left(\frac{1}{z'^*-e^{-i \phi_D }}-\frac{1}{z'^*-e^{-i \phi_S }}\right) + c.c.,
\end{eqnarray}
where again $z'$ was redefined to absorb the $e^{i \phi_R}$ factor. As for $\sigma^{(2)}_{+}$ term, the result is single-harmonic and one-fold symmetric, however, the phase winding (the sign of $\phi_R$ in the exponent) is the opposite.

\subsection{$\sigma^{(2)}_{-}$ basis function (the $C_3$ symmetric term)}

In this and the following subsections we explicitly derive the three basis functions for the case of vanishing anisotropy in the linear conductivity tensor. We start with the basis function for the $\sigma^{(2)}_{-}$ term and place $z$ on the edge of the unit circle while keeping $z_S$ and $z_D$ inside:
\begin{eqnarray}
&&\Phi^{\newtext{(2_{-})}}_{iso}(\br)=\frac{1}{\sigma}\int_{\Omega}d^2\br' \left(\frac{\partial}{\partial
z'}G_{N}(\br,\br')\right)\left( \sigma^{(2)}_{-} e^{i\chi^{(2)}_{-}}\left(E^{\newtext{(1)}}_x (\br')-iE^{\newtext{(1)}}_y (\br') \right)^2 \right) + c.c. = \\
&&=\frac{1}{\sigma}\int_{\Omega}d^2\br' \left( -\frac{1}{4\pi}\left(\frac{2}{z'-z}\right) \right)\left( \sigma^{(2)}_{-} e^{i\chi^{(2)}_{-}} \frac{I^2}{4\pi^2 \sigma^2}\left( \frac{1}{z'-z_S}
+\frac{1}{z'-\frac{1}{z^{*}_S}}-\frac{1}{z'-z_D}-\frac{1}{z'-\frac{1}{z^{*}_D}} \right)^2 \right) + c.c. = \\
&&= -\frac{\sigma^{(2)}_{-} e^{i\chi^{(2)}_{-}} I^2}{8\pi^3\sigma^3} \int_{\Omega}d^2\br' \left(\frac{1}{z'-z}\right) \left(  \frac{1}{z'-z_S}
+\frac{1}{z'-\frac{1}{z^{*}_S}}-\frac{1}{z'-z_D}-\frac{1}{z'-\frac{1}{z^{*}_D}} \right)^2 + c.c.
\end{eqnarray}
This expression consists of base integrals of generic type
\begin{eqnarray}
&&\frac{\sigma^{(2)}_{-} e^{i\chi^{(2)}_{-}}I^2}{(2\pi\sigma)^3}\int_\Omega d^2\br'\frac{1}{z-z'}\frac{1}{z'-z_i}\frac{1}{z'-z_j}=
-\frac{\sigma^{(2)}_{-} e^{i\chi^{(2)}_{-}}I^2}{(2\pi\sigma)^3}\int_\Omega d^2\br'\frac{1}{z'-z}\frac{1}{z'-z_i}\frac{1}{z'-z_j}=\\
&&-\frac{\sigma^{(2)}_{-} e^{i\chi^{(2)}_{-}}I^2}{(2\pi\sigma)^3}
\left(\frac{1}{z-z_i}\frac{1}{z-z_j}\int_\Omega \frac{d^2\br'}{z'-z}
-\frac{1}{z-z_i}\frac{1}{z_i-z_j}\int_\Omega \frac{d^2\br'}{z'-z_i}+
\frac{1}{z-z_j}\frac{1}{z_i-z_j}\int_\Omega \frac{d^2\br'}{z'-z_j}
\right)=\\
&&\frac{\sigma^{(2)}_{-} e^{i\chi^{(2)}_{-}}I^2}{(2\pi\sigma)^3}
\left(\frac{\pi z^{*}}{(z-z_i)(z-z_j)}
-\frac{\frac{\pi}{z_i}\Theta(|z_i|-1)
+\pi z^{*}_i \Theta(1-|z_i|)}{(z-z_i)(z_i-z_j)}+
\frac{\frac{\pi}{z_j}\Theta(|z_j|-1)
+\pi z^{*}_j \Theta(1-|z_j|)}{(z-z_j)(z_i-z_j)}
\right)\nonumber\\
&&\equiv \frac{\sigma^{(2)}_{-} e^{i\chi^{(2)}_{-}}I^2}{(2\pi\sigma)^3} F(z,z_i,z_j).
\end{eqnarray}
where we used partial fractions decomposition
\begin{eqnarray}
&&\frac{1}{z'-z}\frac{1}{z'-z_i}\frac{1}{z'-z_j}=\frac{1}{z-z_i}\frac{1}{z-z_j}\frac{1}{z'-z}-
\frac{1}{z-z_i}\frac{1}{z_i-z_j}\frac{1}{z'-z_i}+\frac{1}{z-z_j}\frac{1}{z_i-z_j}\frac{1}{z'-z_j}
\end{eqnarray}
and a contour integral
\begin{eqnarray}
&&\int d^2\br'\frac{1}{z'-z_k}=\int_0^1 d\rho \rho \oint \frac{d\zeta}{i\zeta}\frac{1}{\rho \zeta-z_k}=
\int_0^1 d\rho \oint \frac{d\zeta}{i\zeta}\frac{1}{\zeta-\frac{z_k}{\rho}}=
2\pi \int_0^1 d\rho\frac{1}{-\frac{z_k}{\rho}}
+2\pi \Theta(1-|z_k|)\int_{|z_k|}^1 d\rho \frac{\rho}{z_k}\nonumber\\
&&=-\frac{\pi}{z_k}
+\pi\frac{\Theta(1-|z_k|)}{z_k}\left(1-|z_k|^2\right)
=-\frac{\pi}{z_k}\Theta(1-|z_k|)-\frac{\pi}{z_k}\Theta(|z_k|-1)
+\pi\frac{\Theta(1-|z_k|)}{z_k}\left(1-|z_k|^2\right)\\
&&=-\frac{\pi}{z_k}\Theta(|z_k|-1)
-\pi z^{*}_k \Theta(1-|z_k|).
\end{eqnarray}
Using the base integrals one can express the basis function as
\begin{eqnarray}
\Phi^{\newtext{(2_{-})}}_{iso}(\br)=&&\frac{\sigma^{(2)}_{-} e^{i\chi^{(2)}_{-}}I^2}{(2\pi\sigma)^3}\sum_{\mu\nu}
\left(\begin{array}{cccc}
F'(z,z_S,\tilde{z}_S) & F'(z,z_S,\frac{1}{\tilde{z}^*_S}) & -F'(z,z_S,\tilde{z}_D) & -F'(z,z_S,\frac{1}{\tilde{z}^*_D}) \\
F'(z,z_S,\frac{1}{\tilde{z}^*_S}) & F'(z,\frac{1}{\tilde{z}^*_S},\frac{1}{z^{*}_S}) & -F'(z,\frac{1}{z^{*}_S},\tilde{z}_D) &
-F'(z,\frac{1}{z^{*}_S},\frac{1}{\tilde{z}^*_D})\\
-F'(z,z_S,\tilde{z}_D) & -F'(z,\frac{1}{z^{*}_S},\tilde{z}_D) & F'(z,z_D,\tilde{z}_D) & F'(z,z_D,\frac{1}{\tilde{z}^*_D}) \\
-F'(z,z_S,\frac{1}{\tilde{z}^*_D}) & -F'(z,\frac{1}{z^{*}_S},\frac{1}{\tilde{z}^*_D}) & F'(z,z_D,\frac{1}{\tilde{z}^*_D}) &
F'(z,\frac{1}{\tilde{z}^*_D},\frac{1}{z^{*}_D})
\end{array}
\right)_{\mu\nu}+c.c.
\end{eqnarray}
The general $F$-function is given by
\begin{eqnarray}
F(z,z_i,z_j)=\frac{\pi z^{*}}{(z-z_i)(z-z_j)}
-\frac{\frac{\pi}{z_i}\Theta(|z_i|-1)
+\pi z^{*}_i \Theta(1-|z_i|)}{(z-z_i)(z_i-z_j)}+
\frac{\frac{\pi}{z_j}\Theta(|z_j|-1)
+\pi z^{*}_j \Theta(1-|z_j|)}{(z-z_j)(z_i-z_j)},
\end{eqnarray}
where Heaviside $\Theta$-functions control whether the source/drain is located inside or outside of the unit disk sample. $F-$functions that depend on both source and drain are given by
\begin{eqnarray}
&&F(z,z_S,z_D)=\frac{\pi z^{*}}{(z-z_S)(z-z_D)}
-\frac{\pi z^{*}_S}{(z-z_S)(z_S-z_D)}+
\frac{\pi z^{*}_D}{(z-z_D)(z_S-z_D)}\\
&&F\left(z,z_S,\frac{1}{z^{*}_D}\right)=
\frac{\pi z^{*}}{(z-z_S)\left(z-\frac{1}{z^{*}_D}\right)}
-\frac{\pi z^{*}_S}{(z-z_S)\left(z_S-\frac{1}{z^{*}_D}\right)}+
\frac{\pi z^{*}_D}{(z-\frac{1}{z^{*}_D})\left(z_S-\frac{1}{z^{*}_D}\right)}\\
&&F\left(z,\frac{1}{z^{*}_S},z_D\right)=
\frac{\pi z^{*}}{(z-z_D)\left(z-\frac{1}{z^{*}_S}\right)}
-\frac{\pi z^{*}_D}{(z-z_D)\left(z_D-\frac{1}{z^{*}_S}\right)}+
\frac{\pi z^{*}_S}{(z-\frac{1}{z^{*}_S})\left(z_D-\frac{1}{z^{*}_S}\right)}\\
&&F\left(z,\frac{1}{z^{*}_S},\frac{1}{z^{*}_D}\right)=
\frac{\pi z^{*}}{(z-\frac{1}{z^{*}_S})\left(z-\frac{1}{z^{*}_D}\right)}
-\frac{\pi z^{*}_S}{(z-
\frac{1}{z^{*}_S})\left(
\frac{1}{z^{*}_S}-\frac{1}{z^{*}_D}\right)}+
\frac{\pi z^{*}_D}{(z-\frac{1}{z^{*}_D})\left(
\frac{1}{z^{*}_S}-\frac{1}{z^{*}_D}\right)}.
\end{eqnarray}
When $z_S$ and $z_D$ are on the boundary or the unit disk, all four of the above functions are the same.

The $F$-functions that depend only on the position of source (or drain) experience divergence and have to be regularized by considering the source/drain being distributed. In our calculations we use a box distribution, see sketch in Fig. \ref{distr_sketch}. It also turns out that the potential at the boundary of the disk does depend on the shape of source/drain: we find different results for circularly-symmetric distribution and for arc distribution. In this work we focus only on arc distribution as it is the case realized in experiments. To regularize the divergent $F-$functions we consider two points $z_A, \tilde{z}_A$ with $A=S,D$ that belong to the distributed source/drain. 
We first match magnitudes of $z_A$ and $\tilde{z}_A$, then bring both to the boundary, and then match their phases. This would correspond to a source without circular symmetry (an arc of a circle). Taking those limits we obtain 
\begin{equation}
\begin{gathered}
F(z,z_A,\tilde{z}_A) = \frac{\pi z^{*}}{(z-z_A)(z-\tilde{z}_A)}
-\frac{\pi z^{*}_A }{(z-z_A)(z_A-\tilde{z}_A)}+
\frac{\pi \tilde{z}^*_A}{(z-\tilde{z}_A)(z_A-\tilde{z}_A)} = \\
= \frac{\pi z^{*}}{(z-z_A)(z-\tilde{z}_A)}
-\frac{\pi e^{-i \phi_A} }{(z-z_A)(e^{i \phi_A} - e^{i \tilde{\phi}_A})}+
\frac{\pi e^{-i \tilde{\phi}_A}}{(z-\tilde{z}_A)(e^{i \phi_A} - e^{i \tilde{\phi}_A})} 
\rightarrow \frac{\pi z^{*}}{(z-z_A)^2} + \frac{\pi e^{- 2i \phi_A} }{(z-z_A)} - \frac{\pi z^{*}_A}{(z-z_A)^2}; \\
F\left(z,z_A,\frac{1}{\tilde{z}_A^*}\right)=\frac{\pi z^*}{(z-z_A)(z-\frac{1}{\tilde{z}_A^*})}
-\frac{\pi z_A^* }{(z-z_A)(z_A-\frac{1}{\tilde{z}_A^*})}+
\frac{\pi \tilde{z}_A^*}{(z-\frac{1}{\tilde{z}_A^*})(z_A-\frac{1}{\tilde{z}_A^*})} \rightarrow
\frac{\pi z^*}{(z-z_A)^2} + \frac{\pi \left( z_A^* \right)^2 \left( -2 z_A + z  \right)}{(z-z_A)^2}; \\
F\left(z,\frac{1}{z_A^*},\frac{1}{\tilde{z}_A^*}\right)=
\frac{\pi z^*}{\left(z-\frac{1}{z_A^*}\right)^2}
-\frac{\pi z_A^*}{\left(z-\frac{1}{z_A^*}\right)\left(\frac{1}{z_A^*}-\frac{1}{\tilde{z}_A^*}\right)}+
\frac{\pi \tilde{z}_A^*}{\left(z-\frac{1}{\tilde{z}_A^*}\right)\left(\frac{1}{z_A^*}-\frac{1}{\tilde{z}_A^*}\right)} \rightarrow \frac{\pi \left( z^* -2 z_A^* + z (z_A^*)^2 \right)}{\left(z-\frac{1}{z_A^*}\right)^2},
\end{gathered}
\end{equation}
all three results are identical when $z_A$ is at the boundary. The basis function then reads
\begin{equation}
\begin{gathered}
\Phi^{\newtext{(2_{-})}}_{iso}(\br)=
\frac{\sigma^{(2)}_{-} e^{i\chi^{(2)}_{-}}I^2}{(2\sigma)^3\pi^2}
\biggr(\sum_{A=S,D}\left[\frac{4}{z z_A^2}
\right]
-8\left(\frac{ z^{*}}{(z-z_S)(z-z_D)}
+\frac{z^{*}_S}{(z_S-z)(z_S-z_D)}+
\frac{z^{*}_D}{(z_D-z)(z_D-z_S)}\right)
\biggr)+c.c.
\end{gathered}
\label{eq:phi21iso_different_limit}
\end{equation}

\subsection{$\sigma^{(2)}_{+}$ basis function (non-divergent $C_3$ breaking term)}

Consider the second basis function given by
\begin{eqnarray}
&&\Phi^{\newtext{(2_{+})}}_{iso}(\br)=\frac{1}{\sigma}\int_{\Omega}d^2\br' \left(\frac{\partial}{\partial
z'}G_{N}(\br,\br')\right)\left( \sigma^{(2)}_{+} e^{i\chi^{(2)}_{+}}\left(E^{\newtext{(1)}}_x (\br')+iE^{\newtext{(1)}}_y (\br') \right)^2 \right) + c.c. = \\
&&=\frac{1}{\sigma}\int_{\Omega}d^2\br' \left( -\frac{1}{4\pi}\left(\frac{2}{z'-z}\right) \right)\left( \sigma^{(2)}_{+} e^{i\chi^{(2)}_{+}} \frac{I^2}{4\pi^2 \sigma^2}\left( \frac{1}{z^{'*}-z^{*}_S}
+\frac{1}{z^{'*}-\frac{1}{z_S}}-\frac{1}{z^{'*}-z^{*}_D}-\frac{1}{z^{'*}-\frac{1}{z_D}} \right)^2 \right) + c.c. = \\
&&= -\frac{\sigma^{(2)}_{+} e^{i\chi^{(2)}_{+}} I^2}{8\pi^3\sigma^3} \int_{\Omega}d^2\br' \left(\frac{1}{z'-z}\right) \left(  \frac{1}{z^{'*}-z^{*}_S}
+\frac{1}{z^{'*}-\frac{1}{z_S}}-\frac{1}{z^{'*}-z^{*}_D}-\frac{1}{z^{'*}-\frac{1}{z_D}} \right)^2 + c.c.
\end{eqnarray}
As for the $\sigma^{(2)}_{-}$-term, the basis function can be expressed through one base integral
\begin{eqnarray}
\frac{\sigma^{(2)}_{+} e^{i\chi^{(2)}_{+}}I^2}{(2\pi\sigma)^3}\int_\Omega d^2\br'\frac{1}{z-z'}\frac{1}{z^{'*}-z^{*}_i}\frac{1}{z^{'*}-z_j}.
\end{eqnarray}
We place $z$ on the edge of the unit circle and evaluate the base integral:
\begin{eqnarray}
&&\frac{\sigma^{(2)}_{+} e^{i\chi^{(2)}_{+}}I^2}{(2\pi\sigma)^3}\int_\Omega d^2\br'\frac{1}{z-z'}\frac{1}{z^{'*}-z^{*}_i}\frac{1}{z^{'*}-z_j}=
-\frac{\sigma^{(2)}_{+} e^{i\chi^{(2)}_{+}}I^2}{(2\pi\sigma)^3}\int_\Omega d^2\br'\frac{1}{z'-z}\frac{1}{z^{'*}-z^{*}_i}\frac{1}{z^{'*}-z^{*}_j}=\\
&&-\frac{\sigma^{(2)}_{+} e^{i\chi^{(2)}_{+}}I^2}{(2\pi\sigma)^3}\frac{1}{z^{*}_i-z^{*}_j}\int_\Omega d^2\br'\frac{1}{z'-z}
\left(
\frac{1}{z^{'*}-z^{*}_i}-\frac{1}{z^{'*}-z^{*}_j}\right)=
-\frac{\sigma^{(2)}_{+} e^{i\chi^{(2)}_{+}}I^2}{(2\pi\sigma)^3}\frac{1}{z^{*}_i-z^{*}_j}\int_0^1 d\rho \rho \oint \frac{d\zeta}{i\zeta}
\frac{1}{\rho\zeta-z}
\left(
\frac{1}{\frac{\rho}{\zeta}-z^{*}_i}-\frac{1}{\frac{\rho}{\zeta}-z^{*}_j}\right)=\nonumber\\
&&-\frac{\sigma^{(2)}_{+} e^{i\chi^{(2)}_{+}}I^2}{(2\pi\sigma)^3}\frac{1}{z^{*}_i-z^{*}_j}\int_0^1 d\rho \oint \frac{d\zeta}{i}
\frac{1}{\zeta-\frac{z}{\rho}}
\left(
\frac{1}{\rho-z^{*}_i \zeta}-\frac{1}{\rho-z^{*}_j \zeta}\right)=
\frac{\sigma^{(2)}_{+} e^{i\chi^{(2)}_{+}}I^2}{(2\pi\sigma)^3}\frac{1}{z^{*}_i-z^{*}_j}\int_0^1 d\rho \oint \frac{d\zeta}{i}
\frac{1}{\zeta-\frac{z}{\rho}}
\left(\frac{1}{z^{*}_i}
\frac{1}{\zeta-\frac{\rho}{z^{*}_i}}-\frac{1}{z^{*}_j}\frac{1}{\zeta-\frac{\rho}{z^{*}_j}}\right)\nonumber
\\
&&=\frac{\sigma^{(2)}_{+} e^{i\chi^{(2)}_{+}}I^2}{(2\pi\sigma)^3}\frac{2\pi}{z^{*}_i-z^{*}_j}\int_0^1 d\rho
\left(\frac{\Theta(|z_i|-\rho)}{z^{*}_i}\frac{1}{\frac{\rho}{z^{*}_i}-\frac{z}{\rho}}
-\frac{\Theta(|z_j|-\rho)}{z^{*}_j}\frac{1}{\frac{\rho}{z^{*}_j}-\frac{z}{\rho}}\right)\nonumber
\\
&&=\frac{\sigma^{(2)}_{+} e^{i\chi^{(2)}_{+}}I^2}{(2\pi\sigma)^3}
\left(\frac{2\pi}{z^{*}_i-z^{*}_j}
\frac{\Theta(|z_i|-1)}{z^{*}_i}
\int_0^1
\frac{d\rho}{\frac{\rho}{z^{*}_i}-\frac{z}{\rho}}
-\frac{2\pi}{z^{*}_i-z^{*}_j}\frac{\Theta(|z_j|-1)}{z^{*}_j}
\int_0^1 \frac{d\rho}{\frac{\rho}{z^{*}_j}-\frac{z}{\rho}}\right)+\nonumber\\
&&+\frac{\sigma^{(2)}_{+} e^{i\chi^{(2)}_{+}}I^2}{(2\pi\sigma)^3}
\left(\frac{2\pi}{z^{*}_i-z^{*}_j}
\frac{\Theta(1-|z_i|)}{z^{*}_i}
\int_0^{|z_i|}
\frac{d\rho}{\frac{\rho}{z^{*}_i}-\frac{z}{\rho}}
-\frac{2\pi}{z^{*}_i-z^{*}_j}\frac{\Theta(1-|z_j|)}{z^{*}_j}
\int_0^{|z_j|} \frac{d\rho}{\frac{\rho}{z^{*}_j}-\frac{z}{\rho}}\right)
\nonumber
\\
&&=\frac{\sigma^{(2)}_{+} e^{i\chi^{(2)}_{+}}I^2}{(2\pi\sigma)^3}
\left(\frac{2\pi}{z^{*}_i-z^{*}_j}
\Theta(|z_i|-1)
\int_0^1
\frac{d\rho\rho}{\rho^2-zz^{*}_i}
-\frac{2\pi}{z^{*}_i-z^{*}_j}\Theta(|z_j|-1)
\int_0^1 \frac{d\rho \rho}{\rho^2-z^{*}_jz}\right)+\nonumber\\
&&+\frac{\sigma^{(2)}_{+} e^{i\chi^{(2)}_{+}}I^2}{(2\pi\sigma)^3}
\left(\frac{2\pi}{z^{*}_i-z^{*}_j}
\Theta(1-|z_i|)
\int_0^{|z_i|}
\frac{d\rho\rho}{\rho^2-zz^{*}_i}
-\frac{2\pi}{z^{*}_i-z^{*}_j}\Theta(1-|z_j|)
\int_0^{|z_j|} \frac{d\rho\rho}{\rho^2-z^{*}_jz}\right)
\nonumber
\\
&&=\frac{\sigma^{(2)}_{+} e^{i\chi^{(2)}_{+}}I^2}{(2\pi\sigma)^3}
\left(\frac{\pi}{z^{*}_i-z^{*}_j}
\Theta(|z_i|-1)\left(\ln[1-zz^{*}_i]-\ln[-zz^{*}_i]\right)
-\frac{\pi}{z^{*}_i-z^{*}_j}\Theta(|z_j|-1)
\left(\ln[1-zz^{*}_j]-\ln[-zz^{*}_j]\right)
\right)+\nonumber\\
&&+\frac{\sigma^{(2)}_{+} e^{i\chi^{(2)}_{+}}I^2}{(2\pi\sigma)^3}
\left(\frac{\pi}{z^{*}_i-z^{*}_j}
\Theta(1-|z_i|)\left(\ln[|z_i|^2-zz^{*}_i]-\ln[-zz^{*}_i]\right)
-\frac{\pi}{z^{*}_i-z^{*}_j}\Theta(1-|z_j|)
\left(\ln[|z_j|^2-zz^{*}_j]-\ln[-zz^{*}_j]\right)
\right)
\nonumber
\\
&&\equiv\frac{\sigma^{(2)}_{+} e^{i\chi^{(2)}_{+}}I^2}{(2\pi\sigma)^3}F'\left(z,z_i,z_j\right),
\end{eqnarray}
where
\begin{eqnarray}
F'\left(z,z_i,z_j\right)&=&\frac{\pi}{z^{*}_i-z^{*}_j}\left(\ln\left[-zz^{*}_j\right]-\ln\left[-zz^{*}_i\right]
\right)\\
&+&\frac{\pi}{z^{*}_i-z^{*}_j}\left(
\Theta\left(|z_i|-1\right)\ln\left[1-zz^{*}_i\right]+\Theta\left(1-|z_i|\right)\ln\left[|z_i|^2-zz^{*}_i\right]
\right)\\
&-&\frac{\pi}{z^{*}_i-z^{*}_j}\left(
\Theta\left(|z_j|-1\right)\ln\left[1-zz^{*}_j\right]+
\Theta\left(1-|z_j|\right)\ln\left[|z_j|^2-zz^{*}_j\right]
\right)
\end{eqnarray}
This allows us to express the second basis function as
\begin{eqnarray}
\Phi^{\newtext{(2_{+})}}_{iso}(\br)=&&\frac{\sigma^{(2)}_{+} e^{i\chi^{(2)}_{+}}I^2}{(2\pi\sigma)^3}\sum_{\mu\nu}
\left(\begin{array}{cccc}
F'(z,z_S,\tilde{z}_S) & F'(z,z_S,\frac{1}{\tilde{z}^*_S}) & -F'(z,z_S,\tilde{z}_D) & -F'(z,z_S,\frac{1}{\tilde{z}^*_D}) \\
F'(z,z_S,\frac{1}{\tilde{z}^*_S}) & F'(z,\frac{1}{\tilde{z}^*_S},\frac{1}{z^{*}_S}) & -F'(z,\frac{1}{z^{*}_S},\tilde{z}_D) &
-F'(z,\frac{1}{z^{*}_S},\frac{1}{\tilde{z}^*_D})\\
-F'(z,z_S,\tilde{z}_D) & -F'(z,\frac{1}{z^{*}_S},\tilde{z}_D) & F'(z,z_D,\tilde{z}_D) & F'(z,z_D,\frac{1}{\tilde{z}^*_D}) \\
-F'(z,z_S,\frac{1}{\tilde{z}^*_D}) & -F'(z,\frac{1}{z^{*}_S},\frac{1}{\tilde{z}^*_D}) & F'(z,z_D,\frac{1}{\tilde{z}^*_D}) &
F'(z,\frac{1}{\tilde{z}^*_D},\frac{1}{z^{*}_D})
\end{array}
\right)_{\mu\nu}+c.c.
\end{eqnarray}
\begin{eqnarray}
F'\left(z,z_i,z_j\right)&=&\frac{\pi}{z^{*}_i-z^{*}_j}\left(\ln\left[-zz^{*}_j\right]-\ln\left[-zz^{*}_i\right]
\right)\\
&+&\frac{\pi}{z^{*}_i-z^{*}_j}\left(
\Theta\left(|z_i|-1\right)\ln\left[1-zz^{*}_i\right]+\Theta\left(1-|z_i|\right)\ln\left[|z_i|^2-zz^{*}_i\right]
\right)\\
&-&\frac{\pi}{z^{*}_i-z^{*}_j}\left(
\Theta\left(|z_j|-1\right)\ln\left[1-zz^{*}_j\right]+
\Theta\left(1-|z_j|\right)\ln\left[|z_j|^2-zz^{*}_j\right]
\right)
\end{eqnarray}
As in the previous case, $F-$functions that depend solely on positions of source/drain need to be regularized. We use the same regularization for arc-shaped contacts: we first match magnitudes of $z_A$ and $\tilde{z}_A$, then take $z_A$ and $\tilde{z}_A$ to the boundary, and then finally match their phases, to obtain
\begin{equation}
    \begin{gathered}
        F'\left(z,z_A,\tilde{z}_A\right)=\frac{\pi}{z^{*}_A-\tilde{z}^*_A}\left(\ln\left[-z\tilde{z}^*_A\right]-\ln\left[-zz^{*}_A\right]
\right)+\frac{\pi}{z^{*}_A-\tilde{z}^*_A}\left(
\ln\left[|z_A|^2-zz^{*}_A\right]-\ln\left[|\tilde{z}_A|^2-z\tilde{z}^*_A\right]
\right) = \\ 
= \frac{\pi}{z^{*}_A-\tilde{z}^*_A} \ln\left[\frac{\tilde{z}^*_A}{z^{*}_A}\right]
+\frac{\pi}{z^{*}_A-\tilde{z}^*_A} 
\ln\left[\frac{|z_A|^2-zz^{*}_A}{|\tilde{z}_A|^2-z\tilde{z}^*_A}\right] = 
\frac{\pi}{z^{*}_A-\tilde{z}^*_A} \ln\left[\frac{\tilde{z}^*_A}{z^{*}_A}\right]
+\frac{\pi}{z^{*}_A-\tilde{z}^*_A} 
\ln\left[\frac{z^{*}_A(z_A-z)}{\tilde{z}^*_A(\tilde{z}_A-z)}\right] = \\=
\frac{\pi}{z^{*}_A-\tilde{z}^*_A} 
\ln\left[\frac{(z_A-z)}{(\tilde{z}_A-z)}\right] = 
\frac{\pi}{|z_A|\left(e^{-i \phi_A}-e^{-i \tilde{\phi}_A}\right)} 
\ln\left[\frac{|z_A| e^{i \phi_A}-z}{|z_A| e^{i \tilde{\phi}_A}-z}\right] = \\ = 
\frac{\pi}{|z_A|\left(e^{-i \phi_A}-e^{-i \tilde{\phi}_A}\right)} 
\ln\left[1 + \frac{|z_A| e^{i \phi_A} - |z_A| e^{i \tilde{\phi}_A}}{|z_A| e^{i \tilde{\phi}_A}-z}\right] = \frac{\pi}{|z_A|\left(e^{-i \phi_A}-e^{-i \tilde{\phi}_A}\right)} 
\ln\left[1 + \frac{|z_A| (e^{i \phi_A} - e^{i \tilde{\phi}_A})}{|z_A| e^{i \tilde{\phi}_A}-z}\right] \simeq \\ \simeq
\frac{\pi}{|z_A|\left(e^{-i \phi_A}-e^{-i \tilde{\phi}_A}\right)} 
 \frac{|z_A| (e^{i \phi_A} - e^{i \tilde{\phi}_A})}{|z_A| e^{i \tilde{\phi}_A}-z} \rightarrow  \frac{\pi e^{2 i \phi_A}}{z - z_A}
    \end{gathered}
\end{equation}

\begin{equation}
    \begin{gathered}
        F'\left(z,z_A,\frac{1}{\tilde{z}^*_A}\right)=\frac{\pi}{z^*_A-\frac{1}{\tilde{z}_A}}\left(\ln\left[-\frac{z}{\tilde{z}_A}\right]-\ln\left[-zz^*_A\right]
+\ln\left[|z_A|^2-zz^*_A\right]-\ln\left[1-\frac{z}{\tilde{z}_A}\right]\right) = \\ =
\frac{\pi}{z^*_A-\frac{1}{\tilde{z}_A}}\left(\ln\left[\frac{1}{\tilde{z}_A z^*_A}\right]
+\ln\left[z^*_A(z_A-z)\right]-\ln\left[\frac{\tilde{z}_A - z}{\tilde{z}_A}\right]\right) = 
\frac{\pi}{z^*_A-\frac{1}{\tilde{z}_A}}\left(\ln\left[\frac{\tilde{z}_A z^*_A(z_A-z)}{\tilde{z}_A z^*_A (\tilde{z}_A - z)}\right]\right) = \\ =
\frac{\pi}{z^*_A-\frac{1}{\tilde{z}_A}}\ln\left[\frac{z_A - \tilde{z}_A + \tilde{z}_A-z}{\tilde{z}_A - z}\right] \simeq \frac{\pi}{z^*_A-\frac{1}{\tilde{z}_A}} \frac{z_A - \tilde{z}_A}{\tilde{z}_A - z} = \frac{\pi |z_A|}{|z_A|e^{- i \phi_A}-\frac{e^{- i \tilde{\phi}_A}}{|z_A|}} \frac{e^{i \phi_A} - e^{i \tilde{\phi}_A}}{\tilde{z}_A - z} = \\ = \frac{\pi }{e^{- i \phi_A}-e^{- i \tilde{\phi}_A}} \frac{e^{i \phi_A} - e^{i \tilde{\phi}_A}}{\tilde{z}_A - z} \rightarrow \frac{\pi e^{2 i \phi_A}}{z - z_A}
    \end{gathered}
\end{equation}

\begin{equation}
    \begin{gathered}
        F'\left(z,\frac{1}{z_A^*},\frac{1}{\tilde{z}^*_A}\right)=
        \frac{\pi}{\frac{1}{z_A}-\frac{1}{\tilde{z}_A}}\left(\ln\left[-\frac{z}{\tilde{z}_A}\right]-\ln\left[-\frac{z}{z_A}\right]
+\ln\left[1-\frac{z}{z_A}\right]-\ln\left[1-\frac{z}{\tilde{z}_A}\right]
\right) = \\ = 
\frac{\pi}{\frac{1}{z_A}-\frac{1}{\tilde{z}_A}} \ln\left[\frac{z_A - z}{\tilde{z}_A -z}\right] \simeq \frac{\pi}{\frac{1}{z_A}-\frac{1}{\tilde{z}_A}} \frac{z_A - \tilde{z}_A}{\tilde{z}_A -z} = \frac{\pi z_A \tilde{z}_A}{\tilde{z}_A-z_A} \frac{z_A - \tilde{z}_A}{\tilde{z}_A -z} \to \frac{\pi e^{2 i \phi_A}}{z-z_A}
    \end{gathered}
\end{equation}
Therefore, the second basis function reads
\begin{eqnarray}\label{eq:phi22iso_different_limit}
\Phi^{\newtext{(2_{+})}}_{iso}(\br)=&&\frac{\sigma^{(2)}_{+} e^{i\chi^{(2)}_{+}}I^2}{(2\pi\sigma)^3}
\left(\frac{4\pi z^2_S}{z-z_S}+\frac{4\pi z^2_D}{z-z_D}-\frac{8\pi}{z^*_S-z^*_D}\left(
\ln\left[-zz^*_D\right]-\ln\left[-zz^*_S\right]+\ln\left[1-zz^*_S\right]-\ln\left[1-zz^*_D\right]
\right)
\right)+c.c.\nonumber\\
\end{eqnarray}

\subsection{A simple proof of logarithmic divergence of the basis function for the $\sigma^{(2)}_{0}$ (divergent $C_3$ breaking) term}

To show the presence of logarithmic divergence we split the gradient of the Green's function into a singular at source part and into a regular part:
\begin{equation}
    \nabla_{\br'}G_{N}(\br, \br') = B(\br, \br') = B(\br, \br_S') + \left( B(\br, \br') - B(\br, \br_S') \right).
\end{equation}
The part in the brackets is regular as the singularity at $\br_S$ is subtracted from the full expression; the first term contains the singularity, however, one can pull it out of the integral as it doesn't depend on the integration variable anymore. So, let us consider a base integral 
\begin{equation}
    \begin{gathered}
        -\int_{\Omega} d^2 \br' \frac{1}{z'-z_S} \frac{1}{{z'}^* - z_S^*} \frac{1}{z' - z} = -\int_{\Omega} d^2 \br' \frac{1}{z'-z_S} \frac{1}{{z'}^* - z_S^*} \left( \frac{1}{z_S - z} +  \frac{1}{z' - z} - \frac{1}{z_S - z} \right)
    \end{gathered}
\end{equation}
and focus only on the singular part:
\begin{equation}
    \begin{gathered}
       -\int_{\Omega} d^2 \br' \frac{1}{z'-z_S} \frac{1}{{z'}^* - z_S^*}  \frac{1}{z_S - z} = - \frac{1}{z_S - z} \int_{\Omega} d^2 \br' \frac{1}{z'-z_S} \frac{1}{{z'}^* - z_S^*} = - \frac{1}{z_S - z} \int_{\Omega} d^2 \br' \frac{1}{|z'-z_S|^2}
    \end{gathered}
\end{equation}
We now puncture a disk around $z_S$ and split the integration region in two parts: (i) a tiny annulus $An$ with an outer radius $\lambda_{out}$ being the radius of the disk surrounding $z_S$ and the inner radius $\lambda_{in}$ being our small parameter $\lambda$, which we eventually will take to zero, and (ii) the rest of the initial unit disk: $\Omega / An $. Then the integral over  $\Omega / An $ gives a correction to the regular part as long as the outer radius of the annulus is finite. For the integral over the annulus we shift the origin of coordinates to $z_S$ and obtain
\begin{equation}
    \begin{gathered}
       - \frac{1}{z_S - z} \int_{An} d^2 \br' \frac{1}{|z'-z_S|^2} = 
       - \frac{1}{z_S - z} \int_{An} d^2 \tilde{\br}' \frac{1}{(\tilde{\br}')^2} = 
       - \frac{2 \pi}{z_S - z} \ln \left( \frac{\lambda_{out}}{\lambda_{in}} \right) = - \frac{2 \pi}{z_S - z} \ln \lambda_{out} + \frac{2 \pi}{z_S - z} \ln \lambda_{in},
    \end{gathered}
\end{equation}
where the log with $\lambda_{out}$ just corrects the regular part and the second log is divergent in the limit of $\lambda_{in} \rightarrow 0$. 

In the next subsection we derive the base integrals and select the regular part of the basis function for the term $\sigma^{(2)}_{0}$.

\subsection{Basis function for the $\sigma^{(2)}_{0}$ term (divergent $C_3$ breaking)}

We finally proceed to the last basis function in the isotropic linear conductivity case. The $\sigma^{(2)}_{0}$ basis function reads
\begin{eqnarray}
&&\Phi^{\newtext{(2_0)}}_{iso}(\br)=
\frac{1}{\sigma}\int_{\Omega}d^2\br' \left[\left(\frac{\partial}{\partial
z'}G_{N}(\br,\br')\right)\left( \sigma^{(2)}_{0} e^{i\chi^{(2)}_{0}}\left({E^{\newtext{(1)}}_x}^2+{E^{\newtext{(1)}}_y}^2\right) \right)+c.c.
\right],
\end{eqnarray}
where
\begin{eqnarray}
&&E^{\newtext{(1)}}_x+iE^{\newtext{(1)}}_y=\frac{I}{2\pi\sigma}\left(\frac{1}{z^*-z^*_S}
+\frac{1}{z^*-\frac{1}{z_S}}-\frac{1}{z^*-z^*_D}-\frac{1}{z^*-\frac{1}{z_D}}
\right)\\
&&E^{\newtext{(1)}}_x-iE^{\newtext{(1)}}_y
=\frac{I}{2\pi\sigma}\left(\frac{1}{z-z_S}
+\frac{1}{z-\frac{1}{z^*_S}}-\frac{1}{z-z_D}-\frac{1}{z-\frac{1}{z^*_D}}
\right)\\
&&\frac{\partial}{\partial z'}G_N(\br,\br')=-\frac{1}{4\pi}\left(\frac{1}{z'-z}+\frac{1}{z'-\frac{1}{z^{*}}}\right).
\end{eqnarray}
As in the two previous cases, we put $z$ on the boundary of the disk and define the base integral
\begin{eqnarray}
F''(z,z_i,z_j)&=&\int_\Omega d^2\br'\frac{1}{z-z'}\frac{1}{z'-z_i}\frac{1}{z'^*-z^*_j}.
\label{F''base_integral}
\end{eqnarray}
While in the above $z$ is on the boundary of the disk, $z_i$ and $z_j$ can be inside or outside of the disk. Then the basis function is given by the following expression
\begin{eqnarray}
\Phi^{\newtext{(2_0)}}_{iso}(\br)=&&\frac{\sigma^{(2)}_{0} e^{i\chi^{(2)}_{0}}I^2}{(2\pi\sigma)^3}\sum_{\mu\nu}
\left(\begin{array}{cccc}
F''(z,z_S,\tilde{z}_S) & F''(z,z_S,\frac{1}{\tilde{z}^*_S}) & -F''(z,z_S,\tilde{z}_D) & -F''(z,z_S,\frac{1}{\tilde{z}^*_D}) \\
F''(z,\frac{1}{z^*_S},\tilde{z}_S) & F''(z,\frac{1}{z^*_S},\frac{1}{\tilde{z}^*_S}) & -F''(z,\frac{1}{z^*_S},\tilde{z}_D) &
-F''(z,\frac{1}{z^*_S},\frac{1}{\tilde{z}^*_D})\\
-F''(z,z_D,\tilde{z}_S) & -F''(z,z_D,\frac{1}{\tilde{z}^*_S}) & F''(z,z_D,\tilde{z}_D) & F''(z,z_D,\frac{1}{\tilde{z}^*_D}) \\
-F''(z,\frac{1}{z^*_D},\frac{1}{\tilde{z}^*_D}) & -F''(z,\frac{1}{z^*_D},\frac{1}{\tilde{z}^*_S}) & F''(z,\frac{1}{z^*_D},\tilde{z}_D) &
F''(z,\frac{1}{z^*_D},\frac{1}{\tilde{z}^*_D})
\end{array}
\right)_{\mu\nu}+c.c.
\end{eqnarray}
We evaluate the base integral for arbitrary $i,j$:
\begin{eqnarray}
F''(z,z_i,z_j)&=&\int_{|z_i|}^1\frac{d\rho \rho}{z_iz^*_j-\rho^2}\frac{2\pi \Theta(1-|z_i|)}{z_i-z}
-\int_0^1\frac{d\rho}{z_iz^*_j-\rho^2}\frac{2\pi\Theta(|z_j|-1)}{\frac{\rho}{z^*_j}-\frac{z}{\rho}}
-\int_0^{|z_j|}\frac{d\rho}{z_iz^*_j-\rho^2}\frac{2\pi \Theta(1-|z_j|)}{\frac{\rho}{z^*_j}-\frac{z}{\rho}}\\
&=&\frac{\pi \Theta(1-|z_i|)}{z-z_i}
\left(\ln\left[1-z_iz^*_j\right]-\ln\left[|z_i|^2-z_iz^*_j\right]\right)\\
&+&z^*_j\pi\Theta(|z_j|-1) \int_0^1\frac{dx}{(x-z_iz^*_j)}\frac{1}{(x-zz^*_j)}
+z^*_j\pi \Theta(1-|z_j|)\int_0^{|z_j|^2}\frac{dx}{(x-z_iz^*_j)}\frac{1}{(x-zz^*_j)}\\
&=&\frac{\pi \Theta(1-|z_i|)}{z-z_i}
\left(\ln\left[1-z_iz^*_j\right]-\ln\left[|z_i|^2-z_iz^*_j\right]\right)\nonumber\\
&+&\frac{\pi\Theta(|z_j|-1)}{z_i-z}\left(\ln\left[1-z_iz^*_j\right]-\ln\left[-z_iz^*_j\right]
-\ln\left[1-zz^*_j\right]+\ln\left[-zz^*_j\right]\right)\nonumber\\
&+&\frac{\pi \Theta(1-|z_j|)}{z_i-z}\left(
\ln\left[|z_j|^2-z_iz^*_j\right]-\ln\left[-z_iz^*_j\right]-\ln\left[|z_j|^2-zz^*_j\right]+\ln\left[-zz^*_j\right]
\right).
\label{F''base_full}
\end{eqnarray}
Again, the base integral is divergent for $i=j=S,D$, therefore, regularization is needed.

We will select the regular contribution to the basis function. To do so we substitute $z_{i} = z_A, 1/z_A^*$ and $z_j = \tilde{z}_A, 1/\tilde{z}_A^*$ into Eq. \eqref{F''base_integral} and keep $z_A, \tilde{z}_A$ explicitly inside the disk, and then collect only the terms that are divergent in the limit of $\tilde{z}_A \rightarrow z_A$ and $|z_A| \rightarrow 1$: 
\begin{eqnarray} \label{F''base_singular}
    F''(z,z_A,\tilde{z}_A)_{singular} = \frac{\pi}{z-z_A}  \left[ \ln \left( 1 - z_A \tilde{z}_A^*  \right) - \ln \left( |z_A|^2 - z_A \tilde{z}_A^*  \right) -  \ln \left( |\tilde{z}_A^*|^2 - z_A \tilde{z}_A^* \right) \right], \\
    F''(z,z_A,1/\tilde{z}_A^*)_{singular} = - \frac{\pi}{z-z_A}   \ln \left( |z_A|^2 - z_A /\tilde{z}_A  \right), \\
    F''(z,1/z_A^*,\tilde{z}_A)_{singular} = - \frac{\pi}{z-1/z_A^*}   \ln \left( 1 - \frac{\tilde{z}_A^*}{ z_A^* } \right), \\
    F''(z,1/z_A^*,1/\tilde{z}_A^*)_{singular} =  - \frac{\pi}{z-1/z_A^*}   \ln \left( 1 - \frac{1}{ z_A^* \tilde{z}_A } \right),
\end{eqnarray}
We can see that every term above diverges either for $\tilde{z}_A \rightarrow z_A$, or for $|z_A| \rightarrow 1$. 
Now we can subtract our result \eqref{F''base_singular} from \eqref{F''base_full} to obtain the regular part. Note that this only applies to terms involving only $z_S$ or only $z_D$ but not the cross terms, as they don't experience any singularities. Subtracting the singular part from the full expression and substituting $z_A=z_S, z_D$ inside the disk we obtain 
\begin{equation}
    \begin{gathered}
        F''(z,z_A,\tilde{z}_A)_{regular} = F''(z,z_A,\tilde{z}_A) - F''(z,z_A,\tilde{z}_A)_{singular} = \frac{\pi}{z-z_A} \left[ \ln \left( - z_A \tilde{z}_A^* \right) + \ln \left( |\tilde{z}_A|^2 - z \tilde{z}_A^* \right) - \ln \left( - z \tilde{z}_A^* \right) \right], \\
        F''(z,z_A,1/\tilde{z}_A^*)_{regular} = \frac{\pi}{z-z_A} \left[ \ln \left( \frac{-z_A}{\tilde{z}_A} \right) + \ln \left( 1-  \frac{z}{\tilde{z}_A} \right) - \ln \left( \frac{-z}{\tilde{z}_A} \right) \right], \\
        F''(z,1/z_A^*,\tilde{z}_A^*)_{regular} = \frac{\pi}{z-1/z_A^*} \left[ \ln \left( \frac{-  \tilde{z}_A^*}{z_A^*} \right) + \ln \left( 1 - z \tilde{z}_A^* \right) - \ln \left( - z \tilde{z}_A^* \right) \right], \\
        F''(z,1/z_A^*,1/\tilde{z}_A^*)_{regular} = \frac{\pi}{z-1/z_A^*} \left[ \ln \left( \frac{ -1}{z_A^* \tilde{z}_A} \right) + \ln \left( 1 - \frac{ z}{\tilde{z}_A} \right) -  \ln \left( \frac{-z}{\tilde{z}_A} \right) \right].
    \end{gathered}
\end{equation}
We consider the limit of $|\tilde{z}_A| \rightarrow |z_A|$ first and then take $|z_A| \rightarrow 1$. Then we get
\begin{equation}
    \begin{gathered}
        F''(z,z_A,\tilde{z}_A)_{regular} = F''(z,z_A,1/\tilde{z}_A^*)_{regular} = F''(z,1/z_A^*,\tilde{z}_A^*)_{regular} = F''(z,1/z_A^*,1/\tilde{z}_A^*)_{regular} = \\ = \frac{\pi}{z-z_A} \left[ \ln \left( - z_A \tilde{z}_A^* \right) + \ln \left( |\tilde{z}_A|^2 - z \tilde{z}_A^* \right) - \ln \left( - z \tilde{z}_A^* \right) \right] = \\ = \frac{\pi}{z-z_A} \left[ \ln \left( - |z_A|^2 e^{i (\phi_A - \tilde{\phi}_A)} \right) + \ln \left( |z_A|^2 - z \tilde{z}_A^* \right) - \ln \left( - z \tilde{z}_A^* \right) \right] = \\ = \frac{\pi}{z-e^{i \phi_A}} \left[ \ln \left( - e^{i (\phi_A - \tilde{\phi}_A)} \right) + \ln \left( 1 - z e^{-i \tilde{\phi}_A} \right) - \ln \left( - z e^{-i \tilde{\phi}_A} \right) \right].
    \end{gathered}
\end{equation}  
We can also get the expression for the singular part of the basis function for the arc contact:
\begin{eqnarray} 
    F''(z,z_A,\tilde{z}_A)_{singular} = F''(z,z_A,1/\tilde{z}_A^*)_{singular} = \\ = F''(z,1/z_A^*,\tilde{z}_A^*)_{singular} = F''(z,1/z_A^*,1/\tilde{z}_A^*)_{singular}  
    = - \frac{\pi}{z - z_A} \ln \left( 1 - e^{i \left( \phi_A - \tilde{\phi}_A \right)} \right).
\end{eqnarray}
Now we will need to integrate both regular and singular contributions with some distribution function corresponding to arc width of the contact. Let us assume a box distribution function with width $\lambda$ (in experimental setting $\lambda$ is given by the angular width of a lead measured in radians). We calculate the sum of two integrals:
\begin{equation}
    \begin{gathered}
        \mathcal{L}_{\lambda}=\frac{1}{\lambda^2} \int_{-\frac{\lambda}{2}}^{\frac{\lambda}{2}} d\phi_A \int_{-\frac{\lambda}{2}}^{\frac{\lambda}{2}} d \tilde{\phi}_A \left[ \ln \left( 1 - e^{i \left( \phi_A - \tilde{\phi}_A \right)} \right) - \ln \left( - e^{i \left( \phi_A - \tilde{\phi}_A \right)} \right) \right] = \\ = \frac{1}{\lambda^2} \left[ \frac{\lambda}{2} \left( \lambda - i \ln \left( - e^{- i \lambda} \right) + i \ln \left( - e^{i \lambda} \right)  \right) \left( \ln \left( - e^{- i \lambda} \right) + \ln \left( - e^{i \lambda} \right) \right) + \mathrm{Li}_3 (e^{i \lambda}) + \mathrm{Li}_3 (e^{-i \lambda}) - 2 \zeta(3) \right].
    \end{gathered}
    \label{A2''arc_integrals}
\end{equation}
This expression is very well approximated by $-\frac{3}{2} + \ln (\lambda)$ in the range of $\lambda \sim 0.5$.
Before we proceed further we also write down the expression for $F''(z,z_S,z_D)$ using an explicit sum of logs:
\begin{equation}
    \begin{gathered}
        F''(z,z_S,z_D) = \frac{\pi}{z - z_S} \left[ \ln \left( 1 - z z_D^* \right) - \ln \left( - z z_D^* \right) - \ln \left( 1 - z_S z_D^* \right) + \ln \left( - z_S z_D^* \right) \right], \\
        F''(z,z_D,z_S) = \frac{\pi}{z - z_D} \left[ \ln \left( 1 - z z_S^* \right) - \ln \left( - z z_S^* \right) - \ln \left( 1 - z_D z_S^* \right) + \ln \left( - z_D z_S^* \right) \right].
    \end{gathered}
\end{equation}
where $z=e^{i \phi}, z_S=e^{i \phi_S}, z_D=e^{i \phi_D}$. The combinations with $1/z_D$ and $1/z_S$ give the same result.

Now we combine all terms together to get the final result for the full (both singular and regular parts) basis function for arc contact:
\begin{eqnarray}
\label{eq:phi23iso}
&&\Phi^{\newtext{(2^{arc}_0)}}_{iso}(\br)=\frac{\sigma^{(2)}_{0} e^{i\chi^{(2)}_{0}}I^2}{(2\pi\sigma)^3}
\left(
\left(\sum_{A=S,D}\frac{4\pi}{z_A-z}\left(\mathcal{L}_\lambda-\ln\left[1-\frac{z}{z_A}\right]+\ln\left[-\frac{z}{z_A}\right]\right)
\right)-\right.\\
&&\left.
\left(\frac{4\pi}{z_S-z}
\left(\ln\left[1-\frac{z_S}{z_D}\right]-\ln\left[-\frac{z_S}{z_D}\right]-\ln\left[1-\frac{z}{z_D}\right]+\ln\left[-\frac{z}{z_D}\right]\right)
+S\leftrightarrow D
\right)
\right)
+c.c.
\end{eqnarray}
Note, that this basis function \underline{does depend on the angular size of source/drain} -- a property that distinguishes it from the other two basis functions. It diverges logarithmically in the limit of valishingly small angular width of contacts. Therefore, one needs to know the size of contacts to properly extract the components of nonlinear tensor.

\section{Neumann Green's function for the ellipse in Zhukovsky's coordinates}

For the more general case of anisotropic linear conductivity, we will need the Neumann Green's function on an ellipse, as the in the anisotropic case the Poisson equation is defined on an ellipse, see \cite{Oskar_solo}. 

Let's try to determine the Green's function in this case.
To do so we proceed the same way as in the linear case: we define the coordinate system such that the coordinate axes coincide with the principal axes of the linear
conductivity tensor and then rescale the original $x-y$ variables into $X-Y$. This turns the boundary region from a circle to an ellipse and transforms the differential
operator into a Laplacian. Then the Green's function satisfies
\begin{eqnarray}
\left(\frac{\partial^2}{\partial^2 u'}+\frac{\partial^2}{\partial^2 v'}\right)G(u,v;u',v')&=&-\delta(u-u')\delta(v-v')
\end{eqnarray}
where (as a result of the Zhukovsky conformal mapping of an ellipse to an annulus \cite{Oskar_solo})
\begin{eqnarray}
w&=&u+iv\\
X+iY=Z&=&\alpha_+w+\frac{\alpha_-}{w}\\
X&=&\frac{x}{\sqrt{1+\frac{\Delta\sigma}{\bar\sigma}}}\\
Y&=&\frac{y}{\sqrt{1-\frac{\Delta\sigma}{\bar\sigma}}}\\
\Omega&=&\sqrt{\frac{\sqrt{1-\frac{\Delta \sigma}{\bar{\sigma}}}-\sqrt{1+\frac{\Delta \sigma}{\bar{\sigma}}}}{\sqrt{1-\frac{\Delta \sigma}{\bar{\sigma}}}+\sqrt{1+\frac{\Delta \sigma}{\bar{\sigma}}}}}.
\end{eqnarray}
Now, we switch to polar coordinates
\begin{eqnarray}
\rho&=&\sqrt{u^2+v^2}\\
\phi&=&\tan^{-1}\frac{v}{u}
\end{eqnarray}

A vector normal to the original circle (boundary of the disk-shaped sample) is $((1+\delta)x-x)\hat{\bx}+((1+\delta)y -y)\hat{\by}$, where $\hat{\bx}$ is a unit vector perpendicular to
the constant $x$ surface and $\hat{\by}$ is a unit vector perpendicular to the constant $y$ surface. Note that a perpendicular vector to the normal,
i.e. tangential to surface vector, is $\hat{\bz}\times((1+\delta)x-x)\hat{\bx}+((1+\delta)y -y)\hat{\by}=((1+\delta)x-x)\hat{\by}-((1+\delta)y
-y)\hat{\bx}$.
Moving by an infinitesimal distance along this tangential vector still places us on the circle.

Rescaling to $X,Y$ does not change the orientation of the unit vectors, but it does change the orientation of the normal vector to
$\be_1=\sqrt{1+\frac{\Delta\sigma}{\bar\sigma}}((1+\delta)X-X)\hat{\bx}+\sqrt{1-\frac{\Delta\sigma}{\bar\sigma}}((1+\delta)Y -Y)\hat{\by}$.
The original tangential vector in turn changes to
$\be_2=\sqrt{1+\frac{\Delta\sigma}{\bar\sigma}}((1+\delta)X-X)\hat{\by}-\sqrt{1-\frac{\Delta\sigma}{\bar\sigma}}((1+\delta)Y -Y)\hat{\bx}$. Clearly
we still have $\be_1\cdot\be_2=0$.
But since points on the circle map onto points on the ellipse, the $\be_2$ must move us along the edge of the ellipse.
So if we have a vector $\nabla \Phi$ and we take its scalar product with the normal of the circle in the $x-y$ coordinate system $\hat{n}\cdot\nabla
\Phi=\frac{\partial \Phi}{\partial n}$, we also take the normal to the ellipse derivative in the $X-Y$ system.

The Green's function for the Laplacian now reads
\begin{eqnarray}
\left(\frac{1}{\rho'}\frac{\partial}{\partial \rho'}\rho'\frac{\partial}{\partial \rho'}+\frac{1}{\rho'^2}\frac{\partial^2}{\partial^2
\phi'}\right)G(\rho,\phi;\rho',\phi')&=&-\frac{1}{\rho'}\delta(\rho-\rho')\delta(\phi-\phi')\\
\Rightarrow\left(\frac{\partial}{\partial \rho'}\rho'\frac{\partial}{\partial \rho'}+\frac{1}{\rho'}\frac{\partial^2}{\partial^2
\phi'}\right)G(\rho,\phi;\rho',\phi')&=&-\delta(\rho-\rho')\delta(\phi-\phi')
\end{eqnarray}
We need to distinguish the two regions, region 1: $\rho'<\rho$ and region 2: $\rho'>\rho$. In each region $G$ must satisfy the Laplace equation.
The boundary conditions in the region 1 are the same as in \cite{Oskar_solo}.

This restriction also applies in the region 2 because we can have $\rho'\rightarrow \Omega+2\eps$ while $\rho\rightarrow\Omega+\eps$, and as
$\eps\rightarrow 0^+$ the solution has to be continuous and differentiable across the line segment joining the foci of the $X-Y$ ellipse. This can
be accomplished by requiring continuity across $\rho=\rho'$ and for $\phi\neq \phi'$ differentiability as well.
In addition, we wish for the normal derivative of $G_N$ at $\rho'=1$ to be independent of $\rho,\phi$. Thus,
\begin{eqnarray}
&&\text{For } \rho'<\rho:\nonumber\\
&&
G(\rho,\phi;\rho',\phi')=A_{0}(\rho,\phi)+B_{0}(\rho,\phi)\ln\rho'+\sum_{m=1}A_{|m|}(\rho,\phi)\left(\frac{\rho'^m}{\Omega^m}+\frac{\Omega^m}{\rho'^m}\right)\cos(m\phi')+
B_{|m|}(\rho,\phi)\left(\frac{\rho'^m}{\Omega^m}-\frac{\Omega^m}{\rho'^m}\right)\sin(m\phi')\nonumber\\ \\
&&\text{For } \rho'>\rho:\nonumber\\
&&G(\rho,\phi;\rho',\phi')=C_{0}(\rho,\phi)+D_{0}\ln\rho'+\sum_{m\in\mathbb{Z}\backslash\{0\}}C_{m}(\rho,\phi)
\left(\rho'^m+\frac{1}{\rho'^m}\right)e^{im\phi'}.\nonumber\\
\end{eqnarray}
The Green's function has to be continuous across $\rho'=\rho$ circle:
\begin{eqnarray}
\lim_{\eps\rightarrow 0^+}G(\rho,\phi;\rho+\eps,\phi')=\lim_{\eps\rightarrow 0^+}G(\rho,\phi;\rho-\eps,\phi').
\end{eqnarray}
We integrate the Poisson equation over a small interval around $\rho$,
\begin{eqnarray}
&&\lim_{\eps\rightarrow 0^+}\int_{\rho-\eps}^{\rho+\eps} d\rho'\left(\frac{\partial}{\partial \rho'}\rho'\frac{\partial}{\partial
\rho'}+\frac{1}{\rho'}\frac{\partial^2}{\partial^2 \phi'}\right)G(\rho,\phi;\rho',\phi')=-\lim_{\eps\rightarrow 0^+}\int_{\rho-\eps}^{\rho+\eps}
d\rho'\delta(\rho-\rho')\delta(\phi-\phi')\nonumber \\ 
&&\Rightarrow \lim_{\eps\rightarrow 0^+}\rho\left(\frac{\partial}{\partial \rho'}G(\rho,\phi;\rho',\phi')|_{\rho'=\rho+\eps}-
\frac{\partial}{\partial \rho'}G(\rho,\phi;\rho',\phi')|_{\rho'=\rho-\eps}
\right)=-\delta(\phi-\phi').
\label{deriv_discont_eq}
\end{eqnarray}
To evaluate the limit one needs to know $\rho'\frac{\partial}{\partial \rho'}G(\rho,\phi;\rho',\phi')$ for $\rho' \lessgtr \rho$. It is given by
\begin{eqnarray}
&&\text{For } \rho'<\rho:\nonumber\\
&& \rho'\frac{\partial}{\partial \rho'}G(\rho,\phi;\rho',\phi')=B_0(\rho,\phi)+\sum_{m=1}m
A_{|m|}(\rho,\phi)\left(\frac{\rho'^{m}}{\Omega^m}-\frac{\Omega^m}{\rho'^{m}}\right)\cos(m\phi')+
m B_{|m|}(\rho,\phi)\left(\frac{\rho'^m}{\Omega^m}+\frac{\Omega^m}{\rho'^m}\right)\sin(m\phi')\nonumber\\
&&\text{For } \rho'>\rho:\nonumber\\
&&\rho'\frac{\partial}{\partial \rho'}G(\rho,\phi;\rho',\phi')=D_{0}+\sum_{m\in\mathbb{Z}\backslash\{0\}}m
C_{m}(\rho,\phi)\left(\rho'^{m}-\frac{1}{\rho'^{m}}\right)e^{im\phi'}.\nonumber
\end{eqnarray}
Substituting into the derivative discontinuity equation \eqref{deriv_discont_eq} we have
\begin{eqnarray}\label{eqn:derivative conditions}
&&D_{0}+\sum_{m\in\mathbb{Z}\backslash\{0\}}m C_{m}(\rho,\phi)
\left(\rho^m-\frac{1}{\rho^m}\right)e^{im\phi'}
\nonumber\\
&&-\left[B_0(\rho,\phi)+
\sum_{m=1}m A_{|m|}(\rho,\phi)\left(\frac{\rho^{m}}{\Omega^m}-\frac{\Omega^m}{\rho^{m}}\right)\cos(m\phi')+
m B_{|m|}(\rho,\phi)\left(\frac{\rho^m}{\Omega^m}+\frac{\Omega^m}{\rho^m}\right)\sin(m\phi')
\right]\nonumber\\
&&=-\frac{1}{2\pi}\sum_{m\in\mathbb{Z}}e^{im\phi}e^{-im\phi'},
\end{eqnarray}
where we used
\begin{eqnarray}
\delta(\phi-\phi')&=&\frac{1}{2\pi}\sum_{m\in\mathbb{Z}}e^{im\phi}e^{-im\phi'}=\frac{1}{2\pi}+\frac{1}{\pi}\sum_{m=1}^\infty \left(\cos m\phi\cos
m\phi'+
\sin m\phi\sin m\phi'\right).
\end{eqnarray}
At the same time, the continuity of $G$ at $\rho=\rho'$ requires
\begin{eqnarray}
&&C_{0}(\rho,\phi)+D_{0}\ln\rho+\sum_{m\in\mathbb{Z}\backslash\{0\}}C_{m}(\rho,\phi)
\left(\rho^m+\frac{1}{\rho^m}\right)e^{im\phi'}\nonumber\\
=&&A_{0}(\rho,\phi)+B_{0}(\rho,\phi)\ln\rho+\sum_{m=1}A_{|m|}(\rho,\phi)\left(\frac{\rho^m}{\Omega^m}+\frac{\Omega^m}{\rho^m}\right)\cos(m\phi')+
B_{|m|}(\rho,\phi)\left(\frac{\rho^m}{\Omega^m}-\frac{\Omega^m}{\rho^m}\right)\sin(m\phi').
\end{eqnarray}
Each $\phi'$ angular harmonic must be matched on the two sides of the Eq. (\ref{eqn:derivative conditions}). Therefore,
$B_0(\rho,\phi)=D_0+\frac{1}{2\pi}$.
For non-zero $m$, by matching the $\phi'$ angular harmonics, we get two linear equations relating $A_{|m|},B_{|m|},C_{|m|}$ and two linear equations
relating $A_{|m|},B_{|m|},C_{-|m|}$. So, in total we get four linear equations of four variables which we investigate below.

\subsection{Fourier coefficients}

To find the harmonic coefficients we match the terms for $\rho'<\rho$ and $\rho'>\rho$
\begin{eqnarray}
&&\sum_{m\in\mathbb{Z}\backslash\{0\}}C_{m}(\rho,\phi)
\left(\rho^m+\frac{1}{\rho^m}\right)e^{im\phi'}
=\sum_{m=1}A_{|m|}(\rho,\phi)\left(\frac{\rho^m}{\Omega^m}+\frac{\Omega^m}{\rho^m}\right)\cos(m\phi')+
B_{|m|}(\rho,\phi)\left(\frac{\rho^m}{\Omega^m}-\frac{\Omega^m}{\rho^m}\right)\sin(m\phi'),\nonumber\\
&&\sum_{m\in\mathbb{Z}\backslash\{0\}}m C_{m}(\rho,\phi)
\left(\rho^m-\frac{1}{\rho^m}\right)e^{im\phi'}
-\left[
\sum_{m=1}m A_{|m|}(\rho,\phi)\left(\frac{\rho^{m}}{\Omega^m}-\frac{\Omega^m}{\rho^{m}}\right)\cos(m\phi')+
m B_{|m|}(\rho,\phi)\left(\frac{\rho^m}{\Omega^m}+\frac{\Omega^m}{\rho^m}\right)\sin(m\phi')
\right]\nonumber\\
&&=-\frac{1}{2\pi}\sum_{m\in\mathbb{Z}\backslash\{0\}}e^{-im\phi}e^{im\phi'}
\end{eqnarray}

Matching the coefficients for harmonics we get the two conditions for different signs of $m$:

\noindent$\bullet$ For $m>0$:
\begin{eqnarray}
&&C_{m}(\rho,\phi)
\left(\rho^m+\frac{1}{\rho^m}\right)
=\frac{1}{2}A_{|m|}(\rho,\phi)\left(\frac{\rho^m}{\Omega^m}+\frac{\Omega^m}{\rho^m}\right)+
\frac{1}{2i}B_{|m|}(\rho,\phi)\left(\frac{\rho^m}{\Omega^m}-\frac{\Omega^m}{\rho^m}\right),\\
&&m C_{m}(\rho,\phi)
\left(\rho^m-\frac{1}{\rho^m}\right)
-\left[
\frac{m}{2} A_{|m|}(\rho,\phi)\left(\frac{\rho^{m}}{\Omega^m}-\frac{\Omega^m}{\rho^{m}}\right)+
\frac{m}{2i} B_{|m|}(\rho,\phi)\left(\frac{\rho^m}{\Omega^m}+\frac{\Omega^m}{\rho^m}\right)
\right]=-\frac{1}{2\pi}e^{-im\phi}. \nonumber
\end{eqnarray}
\noindent $\bullet$ For $m<0$:
\begin{eqnarray}
&&C_{m}(\rho,\phi)
\left(\rho^m+\frac{1}{\rho^m}\right)
=\frac{1}{2}A_{|m|}(\rho,\phi)\left(\frac{\rho^{|m|}}{\Omega^{|m|}}+\frac{\Omega^{|m|}}{\rho^{|m|}}\right)-
\frac{1}{2i}B_{|m|}(\rho,\phi)\left(\frac{\rho^{|m|}}{\Omega^{|m|}}-\frac{\Omega^{|m|}}{\rho^{|m|}}\right),\\
&&m C_{m}(\rho,\phi)
\left(\rho^m-\frac{1}{\rho^m}\right)
-\left[
\frac{|m|}{2} A_{|m|}(\rho,\phi)\left(\frac{\rho^{|m|}}{\Omega^{|m|}}-\frac{\Omega^{|m|}}{\rho^{|m|}}\right)-
\frac{|m|}{2i} B_{|m|}(\rho,\phi)\left(\frac{\rho^{|m|}}{\Omega^{|m|}}+\frac{\Omega^{|m|}}{\rho^{|m|}}\right)
\right]=-\frac{1}{2\pi}e^{-im\phi}. \nonumber
\end{eqnarray}
Eliminating $C_m$ we have for $m>0$:\\
\begin{eqnarray}
&&
\frac{\rho^{2m}-1}{\rho^{2m}+1}
\left(
A_{|m|}(\rho,\phi)\left(\frac{\rho^m}{\Omega^m}+\frac{\Omega^m}{\rho^m}\right)-
iB_{|m|}(\rho,\phi)\left(\frac{\rho^m}{\Omega^m}-\frac{\Omega^m}{\rho^m}\right)
\right)
\nonumber\\
&&-
 A_{|m|}(\rho,\phi)\left(\frac{\rho^{m}}{\Omega^m}-\frac{\Omega^m}{\rho^{m}}\right)+i
 B_{|m|}(\rho,\phi)\left(\frac{\rho^m}{\Omega^m}+\frac{\Omega^m}{\rho^m}\right)
=-\frac{1}{\pi m}e^{-im\phi},
\end{eqnarray}
which can be rewritten as
\begin{eqnarray}
&&\left[\frac{\rho^{2m}-1}{\rho^{2m}+1}\left(\frac{\rho^m}{\Omega^m}+\frac{\Omega^m}{\rho^m}\right)-
\left(\frac{\rho^{m}}{\Omega^m}-\frac{\Omega^m}{\rho^{m}}\right)
\right] A_{|m|}(\rho,\phi)+i\left[\left(\frac{\rho^m}{\Omega^m}+\frac{\Omega^m}{\rho^m}\right)-
\frac{\rho^{2m}-1}{\rho^{2m}+1}\left(\frac{\rho^m}{\Omega^m}-\frac{\Omega^m}{\rho^m}\right)
\right]B_{|m|}(\rho,\phi)=-\frac{e^{-im\phi}}{\pi m}.\nonumber
\end{eqnarray}
Now we eliminate $C_m$ for $m<0$ and get:\\
\begin{eqnarray}
&&
\frac{\left(\rho^m-\frac{1}{\rho^m}\right)}{\left(\rho^m+\frac{1}{\rho^m}\right)
}\left(A_{|m|}(\rho,\phi)\left(\frac{\rho^{|m|}}{\Omega^{|m|}}+\frac{\Omega^{|m|}}{\rho^{|m|}}\right)+i
B_{|m|}(\rho,\phi)\left(\frac{\rho^{|m|}}{\Omega^{|m|}}-\frac{\Omega^{|m|}}{\rho^{|m|}}\right)\right)\nonumber\\
&&
+A_{|m|}(\rho,\phi)\left(\frac{\rho^{|m|}}{\Omega^{|m|}}-\frac{\Omega^{|m|}}{\rho^{{|m|}}}\right)+i
B_{|m|}(\rho,\phi)\left(\frac{\rho^{|m|}}{\Omega^{|m|}}+\frac{\Omega^{|m|}}{\rho^{|m|}}\right)
=-\frac{1}{\pi m}e^{-im\phi}.
\end{eqnarray}
Equivalently we can right the same expression as
\begin{eqnarray}
&&
-\frac{\rho^{2|m|}-1}{\rho^{2|m|}+1}\left(A_{|m|}(\rho,\phi)\left(\frac{\rho^{|m|}}{\Omega^{|m|}}+\frac{\Omega^{|m|}}{\rho^{|m|}}\right)+i
B_{|m|}(\rho,\phi)\left(\frac{\rho^{|m|}}{\Omega^{|m|}}-\frac{\Omega^{|m|}}{\rho^{|m|}}\right)\right)\nonumber\\
&&
+A_{|m|}(\rho,\phi)\left(\frac{\rho^{|m|}}{\Omega^{|m|}}-\frac{\Omega^{|m|}}{\rho^{{|m|}}}\right)+i
B_{|m|}(\rho,\phi)\left(\frac{\rho^{|m|}}{\Omega^{|m|}}+\frac{\Omega^{|m|}}{\rho^{|m|}}\right)
=\frac{1}{\pi |m|}e^{i|m|\phi},
\end{eqnarray}
or
\begin{eqnarray}
&&\left[-\frac{\rho^{2|m|}-1}{\rho^{2|m|}+1}\left(\frac{\rho^{|m|}}{\Omega^{|m|}}+\frac{\Omega^{|m|}}{\rho^{|m|}}\right)+
\left(\frac{\rho^{|m|}}{\Omega^{|m|}}-\frac{\Omega^{|m|}}{\rho^{{|m|}}}\right)
\right]A_{|m|}(\rho,\phi) \nonumber \\
&&+i\left[\left(\frac{\rho^{|m|}}{\Omega^{|m|}}+\frac{\Omega^{|m|}}{\rho^{|m|}}\right)-\frac{\rho^{2|m|}-1}{\rho^{2|m|}+1}
\left(\frac{\rho^{|m|}}{\Omega^{|m|}}-\frac{\Omega^{|m|}}{\rho^{|m|}}\right)\right]B_{|m|}(\rho,\phi)
=\frac{1}{\pi |m|}e^{i|m|\phi}.
\end{eqnarray}
This implies
\begin{eqnarray}
&&\left[\frac{\rho^{2|m|}-1}{\rho^{2|m|}+1}\left(\frac{\rho^{|m|}}{\Omega^{|m|}}+\frac{\Omega^{|m|}}{\rho^{|m|}}\right)-
\left(\frac{\rho^{|m|}}{\Omega^{|m|}}-\frac{\Omega^{|m|}}{\rho^{{|m|}}}\right)
\right]A_{|m|}(\rho,\phi)=-\frac{1}{2\pi |m|}\left(e^{i|m|\phi}+e^{-i|m|\phi}\right),
\end{eqnarray}
which can be transformed into
\begin{eqnarray}
A_{|m|}(\rho,\phi)=-\frac{\rho^{|m|}+\rho^{-|m|}}{4\pi |m|}
\left(\frac{e^{i|m|\phi}+e^{-i|m|\phi}}{\Omega^{|m|}-\Omega^{-|m|}}\right)\label{eq:Am}.
\end{eqnarray}
Performing similar manipulations for 
\begin{eqnarray}
&&2i\left[\left(\frac{\rho^{|m|}}{\Omega^{|m|}}+\frac{\Omega^{|m|}}{\rho^{|m|}}\right)-\frac{\rho^{2|m|}-1}{\rho^{2|m|}+1}
\left(\frac{\rho^{|m|}}{\Omega^{|m|}}-\frac{\Omega^{|m|}}{\rho^{|m|}}\right)\right]B_{|m|}(\rho,\phi)=\frac{1}{\pi
|m|}\left(e^{i|m|\phi}-e^{-i|m|\phi}\right)
\end{eqnarray}
we obtain 
\begin{eqnarray}
B_{|m|}(\rho,\phi)=\frac{\rho^{|m|}+\rho^{-|m|}}{4\pi i
|m|}\frac{\left(e^{i|m|\phi}-e^{-i|m|\phi}\right)}{\Omega^{|m|}+\Omega^{-|m|}}\label{eq:Bm}.
\end{eqnarray}

\noindent Then for $m>0$ we get 
\begin{eqnarray}
&&C_{m}(\rho,\phi)
\left(\rho^m+\frac{1}{\rho^m}\right)
=\frac{1}{2}A_{|m|}(\rho,\phi)\left(\frac{\rho^m}{\Omega^m}+\frac{\Omega^m}{\rho^m}\right)+
\frac{1}{2i}B_{|m|}(\rho,\phi)\left(\frac{\rho^m}{\Omega^m}-\frac{\Omega^m}{\rho^m}\right),
\end{eqnarray}
and for $m<0$ 
\begin{eqnarray}
&&C_{m}(\rho,\phi)
\left(\rho^m+\frac{1}{\rho^m}\right)
=\frac{1}{2}A_{|m|}(\rho,\phi)\left(\frac{\rho^{|m|}}{\Omega^{|m|}}+\frac{\Omega^{|m|}}{\rho^{|m|}}\right)-
\frac{1}{2i}B_{|m|}(\rho,\phi)\left(\frac{\rho^{|m|}}{\Omega^{|m|}}-\frac{\Omega^{|m|}}{\rho^{|m|}}\right).
\end{eqnarray}
This implies that for any sign of $m$ we can write
\begin{eqnarray}
C_{m}(\rho,\phi)&&=\frac{1}{2}\frac{1}{\left(\rho^{|m|}+\rho^{-|m|}\right)}
\left(
A_{|m|}(\rho,\phi)\left(\frac{\rho^{|m|}}{\Omega^{|m|}}+\frac{\Omega^{|m|}}{\rho^{|m|}}\right)-i\text{sign}(m)
B_{|m|}(\rho,\phi)\left(\frac{\rho^{|m|}}{\Omega^{|m|}}-\frac{\Omega^{|m|}}{\rho^{|m|}}\right)
\right)= \nonumber \\
&&=\frac{-1}{8\pi |m|}
\left(
\frac{e^{i|m|\phi}+e^{-i|m|\phi}}{\Omega^{|m|}-\Omega^{-|m|}}
\left(\frac{\rho^{|m|}}{\Omega^{|m|}}+\frac{\Omega^{|m|}}{\rho^{|m|}}\right)+\text{sign}(m)
\frac{e^{i|m|\phi}-e^{-i|m|\phi}}{\Omega^{|m|}+\Omega^{-|m|}}\left(\frac{\rho^{|m|}}{\Omega^{|m|}}-\frac{\Omega^{|m|}}{\rho^{|m|}}\right)
\right)\label{eq:Cm}.
\end{eqnarray}
Using
\begin{eqnarray}
&&\sum_{m=1}^{\infty}\frac{x^m}{m}=-\ln\left(1-x\right)\\
&&\frac{1}{\Omega^{|m|}\pm \Omega^{-|m|}}=\pm\Omega^{|m|}\frac{1}{1\pm \Omega^{2|m|}}=
\pm\Omega^{|m|}\sum_{n=0}^{\infty}(\mp 1)^n\left(\Omega^{2|m|}\right)^n
\end{eqnarray}
and Eqs. (\ref{eq:Am}), (\ref{eq:Bm}), (\ref{eq:Cm})
we can now compute the Neumann Green's function for the ellipse:
\begin{eqnarray}
&&\text{For } \rho'<\rho:\nonumber\\
&&
G(\rho,\phi;\rho',\phi')=\frac{1}{2\pi}\ln\rho'+\sum_{m=1}A_{|m|}(\rho,\phi)\left(\frac{\rho'^m}{\Omega^m}+\frac{\Omega^m}{\rho'^m}\right)\cos(m\phi')+
B_{|m|}(\rho,\phi)\left(\frac{\rho'^m}{\Omega^m}-\frac{\Omega^m}{\rho'^m}\right)\sin(m\phi')\nonumber \\
&&\text{For } \rho'>\rho:\nonumber\\
&&G(\rho,\phi;\rho',\phi')=\frac{1}{2\pi}\ln\rho+\sum_{m\in\mathbb{Z}\backslash\{0\}}C_{m}(\rho,\phi)
\left(\rho'^m+\frac{1}{\rho'^m}\right)e^{im\phi'},
\end{eqnarray}
where
\begin{eqnarray}
&&A_{|m|}(\rho,\phi)=-\frac{\rho^{|m|}+\rho^{-|m|}}{4\pi |m|}
\left(\frac{e^{i|m|\phi}+e^{-i|m|\phi}}{\Omega^{|m|}-\Omega^{-|m|}}\right)\\
&&B_{|m|}(\rho,\phi)=\frac{\rho^{|m|}+\rho^{-|m|}}{4\pi i |m|}\frac{\left(e^{i|m|\phi}-e^{-i|m|\phi}\right)}{\Omega^{|m|}+\Omega^{-|m|}}\\
&&C_{m}(\rho,\phi)=\frac{-1}{8\pi |m|}
\left(
\frac{e^{i|m|\phi}+e^{-i|m|\phi}}{\Omega^{|m|}-\Omega^{-|m|}}
\left(\frac{\rho^{|m|}}{\Omega^{|m|}}+\frac{\Omega^{|m|}}{\rho^{|m|}}\right)+\text{sign}(m)
\frac{e^{i|m|\phi}-e^{-i|m|\phi}}{\Omega^{|m|}+\Omega^{-|m|}}\left(\frac{\rho^{|m|}}{\Omega^{|m|}}-\frac{\Omega^{|m|}}{\rho^{|m|}}\right)
\right).
\end{eqnarray}
Thus, for $\rho'<\rho$:
\begin{eqnarray}
G(\rho,\phi;\rho',\phi')=\frac{1}{2\pi}\ln\rho'-&&\sum_{m=1}
\frac{\rho^{|m|}+\rho^{-|m|}}{4\pi |m|}
\left(\frac{e^{i|m|\phi}+e^{-i|m|\phi}}{\Omega^{|m|}-\Omega^{-|m|}}\right)\left(\frac{\rho'^m}{\Omega^m}+\frac{\Omega^m}{\rho'^m}\right)\cos(m\phi')\nonumber\\
+&&
\sum_{m=1}\frac{\rho^{|m|}+\rho^{-|m|}}{4\pi i
|m|}\frac{\left(e^{i|m|\phi}-e^{-i|m|\phi}\right)}{\Omega^{|m|}+\Omega^{-|m|}}\left(\frac{\rho'^m}{\Omega^m}-\frac{\Omega^m}{\rho'^m}\right)\sin(m\phi'),
\end{eqnarray}
and for $\rho'>\rho$:
\begin{eqnarray}
&&G(\rho,\phi;\rho',\phi')=\frac{1}{2\pi}\ln\rho+\sum_{m=1}^{\infty}C_{m}(\rho,\phi)
\left(\rho'^m+\rho'^{-m}\right)e^{im\phi'}
+\sum_{m=1}^{\infty}C_{-m}(\rho,\phi)
\left(\rho'^m+\rho'^{-m}\right)e^{-im\phi'}=\\
&&\frac{1}{2\pi}\ln\rho+\sum_{m=1}^{\infty}\left(C_{m}(\rho,\phi)+C_{-m}(\rho,\phi)\right)
\left(\rho'^m+\rho'^{-m}\right)\cos(m\phi')+i\sum_{m=1}^{\infty}\left(C_{m}(\rho,\phi)-C_{-m}(\rho,\phi)\right)
\left(\rho'^m+\rho'^{-m}\right)\sin\left(m\phi'\right)\nonumber\\
&&=\frac{1}{2\pi}\ln\rho-\sum_{m=1}^{\infty}
\frac{1}{4\pi |m|}
\frac{e^{i|m|\phi}+e^{-i|m|\phi}}{\Omega^{|m|}-\Omega^{-|m|}}
\left(\frac{\rho^{|m|}}{\Omega^{|m|}}+\frac{\Omega^{|m|}}{\rho^{|m|}}\right)
\left(\rho'^m+\rho'^{-m}\right)\cos(m\phi')\nonumber\\
&&-i\sum_{m=1}^{\infty}\frac{1}{4\pi m}\left(\frac{e^{i|m|\phi}-e^{-i|m|\phi}}{\Omega^{|m|}+\Omega^{-|m|}}
\left(\frac{\rho^{|m|}}{\Omega^{|m|}}-\frac{\Omega^{|m|}}{\rho^{|m|}}\right)\right)
\left(\rho'^m+\rho'^{-m}\right)\sin\left(m\phi'\right).
\end{eqnarray}
Therefore for $\rho'>\rho$:
\begin{eqnarray}
G(\rho,\phi;\rho',\phi')=
\frac{1}{2\pi}\ln\rho -&&\frac{1}{2\pi}\sum_{m=1}^{\infty}
\frac{1}{m}
\frac{\cos(m\phi)}{\Omega^{m}-\Omega^{-m}}
\left(\frac{\rho^{m}}{\Omega^{m}}+\frac{\Omega^{m}}{\rho^{m}}\right)
\left(\rho'^m+\rho'^{-m}\right)\cos(m\phi')\nonumber\\
+&&\frac{1}{2\pi}\sum_{m=1}^{\infty}\frac{1}{m}\frac{\sin(m\phi)}{\Omega^{m}+\Omega^{-m}}
\left(\frac{\rho^{m}}{\Omega^{m}}-\frac{\Omega^{m}}{\rho^{m}}\right)
\left(\rho'^m+\rho'^{-m}\right)\sin\left(m\phi'\right).
\end{eqnarray}
and for $\rho'<\rho$:
\begin{eqnarray}
G(\rho,\phi;\rho',\phi')=\frac{1}{2\pi}\ln\rho'-&&\frac{1}{2\pi}\sum_{m=1}
\frac{\rho^{|m|}+\rho^{-|m|}}{m}
\left(\frac{\cos(m\phi)}{\Omega^{|m|}-\Omega^{-|m|}}\right)\left(\frac{\rho'^m}{\Omega^m}+\frac{\Omega^m}{\rho'^m}\right)\cos(m\phi')\nonumber\\
+&&\frac{1}{2\pi}
\sum_{m=1}\frac{\rho^{|m|}+\rho^{-|m|}}{m}\frac{\sin(m\phi)}{\Omega^{|m|}+\Omega^{-|m|}}\left(\frac{\rho'^m}{\Omega^m}-\frac{\Omega^m}{\rho'^m}\right)\sin(m\phi')
\end{eqnarray}
This makes the symmetry of the Neumann Green's function under $(\rho,\phi)\leftrightarrow (\rho',\phi')$,  explicit.

We can therefore write the Green's function succinctly as
\begin{eqnarray}
&&G(\rho,\phi;\rho',\phi')=
\frac{1}{2\pi}\left(\ln\rho_< -\sum_{m=1}^{\infty}
\frac{1}{m}
\frac{1}{\Omega^{m}-\Omega^{-m}}
\left(\frac{\rho_<^{m}}{\Omega^{m}}+\frac{\Omega^{m}}{\rho_<^{m}}\right)
\left(\rho_>^m+\rho_>^{-m}\right)\cos(m\phi)\cos(m\phi')\right.\nonumber\\
&&+\left.\sum_{m=1}^{\infty}\frac{1}{m}\frac{1}{\Omega^{m}+\Omega^{-m}}
\left(\frac{\rho_<^{m}}{\Omega^{m}}-\frac{\Omega^{m}}{\rho_<^{m}}\right)
\left(\rho_>^m+\rho_>^{-m}\right)\sin(m\phi)\sin\left(m\phi'\right)\right).
\end{eqnarray}
Furthermore, we can use
\begin{eqnarray}
&&\sum_{m=1}^{\infty}\frac{x^m}{m}=-\ln\left(1-x\right)\\
&&\frac{1}{\Omega^{m}\pm \Omega^{-m}}=\pm\Omega^{m}\frac{1}{1\pm \Omega^{2m}}=
\pm\Omega^{m}\sum_{n=0}^{\infty}(\mp 1)^n\left(\Omega^{2m}\right)^n
\end{eqnarray}
to sum over $m$ and write a rapidly convergent series expression for $G$:
\begin{eqnarray}
&&G(\rho,\phi;\rho',\phi')=
\frac{1}{2\pi}\left(\ln\rho_< +\frac{1}{4}\sum_{n=0}^{\infty}\sum_{s_1,s_2,s_3,s_4=\pm1}\sum_{m=1}^{\infty}
\frac{1}{m}
\Omega^{(2n+1)m}
\left(\frac{\rho_<}{\Omega}\right)^{s_1m}
\rho_>^{s_2m}e^{is_3m\phi}e^{is_4m\phi'}\right.\nonumber\\
&&+\left.\frac{-1}{4}\sum_{n=0}^{\infty}(-1)^n\sum_{s_1,s_2,s_3,s_4=\pm1}s_1s_3s_4
\sum_{m=1}^{\infty}\frac{1}{m}\Omega^{(2n+1)m}
\left(\frac{\rho_<}{\Omega}\right)^{s_1m}
\rho_>^{s_2m}e^{is_3m\phi}e^{is_4m\phi'}\right)\\
&&=\frac{1}{2\pi}\left(\ln\rho_< -\frac{1}{4}\sum_{n=0}^{\infty}\sum_{s_1,s_2,s_3,s_4=\pm1}
\ln\left[1-\Omega^{(2n+1)}\left(\frac{\rho_<}{\Omega}\right)^{s_1}\rho_>^{s_2}e^{is_3\phi}e^{is_4\phi'}\right]
\right.\nonumber\\
&&+\left.\frac{1}{4}\sum_{n=0}^{\infty}(-1)^n\sum_{s_1,s_2,s_3,s_4=\pm1}s_1s_3s_4
\ln\left[1-\Omega^{(2n+1)}\left(\frac{\rho_<}{\Omega}\right)^{s_1}\rho_>^{s_2}e^{is_3\phi}e^{is_4\phi'}\right]
\right)\\
&&=\frac{1}{2\pi}\left(\ln\rho_< -\frac{1}{4}\sum_{n=0,2,4,\ldots}^{\infty}\sum_{s_1,s_2,s_3,s_4=\pm1}
\left(1-s_1s_3s_4\right)
\ln\left[1-\Omega^{(2n+1)}\left(\frac{\rho_<}{\Omega}\right)^{s_1}\rho_>^{s_2}e^{is_3\phi}e^{is_4\phi'}\right]
\right.\\
&&-\left.
\frac{1}{4}\sum_{n=1,3,5\ldots}^{\infty}\sum_{s_1,s_2,s_3,s_4=\pm1}
\left(1+s_1s_3s_4\right)\ln\left[1-\Omega^{(2n+1)}\left(\frac{\rho_<}{\Omega}\right)^{s_1}\rho_>^{s_2}e^{is_3\phi}e^{is_4\phi'}\right]
\right)\\
&&=\frac{1}{2\pi}\left(\ln\rho_<\right.\\
&&\left. -\frac{1}{2}\sum_{n=0,2,4,\ldots}^{\infty}\sum_{s_2=\pm1}
\ln\left[\left(1-\Omega^{(2n+1)}\left(\frac{\rho_<}{\Omega}\right)^{-1}\rho_>^{s_2}e^{i\phi}e^{i\phi'}\right)
\left(1-\Omega^{(2n+1)}\left(\frac{\rho_<}{\Omega}\right)\rho_>^{s_2}e^{-i\phi}e^{i\phi'}\right)
\right]
\right.\\
&&\left. -\frac{1}{2}\sum_{n=0,2,4,\ldots}^{\infty}\sum_{s_2=\pm1}
\ln\left[
\left(1-\Omega^{(2n+1)}\left(\frac{\rho_<}{\Omega}\right)\rho_>^{s_2}e^{i\phi}e^{-i\phi'}\right)
\left(1-\Omega^{(2n+1)}\left(\frac{\rho_<}{\Omega}\right)^{-1}\rho_>^{s_2}e^{-i\phi}e^{-i\phi'}\right)
\right]
\right.\\
&&-\left.
\frac{1}{4}\sum_{n=1,3,5\ldots}^{\infty}\sum_{s_1,s_2,s_3,s_4=\pm1}
\left(1+s_1s_3s_4\right)\ln\left[1-\Omega^{(2n+1)}\left(\frac{\rho_<}{\Omega}\right)^{s_1}\rho_>^{s_2}e^{is_3\phi}e^{is_4\phi'}\right]
\right).
\end{eqnarray}
Now, because the expression is symmetric under $\phi$ and $\phi'$, we can assign either one of these with $>$ and $<$.
For the $n$ even case this gives
\begin{eqnarray}
&&G(\rho,\phi;\rho',\phi')=\frac{1}{2\pi}\left(\ln\rho_<\right.\\
&&\left. -\frac{1}{2}\sum_{n=0,2,4,\ldots}^{\infty}\sum_{s_2=\pm1}
\ln\left[\left(1-\Omega^{(2n+1)}\left(\frac{\rho_<}{\Omega}\right)^{-1}\rho_>^{s_2}e^{i\phi_<}e^{i\phi_>}\right)
\left(1-\Omega^{(2n+1)}\left(\frac{\rho_<}{\Omega}\right)\rho_>^{s_2}e^{-i\phi_<}e^{i\phi_>}\right)
\right]
\right.\\
&&\left. -\frac{1}{2}\sum_{n=0,2,4,\ldots}^{\infty}\sum_{s_2=\pm1}
\ln\left[
\left(1-\Omega^{(2n+1)}\left(\frac{\rho_<}{\Omega}\right)\rho_>^{s_2}e^{i\phi_<}e^{-i\phi_>}\right)
\left(1-\Omega^{(2n+1)}\left(\frac{\rho_<}{\Omega}\right)^{-1}\rho_>^{s_2}e^{-i\phi_<}e^{-i\phi_>}\right)
\right]
\right.\\
&&-\left.
\frac{1}{4}\sum_{n=1,3,5\ldots}^{\infty}\sum_{s_1,s_2,s_3,s_4=\pm1}
\left(1+s_1s_3s_4\right)\ln\left[1-\Omega^{(2n+1)}\left(\frac{\rho_<}{\Omega}\right)^{s_1}\rho_>^{s_2}e^{is_3\phi}e^{is_4\phi'}\right]
\right)\\
&&=\frac{1}{2\pi}\left(\ln\rho_<\right.\\
&&\left. -\frac{1}{2}\sum_{n=0,2,4,\ldots}^{\infty}\sum_{s_2=\pm1}
\ln\left[\left(1-\Omega^{(2n+1)}\frac{\Omega}{\bar{w}_<}\rho_>^{s_2}e^{i\phi_>}\right)
\left(1-\Omega^{(2n+1)}\frac{\bar{w}_<}{\Omega}\rho_>^{s_2}e^{i\phi_>}\right)
\right]
\right.\\
&&\left. -\frac{1}{2}\sum_{n=0,2,4,\ldots}^{\infty}\sum_{s_2=\pm1}
\ln\left[
\left(1-\Omega^{(2n+1)}\frac{w_<}{\Omega}\rho_>^{s_2}e^{-i\phi_>}\right)
\left(1-\Omega^{(2n+1)}\frac{\Omega}{w_<}\rho_>^{s_2}e^{-i\phi_>}\right)
\right]
\right.\\
&&-\left.
\frac{1}{4}\sum_{n=1,3,5\ldots}^{\infty}\sum_{s_1,s_2,s_3,s_4=\pm1}
\left(1+s_1s_3s_4\right)\ln\left[1-\Omega^{(2n+1)}\left(\frac{\rho_<}{\Omega}\right)^{s_1}\rho_>^{s_2}e^{is_3\phi}e^{is_4\phi'}\right]
\right)\\
&&=\frac{1}{2\pi}\left(\ln\rho_<\right.\\
&&\left. -\frac{1}{2}\sum_{n=0,2,4,\ldots}^{\infty}\sum_{s_2=\pm1}
\left(\ln\left[
\left(1-\Omega^{(2n+1)}\frac{w_<}{\Omega}\rho_>^{s_2}e^{-i\phi_>}\right)
\left(1-\Omega^{(2n+1)}\frac{\Omega}{w_<}\rho_>^{s_2}e^{-i\phi_>}\right)
\right]+c.c.\right)
\right.\\
&&-\left.
\frac{1}{4}\sum_{n=1,3,5\ldots}^{\infty}\sum_{s_1,s_2,s_3,s_4=\pm1}
\left(1+s_1s_3s_4\right)\ln\left[1-\Omega^{(2n+1)}\left(\frac{\rho_<}{\Omega}\right)^{s_1}\rho_>^{s_2}e^{is_3\phi}e^{is_4\phi'}\right]
\right).
\end{eqnarray}
Therefore,
\begin{eqnarray}
&&G(\rho,\phi;\rho',\phi')=\frac{1}{2\pi}\left(\ln\rho_<\right.\\
&&\left. -\frac{1}{2}\sum_{n=0,2,4,\ldots}^{\infty}\sum_{s_2=\pm1}
\left(\ln\left[
\left(1-\Omega^{2n}\frac{Z_<}{\alpha_+}\rho_>^{s_2}e^{-i\phi_>}+\Omega^{4n+2}\rho_>^{2s_2}e^{-2i\phi_>}\right)
\right]+c.c.\right)
\right.\\
&&-\left.
\frac{1}{4}\sum_{n=1,3,5\ldots}^{\infty}\sum_{s_1,s_2,s_3,s_4=\pm1}
\left(1+s_1s_3s_4\right)\ln\left[1-\Omega^{(2n+1)}\left(\frac{\rho_<}{\Omega}\right)^{s_1}\rho_>^{s_2}e^{is_3\phi}e^{is_4\phi'}\right]
\right)\\
&&=\frac{1}{2\pi}\left(\ln\rho_<\right.\\
&&\left. -\frac{1}{2}\sum_{n=0,2,4,\ldots}^{\infty}
\left(\ln\left[
\left|w_>-\Omega^{2n}\frac{Z_<}{\alpha_+}+\Omega^{4n+2}\frac{1}{w_>}\right|^2
\right]+\ln\left[
\left|\frac{1}{w^*_>}-\Omega^{2n}\frac{Z_<}{\alpha_+}+\Omega^{4n+2}w^*_>\right|^2
\right]\right)
\right.\\
&&-\left.
\frac{1}{4}\sum_{n=1,3,5\ldots}^{\infty}\sum_{s_1,s_2,s_3,s_4=\pm1}
\left(1+s_1s_3s_4\right)\ln\left[1-\Omega^{(2n+1)}\left(\frac{\rho_<}{\Omega}\right)^{s_1}\rho_>^{s_2}e^{is_3\phi}e^{is_4\phi'}\right]
\right).
\end{eqnarray}
For the odd $n$ terms
\begin{eqnarray}
&&G(\rho,\phi;\rho',\phi')=\frac{1}{2\pi}\left(\ln\rho_<\right.\\
&&\left. -\frac{1}{2}\sum_{n=0,2,4,\ldots}^{\infty}
\left(\ln\left[
\left|w_>-\Omega^{2n}\frac{Z_<}{\alpha_+}+\Omega^{4n+2}\frac{1}{w_>}\right|^2
\right]+\ln\left[
\left|\frac{1}{w^*_>}-\Omega^{2n}\frac{Z_<}{\alpha_+}+\Omega^{4n+2}w^*_>\right|^2
\right]\right)
\right.\\
&&-\left.
\frac{1}{2}\sum_{n=1,3,5\ldots}^{\infty}\sum_{s_2=\pm1}
\left(\ln\left[1-\Omega^{(2n+1)}\left(\frac{\rho_<}{\Omega}\right)\rho_>^{s_2}e^{i\phi_<}e^{i\phi_>}\right]+
\ln\left[1-\Omega^{(2n+1)}\left(\frac{\rho_<}{\Omega}\right)\rho_>^{s_2}e^{-i\phi_<}e^{-i\phi_>}\right]+\right.\right.\\
&&\left.\left.\ln\left[1-\Omega^{(2n+1)}\left(\frac{\rho_<}{\Omega}\right)^{-1}\rho_>^{s_2}e^{-i\phi_<}e^{i\phi_>}\right]+
\ln\left[1-\Omega^{(2n+1)}\left(\frac{\rho_<}{\Omega}\right)^{-1}\rho_>^{s_2}e^{i\phi_<}e^{-i\phi_>}\right]
\right)\right)\\
&&=\frac{1}{2\pi}\left(\ln\rho_<\right.\\
&&\left. -\frac{1}{2}\sum_{n=0,2,4,\ldots}^{\infty}
\left(\ln\left[
\left|w_>-\Omega^{2n}\frac{Z_<}{\alpha_+}+\Omega^{4n+2}\frac{1}{w_>}\right|^2
\right]+\ln\left[
\left|\frac{1}{w^*_>}-\Omega^{2n}\frac{Z_<}{\alpha_+}+\Omega^{4n+2}w^*_>\right|^2
\right]\right)
\right.\\
&&-\left.
\frac{1}{2}\sum_{n=1,3,5\ldots}^{\infty}\sum_{s_2=\pm1}
\left(\ln\left[1-\Omega^{2n}\frac{Z_<}{\alpha_+}\rho_>^{s_2}e^{i\phi_>}+\Omega^{4n+2}\rho^{2s_2}_>e^{2i\phi_>}
\right]+c.c.
\right)\right)\\
&&=\frac{1}{2\pi}\left(\ln\rho_<\right.\\
&&\left. -\frac{1}{2}\sum_{n=0,2,4,\ldots}^{\infty}
\left(\ln\left[
\left|w_>-\Omega^{2n}\frac{Z_<}{\alpha_+}+\Omega^{4n+2}\frac{1}{w_>}\right|^2
\right]+\ln\left[
\left|\frac{1}{w^*_>}-\Omega^{2n}\frac{Z_<}{\alpha_+}+\Omega^{4n+2}w^*_>\right|^2
\right]\right)
\right.\\
&&-\left.
\frac{1}{2}\sum_{n=1,3,5\ldots}^{\infty}%\sum_{s_2=\pm1}
\left(\ln\left[\left|1-\Omega^{2n}\frac{Z_<}{\alpha_+}w_>+\Omega^{4n+2}w^2_>
\right|^2\right]+
\ln\left[\left|1-\Omega^{2n}\frac{Z_<}{\alpha_+}\frac{1}{w^*_>}+\Omega^{4n+2}\frac{1}{{w^*}^2_>}
\right|^2\right]
\right)\right).
\end{eqnarray}
Now, notice that if we isolate the $n=0$ term in the sum, which should dominate at low $\Omega$, the Green's function reads
\begin{eqnarray}
&&G(\rho,\phi;\rho',\phi')
=\frac{1}{2\pi}\left(\ln\rho_<
-\frac{1}{2}
\left(\ln\left[
\left|w_>-\frac{Z_<}{\alpha_+}+\Omega^{2}\frac{1}{w_>}\right|^2
\right]+\ln\left[
\left|\frac{1}{w^*_>}-\frac{Z_<}{\alpha_+}+\Omega^{2}w^*_>\right|^2
\right]\right)
\right.\nonumber\\
&&\left. -\frac{1}{2}\sum_{n=2,4,\ldots}^{\infty}
\left(\ln\left[
\left|w_>-\Omega^{2n}\frac{Z_<}{\alpha_+}+\Omega^{4n+2}\frac{1}{w_>}\right|^2
\right]+\ln\left[
\left|\frac{1}{w^*_>}-\Omega^{2n}\frac{Z_<}{\alpha_+}+\Omega^{4n+2}w^*_>\right|^2
\right]\right)
\right.\\
&&-\left.
\frac{1}{2}\sum_{n=1,3,5\ldots}^{\infty}%\sum_{s_2=\pm1}
\left(\ln\left[\left|1-\Omega^{2n}\frac{Z_<}{\alpha_+}w_>+\Omega^{4n+2}w^2_>
\right|^2\right]+
\ln\left[\left|1-\Omega^{2n}\frac{Z_<}{\alpha_+}\frac{1}{w^*_>}+\Omega^{4n+2}\frac{1}{{w^*}^2_>}
\right|^2\right]
\right)\right)
\end{eqnarray}
Using $w+\frac{\Omega^2}{w}=\frac{Z}{\alpha_+}$, we have
\begin{eqnarray}
&&G(\rho,\phi;\rho',\phi')
=\frac{1}{2\pi}\left(\ln\rho_<
-\frac{1}{2}
\left(\ln\left[
\left|\frac{Z_>-Z_<}{\alpha_+}\right|^2
\right]+\ln\left[
\left|\frac{1}{w^*_>}-\frac{Z_<}{\alpha_+}+\Omega^{2}w^*_>\right|^2
\right]\right)
\right.\nonumber\\
&&\left. -\frac{1}{2}\sum_{n=2,4,\ldots}^{\infty}
\left(\ln\left[
\left|1-\Omega^{2n}\frac{Z_<}{\alpha_+ w_>}+\Omega^{4n+2}\frac{1}{w^2_>}\right|^2
\right]+\ln\left[
\left|1-\Omega^{2n}\frac{Z_<}{\alpha_+}w^*_>+\Omega^{4n+2}{w^*}^2_>\right|^2
\right]\right)
\right.\\
&&-\left.
\frac{1}{2}\sum_{n=1,3,5\ldots}^{\infty}%\sum_{s_2=\pm1}
\left(\ln\left[\left|1-\Omega^{2n}\frac{Z_<}{\alpha_+}w_>+\Omega^{4n+2}w^2_>
\right|^2\right]+
\ln\left[\left|1-\Omega^{2n}\frac{Z_<}{\alpha_+}\frac{1}{w^*_>}+\Omega^{4n+2}\frac{1}{{w^*}^2_>}
\right|^2\right]
\right)\right).
\end{eqnarray}
And if the $Z_>$ is on the boundary, then
\begin{eqnarray}
\left.\left(\frac{1}{w^*_>}-\frac{Z_<}{\alpha_+}+\Omega^{2}w^*_>\right)\right|_{w^*_>\in\text{outer boundary}}\rightarrow
\left.\left(w_>-\frac{Z_<}{\alpha_+}+\frac{\Omega^{2}}{w_>}\right)\right|_{w_>\in\text{outer boundary}}=
\left.\left(\frac{Z_>-Z_<}{\alpha_+}\right)\right|_{Z_>\in\text{outer boundary}}
\end{eqnarray}

The $\alpha_+$ in the denominator just corresponds to an overall constant to $G$ which does not change the resulting potential.
So we see that when $Z_>$ is placed on the boundary, and if $\Omega$ is small,
then the leading order term Green's function is the same as in the isotropic case except that the coordinate $z$ is replaced by $Z$, i.e. it is
rescaled.

\section{Analytical solution for the anisotropic linear conductivity tensor case}

In this section we seek for a nonlinear electrostatic potential $\Phi^{\newtext{(2)}}_{ani}(\bR)$ in the case of arbitrary anisotropy of linear conductivity tensor.
As we established in Eq. \eqref{eq:marvelous}, the solution for nonlinear potential in the case of arbitrary anisotropy of the linear conductivity tensor is given by
\begin{eqnarray}
\phi^{\newtext{(2)}}(\bR)=
\frac{1}{\bar{\sigma}}\int_{\Omega}d^2\bR' \left(\nabla_{\bR'}G_{N}(\bR,\bR')\right) \cdot\tilde{\mathcal{E}}^{(2)}(\bR'),
\end{eqnarray}
which can be reexpressed as
\begin{eqnarray}
&&\Phi^{\newtext{(2)}}_{ani}(\bR)=
\frac{1}{\bar{\sigma}}\int_{\Omega}d^2\bR' \left[\left(\frac{\partial}{\partial
Z'}G_{N}(\bR,\bR')\right)\left(\tilde{\mathcal{E}}_x^{(2)}(\bR')+i\tilde{\mathcal{E}}_y^{(2)}(\bR')\right)+
\left(\frac{\partial}{\partial Z'^*}G_{N}(\bR,\bR')\right)\left(\tilde{\mathcal{E}}_x^{(2)}(\bR')-i\tilde{\mathcal{E}}_y^{(2)}(\bR')\right)\right]\nonumber\\
\end{eqnarray}
where the integration region is defined by an \textit{ellipse} and
\begin{eqnarray}
&&\tilde{\mathcal{E}}_x^{(2)}(\bR')+i\tilde{\mathcal{E}}_y^{(2)}(\bR')=
\frac{1}{\sqrt{1+\frac{\Delta\sigma}{\bar{\sigma}}}}\mathcal{E}_x^{(2)}(\br')+
\frac{i}{\sqrt{1-\frac{\Delta\sigma}{\bar{\sigma}}}}\mathcal{E}_y^{(2)}(\br')=\\
&&
\frac{1}{\sqrt{1+\frac{\Delta\sigma}{\bar{\sigma}}}}
\left(\frac{\mathcal{E}_x^{(2)}(\br')+i\mathcal{E}_y^{(2)}(\br')}{2}+
\frac{\mathcal{E}_x^{(2)}(\br')-i\mathcal{E}_y^{(2)}(\br')}{2}
\right)+
\frac{1}{\sqrt{1-\frac{\Delta\sigma}{\bar{\sigma}}}}
\left(\frac{\mathcal{E}_x^{(2)}(\br')
+i\mathcal{E}_y^{(2)}(\br')}{2}-\frac{\mathcal{E}_x^{(2)}(\br')
-i\mathcal{E}_y^{(2)}(\br')}{2}
\right)\nonumber\\
&&=\alpha_+\left(\mathcal{E}_x^{(2)}(\br')+i\mathcal{E}_y^{(2)}(\br')\right)+
\alpha_-\left(\mathcal{E}_x^{(2)}(\br')-i\mathcal{E}_y^{(2)}(\br')\right)= \\
&&=\alpha_+\left(\left(\sigma^{(2)}_{-} e^{i\chi^{(2)}_{-}} + \Omega^2 \sigma^{(2)}_{+} e^{-i\chi^{(2)}_{+}} \right) \left(E^{\newtext{(1)}}_x-iE^{\newtext{(1)}}_y\right)^2+
\left(\sigma^{(2)}_{+} e^{i\chi^{(2)}_{+}} + \Omega^2 \sigma^{(2)}_{-} e^{-i\chi^{(2)}_{-}} \right) \left(E^{\newtext{(1)}}_x+iE^{\newtext{(1)}}_y\right)^2 +\right. \\
&&\left.+ \sigma^{(2)}_{0} \left( e^{i\chi^{(2)}_{0}} + \Omega^2  e^{-i\chi^{(2)}_{0}} \right) \left({E^{\newtext{(1)}}_x}^2+{E^{\newtext{(1)}}_y}^2\right)\right),
\end{eqnarray}
where $E^{\newtext{(1)}}_{x,y}$ are functions of $\br'$. To avoid confusion, we note that $\br'=\br'(\bR')$ via Eq. (\ref{eqn:Zxy}):
\begin{eqnarray}
\label{eqn:Zxy}
Z=\frac{x}{\sqrt{1+\frac{\Delta\sigma}{\bar{\sigma}}}}+i\frac{y}{\sqrt{1-\frac{\Delta\sigma}{\bar{\sigma}}}}.
\end{eqnarray}
The electric field $E^{\newtext{(1)}}$ can be explicitly expressed via the linear electrostatic potential
\begin{eqnarray}
&&E^{\newtext{(1)}}_x(\br)-iE^{\newtext{(1)}}_y(\br)=\left(-\frac{\partial}{\partial x}+i\frac{\partial}{\partial y}\right)\Phi^{\newtext{(1)}}_{ani}(\bR)=
\left(-\frac{\partial Z}{\partial x}\frac{\partial}{\partial Z}
-\frac{\partial Z^*}{\partial x}\frac{\partial}{\partial Z^*}
+i\frac{\partial Z}{\partial y}\frac{\partial}{\partial Z}
+i\frac{\partial Z^*}{\partial y}\frac{\partial}{\partial Z^*}\right)\Phi^{\newtext{(1)}}_{ani}(\bR)= \nonumber \\
&&
\left(-\frac{1}{\sqrt{1+\frac{\Delta\sigma}{\bar{\sigma}}}}\frac{\partial}{\partial Z}
-\frac{1}{\sqrt{1+\frac{\Delta\sigma}{\bar{\sigma}}}}\frac{\partial}{\partial Z^*}
-\frac{1}{\sqrt{1-\frac{\Delta\sigma}{\bar{\sigma}}}}\frac{\partial}{\partial Z}
+\frac{1}{\sqrt{1-\frac{\Delta\sigma}{\bar{\sigma}}}}\frac{\partial}{\partial Z^*}\right)\Phi^{\newtext{(1)}}_{ani}(\bR)=\\
&&
\left(-2\alpha_+\frac{\partial}{\partial Z}
-2\alpha_-\frac{\partial}{\partial Z^*}
\right)\Phi^{\newtext{(1)}}_{ani}(\bR),
\end{eqnarray}
with $a=1$ (we consider unit disk sample).
As we showed in the previous section, Neumann Green's function for the ellipse reads 
\begin{eqnarray}
&&G_N(\rho,\phi;\rho',\phi')
=\frac{\Theta(\rho-\rho')}{2\pi}\left(\ln\rho'
-\frac{1}{2}
\left(\ln\left[
\left|\frac{Z-Z'}{\alpha_+}\right|^2
\right]+\ln\left[
\left|\frac{1}{w^*}-\frac{Z'}{\alpha_+}+\Omega^{2}w^*\right|^2
\right]\right)
\right.\nonumber\\
&&\left. -\frac{1}{2}\sum_{n=2,4,\ldots}^{\infty}
\left(\ln\left[
\left|1-\Omega^{2n}\frac{Z'}{\alpha_+ w}+\Omega^{4n+2}\frac{1}{w^2}\right|^2
\right]+\ln\left[
\left|1-\Omega^{2n}\frac{Z'}{\alpha_+}w^*+\Omega^{4n+2}{w^*}^2\right|^2
\right]\right)
\right.\\
&&-\left.
\frac{1}{2}\sum_{n=1,3,5\ldots}^{\infty}
\left(\ln\left[\left|1-\Omega^{2n}\frac{Z'}{\alpha_+}w+\Omega^{4n+2}w^2
\right|^2\right]+
\ln\left[\left|1-\Omega^{2n}\frac{Z'}{\alpha_+}\frac{1}{w^*}+\Omega^{4n+2}\frac{1}{{w^*}^2}
\right|^2\right]
\right)\right)\\
&&+\frac{\Theta(\rho'-\rho)}{2\pi}\left(\ln\rho
-\frac{1}{2}
\left(\ln\left[
\left|\frac{Z'-Z}{\alpha_+}\right|^2
\right]+\ln\left[
\left|\frac{1}{w'^*}-\frac{Z}{\alpha_+}+\Omega^{2}w'^*\right|^2
\right]\right)
\right.\nonumber\\
&&\left. -\frac{1}{2}\sum_{n=2,4,\ldots}^{\infty}
\left(\ln\left[
\left|1-\Omega^{2n}\frac{Z}{\alpha_+ w'}+\Omega^{4n+2}\frac{1}{w'^2}\right|^2
\right]+\ln\left[
\left|1-\Omega^{2n}\frac{Z}{\alpha_+}w'^*+\Omega^{4n+2}{w'^*}^2\right|^2
\right]\right)
\right.\\
&&-\left.
\frac{1}{2}\sum_{n=1,3,5\ldots}^{\infty}
\left(\ln\left[\left|1-\Omega^{2n}\frac{Z}{\alpha_+}w'+\Omega^{4n+2}w'^2
\right|^2\right]+
\ln\left[\left|1-\Omega^{2n}\frac{Z}{\alpha_+}\frac{1}{w'^*}+\Omega^{4n+2}\frac{1}{{w'^*}^2}
\right|^2\right]
\right)\right)
\end{eqnarray}
with
$$\frac{w}{\Omega}=\frac{Z}{2\alpha_+\Omega}+\text{sign}(\mathcal{R}e Z)\sqrt{\left(\frac{Z}{2\alpha_+\Omega}\right)^2-1}.$$

We can use the expressions above and
\begin{eqnarray}
\nabla\cdot j^{\newtext{(1)}}(\br)&=&-\nabla\cdot\sigma\cdot\nabla \Phi^{\newtext{(1)}}_{ani}(\br)\\
\nabla\cdot j^{\newtext{(1)}}(\br)&=&I\left(\delta(\br-\br_S)-\delta(\br-\br_D)\right)\\
-\bar{\sigma}\left(\frac{\partial^2}{\partial X^2}+\frac{\partial^2}{\partial Y^2}\right)\Phi^{\newtext{(1)}}_{ani} (\bR) &=&\frac{I}{\sqrt{1-\left(\frac{\Delta\sigma}{\bar{\sigma}}\right)^2}}
\left(\delta(\bR-\bR_S)-\delta(\bR-\bR_D)\right)
\end{eqnarray}
to determine the closed form of $\Phi^{\newtext{(1)}}_{ani} (\bR) $ via
\begin{eqnarray}
\Phi^{\newtext{(1)}}_{ani}(\bR)&&=-\int_{\Omega}d^2\bR' G_{N}(\bR,\bR')\nabla^2_{\bR'}\Phi^{\newtext{(1)}}_{ani}(\bR')+\int_C d\ell' G_{N}(\bR,\bR')\hat{\bN}\cdot\nabla_{\bR'}\Phi^{\newtext{(1)}}_{ani}(\bR')\\
\Phi^{\newtext{(1)}}_{ani}(\bR)&&=\frac{I}{\bar{\sigma}\sqrt{1-\left(\frac{\Delta\sigma}{\bar{\sigma}}\right)^2}}
\left(G_{N}(\bR,\bR_S)-G_{N}(\bR,\bR_D)\right)
\end{eqnarray}
It will prove convenient to work in the Zhukovsky coordinates, where Neumann Green's function takes the form
\begin{eqnarray}
&&G_N(\rho,\phi;\rho',\phi')=\frac{1}{2\pi}\left(\ln\rho_<\right.\\
&&\left. -\frac{1}{2}\sum_{n=0,2,4,\ldots}^{\infty}
\left(\ln\left[
\left|w_>-\Omega^{2n}\frac{Z_<}{\alpha_+}+\Omega^{4n+2}\frac{1}{w_>}\right|^2
\right]+\ln\left[
\left|\frac{1}{w^*_>}-\Omega^{2n}\frac{Z_<}{\alpha_+}+\Omega^{4n+2}w^*_>\right|^2
\right]\right)
\right.\\
&&-\left.
\frac{1}{2}\sum_{n=1,3,5\ldots}^{\infty}
\left(\ln\left[\left|1-\Omega^{2n}\frac{Z_<}{\alpha_+}w_>+\Omega^{4n+2}w^2_>
\right|^2\right]+
\ln\left[\left|1-\Omega^{2n}\frac{Z_<}{\alpha_+}\frac{1}{w^*_>}+\Omega^{4n+2}\frac{1}{{w^*}^2_>}
\right|^2\right]
\right)\right).
\end{eqnarray}
For the source and drain at the boundary of the original disk, $w_>$ is tied to $R_{S,D}$, And, as shown in \cite{Oskar_solo}, it is just the original coordinate. So, the linear potential reads
\begin{eqnarray}
\Phi^{\newtext{(1)}}_{ani}(\bR)&&=\frac{I}{2\pi \bar{\sigma}\sqrt{1-\left(\frac{\Delta\sigma}{\bar{\sigma}}\right)^2}}
\left(\sum_{n=0,2,4,\ldots}^{\infty}
\ln
\frac{\left|1+\Omega^{4n+2}{z^*_D}^2-\Omega^{2n}\frac{Z}{\alpha_+}z^*_D\right|^2
}{\left|1+\Omega^{4n+2}{z^*_S}^2-\Omega^{2n}\frac{Z}{\alpha_+}z_S^*\right|^2
}+
\sum_{n=1,3,5,\ldots}^{\infty}
\ln
\frac{\left|1+\Omega^{4n+2}z^2_D-\Omega^{2n}\frac{Z}{\alpha_+}z_D\right|^2
}{\left|1+\Omega^{4n+2}z^2_S-\Omega^{2n}\frac{Z}{\alpha_+}z_S\right|^2
}\right)\nonumber\\
\end{eqnarray}

We will find it helpful to transform the 2D integration in $R$ variables to Zhukovsky variables as well.
For this we will need the Jacobian \cite{Oskar_solo}
\begin{eqnarray}
&&\int_{\text{ellipse}} dXdY (\ldots)=\int_{\text{annulus}} du dv J\left(\frac{X,Y}{u,v}\right)(\ldots)=\alpha^2_+\int_{\text{annulus}} du dv\left(1-\frac{\Omega^2}{w^2}\right)
\left(1-\frac{\Omega^2}{{w^*}^2}\right)(\ldots)\\
&&=\alpha^2_+\int_{\Omega}^1 d\rho \rho \int_0^{2\pi}d\theta\left(1-\frac{\Omega^2}{\rho^2}e^{-2i\theta}\right)
\left(1-\frac{\Omega^2}{\rho^2}e^{2i\theta}\right)(\ldots).
\end{eqnarray}
Knowing the linear potential $\Phi^{\newtext{(1)}}_{ani}(\bR)$ allows us to find an explicit expression for the electric field $E^{\newtext{(1)}}$:
\begin{eqnarray}
&&E^{\newtext{(1)}}_x(\br)-iE^{\newtext{(1)}}_y(\br)=
\left(-2\alpha_+\frac{\partial}{\partial Z}
-2\alpha_-\frac{\partial}{\partial Z^*}
\right)\Phi^{\newtext{(1)}}_{ani}(\bR)\\
&&=\frac{\left(-2\alpha_+\right)I}{2\pi \bar{\sigma}\sqrt{1-\left(\frac{\Delta\sigma}{\bar{\sigma}}\right)^2}}
\frac{\partial}{\partial Z}\left(\sum_{n=0,2,4,\ldots}^{\infty}
\ln
\frac{\left|1+\Omega^{4n+2}{z^*_D}^2-\Omega^{2n}\frac{Z}{\alpha_+}z^*_D\right|^2
}{\left|1+\Omega^{4n+2}{z^*_S}^2-\Omega^{2n}\frac{Z}{\alpha_+}z_S^*\right|^2
}+
\sum_{n=1,3,5,\ldots}^{\infty}
\ln
\frac{\left|1+\Omega^{4n+2}z^2_D-\Omega^{2n}\frac{Z}{\alpha_+}z_D\right|^2
}{\left|1+\Omega^{4n+2}z^2_S-\Omega^{2n}\frac{Z}{\alpha_+}z_S\right|^2
}\right)\nonumber\\
&&+\frac{\left(-2\alpha_-\right)I}{2\pi \bar{\sigma}\sqrt{1-\left(\frac{\Delta\sigma}{\bar{\sigma}}\right)^2}}
\frac{\partial}{\partial Z^*}\left(\sum_{n=0,2,4,\ldots}^{\infty}
\ln
\frac{\left|1+\Omega^{4n+2}{z^*_D}^2-\Omega^{2n}\frac{Z}{\alpha_+}z^*_D\right|^2
}{\left|1+\Omega^{4n+2}{z^*_S}^2-\Omega^{2n}\frac{Z}{\alpha_+}z_S^*\right|^2
}+
\sum_{n=1,3,5,\ldots}^{\infty}
\ln
\frac{\left|1+\Omega^{4n+2}z^2_D-\Omega^{2n}\frac{Z}{\alpha_+}z_D\right|^2
}{\left|1+\Omega^{4n+2}z^2_S-\Omega^{2n}\frac{Z}{\alpha_+}z_S\right|^2
}\right)\nonumber\\
&&=\frac{\left(-2\alpha_+\right)I}{2\pi \bar{\sigma}\sqrt{1-\left(\frac{\Delta\sigma}{\bar{\sigma}}\right)^2}}
\left(\sum_{n=0,2,4,\ldots}^{\infty}
\frac{z^*_S}{\alpha_+}
\frac{\Omega^{2n}}{1+\Omega^{4n+2}{z^*_S}^2-\Omega^{2n}\frac{Z}{\alpha_+}z_S^*}
+
\sum_{n=1,3,5,\ldots}^{\infty}
\frac{z_S}{\alpha_+}\frac{\Omega^{2n}}{1+\Omega^{4n+2}z^2_S-\Omega^{2n}\frac{Z}{\alpha_+}z_S}
-(S\rightarrow D)\right)
\nonumber\\
&&+\frac{\left(-2\alpha_-\right)I}{2\pi \bar{\sigma}\sqrt{1-\left(\frac{\Delta\sigma}{\bar{\sigma}}\right)^2}}
\left(\sum_{n=0,2,4,\ldots}^{\infty}
\frac{z_S}{\alpha_+}
\frac{\Omega^{2n}}{1+\Omega^{4n+2}{z_S}^2-\Omega^{2n}\frac{Z^*}{\alpha_+}z_S}
+
\sum_{n=1,3,5,\ldots}^{\infty}
\frac{z^*_S}{\alpha_+}\frac{\Omega^{2n}}{1+\Omega^{4n+2}{z^*_S}^2-\Omega^{2n}\frac{Z^*}{\alpha_+}z^*_S}
-(S\rightarrow D)
\right)\nonumber\\
&&=\frac{-I}{\pi \bar{\sigma}\sqrt{1-\left(\frac{\Delta\sigma}{\bar{\sigma}}\right)^2}}
\left(\sum_{n=0,2,4,\ldots}^{\infty}
\frac{z^*_S\Omega^{2n}}{1+\Omega^{4n+2}{z^*_S}^2-\Omega^{2n}\frac{Z}{\alpha_+}z_S^*}
+
\sum_{n=1,3,5,\ldots}^{\infty}
\frac{z_S\Omega^{2n}}{1+\Omega^{4n+2}z^2_S-\Omega^{2n}\frac{Z}{\alpha_+}z_S}
-(S\rightarrow D)\right)
\nonumber\\
&&+\frac{-I}{\pi \bar{\sigma}\sqrt{1-\left(\frac{\Delta\sigma}{\bar{\sigma}}\right)^2}}
\left(\sum_{n=0,2,4,\ldots}^{\infty}
\frac{z_S\Omega^{2n+2}}{1+\Omega^{4n+2}{z_S}^2-\Omega^{2n}\frac{Z^*}{\alpha_+}z_S}
+
\sum_{n=1,3,5,\ldots}^{\infty}
\frac{z^*_S\Omega^{2n+2}}{1+\Omega^{4n+2}{z^*_S}^2-\Omega^{2n}\frac{Z^*}{\alpha_+}z^*_S}
-(S\rightarrow D)
\right)\nonumber\\
&&=\frac{-I}{\pi \bar{\sigma}\sqrt{1-\left(\frac{\Delta\sigma}{\bar{\sigma}}\right)^2}}
\left(\sum_{n=0,2,4,\ldots}^{\infty}
\frac{\Omega^{2n}}{z_S+\Omega^{4n+2}z^*_S-\Omega^{2n}\frac{Z}{\alpha_+}}
+
\sum_{n=1,3,5,\ldots}^{\infty}
\frac{\Omega^{2n}}{z^*_S+\Omega^{4n+2}z_S-\Omega^{2n}\frac{Z}{\alpha_+}}
-(S\rightarrow D)\right)
\nonumber\\
&&+\frac{-I}{\pi \bar{\sigma}\sqrt{1-\left(\frac{\Delta\sigma}{\bar{\sigma}}\right)^2}}
\left(\sum_{n=0,2,4,\ldots}^{\infty}
\frac{\Omega^{2n+2}}{z^*_S+\Omega^{4n+2}z_S-\Omega^{2n}\frac{Z^*}{\alpha_+}}
+
\sum_{n=1,3,5,\ldots}^{\infty}
\frac{\Omega^{2n+2}}{z_S+\Omega^{4n+2}z^*_S-\Omega^{2n}\frac{Z^*}{\alpha_+}}
-(S\rightarrow D)
\right)\nonumber
\end{eqnarray}
In Zhukovsky variables $Z=\alpha_+w+\frac{\alpha_-}{w}$ and $\frac{\alpha_-}{\alpha_+}=\Omega^2$. Therefore, the electric field can be cast into a convenient form
\begin{eqnarray}
&&E^{\newtext{(1)}}_x(\br)-iE^{\newtext{(1)}}_y(\br)\\
&&=\frac{-I}{\pi \bar{\sigma}\sqrt{1-\left(\frac{\Delta\sigma}{\bar{\sigma}}\right)^2}}
\left(\sum_{n=0,2,4,\ldots}^{\infty}
\frac{\Omega^{2n}}{z_S+\Omega^{4n+2}z^*_S-\Omega^{2n}\left(w+\frac{\Omega^2}{w}\right)}
+
\sum_{n=1,3,5,\ldots}^{\infty}
\frac{\Omega^{2n}}{z^*_S+\Omega^{4n+2}z_S-\Omega^{2n}\left(w+\frac{\Omega^2}{w}\right)}
-(S\rightarrow D)\right)
\nonumber\\
&&+\frac{-I}{\pi \bar{\sigma}\sqrt{1-\left(\frac{\Delta\sigma}{\bar{\sigma}}\right)^2}}
\left(\sum_{n=0,2,4,\ldots}^{\infty}
\frac{\Omega^{2n+2}}{z^*_S+\Omega^{4n+2}z_S-\Omega^{2n}\left(w^*+\frac{\Omega^2}{w^*}\right)}
+
\sum_{n=1,3,5,\ldots}^{\infty}
\frac{\Omega^{2n+2}}{z_S+\Omega^{4n+2}z^*_S-\Omega^{2n}\left(w^*+\frac{\Omega^2}{w^*}\right)}
-(S\rightarrow D)
\right)\nonumber
\end{eqnarray}

For the ``observation'' point $\bR$ on the boundary, $\bR'$ is inside and we find
\begin{eqnarray}
&&\frac{\partial}{\partial Z'}G_{N}(\bR,\bR')=\nonumber\\
&&\frac{\partial}{\partial Z'}\frac{1}{2\pi}\left(\ln|Z'| -\sum_{n=0,2,4,\ldots}^{\infty}
\ln\left[
\left|1-\Omega^{2n}\frac{Z'}{\alpha_+}z^*+\Omega^{4n+2}{z^*}^2\right|^2
\right]
-
\sum_{n=1,3,5\ldots}^{\infty}
\ln\left[\left|1-\Omega^{2n}\frac{Z'}{\alpha_+}z+\Omega^{4n+2}z^2
\right|^2\right]
\right).
\end{eqnarray}
The first term in the expression above can be ignored as it does not depend on $\bR$ and, therefore, only contributes to an overall shift of the potential. Thus,
\begin{eqnarray}
&&\frac{\partial}{\partial Z'}G_{N}(\bR,\bR')\rightarrow\frac{1}{2\pi}\left(\sum_{n=0,2,4,\ldots}^{\infty}
\frac{z^*}{\alpha_+}\frac{\Omega^{2n}}{\left(1-\Omega^{2n}\frac{Z'}{\alpha_+}z^*+\Omega^{4n+2}{z^*}^2\right)}
+
\sum_{n=1,3,5\ldots}^{\infty}
\frac{z}{\alpha_+}\frac{\Omega^{2n}}{1-\Omega^{2n}\frac{Z'}{\alpha_+}z+\Omega^{4n+2}z^2}
\right)\\
&&=\frac{1}{2\pi \alpha_+}\left(\sum_{n=0,2,4,\ldots}^{\infty}
\frac{\Omega^{2n}}{z+\Omega^{4n+2}{z^*}-\Omega^{2n}\frac{Z'}{\alpha_+}}
+
\sum_{n=1,3,5\ldots}^{\infty}
\frac{\Omega^{2n}}{z^*-\Omega^{2n}\frac{Z'}{\alpha_+}+\Omega^{4n+2}z}
\right)\\
&&=\frac{1}{2\pi \alpha_+}\left(\sum_{n=0,2,4,\ldots}^{\infty}
\frac{\Omega^{2n}}{z+\Omega^{4n+2}{z^*}-\Omega^{2n}\left(w'+\frac{\Omega^2}{w'}\right)}
+
\sum_{n=1,3,5\ldots}^{\infty}
\frac{\Omega^{2n}}{z^*+\Omega^{4n+2}z-\Omega^{2n}\left(w'+\frac{\Omega^2}{w'}\right)}
\right).
\end{eqnarray}
To evaluate relevant integrals in Zhukovsky variables we need to know the structure of poles of the integrand. We study it below.

\subsection{Analysis of the poles}

When performing the integrals over $w'$, we encounter terms in the denominator whose factors we will need.
They can be brought to the form
\begin{eqnarray}
\frac{w}{\Omega}+\frac{\Omega}{w}=Q,
\end{eqnarray}
where $Q$ is a complex number.
Decomposing into the amplitude and the phase, $w=\rho e^{i\theta}$ and setting $\zeta=e^{i\theta}$, we have
\begin{eqnarray}
\frac{\rho}{\Omega}\zeta+\frac{\Omega}{\rho}\frac{1}{\zeta}=Q.
\end{eqnarray}
Let
\begin{eqnarray}
\zeta'=\frac{\rho}{\Omega}\zeta.\end{eqnarray}
Then
\begin{eqnarray}
\zeta'+\frac{1}{\zeta'}=Q.
\end{eqnarray}
Therefore, if $\zeta'$ is a solution then so is $1/\zeta'$.
Moreover, unless $Q$ is real and $-2<Q<2$, one of the $\zeta'$ solutions has an amplitude larger than $1$ and the other smaller than $1$, because the solution of the above quadratic equation is 
\begin{eqnarray}
\zeta'_{\pm}=\frac{Q}{2}\pm \sqrt{\frac{Q^2}{4}-1}.
\end{eqnarray}
There are two possibilities then

\noindent $\bullet$ If $\mathcal{R}eQ>0$ then $\zeta'_>=\frac{Q}{2}+\sqrt{\frac{Q^2}{4}-1}$ is the root with the larger amplitude, and 
$\zeta'_<=\frac{Q}{2}-\sqrt{\frac{Q^2}{4}-1}$ is the root with the smaller amplitude. $\zeta'_>$ must be outsize unit complex circle and $\zeta'_<$ inside.\\
\noindent $\bullet$ If $\mathcal{R}eQ<0$ then $\zeta'_>=\frac{Q}{2}-\sqrt{\frac{Q^2}{4}-1}$ is the root with the larger amplitude, and 
$\zeta'_<=\frac{Q}{2}+\sqrt{\frac{Q^2}{4}-1}$ is the root with the smaller amplitude.
Again, $\zeta'_>$ must be outsize unit complex circle and $\zeta'_<$ inside.
\\

The smaller root then leads to $\zeta_<=\frac{\Omega}{\rho}\zeta'_<$, which is guaranteed to be inside the unit circle because 
$\frac{\Omega}{\rho}< 1$.
The larger root leads to $\zeta_>=\frac{\Omega}{\rho}\zeta'_>$. Although $\zeta'_>$ is outside the unit circle, the parameter 
$\frac{\Omega}{\rho}$ scales down its radius. Therefore, depending on the value of $\left|\zeta'_>\right|$ and the value of $\frac{\Omega}{\rho}$, the variable $\zeta_>$ may be either inside or outside the unit circle.
It is outside if $\frac{\rho}{\Omega}<|\zeta'_{>}|$ and inside otherwise.

The expression for $\zeta'_{\pm}$ can be significantly simplified for $|z|=1$:
\begin{eqnarray}
&&\zeta'_{\pm}=\frac{Q}{2}\pm \sqrt{\frac{Q^2}{4}-1} = \frac{z+\Omega^{4n+2}{z^*}}{2\Omega^{2n+1}}\pm \sqrt{ \frac{(z+\Omega^{4n+2}{z^*})^2}{4\Omega^{4n+2}} -1} = 
\frac{z+\Omega^{4n+2}{z^*}}{2\Omega^{2n+1}}\pm \sqrt{ \frac{z^2+2 \Omega^{4n+2} z^* z +\Omega^{8n+4}(z^*)^2 - 4\Omega^{4n+2}}{4\Omega^{4n+2}} } = \nonumber \\ 
&&=\frac{z+\Omega^{4n+2}{z^*}}{2\Omega^{2n+1}}\pm \sqrt{ \frac{z^2-2 \Omega^{4n+2} z^* z +\Omega^{8n+4}(z^*)^2}{4\Omega^{4n+2}} } = 
\frac{z+\Omega^{4n+2}{z^*}}{2\Omega^{2n+1}}\pm \sqrt{ \left(\frac{z-\Omega^{4n+2}z^*}{2\Omega^{2n+1}} \right)^2 } = 
\frac{z+\Omega^{4n+2}{z^*}}{2\Omega^{2n+1}}\pm \frac{z-\Omega^{4n+2}z^*}{2\Omega^{2n+1}}. \nonumber
\end{eqnarray}
Therefore, 
\begin{eqnarray}
    \zeta'_+ = \frac{z}{\Omega^{2n+1}}, \\
    \zeta'_- = z^* \Omega^{2n+1}.
\end{eqnarray}
This means that the larger root is always $\zeta'_+$ as $0<\Omega<1$ and $|z|=1$.

We also encounter terms of the form
\begin{eqnarray}
\frac{w^*}{\Omega}+\frac{\Omega}{w^*}=P,
\end{eqnarray}
where $P$ is a complex number. With $w^*=\rho e^{-i\theta}$ and $\zeta=e^{i\theta}$ this expression becomes
\begin{eqnarray}
\frac{\rho}{\Omega}\frac{1}{\zeta}+\frac{\Omega}{\rho}\zeta=P
\end{eqnarray}
As in the previous case we introduce
\begin{eqnarray}
\xi'=\frac{\Omega}{\rho}\zeta.
\end{eqnarray}
Then,
\begin{eqnarray}
\frac{1}{\xi'}+\xi'=P.
\end{eqnarray}
Therefore if $\xi'$ is a solution then so is $1/\xi'$.
Similar to the above analysis, 
\begin{eqnarray}
\xi'_{\pm}=\frac{P}{2}\pm \sqrt{\frac{P^2}{4}-1},
\end{eqnarray}
and we again encounter the two possibilities: \\
\noindent $\bullet$ If $\mathcal{R}eP>0$, then $\xi'_>=\frac{P}{2}+\sqrt{\frac{P^2}{4}-1}$ is the root with the larger amplitude, and
$\xi'_<=\frac{P}{2}-\sqrt{\frac{P^2}{4}-1}$ is the root with the smaller amplitude. $\xi'_>$ must be outsize unit complex circle and $\xi'_<$ inside.\\
\noindent $\bullet$ If $\mathcal{R}eP<0$, then $\xi'_>=\frac{P}{2}-\sqrt{\frac{P^2}{4}-1}$ is the root with the larger amplitude, and
$\xi'_<=\frac{P}{2}+\sqrt{\frac{P^2}{4}-1}$ is the root with the smaller amplitude.
Again, $\xi'_>$ must be outsize unit complex circle and $\xi'_<$ inside; and $\xi'_<\xi'_>=1$.
\\

The larger root leads to $\zeta_>=\frac{\rho}{\Omega}\xi'_>$. Since $\Omega<\rho<1$ the ratio is $\frac{\rho}{\Omega}$ larger than $1$, and since $\xi'_>$ is guaranteed to be outside the unit complex circle, $\zeta_>$ is guaranteed to be outside the unit circle. The smaller root leads to $\zeta_<=\frac{\rho}{\Omega}\xi'_<$. Although $\xi'_<$ is guaranteed to be inside the unit complex circle, $\zeta_<$ is not guaranteed to be inside the unit circle because the factor $\frac{\rho}{\Omega}>1$ can dilate it outside. 
$\zeta_<$ is inside if $\rho<\frac{\Omega}{|\xi'_<|}$. The two roots can be simplified in the same way as previously.

\subsection{Base integral 1}

The nonlinear electrostatic potential can be expressed in terms of the base integrals. In this and the following subsection we evaluate the two base integrals. 

The first base integral is given by
\begin{eqnarray}
&&\mathcal{I}_1(Q)\equiv\alpha^2_+\int_{\text{annulus}}dudv\left(1-\frac{\Omega^2}{w^2}\right)
\left(1-\frac{\Omega^2}{{w^*}^2}\right)
\frac{1}{Q-\left(\frac{w}{\Omega}+\frac{\Omega}{w}\right)}=\Omega\Upsilon(Q)
\end{eqnarray}
and let $$Q=\frac{z+\Omega^{4n+2}{z^*}}{\Omega^{2n+1}}.$$
Then integration can be split into three independent terms as follows:
\begin{eqnarray}
&&\Upsilon=-\frac{\alpha^2_+}{\Omega}\int_{\text{annulus}}dudv\left(1-\frac{\Omega^2}{w^2}\right)
\left(1-\frac{\Omega^2}{{w^*}^2}\right)
\frac{1}{\frac{w}{\Omega}+\frac{\Omega}{w}-Q}\\
&&=-\frac{\alpha^2_+}{\Omega}\int_\Omega^1 d\rho \rho \oint \frac{d\zeta}{i\zeta}\left(1-\frac{\Omega^2}{\rho^2\zeta^2}\right)
\left(1-\frac{\Omega^2\zeta^2}{\rho^2}\right)
\frac{1}{\frac{\rho}{\Omega}\zeta+\frac{\Omega}{\rho}\frac{1}{\zeta}-Q}\\
&&=i\frac{\alpha^2_+}{\Omega^2}\int_\Omega^1 d\rho \rho^2 \oint d\zeta\left(1-\frac{\Omega^2}{\rho^2\zeta^2}\right)
\left(1-\frac{\Omega^2\zeta^2}{\rho^2}\right)\frac{\Omega}{\rho \zeta}
\frac{1}{\frac{\rho}{\Omega}\zeta+\frac{\Omega}{\rho}\frac{1}{\zeta}-Q}\\
&&=i\frac{\alpha^2_+}{\Omega^2}\int_\Omega^1 d\rho \rho^2 \oint d\zeta\left(1-\frac{\Omega^2}{\rho^2\zeta^2}\right)
\left(1-\frac{\Omega^2\zeta^2}{\rho^2}\right)
\frac{1}{\left(\frac{\rho}{\Omega}\zeta-\zeta'_+\right)\left(\frac{\rho}{\Omega}\zeta-\zeta'_-\right)}\\
&&=i\alpha^2_+\int_\Omega^1 d\rho \oint d\zeta\left(1-\frac{\Omega^2}{\rho^2\zeta^2}\right)
\left(1-\frac{\Omega^2\zeta^2}{\rho^2}\right)
\frac{1}{\left(\zeta-\frac{\Omega}{\rho}\zeta'_<\right)\left(\zeta-\frac{\Omega}{\rho}\zeta'_>\right)}\\
&&=i\alpha^2_+\int_\Omega^1 d\rho 
\left(1+\frac{\Omega^4}{\rho^4}\right)\oint 
\frac{d\zeta}{\left(\zeta-\frac{\Omega}{\rho}\zeta'_<\right)\left(\zeta-\frac{\Omega}{\rho}\zeta'_>\right)}\\
&&-i\alpha^2_+\int_\Omega^1 d\rho \oint d\zeta\left(\frac{\Omega^2}{\rho^2\zeta^2}
+\frac{\Omega^2\zeta^2}{\rho^2}\right)
\frac{1}{\left(\zeta-\frac{\Omega}{\rho}\zeta'_<\right)\left(\zeta-\frac{\Omega}{\rho}\zeta'_>\right)}\\
&&=\Upsilon_1+\Upsilon_2+\Upsilon_3.
\end{eqnarray}
Below we evaluate the three terms one by one using $\zeta'_<\zeta'_>=1$. The first component is given by
\begin{eqnarray}
&&\Upsilon_1=i\alpha^2_+\int_\Omega^1 d\rho
\left(1+\frac{\Omega^4}{\rho^4}\right)\oint
\frac{d\zeta}{\left(\zeta-\frac{\Omega}{\rho}\zeta'_<\right)\left(\zeta-\frac{\Omega}{\rho}\zeta'_>\right)}
=i\alpha^2_+\int_\Omega^1 d\rho
\left(1+\frac{\Omega^4}{\rho^4}\right)\Theta\left(\frac{\Omega}{\rho}|\zeta'_>|-1\right)
\frac{2\pi i}{\left(\frac{\Omega}{\rho}\zeta'_<-\frac{\Omega}{\rho}\zeta'_>\right)} \nonumber \\
&&=-
\frac{2\pi \alpha^2_+}{\Omega\left(\zeta'_<-\zeta'_>\right)}
\int_\Omega^1 d\rho\rho
\left(1+\frac{\Omega^4}{\rho^4}\right)\Theta\left(\Omega|\zeta'_>|-\rho\right)
=-
\frac{2\pi \alpha^2_+}{\Omega\left(\zeta'_<-\zeta'_>\right)}
\int_\Omega^{\text{min}\left(1,\Omega|\zeta'_>|\right)} d\rho\rho
\left(1+\frac{\Omega^4}{\rho^4}\right)
\\
&&=-
\frac{\pi \alpha^2_+}{\Omega\left(\zeta'_<-\zeta'_>\right)}
\left(\text{min}\left(1,\Omega^2|\zeta'_>|^2\right)-\frac{\Omega^4}{\text{min}\left(1,\Omega^2|\zeta'_>|^2\right)}\right)
\end{eqnarray}
The second component reads
\begin{eqnarray}
&&\Upsilon_2=-i\alpha^2_+\Omega^2\int_\Omega^1 \frac{d\rho}{\rho^2} \oint \frac{d\zeta}{\zeta^2}
\frac{1}{\left(\zeta-\frac{\Omega}{\rho}\zeta'_<\right)\left(\zeta-\frac{\Omega}{\rho}\zeta'_>\right)}\\
&&=2\pi\alpha^2_+\Omega^2\int_\Omega^1 \frac{d\rho}{\rho^2}\left(
\frac{-1}{\left(-\frac{\Omega}{\rho}\zeta'_<\right)^2\left(-\frac{\Omega}{\rho}\zeta'_>\right)}
+\frac{-1}{\left(-\frac{\Omega}{\rho}\zeta'_<\right)\left(-\frac{\Omega}{\rho}\zeta'_>\right)^2}\right)\\
&&+2\pi \alpha^2_+\Omega^2\int_\Omega^1 \frac{d\rho}{\rho^2} \frac{1}{\left(\frac{\Omega}{\rho}\zeta'_<\right)^2}
\frac{1}{\left(\frac{\Omega}{\rho}\zeta'_<-\frac{\Omega}{\rho}\zeta'_>\right)}\\
&&+2\pi \alpha^2_+\Omega^2\int_\Omega^1 \frac{d\rho}{\rho^2} 
\frac{\Theta\left(\rho-\Omega|\zeta'_>|\right)
}{\left(\frac{\Omega}{\rho}\zeta'_>\right)^2}
\frac{1}{\left(\frac{\Omega}{\rho}\zeta'_>-\frac{\Omega}{\rho}\zeta'_<\right)} = \\
&=&\frac{\pi \alpha^2_+}{\Omega} \biggr[ \frac{1-\Omega^2}{(\zeta'_{<})^2 (\zeta'_{<} - \zeta'_{>})} + \frac{\zeta'_{<} + \zeta'_{>}}{(\zeta'_{<})^2 (\zeta'_{>})^2} (1 - \Omega^2) -  \Theta(1-\Omega |\zeta'_{>}|) \frac{1 - \max(\Omega^2, \Omega^2 |\zeta'_{>}|^2)}{(\zeta'_{>})^2 (\zeta'_{<} - \zeta'_{>})} \biggr] = \\
&=& \frac{\pi \alpha^2_+}{\Omega} (1 - \Omega^2) \left[ \frac{1}{(\zeta'_{<})^2 (\zeta'_{<} - 1/\zeta'_{<})} + \zeta'_{<} + \zeta'_{>} \right] - \frac{\pi \alpha^2_+}{\Omega} \Theta(1-\Omega |\zeta'_{>}|) \frac{1 - \Omega^2 |\zeta'_{>}|^2}{(1/\zeta'_{<})^2 (\zeta'_{<} - 1/\zeta'_{<})} = \\
&=& \frac{\pi \alpha^2_+}{\Omega}  \left[ \frac{(\zeta'_{<})^3}{(\zeta'_{<})^2 -1} (1 - \Omega^2) 
- \Theta(1-\Omega |\zeta'_{>}|) \left( 1 - \Omega^2 |\zeta'_{>}|^2 \right) \frac{(\zeta'_{<})^3}{(\zeta'_{<})^2 - 1} \right],
\end{eqnarray}
where we used $\zeta'_{>} = 1/\zeta'_{<}, \; \zeta'_{<} \zeta'_{>} = 1$, and the fact that $\Omega^2 |\zeta'_{>}|^2 > \Omega^2$. 
The last component is evaluated in the similar fashion:
\begin{eqnarray}
&&\Upsilon_3=-i\alpha^2_+\Omega^2\int_\Omega^1 \frac{d\rho}{\rho^2} \oint 
\frac{d\zeta
\zeta^2}{\left(\zeta-\frac{\Omega}{\rho}\zeta'_<\right)\left(\zeta-\frac{\Omega}{\rho}\zeta'_>\right)} = \\
&-&i\alpha^2_+\Omega^2\int_\Omega^1 \frac{d\rho}{\rho^2} \oint 
\frac{d\zeta
\zeta^2}{\frac{\Omega}{\rho } \zeta'_{<} - \frac{\Omega}{\rho}\zeta'_{>} } \left( \frac{1}{\zeta-\frac{\Omega}{\rho}\zeta'_<} - \frac{1}{\zeta-\frac{\Omega}{\rho}\zeta'_>} \right) = \\
&=& \frac{2 \pi \alpha^2_+ \Omega^3}{\zeta'_{<} - \zeta'_{>}} \left[ \int_{\Omega}^1 \frac{d \rho' (\zeta'_{<})^2}{(\rho')^3} - \Theta(1 - \Omega |\zeta'_{>}|) \int_{\max(\Omega,\Omega |\zeta'_{>}|)}^1 \frac{d \rho' (\zeta'_{>})^2}{(\rho')^3} \right] = \\
&=& \frac{\pi \alpha^2_+ \Omega^3}{\zeta'_{<} - \zeta'_{>}} \left[ (\zeta'_{<})^2 \left( \frac{1}{\Omega^2} - 1 \right) - \Theta(1 - \Omega |\zeta'_{>}|) (\zeta'_{>})^2 \left( \frac{1}{\max(\Omega^2,\Omega^2 |\zeta'_{>}|^2)} - 1 \right) \right] = \\
&=& \frac{\pi \alpha^2_+ \Omega^3}{\zeta'_{<} - \zeta'_{>}} \left[ (\zeta'_{<})^2 \left( \frac{1}{\Omega^2} - 1 \right) - \Theta(1 - \Omega |\zeta'_{>}|) (\zeta'_{>})^2 \left( \frac{1}{\Omega^2 |\zeta'_{>}|^2} - 1 \right) \right] = \\
&=& \frac{\pi \alpha^2_+ \Omega}{\zeta'_{<} - \zeta'_{>}} \left[ (\zeta'_{<})^2 \left( 1 - \Omega^2 \right) - \Theta(1 - \Omega |\zeta'_{>}|) (\zeta'_{>})^2 \left( \frac{1}{|\zeta'_{>}|^2} - \Omega^2 \right) \right].
\end{eqnarray}

Using 
\begin{eqnarray}
    \zeta'_+ = \frac{z}{\Omega^{2n+1}}, \\
    \zeta'_- = z^* \Omega^{2n+1}.
\end{eqnarray}
allows to significantly simplify the base integral. Note that $\zeta'_> = \zeta'_+$ and $\zeta'_< = \zeta'_-$. Then it is obvious that $\Omega |\zeta'_>| = \frac{1}{\Omega^{2n}}$, hence 
\begin{eqnarray}
    \Theta(1-\Omega |\zeta'_{>}|) = 0, \\
    \text{min}\left(1,\Omega^2|\zeta'_>|^2\right) = 1.
\end{eqnarray}
Therefore,
\begin{eqnarray}
    &&\Upsilon = \Upsilon_1 + \Upsilon_2 + \Upsilon_3 = -
\frac{\pi \alpha^2_+}{\Omega\left(\zeta'_<-\zeta'_>\right)}
\left( 1-\Omega^4 \right) + \frac{\pi \alpha^2_+}{\Omega} \frac{(\zeta'_{<})^3}{(\zeta'_{<})^2 -1} (1 - \Omega^2) + \frac{\pi \alpha^2_+ \Omega}{\zeta'_{<} - \zeta'_{>}}  (\zeta'_{<})^2 \left( 1 - \Omega^2 \right) = \\
&&= \frac{\pi \alpha_+^2 \left( 1 - \Omega^4 \right) \zeta_{<}'}{\Omega}.
\end{eqnarray}
Employing the expressions for $\Upsilon$ and for $\zeta'_{<}$ we can present an explicit expression for
$\mathcal{I}_1(Q)$:
\begin{equation}
    \mathcal{I}_1(Q) = \Omega\Upsilon(Q) =  \pi \alpha_+^2 \left( 1 - \Omega^4 \right) \zeta_{<}' = \pi \alpha_+^2 \left( 1 - \Omega^4 \right) z^* \Omega^{2n+1}.
\end{equation}

\subsection{Base integral 2}

To find the full expression for the potential we also need to evaluate the second base integral:
\begin{eqnarray}
&&\mathcal{I}_{2}(P,Q)\equiv\alpha^2_+\int_{\text{annulus}}dudv\left(1-\frac{\Omega^2}{w^2}\right)
\left(1-\frac{\Omega^2}{{w^*}^2}\right)
\frac{1}{Q-\left(\frac{w}{\Omega}+\frac{\Omega}{w}\right)}
\frac{1}{P-\left(\frac{w^*}{\Omega}+\frac{\Omega}{w^*}\right)}.
\end{eqnarray}
For brevity, let us introduce auxiliary functions:
\begin{eqnarray}
    F_1 (\rho) = \frac{1}{ \left( \frac{\Omega}{\rho} \zeta'_{<} - \frac{\Omega}{\rho} \zeta'_{>} \right) \left( \frac{\Omega}{\rho} \zeta'_{<} - \frac{\rho}{\Omega} \xi'_{>} \right) \left( \frac{\Omega}{\rho} \zeta'_{<} - \frac{\rho}{\Omega} \xi'_{<} \right) }, \\
    F_2 (\rho) = \frac{1}{ \left( \frac{\Omega}{\rho} \zeta'_{<} - \frac{\Omega}{\rho} \zeta'_{>} \right) \left( \frac{\Omega}{\rho} \zeta'_{>} - \frac{\rho}{\Omega} \xi'_{>} \right) \left( \frac{\Omega}{\rho} \zeta'_{>} - \frac{\rho}{\Omega} \xi'_{<} \right) }, \\
    F_3 (\rho) = \frac{1}{ \left( \frac{\Omega}{\rho} \zeta'_{<} - \frac{\rho}{\Omega} \xi'_{>} \right) \left( \frac{\Omega}{\rho} \zeta'_{>} - \frac{\rho}{\Omega} \xi'_{>} \right) \left( \frac{\rho}{\Omega} \xi'_{>} - \frac{\rho}{\Omega} \xi'_{<} \right) }, \\
    F_4 (\rho) = \frac{1}{ \left( \frac{\Omega}{\rho} \zeta'_{<} - \frac{\rho}{\Omega} \xi'_{<} \right) \left( \frac{\Omega}{\rho} \zeta'_{>} - \frac{\rho}{\Omega} \xi'_{<} \right) \left( \frac{\rho}{\Omega} \xi'_{>} - \frac{\rho}{\Omega} \xi'_{<} \right) }.
\end{eqnarray}
Using these auxiliary functions we can express the base integral 2 as
\begin{eqnarray}
&&\mathcal{I}_{2}(P,Q)\equiv\alpha^2_+\int_{\text{annulus}}dudv\left(1-\frac{\Omega^2}{w^2}\right)
\left(1-\frac{\Omega^2}{{w^*}^2}\right)
\frac{1}{Q-\left(\frac{w}{\Omega}+\frac{\Omega}{w}\right)}
\frac{1}{P-\left(\frac{w^*}{\Omega}+\frac{\Omega}{w^*}\right)} = \\
&&=\alpha^2_+\int_{\Omega}^1 d\rho \rho \oint d\zeta \frac{\zeta}{i} \left(1-\frac{\Omega^2}{\rho^2}\frac{1}{\zeta^2}\right)
\left(1-\frac{\Omega^2}{\rho^2}\zeta^2\right)
\frac{1}{\left(\zeta-\frac{\Omega}{\rho}\zeta'_<\right)\left(\zeta-\frac{\Omega}{\rho}\zeta'_>\right)}
\frac{1}{\left(\zeta-\frac{\rho}{\Omega}\xi'_>\right)\left(\zeta-\frac{\rho}{\Omega}\xi'_<\right)}= \\
&&=\alpha^2_+\int_{\Omega}^1 d\rho \rho \oint d\zeta \frac{\zeta}{i} \left(1-\frac{\Omega^2}{\rho^2}\frac{1}{\zeta^2}\right)
\left(1-\frac{\Omega^2}{\rho^2}\zeta^2\right) \left[ \frac{F_1 (\rho)}{\left(\zeta-\frac{\Omega}{\rho}\zeta'_<\right)} - \frac{F_2 (\rho)}{\left(\zeta-\frac{\Omega}{\rho}\zeta'_>\right)} + \frac{F_3 (\rho)}{\left(\zeta-\frac{\rho}{\Omega}\xi'_>\right)} - \frac{F_4 (\rho)}{\left(\zeta-\frac{\rho}{\Omega}\xi'_<\right)} \right] \\
&&=\alpha^2_+\int_{\Omega}^1 d\rho \rho \oint d\zeta \frac{\zeta}{i} \left(1+\frac{\Omega^4}{\rho^4} -\frac{\Omega^2}{\rho^2}\zeta^2 -\frac{\Omega^2}{\rho^2} \frac{1}{\zeta^2}\right) \left[ \frac{F_1 (\rho)}{\left(\zeta-\frac{\Omega}{\rho}\zeta'_<\right)} - \frac{F_2 (\rho)}{\left(\zeta-\frac{\Omega}{\rho}\zeta'_>\right)} + \frac{F_3 (\rho)}{\left(\zeta-\frac{\rho}{\Omega}\xi'_>\right)} - \frac{F_4 (\rho)}{\left(\zeta-\frac{\rho}{\Omega}\xi'_<\right)} \right] \\
&&= \aleph_{1,1} + \aleph_{1,2} + \aleph_{1,3} + \aleph_{1,4} - \aleph_{2,1} - \aleph_{2,2} - \aleph_{2,3} - \aleph_{2,4} - \aleph_{3,1} - \aleph_{3,2} - \aleph_{3,3} - \aleph_{3,4}
\end{eqnarray}

Then the base integral 2 can be evaluated term by term:
\begin{eqnarray}
    \aleph_{1,1} &=& \alpha_{+}^2 \int_{\Omega}^{1} \rho d \rho \left( 1 + \frac{\Omega^4}{\rho^4} \right) F_1 (\rho) \oint \frac{d \zeta}{i} \frac{\zeta}{\zeta - \frac{\Omega}{\rho} \zeta'_{<}} = \\
    &=& 2 \pi \alpha_+^2 \zeta'_{<} \Omega \int_{\Omega}^1 d \rho \left( 1 + \frac{\Omega^4}{\rho^4} \right) F_1 (\rho) = \\
    &=& \frac{2 \pi \alpha_+^2 \zeta'_{<} \Omega^2}{\left( \zeta'_{<} - \zeta'_{>} \right)} \int_{\Omega}^1 \rho^3 d \rho \left( 1 + \frac{\Omega^4}{\rho^4} \right) \frac{1}{  \left( \rho^2 - \frac{\Omega^2 \zeta'_{<}}{\xi'_{>}} \right) \left( \rho^2 - \frac{\Omega^2 \zeta'_{<}}{\xi'_{<}} \right) } = \\
    &=& \frac{\pi \alpha_+^2 \Omega^2 \zeta'_{<}}{\left( \zeta'_{<} - \zeta'_{>} \right)  \xi'_{<} \xi'_{>} } \int_{\Omega^2}^{1} dx \left( x + \frac{\Omega^4}{x} \right) \left[ \frac{1}{\left( x - \frac{\Omega^2 \zeta'_{<}}{\xi'_{>}} \right) } \frac{1}{\left( x - \frac{\Omega^2 \zeta'_{<}}{\xi'_{<}} \right)} \right] = \\
    &=& - \frac{\pi \alpha_+^2 \Omega^2}{\zeta'_{<} \left( \zeta'_{<} - \zeta'_{>} \right)} \ln \left( \Omega^2 \right) + 
    \frac{\pi \alpha_+^2 }{\left( \zeta'_{<} - \zeta'_{>} \right) \left( \xi'_{<} - \xi'_{>} \right)} \biggr[
    \frac{\left( \frac{\Omega^2 \zeta'_{<}}{\xi'_{>}} \right)^2 + \Omega^4}{\frac{\Omega^2 \zeta'_{<}}{\xi'_{>}}} \left( \ln \left[ 1 - \frac{\Omega^2 \zeta'_{<}}{\xi'_{>}} \right] - \ln \left[ \Omega^2 - \frac{\Omega^2 \zeta'_{<}}{\xi'_{>}} \right] \right)  \\ 
    &-&  \frac{\left( \frac{\Omega^2 \zeta'_{<}}{\xi'_{<}} \right)^2 + \Omega^4}{\frac{\Omega^2 \zeta'_{<}}{\xi'_{<}}} \left( \ln \left[ 1 - \frac{\Omega^2 \zeta'_{<}}{\xi'_{<}} \right] - \ln \left[ \Omega^2 - \frac{\Omega^2 \zeta'_{<}}{\xi'_{<}} \right] \right)  \biggr]
\end{eqnarray}

\begin{eqnarray}
    \aleph_{1,2} &=& -\alpha_{+}^2 \int_{\Omega}^{1} \rho d \rho \left( 1 + \frac{\Omega^4}{\rho^4} \right) F_2 (\rho) \oint \frac{d \zeta}{i} \frac{\zeta}{\zeta - \frac{\Omega}{\rho} \zeta'_{>}} = \\
    &=& -2 \pi \alpha_{+}^2 \zeta'_{>} \Omega \int_{\Omega |\zeta'_{>}|}^{1} d \rho \left( 1 + \frac{\Omega^4}{\rho^4} \right) F_2 (\rho) \Theta \left( 1 - \Omega |\zeta'_{>}| \right) = \\
    &=& -\frac{2 \pi \alpha_{+}^2 \zeta'_{>} \Omega^2}{\left(  \zeta'_{<} -  \zeta'_{>} \right)} \int_{\Omega |\zeta'_{>}|}^{1} \rho d \rho \left( 1 + \frac{\Omega^4}{\rho^4} \right) \frac{\rho^2}{ \left( \rho^2 \xi'_{>} - \Omega^2 \zeta'_{>} \right) \left( \rho^2 \xi'_{<} - \Omega^2 \zeta'_{>} \right) } \Theta \left( 1 - \Omega |\zeta'_{>}| \right) = \\
    &=& -\frac{\pi \alpha_{+}^2 \zeta'_{>} \Omega^2}{\left(  \zeta'_{<} -  \zeta'_{>} \right) \xi'_{>} \xi'_{<}} \int_{\Omega^2 |\zeta'_{>}|^2}^{1} d x \left( x + \frac{\Omega^4}{x} \right) \frac{1}{ \left( x - \frac{\Omega^2 \zeta'_{>}}{\xi'_{>}} \right) \left( x  - \frac{\Omega^2 \zeta'_{>}}{\xi'_{<}} \right) } \Theta \left( 1 - \Omega |\zeta'_{>}| \right) = \\
    &=& -\frac{\pi \alpha_{+}^2 }{ \left(  \zeta'_{<} -  \zeta'_{>} \right) \left( \xi'_{<} - \xi'_{>} \right) } 
    \biggr[
    \frac{\left( \frac{\Omega^2 \zeta'_{>}}{\xi'_{>}} \right)^2 + \Omega^4}{\frac{\Omega^2 \zeta'_{>}}{\xi'_{>}}} \left( \ln \left[ 1 - \frac{\Omega^2 \zeta'_{>}}{\xi'_{>}} \right] - \ln \left[ \Omega^2 |\zeta'_{>}|^2 - \frac{\Omega^2 \zeta'_{>}}{\xi'_{>}} \right] \right) - \\ 
    &-&  \frac{\left( \frac{\Omega^2 \zeta'_{>}}{\xi'_{<}} \right)^2 + \Omega^4}{\frac{\Omega^2 \zeta'_{>}}{\xi'_{<}}} \left( \ln \left[ 1 - \frac{\Omega^2 \zeta'_{>}}{\xi'_{<}} \right] - \ln \left[ \Omega^2 |\zeta'_{>}|^2 - \frac{\Omega^2 \zeta'_{>}}{\xi'_{<}} \right] \right)  \biggr] \Theta \left( 1 - \Omega |\zeta'_{>}| \right) + \\
    &+& \frac{\pi \alpha_{+}^2 \Omega^2}{\zeta'_{>} \left(  \zeta'_{<} -  \zeta'_{>} \right)} \ln \left( \Omega^2 |\zeta'_{>}|^2  \right) \Theta \left( 1 - \Omega |\zeta'_{>}| \right)
\end{eqnarray}

\begin{eqnarray}
    \aleph_{1,3} &=& \alpha_{+}^2 \int_{\Omega}^{1} \rho d \rho \left( 1 + \frac{\Omega^4}{\rho^4} \right) F_3 (\rho) \oint \frac{d \zeta}{i} \frac{\zeta}{\underbrace{\zeta - \frac{\rho}{\Omega} \xi'_{>}}_{outside}} = 0
\end{eqnarray}

\begin{eqnarray}
    \aleph_{1,4} &=& -\alpha_{+}^2 \int_{\Omega}^{1} \rho d \rho \left( 1 + \frac{\Omega^4}{\rho^4} \right) F_4 (\rho) \oint \frac{d \zeta}{i} \frac{\zeta}{\zeta - \frac{\rho}{\Omega} \xi'_{<}} = \\
    &=& -\frac{2 \pi \alpha_{+}^2 \xi'_{<}}{\Omega} \int_{\Omega}^{\min \left(1, \frac{\Omega}{|\xi'_{<}|}\right)} \rho^2 d \rho \left( 1 + \frac{\Omega^4}{\rho^4} \right) F_4 (\rho) = \\ 
    &=&  -\frac{2 \pi \alpha_{+}^2 \Omega^2}{\left( \xi'_{>} - \xi'_{<} \right) \xi'_{<} } \int_{\Omega}^{\min \left(1, \frac{\Omega}{|\xi'_{<}|}\right)} \rho^3 d \rho \left( 1 + \frac{\Omega^4}{\rho^4} \right) \frac{1}{\left( \rho^2 - \frac{\Omega^2 \zeta'_<}{\xi'_{<}} \right) \left( \rho^2 - \frac{\Omega^2 \zeta'_>}{\xi'_{<}} \right)} = \\
    &=& -\frac{\pi \alpha_{+}^2 \Omega^2}{\left( \xi'_{>} - \xi'_{<} \right) \xi'_{<}} \int_{\Omega^2}^{\min \left(1, \frac{\Omega^2}{|\xi'_{<}|^2}\right)} dx
    \left( x + \frac{\Omega^4}{x} \right) \left( \frac{1}{\left( x - \frac{\Omega^2 \zeta'_{<}}{\xi'_{<}}\right)}  
    \frac{1}{\left( x - \frac{\Omega^2 \zeta'_{>}}{\xi'_{<}} \right)} \right) = \\
    &=& \frac{\pi \alpha_{+}^2}{\left( \xi'_{>} - \xi'_{<} \right) \left( \zeta'_{<} - \zeta'_{>} \right) } \biggr[ \frac{\left( \frac{\Omega^2 \zeta'_{>}}{\xi'_{<}} \right)^2 + \Omega^4}{\frac{\Omega^2 \zeta'_{>}}{\xi'_{<}}} \left( \ln \left[ \min \left(1, \frac{\Omega^2}{|\xi'_{<}|^2}\right) - \frac{\Omega^2 \zeta'_{>}}{\xi'_{<}} \right] - \ln \left[ \Omega^2 - \frac{\Omega^2 \zeta'_{>}}{\xi'_{<}} \right] \right) - \\
    &-& \frac{\left( \frac{\Omega^2 \zeta'_{<}}{\xi'_{<}} \right)^2 + \Omega^4}{\frac{\Omega^2 \zeta'_{<}}{\xi'_{<}}} \left( \ln \left[ \min \left(1, \frac{\Omega^2}{|\xi'_{<}|^2}\right) - \frac{\Omega^2 \zeta'_{<}}{\xi'_{<}} \right] - \ln \left[ \Omega^2 - \frac{\Omega^2 \zeta'_{<}}{\xi'_{<}} \right] \right)
    \biggr] - \\
    &-& \frac{\pi \alpha_{+}^2 \Omega^2 \xi'_{<}}{\left( \xi'_{>} - \xi'_{<} \right) \zeta'_{<} \zeta'_{>}} \left( \ln \left[ \min \left(1, \frac{\Omega^2}{|\xi'_{<}|^2}\right) \right]
    - \ln \left[ \Omega^2 \right] \right)
\end{eqnarray}

\begin{eqnarray}
    \aleph_{2,1} &=& \alpha_{+}^2 \int_{\Omega}^{1} \rho d \rho F_1 (\rho) \oint \frac{d \zeta}{i} \frac{\Omega^2 \zeta^2}{\rho^2} \frac{\zeta}{\zeta - \frac{\Omega}{\rho} \zeta'_{<}} = \\
    &=& \frac{2 \pi \alpha_+^2 \Omega^4 \left( \zeta'_{<} \right)^3}{\left( \zeta'_{<} - \zeta'_{>} \right)} \int_{\Omega}^1 \frac{\rho d \rho}{\rho^4} \frac{1}{ \left( \frac{\Omega}{\rho} \zeta'_{<} - \frac{\rho}{\Omega} \xi'_{>} \right) \left( \frac{\Omega}{\rho} \zeta'_{<} - \frac{\rho}{\Omega} \xi'_{<} \right) } = \\
    &=& \frac{\pi \alpha_+^2 \Omega^6 \left( \zeta'_{<} \right)^3}{\left( \zeta'_{<} - \zeta'_{>} \right) \xi'_{<} \xi'_{>} } \int_{\Omega^2}^1 dx \left[ \frac{1}{ x \left( x - \frac{\Omega^2 \zeta'_{<}}{\xi'_{>}} \right) \left( x - \frac{\Omega^2 \zeta'_{<}}{\xi'_{<}} \right)}  \right] = \\
    &=& \frac{\pi \alpha_+^2 \Omega^4 \left( \zeta'_{<} \right)^2}{\left( \zeta'_{<} - \zeta'_{>} \right) \left( \xi'_{<} - \xi'_{>} \right)} \biggr[ \left( \frac{\xi'_{>}}{\Omega^2 \zeta'_{<}} \right) \left( \ln \left[ 1 - \frac{\Omega^2 \zeta'_{<}}{\xi'_{>}} \right] - \ln \left[ \Omega^2 - \frac{\Omega^2 \zeta'_{<}}{\xi'_{>}} \right] \right) - \\
    &-& \left( \frac{\xi'_{<}}{\Omega^2 \zeta'_{<}} \right) \left( \ln \left[ 1 - \frac{\Omega^2 \zeta'_{<}}{\xi'_{<}} \right] - \ln \left[ \Omega^2 - \frac{\Omega^2 \zeta'_{<}}{\xi'_{<}} \right] \right)
    \biggr] - \frac{\pi \alpha_+^2 \Omega^2 \zeta'_{<} }{\left( \zeta'_{<} - \zeta'_{>} \right)} \ln \left[ \Omega^2 \right]
\end{eqnarray}

\begin{eqnarray}
    \aleph_{2,2} &=& -\alpha_{+}^2 \int_{\Omega}^{1} \rho d \rho F_2 (\rho) \oint \frac{d \zeta}{i} \frac{\Omega^2 \zeta^2}{\rho^2} \frac{\zeta}{\zeta - \frac{\Omega}{\rho} \zeta'_{>}} = \\
    &=& -\Theta \left( 1 - \Omega |\zeta'_{>}| \right) \frac{2 \pi \alpha_+^2 \Omega^6 \left( \zeta'_{>}  \right)^3}{\left( \zeta'_{<} - \zeta'_{>} \right) \xi'_{<} \xi'_{>} } \int_{\Omega |\zeta'_{>}|}^1 \rho d \rho \left[ \frac{1}{\rho^2 \left( \rho^2 - \frac{\Omega^2 \zeta'_{>}}{\xi'_{>}} \right) \left( \rho^2 - \frac{\Omega^2 \zeta'_{>}}{\xi'_{<}} \right)} \right] = \\
    &=& -\Theta \left( 1 - \Omega |\zeta'_{>}| \right) \frac{\pi \alpha_+^2 \Omega^6 \left( \zeta'_{>}  \right)^3}{\left( \zeta'_{<} - \zeta'_{>} \right) \xi'_{<} \xi'_{>} } \int_{\Omega^2 |\zeta'_{>}|^2}^1 dx \left[ \frac{1}{x \left( x - \frac{\Omega^2 \zeta'_{>}}{\xi'_{>}} \right) \left( x - \frac{\Omega^2 \zeta'_{>}}{\xi'_{<}} \right)} \right] = \\
    &=& -\Theta \left( 1 - \Omega |\zeta'_{>}| \right) \frac{\pi \alpha_+^2 \Omega^4 \left( \zeta'_{>}  \right)^2}{\left( \zeta'_{<} - \zeta'_{>} \right) \left( \xi'_{<} - \xi'_{>} \right)} \biggr[ \left( \frac{\xi'_{>}}{\Omega^2 \zeta'_{>}} \right) \left( \ln \left[ 1 - \frac{\Omega^2 \zeta'_{>}}{\xi'_{>}} \right] - \ln \left[ \Omega^2 |\zeta'_{>}|^2 - \frac{\Omega^2 \zeta'_{>}}{\xi'_{>}} \right] \right) - \\
    &-& \left( \frac{\xi'_{<}}{\Omega^2 \zeta'_{>}} \right) \left( \ln \left[ 1 - \frac{\Omega^2 \zeta'_{>}}{\xi'_{<}} \right] - \ln \left[ \Omega^2 |\zeta'_{>}|^2 - \frac{\Omega^2 \zeta'_{>}}{\xi'_{>}} \right] \right)
    \biggr] + \Theta \left( 1 - \Omega |\zeta'_{>}| \right) \frac{\pi \alpha_+^2 \Omega^2  \zeta'_{>} }{\left( \zeta'_{<} - \zeta'_{>} \right)} \ln \left[ \Omega^2 |\zeta'_{>}|^2 \right]
\end{eqnarray}

\begin{eqnarray}
    \aleph_{2,3} &=& \alpha_{+}^2 \int_{\Omega}^{1} \rho d \rho F_3 (\rho) \oint \frac{d \zeta}{i} \frac{\Omega^2 \zeta^2}{\rho^2} \frac{\zeta}{\underbrace{\zeta - \frac{\rho}{\Omega} \xi'_{>}}_{outside}} = 0
\end{eqnarray}

\begin{eqnarray}
    \aleph_{2,4} &=& -\alpha_{+}^2 \int_{\Omega}^{1} \rho d \rho F_4 (\rho) \oint \frac{d \zeta}{i} \frac{\Omega^2 \zeta^2}{\rho^2} \frac{\zeta}{\zeta - \frac{\rho}{\Omega} \xi'_{<}} = \\
    &=& -\frac{2 \pi \alpha_+^2 \Omega^2 \left( \xi'_{<}  \right)}{\left( \xi'_{>} - \xi'_{<} \right) } \int_{\Omega}^{\min \left(1, \frac{\Omega}{|\xi'_{<}|} \right)} \rho^3 d \rho \left[ \frac{1}{\left(\rho^2 - \frac{\Omega^2 \zeta'_{<}}{\xi'_{<}} \right) \left( \rho^2 - \frac{\Omega^2 \zeta'_{>}}{\xi'_{<}} \right)} \right] = \\
    &=& -\frac{\pi \alpha_+^2 \Omega^2 \left( \xi'_{<}  \right)}{\left( \xi'_{>} - \xi'_{<} \right) } \int_{\Omega^2}^{\min \left(1, \frac{\Omega^2}{|\xi'_{<}|^2} \right)} x dx \left[ \frac{1}{ \left( x - \frac{\Omega^2 \zeta'_{<}}{\xi'_{<}} \right)\left( x - \frac{\Omega^2 \zeta'_{>}}{\xi'_{<}} \right)} \right] = \\
    &=& -\frac{\pi \alpha_+^2 \left( \xi'_{<}  \right) \Omega^2}{\left( \xi'_{>} - \xi'_{<} \right) \left( \zeta'_{<} - \zeta'_{>} \right)} \biggr[
    \zeta'_{<}  \left( \ln \left[ \min \left(1, \frac{\Omega^2}{|\xi'_{<}|^2} \right) - \frac{\Omega^2 \zeta'_{<}}{\xi'_{<}} \right] - \ln \left[ \Omega^2 - \frac{\Omega^2 \zeta'_{<}}{\xi'_{<}} \right] \right) - \\
    &-& \zeta'_{>} \left( \ln \left[ \min \left(1, \frac{\Omega^2}{|\xi'_{<}|^2} \right) - \frac{\Omega^2 \zeta'_{>}}{\xi'_{<}} \right] - \ln \left[ \Omega^2 - \frac{\Omega^2 \zeta'_{>}}{\xi'_{<}} \right] \right)
    \biggr]
\end{eqnarray}

\begin{eqnarray}
    \aleph_{3,1} &=& \alpha_{+}^2 \int_{\Omega}^{1} \rho d \rho F_1 (\rho) \oint \frac{d \zeta}{i} \frac{\Omega^2}{\rho^2 \zeta^2} \frac{\zeta}{\zeta - \frac{\Omega}{\rho} \zeta'_{<}} = \alpha_{+}^2 \int_{\Omega}^{1} \rho d \rho F_1 (\rho) \frac{\Omega^2}{\rho^2} \oint \frac{d \zeta}{i \zeta} \frac{1}{\zeta - \frac{\Omega}{\rho} \zeta'_{<}} = 0
\end{eqnarray}

\begin{eqnarray}
    \aleph_{3,2} &=& -\alpha_{+}^2 \int_{\Omega}^{1} \rho d \rho F_2 (\rho) \oint \frac{d \zeta}{i} \frac{\Omega^2}{\rho^2 \zeta^2} \frac{\zeta}{\zeta - \frac{\Omega}{\rho} \zeta'_{>}} = \\
    &=& \frac{2\pi \alpha_+^2 \Omega^2}{ \left( \zeta'_{<} - \zeta'_{>} \right) \xi'_{<} \xi'_{>} \zeta'_{>}  } \int_{\Omega}^{\min \left( 1, \Omega |\zeta'_{>}| \right)} \rho^3 d \rho \left[ \frac{1}{\left( \rho^2 - \frac{\Omega^2 \zeta'_{>}}{\xi'_{>}} \right)\left( \rho^2 - \frac{\Omega^2 \zeta'_{>}}{\xi'_{<}} \right)} \right] = \\
    &=& \frac{\pi \alpha_+^2 \Omega^2}{ \left( \zeta'_{<} - \zeta'_{>} \right) \xi'_{<} \xi'_{>} \zeta'_{>}  } \int_{\Omega^2}^{\min \left( 1, \Omega^2 |\zeta'_{>}|^2 \right)} x dx \left[ \frac{1}{\left( x - \frac{\Omega^2 \zeta'_{>}}{\xi'_{>}} \right)\left( x - \frac{\Omega^2 \zeta'_{>}}{\xi'_{<}} \right)} \right] = \\
    &=& \frac{\pi \alpha_+^2 \Omega^2}{ \left( \zeta'_{<} - \zeta'_{>} \right) \left( \xi'_{<} - \xi'_{>} \right) \left( \zeta'_{>}  \right)} \biggr[ 
    \frac{1}{\xi'_{>}} \left( \ln \left[ \min \left( 1, \Omega^2 |\zeta'_{>}|^2 \right) - \frac{\Omega^2 \zeta'_{>}}{\xi'_{>}} \right] - \ln \left[ \Omega^2 - \frac{\Omega^2 \zeta'_{>}}{\xi'_{>}} \right] \right) - \\
    &-& \frac{1}{\xi'_{<}} \left( \ln \left[ \min \left( 1, \Omega^2 |\zeta'_{>}|^2 \right) - \frac{\Omega^2 \zeta'_{>}}{\xi'_{<}} \right] - \ln \left[ \Omega^2 - \frac{\Omega^2 \zeta'_{>}}{\xi'_{<}} \right] \right)
    \biggr]
\end{eqnarray}

\begin{eqnarray}
    \aleph_{3,3} &=& \alpha_{+}^2 \int_{\Omega}^{1} \rho d \rho F_3 (\rho) \oint \frac{d \zeta}{i} \frac{\Omega^2}{\rho^2 \zeta^2} \frac{\zeta}{\zeta - \frac{\rho}{\Omega} \xi'_{>}} = \\
    &=& -\frac{2 \pi \alpha_+^2 \Omega^6 }{\left( \xi'_{>} - \xi'_{<} \right) \left( \xi'_{>}  \right)^3} \int_{\Omega}^1 \frac{\rho d \rho}{\rho^2 \left( \rho^2 - \frac{\Omega^2 \zeta'_{<}}{\xi'_{>}} \right) \left( \rho^2 - \frac{\Omega^2 \zeta'_{>}}{\xi'_{>}} \right)} = \\
    &=& -\frac{\pi \alpha_+^2 \Omega^6 }{\left( \xi'_{>} - \xi'_{<} \right) \left( \xi'_{>}  \right)^3} \int_{\Omega^2}^1 dx \left[ \frac{1}{x \left( x - \frac{\Omega^2 \zeta'_{<}}{\xi'_{>}} \right)\left( x - \frac{\Omega^2 \zeta'_{>}}{\xi'_{>}} \right)} \right] = \\
    &=& -\frac{\pi \alpha_+^2 \Omega^4 }{\left( \xi'_{>} - \xi'_{<} \right) \left( \zeta'_{<} - \zeta'_{>} \right) \left( \xi'_{>}  \right)^2} \biggr[ \left( \frac{\xi'_{>}}{\Omega^2 \zeta'_{<}} \right) \left( \ln \left[ 1 - \frac{\Omega^2 \zeta'_{<}}{\xi'_{>}} \right] - \ln \left[ \Omega^2 - \frac{\Omega^2 \zeta'_{<}}{\xi'_{>}} \right] \right) - \\
    &-& \left( \frac{\xi'_{>}}{\Omega^2 \zeta'_{>}} \right) \left( \ln \left[ 1 - \frac{\Omega^2 \zeta'_{>}}{\xi'_{>}} \right] - \ln \left[ \Omega^2 - \frac{\Omega^2 \zeta'_{>}}{\xi'_{>}} \right] \right)
    \biggr] + \frac{\pi \alpha_{+}^2 \Omega^2}{\left( \xi'_{>} - \xi'_{<} \right) \zeta'_{<} \zeta'_{>} \xi'_{>}} \ln \left[ \Omega^2 \right]
\end{eqnarray}

\begin{eqnarray}
    \aleph_{3,4} &=& -\alpha_{+}^2 \int_{\Omega}^{1} \rho d \rho F_4 (\rho) \oint \frac{d \zeta}{i} \frac{\Omega^2}{\rho^2 \zeta^2} \frac{\zeta}{\zeta - \frac{\rho}{\Omega} \xi'_{<}} = \\
    &=& \frac{2 \pi \alpha_+^2 \Omega^6 }{\left( \xi'_{>} - \xi'_{<} \right) \left( \xi'_{<} \right)^3 } \int_{\frac{\Omega}{|\xi'_{<}|}}^1 \frac{\rho d \rho}{\rho^2 \left( \rho^2 - \frac{\Omega^2 \zeta'_{<}}{\xi'_{<}} \right) \left( \rho^2 - \frac{\Omega^2 \zeta'_{>}}{\xi'_{<}} \right)} \Theta \left( \frac{|\xi'_{<}|}{\Omega} - 1 \right)= \\
    &=& \frac{\pi \alpha_+^2 \Omega^6 }{\left( \xi'_{>} - \xi'_{<} \right) \left( \xi'_{<} \right)^3 } \int_{\frac{\Omega^2}{|\xi'_{<}|^2}}^1 \frac{d x}{x \left( x - \frac{\Omega^2 \zeta'_{<}}{\xi'_{<}} \right) \left( x - \frac{\Omega^2 \zeta'_{>}}{\xi'_{<}} \right)} \Theta \left( \frac{|\xi'_{<}|}{\Omega} - 1 \right) = \\
    &=& \frac{\pi \alpha_+^2 \Omega^4 }{\left( \xi'_{>} - \xi'_{<} \right) \left( \zeta'_{<} - \zeta'_{>} \right) \left( \xi'_{<}  \right)^2} \biggr[ \left( \frac{\xi'_{<}}{\Omega^2 \zeta'_{<}} \right) \left( \ln \left[ 1 - \frac{\Omega^2 \zeta'_{<}}{\xi'_{<}} \right] - \ln \left[ \frac{\Omega^2}{|\xi'_{<}|^2} - \frac{\Omega^2 \zeta'_{<}}{\xi'_{<}} \right] \right) - \\
    &-& \left( \frac{\xi'_{<}}{\Omega^2 \zeta'_{>}} \right) \left( \ln \left[ 1 - \frac{\Omega^2 \zeta'_{>}}{\xi'_{<}} \right] - \ln \left[ \frac{\Omega^2}{|\xi'_{<}|^2} - \frac{\Omega^2 \zeta'_{>}}{\xi'_{<}} \right] \right)
    \biggr] \Theta \left( \frac{|\xi'_{<}|}{\Omega} - 1 \right)- \\
    &-& \frac{\pi \alpha_+^2 \Omega^2}{\left( \xi'_{>} - \xi'_{<} \right) \zeta'_{<} \zeta'_{>} \xi'_{<} } \ln \left[ \frac{\Omega^2}{|\xi'_{<}|^2} \right] \Theta \left( \frac{|\xi'_{<}|}{\Omega} - 1 \right)
\end{eqnarray}

Using our explicit expressions for $\zeta'_{<,>}$ and $\xi'_{<,>}$ we can simplify $\aleph$-integrals as well.
We make use of the expressions 
\begin{eqnarray}
    \Omega^2 |\zeta'_{>}|^2 = \frac{1}{\Omega^{4n}} \geq 1, \;\;\;
    \frac{\Omega^2}{|\xi'_{<}|^2} = \frac{1}{\Omega^{4n}} \geq 1,
\end{eqnarray}
therefore,
\begin{eqnarray}
    \Theta \left( \frac{|\xi'_{<}|}{\Omega} - 1 \right) = \Theta \left( 1 - \Omega |\zeta'_{>}| \right) = 0, \\
    \min \left(1, \frac{\Omega^2}{|\xi'_{<}|^2} \right) = \min \left( 1, \Omega^2 |\zeta'_{>}|^2 \right) = 1.
\end{eqnarray}
Then it is easy to see that 
\begin{eqnarray}
    \aleph_{1,2} = \aleph_{2,2} = \aleph_{3,4} = 0.
\end{eqnarray}
We employ useful identities
\begin{eqnarray}
    \frac{\left( \frac{\zeta'_{<}}{\xi'_{>}} \right)^2 + 1}{\frac{\zeta'_{<}}{\xi'_{>}}} = \frac{\zeta'_{<}}{\xi'_{>}} + \frac{\xi'_{>}}{\zeta'_{<}}; \; \; \;
    \frac{\left( \frac{\zeta'_{<}}{\xi'_{<}} \right)^2 + 1}{\frac{\zeta'_{<}}{\xi'_{<}}} = \frac{\zeta'_{<}}{\xi'_{<}} + \frac{\xi'_{<}}{\zeta'_{<}}; \; \; \;
    \frac{\left( \frac{\zeta'_{>}}{\xi'_{<}} \right)^2 + 1}{\frac{ \zeta'_{>}}{\xi'_{<}}} = \frac{\zeta'_{>}}{\xi'_{<}} + \frac{\xi'_{<}}{\zeta'_{>}}; \; \; \;
    \frac{\left( \frac{\zeta'_{<}}{\xi'_{<}} \right)^2 + 1}{\frac{\zeta'_{<}}{\xi'_{<}}} = \frac{\zeta'_{<}}{\xi'_{<}} + \frac{\xi'_{<}}{\zeta'_{<}}
\end{eqnarray}
to simplify other $\aleph$-integrals:
\begin{eqnarray}
    \aleph_{1,1} &=&  
    \frac{\pi \alpha_+^2 \Omega^2}{\left( \zeta'_{<} - \zeta'_{>} \right) \left( \xi'_{<} - \xi'_{>} \right)} \biggr[
    \left( \frac{\zeta'_{<}}{\xi'_{>}} + \frac{\xi'_{>}}{\zeta'_{<}} \right) \left( \ln \left[ 1 - \frac{\Omega^2 \zeta'_{<}}{\xi'_{>}} \right] - \ln \left[ \Omega^2 - \frac{\Omega^2 \zeta'_{<}}{\xi'_{>}} \right] \right)  \\ 
    &-&  \left( \frac{\zeta'_{<}}{\xi'_{<}} + \frac{\xi'_{<}}{\zeta'_{<}} \right) \left( \ln \left[ 1 - \frac{\Omega^2 \zeta'_{<}}{\xi'_{<}} \right] - \ln \left[ \Omega^2 - \frac{\Omega^2 \zeta'_{<}}{\xi'_{<}} \right] \right)  \biggr] - \frac{\pi \alpha_+^2 \Omega^2}{\zeta'_{<} \left( \zeta'_{<} - \zeta'_{>} \right)} \ln \left( \Omega^2 \right)
\end{eqnarray}
\begin{eqnarray}
    \aleph_{1,4} &=& \frac{\pi \alpha_{+}^2 \Omega^2}{\left( \xi'_{<} - \xi'_{>} \right) \left( \zeta'_{<} - \zeta'_{>} \right) } \biggr[ \left( \frac{\zeta'_{<}}{\xi'_{<}} + \frac{\xi'_{<}}{\zeta'_{<}} \right) \left( \ln \left[ 1 - \frac{\Omega^2 \zeta'_{<}}{\xi'_{<}} \right] - \ln \left[ \Omega^2 - \frac{\Omega^2 \zeta'_{<}}{\xi'_{<}} \right] \right) - \\
    &-& \left( \frac{\zeta'_{>}}{\xi'_{<}} + \frac{\xi'_{<}}{\zeta'_{>}} \right) \left( \ln \left[ 1 - \frac{\Omega^2 \zeta'_{>}}{\xi'_{<}} \right] - \ln \left[ \Omega^2 - \frac{\Omega^2 \zeta'_{>}}{\xi'_{<}} \right] \right) 
    \biggr] + \frac{\pi \alpha_{+}^2 \Omega^2 \xi'_{<}}{\left( \xi'_{>} - \xi'_{<} \right)} \ln \left[ \Omega^2 \right]
\end{eqnarray}
\begin{eqnarray}
    \aleph_{2,1} &=& \frac{\pi \alpha_+^2 \Omega^2 }{\left( \zeta'_{<} - \zeta'_{>} \right) \left( \xi'_{<} - \xi'_{>} \right)} \biggr[  \frac{\xi'_{>}}{\zeta'_{>}}  \left( \ln \left[ 1 - \frac{\Omega^2 \zeta'_{<}}{\xi'_{>}} \right] - \ln \left[ \Omega^2 - \frac{\Omega^2 \zeta'_{<}}{\xi'_{>}} \right] \right) - \\
    &-&  \frac{\xi'_{<}}{\zeta'_{>}}  \left( \ln \left[ 1 - \frac{\Omega^2 \zeta'_{<}}{\xi'_{<}} \right] - \ln \left[ \Omega^2 - \frac{\Omega^2 \zeta'_{<}}{\xi'_{<}} \right] \right)
    \biggr] - \frac{\pi \alpha_+^2 \Omega^2 \zeta'_{<} }{\left( \zeta'_{<} - \zeta'_{>} \right)} \ln \left[ \Omega^2 \right]
\end{eqnarray}
\begin{eqnarray}
    \aleph_{2,4} &=& \frac{\pi \alpha_+^2 \Omega^2}{\left( \xi'_{<} - \xi'_{>} \right) \left( \zeta'_{<} - \zeta'_{>} \right)} \biggr[
    \frac{\zeta'_{<}}{\xi'_{>}}  \left( \ln \left[ 1 - \frac{\Omega^2 \zeta'_{<}}{\xi'_{<}} \right] - \ln \left[ \Omega^2 - \frac{\Omega^2 \zeta'_{<}}{\xi'_{<}} \right] \right) - \\
    &-& \frac{\zeta'_{>}}{\xi'_{>}} \left( \ln \left[ 1 - \frac{\Omega^2 \zeta'_{>}}{\xi'_{<}} \right] - \ln \left[ \Omega^2 - \frac{\Omega^2 \zeta'_{>}}{\xi'_{<}} \right] \right)
    \biggr]
\end{eqnarray}
\begin{eqnarray}
    \aleph_{3,2} &=& \frac{\pi \alpha_+^2 \Omega^2}{ \left( \zeta'_{<} - \zeta'_{>} \right) \left( \xi'_{<} - \xi'_{>} \right) } \biggr[ 
    \frac{\zeta'_{<}}{\xi'_{>}} \left( \ln \left[ 1 - \frac{\Omega^2 \zeta'_{>}}{\xi'_{>}} \right] - \ln \left[ \Omega^2 - \frac{\Omega^2 \zeta'_{>}}{\xi'_{>}} \right] \right) - \\
    &-& \frac{\zeta'_{<}}{\xi'_{<}} \left( \ln \left[ 1 - \frac{\Omega^2 \zeta'_{>}}{\xi'_{<}} \right] - \ln \left[ \Omega^2 - \frac{\Omega^2 \zeta'_{>}}{\xi'_{<}} \right] \right)
    \biggr]
\end{eqnarray}
\begin{eqnarray}
    \aleph_{3,3} &=& \frac{\pi \alpha_+^2 \Omega^2 }{\left( \xi'_{<} - \xi'_{>} \right) \left( \zeta'_{<} - \zeta'_{>} \right) } \biggr[ \left( \frac{\xi'_{<}}{\zeta'_{<}} \right) \left( \ln \left[ 1 - \frac{\Omega^2 \zeta'_{<}}{\xi'_{>}} \right] - \ln \left[ \Omega^2 - \frac{\Omega^2 \zeta'_{<}}{\xi'_{>}} \right] \right) - \\
    &-& \left( \frac{\xi'_{<}}{\zeta'_{>}} \right) \left( \ln \left[ 1 - \frac{\Omega^2 \zeta'_{>}}{\xi'_{>}} \right] - \ln \left[ \Omega^2 - \frac{\Omega^2 \zeta'_{>}}{\xi'_{>}} \right] \right)
    \biggr] + \frac{\pi \alpha_{+}^2 \Omega^2}{\left( \xi'_{>} - \xi'_{<} \right) \xi'_{>}} \ln \left[ \Omega^2 \right]
\end{eqnarray}
We take sums of $\aleph-$integrals separately:
\begin{eqnarray}
    &&\aleph_{1,1} + \aleph_{1,4} =  
    \frac{\pi \alpha_+^2 \Omega^2}{\left( \zeta'_{<} - \zeta'_{>} \right) \left( \xi'_{<} - \xi'_{>} \right)} \left( \frac{\zeta'_{<}}{\xi'_{>}} + \frac{\xi'_{>}}{\zeta'_{<}} \right) \biggr[
     \left( \ln \left[ 1 - \frac{\Omega^2 \zeta'_{<}}{\xi'_{>}} \right] - \ln \left[ \Omega^2 - \frac{\Omega^2 \zeta'_{<}}{\xi'_{>}} \right] \right)  \\ 
    &&-   \left( \ln \left[ 1 - \frac{\Omega^2 \zeta'_{>}}{\xi'_{<}} \right] - \ln \left[ \Omega^2 - \frac{\Omega^2 \zeta'_{>}}{\xi'_{<}} \right] \right)  \biggr] - \frac{\pi \alpha_+^2 \Omega^2}{\zeta'_{<} \left( \zeta'_{<} - \zeta'_{>} \right)} \ln \left( \Omega^2 \right) + \frac{\pi \alpha_{+}^2 \Omega^2 \xi'_{<}}{\left( \xi'_{>} - \xi'_{<} \right)} \ln \left[ \Omega^2 \right]
\end{eqnarray}
\begin{eqnarray}
    &&\aleph_{2,1} + \aleph_{2,4} + \aleph_{3,2} + \aleph_{3,3} =  
    \frac{\pi \alpha_+^2 \Omega^2}{\left( \zeta'_{<} - \zeta'_{>} \right) \left( \xi'_{<} - \xi'_{>} \right)} \left( \frac{\zeta'_{>}}{\xi'_{>}} + \frac{\xi'_{>}}{\zeta'_{>}} \right) \biggr[
     \left( \ln \left[ 1 - \frac{\Omega^2 \zeta'_{<}}{\xi'_{>}} \right] - \ln \left[ \Omega^2 - \frac{\Omega^2 \zeta'_{<}}{\xi'_{>}} \right] \right)  \\ 
    &&-   \left( \ln \left[ 1 - \frac{\Omega^2 \zeta'_{>}}{\xi'_{<}} \right] - \ln \left[ \Omega^2 - \frac{\Omega^2 \zeta'_{>}}{\xi'_{<}} \right] \right)  \biggr] 
    - \frac{\pi \alpha_+^2 \Omega^2 \zeta'_{<} }{\left( \zeta'_{<} - \zeta'_{>} \right)} \ln \left[ \Omega^2 \right] + \frac{\pi \alpha_{+}^2 \Omega^2}{\left( \xi'_{>} - \xi'_{<} \right) \xi'_{>}} \ln \left[ \Omega^2 \right]
\end{eqnarray}
Thus we get an expression for the base integral 2:
\begin{eqnarray}
    &&\mathcal{I}_{2}(P,Q) = \aleph_{1,1} + \aleph_{1,2} + \aleph_{1,3} + \aleph_{1,4} - \aleph_{2,1} - \aleph_{2,2} - \aleph_{2,3} - \aleph_{2,4} - \aleph_{3,1} - \aleph_{3,2} - \aleph_{3,3} - \aleph_{3,4} = \\
    &&= \frac{\pi \alpha_+^2 \Omega^2}{\left( \zeta'_{<} - \zeta'_{>} \right) \left( \xi'_{<} - \xi'_{>} \right)} \left( \frac{\zeta'_{<}}{\xi'_{>}} + \frac{\xi'_{>}}{\zeta'_{<}} - \frac{\zeta'_{>}}{\xi'_{>}} - \frac{\xi'_{>}}{\zeta'_{>}} \right) \biggr[
     \left( \ln \left[ 1 - \frac{\Omega^2 \zeta'_{<}}{\xi'_{>}} \right] - \ln \left[ \Omega^2 - \frac{\Omega^2 \zeta'_{<}}{\xi'_{>}} \right] \right)  \\ 
    &&-   \left( \ln \left[ 1 - \frac{\Omega^2 \zeta'_{>}}{\xi'_{<}} \right] - \ln \left[ \Omega^2 - \frac{\Omega^2 \zeta'_{>}}{\xi'_{<}} \right] \right)  \biggr] - \frac{\pi \alpha_+^2 \Omega^2}{\zeta'_{<} \left( \zeta'_{<} - \zeta'_{>} \right)} \ln \left( \Omega^2 \right) - \frac{\pi \alpha_+^2 \Omega^2 \zeta'_{<} }{\left( \zeta'_{<} - \zeta'_{>} \right)} \ln \left[ \Omega^2 \right] = \\
    &&= \pi \alpha_+^2 \Omega^2  \left(
     \ln \left[ 1 - \frac{\Omega^2 \zeta'_{<}}{\xi'_{>}} \right] - \ln \left[ \Omega^2 - \frac{\Omega^2 \zeta'_{<}}{\xi'_{>}} \right]
    -   \ln \left[ 1 - \frac{\Omega^2 \zeta'_{>}}{\xi'_{<}} \right] + \ln \left[ \Omega^2 - \frac{\Omega^2 \zeta'_{>}}{\xi'_{<}} \right]  \right) - \\
    &&- \frac{\pi \alpha_+^2 \Omega^2}{\zeta'_{<} \left( \zeta'_{<} - \zeta'_{>} \right)} \ln \left( \Omega^2 \right) + \frac{\pi \alpha_+^2 \Omega^2 \zeta'_{<} }{\left( \zeta'_{<} - \zeta'_{>} \right)} \ln \left[ \Omega^2 \right] = \\
    &&=\pi \alpha_+^2 \Omega^2  \left(
     \ln \left[ 1 - \frac{\Omega^2 \zeta'_{<}}{\xi'_{>}} \right] - \ln \left[ \Omega^2 - \frac{\Omega^2 \zeta'_{<}}{\xi'_{>}} \right]
    -   \ln \left[ 1 - \frac{\Omega^2 \zeta'_{>}}{\xi'_{<}} \right] + \ln \left[ \Omega^2 - \frac{\Omega^2 \zeta'_{>}}{\xi'_{<}} \right] + \ln \left[ \Omega^2 \right]  \right). 
\end{eqnarray}
We can simplify the expression for $\mathcal{I}_{2}(P,Q)$ even further by involving explicit expressions for the roots:
\begin{eqnarray}
    &&\zeta'_{<} = z_1^* \Omega^{2n+1}, \; \; \; 
    \zeta'_{>} = \frac{z_2}{\Omega^{2m+1}}\\
    &&\frac{\Omega^2 \zeta'_{<}}{\xi'_{>}} = z_1^* z_2^* \Omega^{2n + 2m + 4}, \; \; \;
    \frac{\Omega^2 \zeta'_{>}}{\xi'_{<}} = \frac{1}{z_1^* z_2^* \Omega^{2n + 2m}}.
\end{eqnarray}
Then
\begin{eqnarray}
    &&\mathcal{I}_{2}(P_m (z_2), Q_n (z_1)) = \pi \alpha_+^2 \Omega^2  \left(
     \ln \left[ 1 - z_1^* z_2^* \Omega^{2n + 2m + 4} \right] - \ln \left[ \Omega^2 - z_1^* z_2^* \Omega^{2n + 2m + 4} \right] - \right. \\ 
    &&- \left.  \ln \left[ 1 - \frac{1}{z_1^* z_2^* \Omega^{2n + 2m}} \right] + \ln \left[ \Omega^2 - \frac{1}{z_1^* z_2^* \Omega^{2n + 2m}} \right] + \ln \left[ \Omega^2 \right]  \right). 
\end{eqnarray}
Note that for $m=n=0$ and $z_1=z_2^*$ this integral diverges because of the third log term in the brackets (also keep in mind that $Q^*_n (z) = Q_n (z^*)$). Importantly, this divergence is the same as in the isotropic case, see arc integrals in \eqref{A2''arc_integrals}. Therefore, one needs to regularize this integral by considering the arc contact limit. 
Later we encounter integrals of the kind $\mathcal{I}_{2}(Q_m^* (z_2), Q_n (z_1))$, so we will consider
\begin{eqnarray}
    &&\mathcal{I}_{2}(Q_0^* (z_2), Q_0 (z_1)) = \pi \alpha_+^2 \Omega^2  \left(
     \ln \left[ 1 - z_1^* z_2 \Omega^{4} \right] - \ln \left[ \Omega^2 - z_1^* z_2 \Omega^{4} \right] -  \ln \left[ 1 - \frac{1}{z_1^* z_2} \right] + \ln \left[ \Omega^2 - \frac{1}{z_1^* z_2} \right] + \ln \left[ \Omega^2 \right]  \right). \nonumber
\end{eqnarray}
To keep an exact analogy with the isotropic case we regularize the two logs (third and fourth) for $z_1 = e^{i \phi_A}, z_2 = e^{i \tilde{\phi}_A}$.
To evaluate regularized integrals we change variables to $u=\phi_A-\tilde{\phi}_A$ and $v=\frac{\phi_A+\tilde{\phi}_A}{2}$ and Taylor expand $\cos$ over small $\lambda$ inside the logs up to quadratic order:
\begin{eqnarray}
    &&\frac{1}{\lambda^2} \int_{-\frac{\lambda}{2}}^{\frac{\lambda}{2}} d\phi_A \int_{-\frac{\lambda}{2}}^{\frac{\lambda}{2}} d \tilde{\phi}_A \left[ \ln \left( 1 - e^{i \left( \phi_A - \tilde{\phi}_A \right)} \right) - \ln \left( \Omega^2 - e^{i \left( \phi_A - \tilde{\phi}_A \right)} \right) \right]=\\ 
    &&\frac{1}{\lambda^2} \int_{-\lambda}^{0} du \int_{-\frac{u + \lambda}{2}}^{\frac{u+\lambda}{2}} d v \left[ \ln \left( 1 - e^{i u } \right) - \ln \left( \Omega^2 - e^{i u} \right) \right]+ \frac{1}{\lambda^2} \int_{0}^{\lambda} du \int_{\frac{u-\lambda}{2}}^{\frac{\lambda-u}{2}} d v \left[ \ln \left( 1 - e^{i u } \right) - \ln \left( \Omega^2 - e^{i u} \right) \right] = \\ 
    &&= \frac{1}{\lambda^2} \int_{0}^{\lambda} du \left( \lambda - u \right) \left[ \ln \left( 2 - 2 \cos u \right) - \ln \left( 1 + \Omega^4 - 2 \Omega^2 \cos u \right) \right] \simeq \\
    &&\simeq \frac{1}{\lambda^2} \int_{0}^{\lambda} du \left( \lambda - u \right) \left[ \ln \left( u^2 \right) - \ln \left( \left( 1 - \Omega^2 \right)^2 + \Omega^2 u^2 \right) \right] = \\
    &&= \ln (\lambda) - 2\frac{\Omega^2 -1 }{\lambda \Omega} \arctan \left( \frac{\lambda \Omega}{\Omega^2 - 1} \right) - \left( \frac{(\Omega^2 - 1)^2}{\lambda^2  \Omega^2 } \right) \ln (1 - \Omega^2)  + \frac{(1-\Omega^2)^2 - \lambda^2 \Omega^2}{2\lambda^2  \Omega^2 } \ln \left( (1-\Omega^2)^2 + \lambda^2 \Omega^2 \right) \nonumber \\
    &&= \ln (\lambda) + 2\gamma \arctan \left( \frac{-1}{\gamma} \right) - \gamma^2 \ln ( 1 - \Omega^2)  + \frac{(1-\Omega^2)^2 - \lambda^2 \Omega^2}{2\lambda^2  \Omega^2 } \ln \left( (1-\Omega^2)^2 + \lambda^2 \Omega^2 \right) = \\
    &&= \ln (\lambda) - 2\gamma \; \mathrm{arccot} \left( \gamma \right) - \gamma^2 \ln (1 - \Omega^2)  + \frac{\gamma^2 - 1}{2} \ln \left( (1 + \gamma^2) \lambda^2 \Omega^2 \right) = \\
    &&= \ln (\lambda) - 2\gamma \; \mathrm{arccot} \left( \gamma \right) - \gamma^2 \ln ( 1 - \Omega^2)  + \frac{\gamma^2 - 1}{2} \left( 2 \ln \left( \lambda \Omega \right) + \ln \left( 1 + \gamma^2\right) \right),
    \\
&&\text{where} \; \; \; \gamma=\frac{1}{\lambda\Omega}-\frac{\Omega}{\lambda}.
\end{eqnarray}

\subsection{Anisotropic case -- combining terms together}

As we mentioned above, the nonlinear electrostatic potential is given by
\begin{eqnarray}
&&\Phi^{\newtext{(2)}}_{ani}(\bR)=
\frac{1}{\bar{\sigma}}\int_{\Omega}d^2\bR' \left[\left(\frac{\partial}{\partial
Z'}G_{N}(\bR,\bR')\right)\left(\tilde{\mathcal{E}}_x^{(2)}(\bR')+i\tilde{\mathcal{E}}_y^{(2)}(\bR')\right)+ \mathrm{c.c.} \right]
\end{eqnarray}
with $x-$axis along the higher resistivity principal axis $(\Delta \sigma<0)$ and
\begin{eqnarray}
&&\tilde{\mathcal{E}}_x^{(2)}(\bR')+i\tilde{\mathcal{E}}_y^{(2)}(\bR')= \\
&&\alpha_+\left(\left(\sigma^{(2)}_{-} e^{i\chi^{(2)}_{-}} + \Omega^2 \sigma^{(2)}_{+} e^{-i\chi^{(2)}_{+}} \right) \left(E^{\newtext{(1)}}_x-iE^{\newtext{(1)}}_y\right)^2+
\left(\sigma^{(2)}_{+} e^{i\chi^{(2)}_{+}} + \Omega^2 \sigma^{(2)}_{-} e^{-i\chi^{(2)}_{-}} \right) \left(E^{\newtext{(1)}}_x+iE^{\newtext{(1)}}_y\right)^2 +\right. \\
&&\left.+ \sigma^{(2)}_{0} \left( e^{i\chi^{(2)}_{0}} + \Omega^2  e^{-i\chi^{(2)}_{0}} \right) \left({E^{\newtext{(1)}}_x}^2+{E^{\newtext{(1)}}_y}^2\right)\right).
\end{eqnarray}
Combining the two expressions together we obtain 
\begin{eqnarray}
&&\Phi^{\newtext{(2)}}_{ani}(\bR)=
\frac{1}{\bar{\sigma}}\int_{\Omega}d^2\bR' \left[\left(\frac{\partial}{\partial
Z'}G_{N}(\bR,\bR')\right)\left(\tilde{\mathcal{E}}_x^{(2)}(\bR')+i\tilde{\mathcal{E}}_y^{(2)}(\bR')\right)+ \mathrm{c.c.} \right] = \\
&&= \frac{\alpha_+}{\bar{\sigma}}\int_{\Omega}d^2\bR' \left[\left(\frac{\partial}{\partial
Z'}G_{N}(\bR,\bR')\right)\left(\left(\sigma^{(2)}_{-} e^{i\chi^{(2)}_{-}} + \Omega^2 \sigma^{(2)}_{+} e^{-i\chi^{(2)}_{+}} \right) \left(E^{\newtext{(1)}}_x-iE^{\newtext{(1)}}_y\right)^2+ \right. \right. \\ &&\left. \left. + 
\left(\sigma^{(2)}_{+} e^{i\chi^{(2)}_{+}} + \Omega^2 \sigma^{(2)}_{-} e^{-i\chi^{(2)}_{-}} \right) \left(E^{\newtext{(1)}}_x+iE^{\newtext{(1)}}_y\right)^2 
+ \sigma^{(2)}_{0} \left( e^{i\chi^{(2)}_{0}} + \Omega^2  e^{-i\chi^{(2)}_{0}} \right) \left({E^{\newtext{(1)}}_x}^2+{E^{\newtext{(1)}}_y}^2\right)\right)+ \mathrm{c.c.} \right].
\end{eqnarray}

For brevity let us introduce the following notation 
\begin{eqnarray}
    &&T^e (n, z) = \frac{\Omega^{2n}}{z + \Omega^{4n+2} z^* - \Omega^{2n} \left( w + \frac{\Omega^2}{w} \right)} = \frac{1}{\frac{z + \Omega^{4n+2} z^*}{\Omega^{2n+1}} - \left( \frac{w}{\Omega} + \frac{\Omega}{w} \right)} \frac{1}{\Omega} = \frac{1}{Q_n (z) - \left( \frac{w}{\Omega} + \frac{\Omega}{w} \right)} \frac{1}{\Omega} = T^e (n, z, w), \\
    &&T^o (n, z) = \frac{\Omega^{2n}}{z^* + \Omega^{4n+2} z - \Omega^{2n} \left( w + \frac{\Omega^2}{w} \right)} = \frac{1}{\frac{z^* + \Omega^{4n+2} z}{\Omega^{2n+1}} - \left( \frac{w}{\Omega} + \frac{\Omega}{w} \right)} \frac{1}{\Omega} = \frac{1}{Q^*_n (z) - \left( \frac{w}{\Omega} + \frac{\Omega}{w} \right)} \frac{1}{\Omega} = T^o (n, z, w),  \\
    &&\tilde{T}^e (n, z) = \frac{\Omega^{2n+2}}{z^* + \Omega^{4n+2} z - \Omega^{2n} \left( w^* + \frac{\Omega^2}{w^*} \right)} = \Omega^2 T^o (n, z, w^*), \\
    &&\tilde{T}^o (n, z) = \frac{\Omega^{2n+2}}{z + \Omega^{4n+2} z^* - \Omega^{2n} \left( w^* + \frac{\Omega^2}{w^*} \right)} = \Omega^2 T^e (n, z, w^*), 
\end{eqnarray}
to rewrite the electric field
\begin{eqnarray}
&&E^{\newtext{(1)}}_x(\br)-iE^{\newtext{(1)}}_y(\br)\\
&&=\frac{-I}{\pi \bar{\sigma}\sqrt{1-\left(\frac{\Delta\sigma}{\bar{\sigma}}\right)^2}}
\left(\sum_{n=0,2,4,\ldots}^{\infty}
\frac{\Omega^{2n}}{z_S+\Omega^{4n+2}z^*_S-\Omega^{2n}\left(w+\frac{\Omega^2}{w}\right)}
+
\sum_{n=1,3,5,\ldots}^{\infty}
\frac{\Omega^{2n}}{z^*_S+\Omega^{4n+2}z_S-\Omega^{2n}\left(w+\frac{\Omega^2}{w}\right)}
-(S\rightarrow D)\right)
\nonumber\\
&&+\frac{-I}{\pi \bar{\sigma}\sqrt{1-\left(\frac{\Delta\sigma}{\bar{\sigma}}\right)^2}}
\left(\sum_{n=0,2,4,\ldots}^{\infty}
\frac{\Omega^{2n+2}}{z^*_S+\Omega^{4n+2}z_S-\Omega^{2n}\left(w^*+\frac{\Omega^2}{w^*}\right)}
+
\sum_{n=1,3,5,\ldots}^{\infty}
\frac{\Omega^{2n+2}}{z_S+\Omega^{4n+2}z^*_S-\Omega^{2n}\left(w^*+\frac{\Omega^2}{w^*}\right)}
-(S\rightarrow D)
\right)\nonumber = \\
&&= \frac{-I}{\pi \bar{\sigma}\sqrt{1-\left(\frac{\Delta\sigma}{\bar{\sigma}}\right)^2}} 
\left( \sum_{n=0,2,4,\ldots}^{\infty} T^e (n, z_S) + \sum_{n=1,3,5,\ldots}^{\infty} T^o (n, z_S) 
 + \sum_{n=0,2,4,\ldots}^{\infty} \tilde{T}^e (n, z_S) + \sum_{n=1,3,5,\ldots}^{\infty} \tilde{T}^o (n, z_S) -(S\rightarrow D) \right)
\end{eqnarray}
and the derivative of the Green's function 
\begin{eqnarray}
&&\frac{\partial}{\partial Z'}G_{N}(\bR,\bR')=\frac{1}{2\pi \alpha_+}\left(\sum_{n=0,2,4,\ldots}^{\infty}
\frac{\Omega^{2n}}{z+\Omega^{4n+2}{z^*}-\Omega^{2n}\left(w'+\frac{\Omega^2}{w'}\right)}
+
\sum_{n=1,3,5\ldots}^{\infty}
\frac{\Omega^{2n}}{z^*+\Omega^{4n+2}z-\Omega^{2n}\left(w'+\frac{\Omega^2}{w'}\right)}
\right) = \\ 
&& = \frac{1}{2\pi \alpha_+} \left( \sum_{n=0,2,4,\ldots}^{\infty} T^e (n, z) +
\sum_{n=1,3,5\ldots}^{\infty} T^o (n, z) \right)
= \frac{1}{2\pi \alpha_+} \left( \sum_{n=0,2,4,\ldots}^{\infty} T^e (n, z, w') +
\sum_{n=1,3,5\ldots}^{\infty} T^o (n, z, w') \right).
\end{eqnarray}
We will also find useful the expression for the complex conjugated field:
\begin{eqnarray}
&&E^{\newtext{(1)}}_x(\br)+iE^{\newtext{(1)}}_y(\br)\\
&&=\frac{-I}{\pi \bar{\sigma}\sqrt{1-\left(\frac{\Delta\sigma}{\bar{\sigma}}\right)^2}}
\left(\sum_{n=0,2,4,\ldots}^{\infty}
\frac{\Omega^{2n}}{z_S^*+\Omega^{4n+2}z_S-\Omega^{2n}\left(w^*+\frac{\Omega^2}{w^*}\right)}
+
\sum_{n=1,3,5,\ldots}^{\infty}
\frac{\Omega^{2n}}{z_S+\Omega^{4n+2}z^*_S-\Omega^{2n}\left(w^*+\frac{\Omega^2}{w^*}\right)}
-(S\rightarrow D)\right)
\nonumber\\
&&+\frac{-I}{\pi \bar{\sigma}\sqrt{1-\left(\frac{\Delta\sigma}{\bar{\sigma}}\right)^2}}
\left(\sum_{n=0,2,4,\ldots}^{\infty}
\frac{\Omega^{2n+2}}{z_S+\Omega^{4n+2}z^*_S-\Omega^{2n}\left(w+\frac{\Omega^2}{w}\right)}
+
\sum_{n=1,3,5,\ldots}^{\infty}
\frac{\Omega^{2n+2}}{z^*_S+\Omega^{4n+2}z_S-\Omega^{2n}\left(w+\frac{\Omega^2}{w}\right)}
-(S\rightarrow D)
\right)\nonumber = \\
&&= \frac{-I}{\pi \bar{\sigma}\sqrt{1-\left(\frac{\Delta\sigma}{\bar{\sigma}}\right)^2}} 
\left( \sum_{n=0,2,4,\ldots}^{\infty} \frac{\tilde{T}^e (n, z_S)}{\Omega^2} + \sum_{n=1,3,5,\ldots}^{\infty} \frac{\tilde{T}^o (n, z_S)}{\Omega^2} 
 + \sum_{n=0,2,4,\ldots}^{\infty} \Omega^2 T^e (n, z_S) + \sum_{n=1,3,5,\ldots}^{\infty} \Omega^2 T^o (n, z_S) -(S\rightarrow D) \right) \nonumber
\end{eqnarray}

Now we use partial fraction decomposition on the integral of the kind:
\begin{eqnarray}
    &&\alpha^2_+\int_{\text{annulus}}dudv\left(1-\frac{\Omega^2}{w^2}\right)
\left(1-\frac{\Omega^2}{{w^*}^2}\right) \frac{1}{\Omega^3} \frac{1}{Q_1 - \left( \frac{w}{\Omega} + \frac{\Omega}{w} \right)} \frac{1}{Q_2 - \left( \frac{w}{\Omega} + \frac{\Omega}{w} \right)} \frac{1}{Q_3 - \left( \frac{w}{\Omega} + \frac{\Omega}{w} \right)} = \\
    &&= \frac{1}{\Omega^3} \left[ \frac{\mathcal{I}_1(Q_1)}{(Q_1 - Q_2)(Q_1 - Q_3)} + \frac{\mathcal{I}_1(Q_2)}{(Q_2 - Q_1)(Q_2 - Q_3)} + \frac{\mathcal{I}_1(Q_3)}{(Q_3 - Q_1)(Q_3 - Q_2)} \right], \\
    &&\alpha^2_+\int_{\text{annulus}}dudv\left(1-\frac{\Omega^2}{w^2}\right)
\left(1-\frac{\Omega^2}{{w^*}^2}\right) \frac{1}{\Omega^3} \frac{1}{Q_1 - \left( \frac{w}{\Omega} + \frac{\Omega}{w} \right)} \frac{1}{Q_2 - \left( \frac{w}{\Omega} + \frac{\Omega}{w} \right)} \frac{1}{Q_3 - \left( \frac{w^*}{\Omega} + \frac{\Omega}{w^*} \right)} = \\ 
&& = \frac{1}{\Omega^3} \left[ \frac{\mathcal{I}_{2} (Q_3, Q_1)}{Q_2 - Q_1} + \frac{\mathcal{I}_{2} (Q_3, Q_2)}{Q_1 - Q_2} \right], \\
&&\alpha^2_+\int_{\text{annulus}}dudv\left(1-\frac{\Omega^2}{w^2}\right)
\left(1-\frac{\Omega^2}{{w^*}^2}\right) \frac{1}{\Omega^3} \frac{1}{Q_1 - \left( \frac{w}{\Omega} + \frac{\Omega}{w} \right)} \frac{1}{Q_2 - \left( \frac{w^*}{\Omega} + \frac{\Omega}{w^*} \right)} \frac{1}{Q_3 - \left( \frac{w}{\Omega} + \frac{\Omega}{w} \right)} = \\ 
&& = \frac{1}{\Omega^3} \left[ \frac{\mathcal{I}_{2} (Q_2, Q_1)}{Q_3 - Q_1} + \frac{\mathcal{I}_{2} (Q_2, Q_3)}{Q_1 - Q_3} \right], \\
&&\alpha^2_+\int_{\text{annulus}}dudv\left(1-\frac{\Omega^2}{w^2}\right)
\left(1-\frac{\Omega^2}{{w^*}^2}\right) \frac{1}{\Omega^3} \frac{1}{Q_1 - \left( \frac{w}{\Omega} + \frac{\Omega}{w} \right)} \frac{1}{Q_2 - \left( \frac{w^*}{\Omega} + \frac{\Omega}{w^*} \right)} \frac{1}{Q_3 - \left( \frac{w^*}{\Omega} + \frac{\Omega}{w^*} \right)} = \\ 
&& = \frac{1}{\Omega^3} \left[ \frac{\mathcal{I}_{2} (Q_2, Q_1)}{Q_3 - Q_2} + \frac{\mathcal{I}_{2} (Q_3, Q_1)}{Q_2 - Q_3} \right].
\end{eqnarray}
Then all possible combinations of integrals of a product of three $T-$terms can be written explicitly.
We will find useful the general notation with dependence on two generic $z_1, z_2$ located at the boundary:
\begin{eqnarray}
    &&\int_{\Omega}d^2\bR' T^e (n, z, w') T^e (n', z_1) T^e (m, z_2) = \frac{1}{\Omega^3} \left[ \frac{\mathcal{I}_1(Q_{n} (z))}{(Q_{n} (z) - Q_{n'} (z_1) )(Q_{n} (z) - Q_{m} (z_2) )} + \right. \\
    &&\left. + \frac{\mathcal{I}_1(Q_{n'} (z_1))}{(Q_{n'} (z_1) - Q_{n} (z))(Q_{n'} (z_1) - Q_{m} (z_2) )} + \frac{\mathcal{I}_1 (Q_{m} (z_2) )}{(Q_{m} (z_2) - Q_{n} (z) )(Q_{m} (z_2) - Q_{n'} (z_1) )} \right], \\
    &&\int_{\Omega}d^2\bR' T^e (n, z, w') T^e (n', z_1) T^o (m, z_2) = \frac{1}{\Omega^3} \left[ \frac{\mathcal{I}_1(Q_{n} (z))}{(Q_{n} (z) - Q_{n'} (z_1) )(Q_{n} (z) - Q_{m}^* (z_2) )} + \right. \\
    &&\left. + \frac{\mathcal{I}_1(Q_{n'} (z_1))}{(Q_{n'} (z_1) - Q_{n} (z))(Q_{n'} (z_1) - Q_{m}^* (z_2) )} + \frac{\mathcal{I}_1 (Q_{m}^* (z_2) )}{(Q_{m}^* (z_2) - Q_{n} (z) )(Q_{m}^* (z_2) - Q_{n'} (z_1) )} \right], \\
    &&\int_{\Omega}d^2\bR' T^e (n, z, w') T^e (n', z_1) \tilde{T}^e (m, z_2) = \frac{1}{\Omega} \left[ \frac{\mathcal{I}_{2} (Q_{m}^* (z_2), Q_{n} (z))}{Q_{n'} (z_1) - Q_{n} (z)} + \frac{\mathcal{I}_{2} (Q_{m}^* (z_2), Q_{n'} (z_1))}{Q_{n} (z) - Q_{n'} (z_1)} \right] \\
    &&\int_{\Omega}d^2\bR' T^e (n, z, w') T^e (n', z_1) \tilde{T}^o (m, z_2) = \frac{1}{\Omega} \left[ \frac{\mathcal{I}_{2} (Q_{m} (z_2), Q_{n} (z))}{Q_{n'} (z_1) - Q_{n} (z)} + \frac{\mathcal{I}_{2} (Q_{m} (z_2), Q_{n'} (z_1))}{Q_{n} (z) - Q_{n'} (z_1)} \right], \\
    &&\int_{\Omega}d^2\bR' T^e (n, z, w') T^o (n', z_1) T^e (m, z_2) = \frac{1}{\Omega^3} \left[ \frac{\mathcal{I}_1(Q_{n} (z))}{(Q_{n} (z) - Q_{n'}^* (z_1) )(Q_{n} (z) - Q_{m} (z_2) )} + \right. \\
    &&\left. + \frac{\mathcal{I}_1(Q_{n'}^* (z_1))}{(Q_{n'}^* (z_1) - Q_{n} (z))(Q_{n'}^* (z_1) - Q_{m} (z_2) )} + \frac{\mathcal{I}_1 (Q_{m} (z_2) )}{(Q_{m} (z_2) - Q_{n} (z) )(Q_{m} (z_2) - Q_{n'}^* (z_1) )} \right], \\
    &&\int_{\Omega}d^2\bR' T^e (n, z, w') T^o (n', z_1) T^o (m, z_2) = \frac{1}{\Omega^3} \left[ \frac{\mathcal{I}_1(Q_{n} (z))}{(Q_{n} (z) - Q_{n'}^* (z_1) )(Q_{n} (z) - Q_{m}^* (z_2) )} + \right. \\
    &&\left. + \frac{\mathcal{I}_1(Q_{n'}^* (z_1))}{(Q_{n'}^* (z_1) - Q_{n} (z))(Q_{n'}^* (z_1) - Q_{m}^* (z_2) )} + \frac{\mathcal{I}_1 (Q_{m}^* (z_2) )}{(Q_{m}^* (z_2) - Q_{n} (z) )(Q_{m}^* (z_2) - Q_{n'}^* (z_1) )} \right], \\
    &&\int_{\Omega}d^2\bR' T^e (n, z, w') T^o (n', z_1) \tilde{T}^e (m, z_2) = \frac{1}{\Omega} \left[ \frac{\mathcal{I}_{2} (Q_{m}^* (z_2), Q_{n} (z))}{Q_{n'}^* (z_1) - Q_{n} (z)} + \frac{\mathcal{I}_{2} (Q_{m}^* (z_2), Q_{n'}^* (z_1))}{Q_{n} (z) - Q_{n'}^* (z_1)} \right] \\
    &&\int_{\Omega}d^2\bR' T^e (n, z, w') T^o (n', z_1) \tilde{T}^o (m, z_2) = \frac{1}{\Omega} \left[ \frac{\mathcal{I}_{2} (Q_{m} (z_2), Q_{n} (z))}{Q_{n'}^* (z_1) - Q_{n} (z)} + \frac{\mathcal{I}_{2} (Q_{m} (z_2), Q_{n'}^* (z_1))}{Q_{n} (z) - Q_{n'}^* (z_1)} \right], \\
    &&\int_{\Omega}d^2\bR' T^e (n, z, w') \tilde{T}^e (n', z_1) T^e (m, z_2) = \frac{1}{\Omega} \left[ \frac{\mathcal{I}_{2} (Q_{n'}^* (z_1), Q_{n} (z))}{Q_{m} (z_2) - Q_{n} (z)} + \frac{\mathcal{I}_{2} (Q_{n'}^* (z_1), Q_{m} (z_2))}{Q_{n} (z) - Q_{m} (z_2)} \right], \\
    &&\int_{\Omega}d^2\bR' T^e (n, z, w') \tilde{T}^e (n', z_1) T^o (m, z_2) = \frac{1}{\Omega} \left[ \frac{\mathcal{I}_{2} (Q_{n'}^* (z_1), Q_{n} (z))}{Q_{m}^* (z_2) - Q_{n} (z)} + \frac{\mathcal{I}_{2} (Q_{n'}^* (z_1), Q_{m}^* (z_2))}{Q_{n} (z) - Q_{m}^* (z_2)} \right], \\
    &&\int_{\Omega}d^2\bR' T^e (n, z, w') \tilde{T}^e (n', z_1) \tilde{T}^e (m, z_2) = \Omega \left[ \frac{\mathcal{I}_{2} (Q_{n'}^* (z_1), Q_{n} (z))}{Q_{m}^* (z_2) - Q_{n'}^* (z_1)} + \frac{\mathcal{I}_{2} (Q_{m}^* (z_2), Q_{n} (z))}{Q_{n'}^* (z_1) - Q_{m}^* (z_2)} \right], \\
    &&\int_{\Omega}d^2\bR' T^e (n, z, w') \tilde{T}^e (n', z_1) \tilde{T}^o (m, z_2) = \Omega \left[ \frac{\mathcal{I}_{2} (Q_{n'}^* (z_1), Q_{n} (z))}{Q_{m} (z_2) - Q_{n'}^* (z_1)} + \frac{\mathcal{I}_{2} (Q_{m} (z_2), Q_{n} (z))}{Q_{n'}^* (z_1) - Q_{m} (z_2)} \right], 
\end{eqnarray}

\begin{eqnarray}
    &&\int_{\Omega}d^2\bR' T^e (n, z, w') \tilde{T}^o (n', z_1) T^e (m, z_2) = \frac{1}{\Omega} \left[ \frac{\mathcal{I}_{2} (Q_{n'} (z_1), Q_{n} (z))}{Q_{m} (z_2) - Q_{n} (z)} + \frac{\mathcal{I}_{2} (Q_{n'} (z_1), Q_{m} (z_2))}{Q_{n} (z) - Q_{m} (z_2)} \right], \\
    &&\int_{\Omega}d^2\bR' T^e (n, z, w') \tilde{T}^o (n', z_1) T^o (m, z_2) = \frac{1}{\Omega} \left[ \frac{\mathcal{I}_{2} (Q_{n'} (z_1), Q_{n} (z))}{Q_{m}^* (z_2) - Q_{n} (z)} + \frac{\mathcal{I}_{2} (Q_{n'} (z_1), Q_{m}^* (z_2))}{Q_{n} (z) - Q_{m}^* (z_2)} \right], \\
    &&\int_{\Omega}d^2\bR' T^e (n, z, w') \tilde{T}^o (n', z_1) \tilde{T}^e (m, z_2) = \Omega \left[ \frac{\mathcal{I}_{2} (Q_{n'} (z_1), Q_{n} (z))}{Q_{m}^* (z_2) - Q_{n'} (z_1)} + \frac{\mathcal{I}_{2} (Q_{m}^* (z_2), Q_{n} (z))}{Q_{n'} (z_1) - Q_{m}^* (z_2)} \right], \\
    &&\int_{\Omega}d^2\bR' T^e (n, z, w') \tilde{T}^o (n', z_1) \tilde{T}^o (m, z_2) = \Omega \left[ \frac{\mathcal{I}_{2} (Q_{n'} (z_1), Q_{n} (z))}{Q_{m} (z_2) - Q_{n'} (z_1)} + \frac{\mathcal{I}_{2} (Q_{m} (z_2), Q_{n} (z))}{Q_{n'} (z_1) - Q_{m} (z_2)} \right],
\end{eqnarray}

\begin{eqnarray}
    &&\int_{\Omega}d^2\bR' T^o (n, z, w') T^e (n', z_1) T^e (m, z_2) = \frac{1}{\Omega^3} \left[ \frac{\mathcal{I}_1(Q_{n}^* (z))}{(Q_{n}^* (z) - Q_{n'} (z_1) )(Q_{n}^* (z) - Q_{m} (z_2) )} + \right. \\
    &&\left. + \frac{\mathcal{I}_1(Q_{n'} (z_1))}{(Q_{n'} (z_1) - Q_{n}^* (z))(Q_{n'} (z_1) - Q_{m} (z_2) )} + \frac{\mathcal{I}_1 (Q_{m} (z_2) )}{(Q_{m} (z_2) - Q_{n}^* (z) )(Q_{m} (z_2) - Q_{n'} (z_1) )} \right], \\
    &&\int_{\Omega}d^2\bR' T^o (n, z, w') T^e (n', z_1) T^o (m, z_2) = \frac{1}{\Omega^3} \left[ \frac{\mathcal{I}_1(Q_{n}^* (z))}{(Q_{n}^* (z) - Q_{n'} (z_1) )(Q_{n}^* (z) - Q_{m}^* (z_2) )} + \right. \\
    &&\left. + \frac{\mathcal{I}_1(Q_{n'} (z_1))}{(Q_{n'} (z_1) - Q_{n}^* (z))(Q_{n'} (z_1) - Q_{m}^* (z_2) )} + \frac{\mathcal{I}_1 (Q_{m}^* (z_2) )}{(Q_{m}^* (z_2) - Q_{n}^* (z) )(Q_{m}^* (z_2) - Q_{n'} (z_1) )} \right], \\
    &&\int_{\Omega}d^2\bR' T^o (n, z, w') T^e (n', z_1) \tilde{T}^e (m, z_2) = \frac{1}{\Omega} \left[ \frac{\mathcal{I}_{2} (Q_{m}^* (z_2), Q_{n}^* (z))}{Q_{n'} (z_1) - Q_{n}^* (z)} + \frac{\mathcal{I}_{2} (Q_{m}^* (z_2), Q_{n'} (z_1))}{Q_{n}^* (z) - Q_{n'} (z_1)} \right] \\
    &&\int_{\Omega}d^2\bR' T^o (n, z, w') T^e (n', z_1) \tilde{T}^o (m, z_2) = \frac{1}{\Omega} \left[ \frac{\mathcal{I}_{2} (Q_{m} (z_2), Q_{n}^* (z))}{Q_{n'} (z_1) - Q_{n}^* (z)} + \frac{\mathcal{I}_{2} (Q_{m} (z_2), Q_{n'} (z_1))}{Q_{n}^* (z) - Q_{n'} (z_1)} \right], \\
    &&\int_{\Omega}d^2\bR' T^o (n, z, w') T^o (n', z_1) T^e (m, z_2) = \frac{1}{\Omega^3} \left[ \frac{\mathcal{I}_1(Q_{n}^* (z))}{(Q_{n}^* (z) - Q_{n'}^* (z_1) )(Q_{n}^* (z) - Q_{m} (z_2) )} + \right. \\
    &&\left. + \frac{\mathcal{I}_1(Q_{n'}^* (z_1))}{(Q_{n'}^* (z_1) - Q_{n}^* (z))(Q_{n'}^* (z_1) - Q_{m} (z_2) )} + \frac{\mathcal{I}_1 (Q_{m} (z_2) )}{(Q_{m} (z_2) - Q_{n}^* (z) )(Q_{m} (z_2) - Q_{n'}^* (z_1) )} \right], \\
    &&\int_{\Omega}d^2\bR' T^o (n, z, w') T^o (n', z_1) T^o (m, z_2) = \frac{1}{\Omega^3} \left[ \frac{\mathcal{I}_1(Q_{n}^* (z))}{(Q_{n}^* (z) - Q_{n'}^* (z_1) )(Q_{n}^* (z) - Q_{m}^* (z_2) )} + \right. \\
    &&\left. + \frac{\mathcal{I}_1(Q_{n'}^* (z_1))}{(Q_{n'}^* (z_1) - Q_{n}^* (z))(Q_{n'}^* (z_1) - Q_{m}^* (z_2) )} + \frac{\mathcal{I}_1 (Q_{m}^* (z_2) )}{(Q_{m}^* (z_2) - Q_{n}^* (z) )(Q_{m}^* (z_2) - Q_{n'}^* (z_1) )} \right], \\
    &&\int_{\Omega}d^2\bR' T^o (n, z, w') T^o (n', z_1) \tilde{T}^e (m, z_2) = \frac{1}{\Omega} \left[ \frac{\mathcal{I}_{2} (Q_{m}^* (z_2), Q_{n}^* (z))}{Q_{n'}^* (z_1) - Q_{n}^* (z)} + \frac{\mathcal{I}_{2} (Q_{m}^* (z_2), Q_{n'}^* (z_1))}{Q_{n}^* (z) - Q_{n'}^* (z_1)} \right] \\
    &&\int_{\Omega}d^2\bR' T^o (n, z, w') T^o (n', z_1) \tilde{T}^o (m, z_2) = \frac{1}{\Omega} \left[ \frac{\mathcal{I}_{2} (Q_{m} (z_2), Q_{n}^* (z))}{Q_{n'}^* (z_1) - Q_{n}^* (z)} + \frac{\mathcal{I}_{2} (Q_{m} (z_2), Q_{n'}^* (z_1))}{Q_{n}^* (z) - Q_{n'}^* (z_1)} \right], \\
    &&\int_{\Omega}d^2\bR' T^o (n, z, w') \tilde{T}^e (n', z_1) T^e (m, z_2) = \frac{1}{\Omega} \left[ \frac{\mathcal{I}_{2} (Q_{n'}^* (z_1), Q_{n}^* (z))}{Q_{m} (z_2) - Q_{n}^* (z)} + \frac{\mathcal{I}_{2} (Q_{n'}^* (z_1), Q_{m} (z_2))}{Q_{n}^* (z) - Q_{m} (z_2)} \right], \\
    &&\int_{\Omega}d^2\bR' T^o (n, z, w') \tilde{T}^e (n', z_1) T^o (m, z_2) = \frac{1}{\Omega} \left[ \frac{\mathcal{I}_{2} (Q_{n'}^* (z_1), Q_{n}^* (z))}{Q_{m}^* (z_2) - Q_{n}^* (z)} + \frac{\mathcal{I}_{2} (Q_{n'}^* (z_1), Q_{m}^* (z_2))}{Q_{n}^* (z) - Q_{m}^* (z_2)} \right], \\
    &&\int_{\Omega}d^2\bR' T^o (n, z, w') \tilde{T}^e (n', z_1) \tilde{T}^e (m, z_2) = \Omega \left[ \frac{\mathcal{I}_{2} (Q_{n'}^* (z_1), Q_{n}^* (z))}{Q_{m}^* (z_2) - Q_{n'}^* (z_1)} + \frac{\mathcal{I}_{2} (Q_{m}^* (z_2), Q_{n}^* (z))}{Q_{n'}^* (z_1) - Q_{m}^* (z_2)} \right], \\
    &&\int_{\Omega}d^2\bR' T^o (n, z, w') \tilde{T}^e (n', z_1) \tilde{T}^o (m, z_2) = \Omega \left[ \frac{\mathcal{I}_{2} (Q_{n'}^* (z_1), Q_{n}^* (z))}{Q_{m} (z_2) - Q_{n'}^* (z_1)} + \frac{\mathcal{I}_{2} (Q_{m} (z_2), Q_{n}^* (z))}{Q_{n'}^* (z_1) - Q_{m} (z_2)} \right], 
\end{eqnarray}

\begin{eqnarray}
    &&\int_{\Omega}d^2\bR' T^o (n, z, w') \tilde{T}^o (n', z_1) T^e (m, z_2) = \frac{1}{\Omega} \left[ \frac{\mathcal{I}_{2} (Q_{n'} (z_1), Q_{n}^* (z))}{Q_{m} (z_2) - Q_{n}^* (z)} + \frac{\mathcal{I}_{2} (Q_{n'} (z_1), Q_{m} (z_2))}{Q_{n}^* (z) - Q_{m} (z_2)} \right], \\
    &&\int_{\Omega}d^2\bR' T^o (n, z, w') \tilde{T}^o (n', z_1) T^o (m, z_2) = \frac{1}{\Omega} \left[ \frac{\mathcal{I}_{2} (Q_{n'} (z_1), Q_{n}^* (z))}{Q_{m}^* (z_2) - Q_{n}^* (z)} + \frac{\mathcal{I}_{2} (Q_{n'} (z_1), Q_{m}^* (z_2))}{Q_{n}^* (z) - Q_{m}^* (z_2)} \right], \\
    &&\int_{\Omega}d^2\bR' T^o (n, z, w') \tilde{T}^o (n', z_1) \tilde{T}^e (m, z_2) = \Omega \left[ \frac{\mathcal{I}_{2} (Q_{n'} (z_1), Q_{n}^* (z))}{Q_{m}^* (z_2) - Q_{n'} (z_1)} + \frac{\mathcal{I}_{2} (Q_{m}^* (z_2), Q_{n}^* (z))}{Q_{n'} (z_1) - Q_{m}^* (z_2)} \right], \\
    &&\int_{\Omega}d^2\bR' T^o (n, z, w') \tilde{T}^o (n', z_1) \tilde{T}^o (m, z_2) = \Omega \left[ \frac{\mathcal{I}_{2} (Q_{n'} (z_1), Q_{n}^* (z))}{Q_{m} (z_2) - Q_{n'} (z_1)} + \frac{\mathcal{I}_{2} (Q_{m} (z_2), Q_{n}^* (z))}{Q_{n'} (z_1) - Q_{m} (z_2)} \right].
\end{eqnarray}
Terms of the type $T^i (n,z) T^j (n',z_1) T^j (m,z_2)$ (also true for terms with tildas)  experience divergences for $z_1=z_2$.  
Therefore, we need to carefully take limits. We do that in the next section.

\subsubsection{Resolving divergences by taking limits}

We need to cure only the divergent terms, so we will focus solely on them. There are 8 such terms:
\begin{eqnarray}
    &&\int_{\Omega}d^2\bR' T^e (n, z, w') T^e (n', z_1) T^e (m, z_2) \\
    &&\int_{\Omega}d^2\bR' T^e (n, z, w') T^o (n', z_1) T^o (m, z_2) \\
    &&\int_{\Omega}d^2\bR' T^e (n, z, w') \tilde{T}^e (n', z_1) \tilde{T}^e (m, z_2) \\
    &&\int_{\Omega}d^2\bR' T^e (n, z, w') \tilde{T}^o (n', z_1) \tilde{T}^o (m, z_2) \\
    &&\int_{\Omega}d^2\bR' T^o (n, z, w') T^e (n', z_1) T^e (m, z_2) \\
    &&\int_{\Omega}d^2\bR' T^o (n, z, w') T^o (n', z_1) T^o (m, z_2) \\
    &&\int_{\Omega}d^2\bR' T^o (n, z, w') \tilde{T}^e (n', z_1) \tilde{T}^e (m, z_2) \\
    &&\int_{\Omega}d^2\bR' T^o (n, z, w') \tilde{T}^o (n', z_1) \tilde{T}^o (m, z_2)
\end{eqnarray}
which diverge for $z_1 = z_2$ \underline{and} for $n'=m$. Further we will only focus on $|z_1|=|z_2|=1$ with $z_{1,2} = e^{i \phi_{1,2}}$. Considering all terms one by one for $\phi_2 \rightarrow \phi_1$ we obtain
\begin{eqnarray}
    &&\int_{\Omega}d^2\bR' T^e (n, z, w') T^e (m, z_1) T^e (m, z_2) = \frac{1}{\Omega^3} \left[ \frac{\mathcal{I}_1(Q_{n} (z))}{(Q_{n} (z) - Q_{m} (z_1) )(Q_{n} (z) - Q_{m} (z_2) )} + \right. \\
    &&\left. + \frac{\mathcal{I}_1(Q_{m} (z_1))}{(Q_{m} (z_1) - Q_{n} (z))(Q_{m} (z_1) - Q_{m} (z_2) )} + \frac{\mathcal{I}_1(Q_{m} (z_2) )}{(Q_{m} (z_2) - Q_{n} (z) )(Q_{m} (z_2) - Q_{m} (z_1) )} \right] = \\
    &&= \frac{1}{\Omega^3} \left[ \frac{\pi \alpha_+^2 \left( 1-\Omega^4 \right) z^* \Omega^{2n+1}}{\left( \frac{z+\Omega^{4n+2}{z^*}}{\Omega^{2n+1}} - \frac{z_1+\Omega^{4m+2}{z_1^*}}{\Omega^{2m+1}} \right) \left(\frac{z+\Omega^{4n+2}{z^*}}{\Omega^{2n+1}} - \frac{z_2+\Omega^{4m+2}{z_2^*}}{\Omega^{2m+1}} \right)} + \right. \\
    &&\left. + \frac{\pi \alpha_+^2 \left( 1-\Omega^4 \right) z_1^* \Omega^{2m+1}}{\left( \frac{z_1+\Omega^{4m+2}{z_1^*}}{\Omega^{2m+1}} - \frac{z+\Omega^{4n+2}{z^*}}{\Omega^{2n+1}} \right) \left( \frac{z_1+\Omega^{4m+2}{z_1^*}}{\Omega^{2m+1}} - \frac{z_2+\Omega^{4m+2}{z_2^*}}{\Omega^{2m+1}} \right)} + \frac{\pi \alpha_+^2 \left( 1-\Omega^4 \right) z_2^* \Omega^{2m+1}}{\left( \frac{z_2+\Omega^{4m+2}{z_2^*}}{\Omega^{2m+1}} - \frac{z+\Omega^{4n+2}{z^*}}{\Omega^{2n+1}} \right) \left( \frac{z_2+\Omega^{4m+2}{z_2^*}}{\Omega^{2m+1}} - \frac{z_1+\Omega^{4m+2}{z_1^*}}{\Omega^{2m+1}} \right)} \right] \nonumber \\
    &&= \frac{\pi \alpha_+^2 (1-\Omega^4)}{\Omega^3} \frac{e^{i \left( \phi + 2 \phi_1 \right)} \Omega^{3+4m+2n}}{\left( e^{2 i \phi_1} - \Omega^{2+4m} \right) \left( e^{ i \left( \phi + \phi_1 \right)} - \Omega^{2+2m+2n} \right)^2} = \frac{\pi \alpha_+^2 (1-\Omega^4) e^{i \left( \phi + 2 \phi_1 \right)} \Omega^{4m+2n}}{\left( e^{2 i \phi_1} - \Omega^{2+4m} \right) \left( e^{ i \left( \phi + \phi_1 \right)} - \Omega^{2+2m+2n} \right)^2}
\end{eqnarray}

\begin{eqnarray}
    &&\int_{\Omega}d^2\bR' T^e (n, z, w') T^o (m, z_1) T^o (m, z_2) = \frac{1}{\Omega^3} \left[ \frac{\mathcal{I}_1(Q_{n} (z))}{(Q_{n} (z) - Q_{m}^* (z_1) )(Q_{n} (z) - Q_{m}^* (z_2) )} + \right. \\
    &&\left. + \frac{\mathcal{I}_1(Q_{m}^* (z_1))}{(Q_{m}^* (z_1) - Q_{n} (z))(Q_{m}^* (z_1) - Q_{m}^* (z_2) )} + \frac{\mathcal{I}_1 (Q_{m}^* (z_2) )}{(Q_{m}^* (z_2) - Q_{n} (z) )(Q_{m}^* (z_2) - Q_{m}^* (z_1) )} \right] = \\
    &&= \frac{1}{\Omega^3} \left[ \frac{\pi \alpha_+^2 \left( 1-\Omega^4 \right) z^* \Omega^{2n+1}}{\left( \frac{z+\Omega^{4n+2}{z^*}}{\Omega^{2n+1}} - \frac{z_1^*+\Omega^{4m+2}{z_1}}{\Omega^{2m+1}} \right) \left(\frac{z+\Omega^{4n+2}{z^*}}{\Omega^{2n+1}} - \frac{z_2^*+\Omega^{4m+2}{z_2}}{\Omega^{2m+1}} \right)} + \right. \\
    &&\left. + \frac{\pi \alpha_+^2 \left( 1-\Omega^4 \right) z_1 \Omega^{2m+1}}{\left( \frac{z_1^*+\Omega^{4m+2}{z_1}}{\Omega^{2m+1}} - \frac{z+\Omega^{4n+2}{z^*}}{\Omega^{2n+1}} \right) \left( \frac{z_1^*+\Omega^{4m+2}{z_1}}{\Omega^{2m+1}} - \frac{z_2^*+\Omega^{4m+2}{z_2}}{\Omega^{2m+1}} \right)} + \frac{\pi \alpha_+^2 \left( 1-\Omega^4 \right) z_2 \Omega^{2m+1}}{\left( \frac{z_2^*+\Omega^{4m+2}{z_2}}{\Omega^{2m+1}} - \frac{z+\Omega^{4n+2}{z^*}}{\Omega^{2n+1}} \right) \left( \frac{z_2^*+\Omega^{4m+2}{z_2}}{\Omega^{2m+1}} - \frac{z_1^*+\Omega^{4m+2}{z_1}}{\Omega^{2m+1}} \right)} \right] \nonumber \\
    &&=- \frac{\pi \alpha_+^2 (1-\Omega^4) e^{i \left( \phi + 2 \phi_1 \right)} \Omega^{4m+2n}}{\left( e^{2 i \phi_1} \Omega^{2+4m} - 1 \right) \left( e^{ i \phi } - e^{ i \phi_1 } \Omega^{2+2m+2n} \right)^2}
\end{eqnarray}

\begin{eqnarray}
    &&\int_{\Omega}d^2\bR' T^o (n, z, w') T^e (m, z_1) T^e (m, z_2) = \frac{1}{\Omega^3} \left[ \frac{\mathcal{I}_1(Q_{n}^* (z))}{(Q_{n}^* (z) - Q_{m} (z_1) )(Q_{n}^* (z) - Q_{m} (z_2) )} + \right. \\
    &&\left. + \frac{\mathcal{I}_1(Q_{m} (z_1))}{(Q_{m} (z_1) - Q_{n}^* (z))(Q_{m} (z_1) - Q_{m} (z_2) )} + \frac{\mathcal{I}_1 (Q_{m} (z_2) )}{(Q_{m} (z_2) - Q_{n}^* (z) )(Q_{m} (z_2) - Q_{m} (z_1) )} \right] = \\ 
    &&= \frac{1}{\Omega^3} \left[ \frac{\pi \alpha_+^2 \left( 1-\Omega^4 \right) z \Omega^{2n+1}}{\left( \frac{z^*+\Omega^{4n+2}{z}}{\Omega^{2n+1}} - \frac{z_1+\Omega^{4m+2}{z_1^*}}{\Omega^{2m+1}} \right) \left(\frac{z^*+\Omega^{4n+2}{z}}{\Omega^{2n+1}} - \frac{z_2+\Omega^{4m+2}{z_2^*}}{\Omega^{2m+1}} \right)} + \right. \\
    &&\left. + \frac{\pi \alpha_+^2 \left( 1-\Omega^4 \right) z_1^* \Omega^{2m+1}}{\left( \frac{z_1+\Omega^{4m+2}{z_1^*}}{\Omega^{2m+1}} - \frac{z^*+\Omega^{4n+2}{z}}{\Omega^{2n+1}} \right) \left( \frac{z_1+\Omega^{4m+2}{z_1^*}}{\Omega^{2m+1}} - \frac{z_2+\Omega^{4m+2}{z_2^*}}{\Omega^{2m+1}} \right)} + \frac{\pi \alpha_+^2 \left( 1-\Omega^4 \right) z_2^* \Omega^{2m+1}}{\left( \frac{z_2+\Omega^{4m+2}{z_2^*}}{\Omega^{2m+1}} - \frac{z^*+\Omega^{4n+2}{z}}{\Omega^{2n+1}} \right) \left( \frac{z_2+\Omega^{4m+2}{z_2^*}}{\Omega^{2m+1}} - \frac{z_1+\Omega^{4m+2}{z_1^*}}{\Omega^{2m+1}} \right)} \right] \nonumber \\
    &&= \frac{\pi \alpha_+^2 (1-\Omega^4) e^{i \left( \phi + 2 \phi_1 \right)} \Omega^{4m+2n}}{\left( e^{2 i \phi_1} - \Omega^{2+4m} \right) \left( e^{ i \phi_1 } - e^{ i \phi } \Omega^{2+2m+2n} \right)^2}
\end{eqnarray}

\begin{eqnarray}
    &&\int_{\Omega}d^2\bR' T^o (n, z, w') T^o (m, z_1) T^o (m, z_2) = \frac{1}{\Omega^3} \left[ \frac{\mathcal{I}_1(Q_{n}^* (z))}{(Q_{n}^* (z) - Q_{m}^* (z_1) )(Q_{n}^* (z) - Q_{m}^* (z_2) )} + \right. \\
    &&\left. + \frac{\mathcal{I}_1(Q_{m}^* (z_1))}{(Q_{m}^* (z_1) - Q_{n}^* (z))(Q_{m}^* (z_1) - Q_{m}^* (z_2) )} + \frac{\mathcal{I}_1 (Q_{m}^* (z_2) )}{(Q_{m}^* (z_2) - Q_{n}^* (z) )(Q_{m}^* (z_2) - Q_{m}^* (z_1) )} \right] = \\
    &&= \frac{1}{\Omega^3} \left[ \frac{\pi \alpha_+^2 \left( 1-\Omega^4 \right) z \Omega^{2n+1}}{\left( \frac{z^*+\Omega^{4n+2}{z}}{\Omega^{2n+1}} - \frac{z_1^*+\Omega^{4m+2}{z_1}}{\Omega^{2m+1}} \right) \left(\frac{z^*+\Omega^{4n+2}{z}}{\Omega^{2n+1}} - \frac{z_2^*+\Omega^{4m+2}{z_2}}{\Omega^{2m+1}} \right)} + \right. \\
    &&\left. + \frac{\pi \alpha_+^2 \left( 1-\Omega^4 \right) z_1 \Omega^{2m+1}}{\left( \frac{z_1^*+\Omega^{4m+2}{z_1}}{\Omega^{2m+1}} - \frac{z^*+\Omega^{4n+2}{z}}{\Omega^{2n+1}} \right) \left( \frac{z_1^*+\Omega^{4m+2}{z_1}}{\Omega^{2m+1}} - \frac{z_2^*+\Omega^{4m+2}{z_2}}{\Omega^{2m+1}} \right)} + \frac{\pi \alpha_+^2 \left( 1-\Omega^4 \right) z_2 \Omega^{2m+1}}{\left( \frac{z_2^*+\Omega^{4m+2}{z_2}}{\Omega^{2m+1}} - \frac{z^*+\Omega^{4n+2}{z}}{\Omega^{2n+1}} \right) \left( \frac{z_2^*+\Omega^{4m+2}{z_2}}{\Omega^{2m+1}} - \frac{z_1^*+\Omega^{4m+2}{z_1}}{\Omega^{2m+1}} \right)} \right] \nonumber \\
    &&=- \frac{\pi \alpha_+^2 (1-\Omega^4) e^{i \left( \phi + 2 \phi_1 \right)} \Omega^{4m+2n}}{\left( e^{2 i \phi_1} \Omega^{2+4m} - 1 \right) \left( 1 - e^{ i \left( \phi + \phi_1 \right) } \Omega^{2+2m+2n} \right)^2}
\end{eqnarray}

\begin{eqnarray}
    &&\int_{\Omega}d^2\bR' T^e (n, z, w') \tilde{T}^e (m, z_1) \tilde{T}^e (m, z_2) = \Omega \left[ \frac{\mathcal{I}_{2} (Q_{m}^* (z_1), Q_{n} (z))}{Q_{m}^* (z_2) - Q_{m}^* (z_1)} + \frac{\mathcal{I}_{2} (Q_{m}^* (z_2), Q_{n} (z))}{Q_{m}^* (z_1) - Q_{m}^* (z_2)} \right] = \\ &&=
    \Omega \frac{\mathcal{I}_{2} (Q_{m}^* (z_1), Q_{n} (z)) - \mathcal{I}_{2} (Q_{m}^* (z_2), Q_{n} (z))}{Q_{m}^* (z_2) - Q_{m}^* (z_1)} = \\
    &&=  \frac{\pi \alpha_+^2 \Omega^3}{\frac{z_2^*+\Omega^{4m+2}{z_2}}{\Omega^{2m+1}} - \frac{z_1^*+\Omega^{4m+2}{z_1}}{\Omega^{2m+1}}} \left(
     \ln \left[ 1 - z^* z_1 \Omega^{2n + 2m + 4} \right] - \ln \left[ \Omega^2 - z^* z_1 \Omega^{2n + 2m + 4} \right] - \right. \\ 
    &&- \left.  \ln \left[ 1 - \frac{1}{z^* z_1 \Omega^{2n + 2m}} \right] + \ln \left[ \Omega^2 - \frac{1}{z^* z_1 \Omega^{2n + 2m}} \right]
    - \ln \left[ 1 - z^* z_2 \Omega^{2n + 2m + 4} \right] + \right. \\ &&+ \left.  \ln \left[ \Omega^2 + z^* z_2 \Omega^{2n + 2m + 4} \right] + 
     \ln \left[ 1 - \frac{1}{z^* z_2 \Omega^{2n + 2m}} \right] - \ln \left[ \Omega^2 - \frac{1}{z^* z_2 \Omega^{2n + 2m}} \right]\right) = \\
     &&= \frac{\pi \alpha_+^2 \left( \Omega^4 - 1 \right) \Omega^{4+4m+2n} e^{i \left( \phi + 2 \phi_1 \right)}}{\left( e^{ 2 i \phi_1 } \Omega^{2 + 4m} -1 \right) \left( e^{ i \phi } - e^{ i \phi_1 } \Omega^{2(m+n)} \right) \left( e^{ i \phi } - e^{ i \phi_1 } \Omega^{2(2+m+n)} \right)}
\end{eqnarray}

\begin{eqnarray}
    &&\int_{\Omega}d^2\bR' T^e (n, z, w') \tilde{T}^o (m, z_1) \tilde{T}^o (m, z_2) = \Omega \left[ \frac{\mathcal{I}_{2} (Q_{m} (z_1), Q_{n} (z))}{Q_{m} (z_2) - Q_{m} (z_1)} + \frac{\mathcal{I}_{2} (Q_{m} (z_2), Q_{n} (z))}{Q_{m} (z_1) - Q_{m} (z_2)} \right] = \\ &&=
    \Omega \frac{\mathcal{I}_{2} (Q_{m} (z_1), Q_{n} (z)) - \mathcal{I}_{2} (Q_{m} (z_2), Q_{n} (z))}{Q_{m} (z_2) - Q_{m} (z_1)} = \\
    &&=  \frac{\pi \alpha_+^2 \Omega^3}{\frac{z_2+\Omega^{4m+2}{z_2^*}}{\Omega^{2m+1}} - \frac{z_1+\Omega^{4m+2}{z_1^*}}{\Omega^{2m+1}}} \left(
     \ln \left[ 1 - z^* z_1^* \Omega^{2n + 2m + 4} \right] - \ln \left[ \Omega^2 - z^* z_1^* \Omega^{2n + 2m + 4} \right] - \right. \\ 
    &&- \left.  \ln \left[ 1 - \frac{1}{z^* z_1^* \Omega^{2n + 2m}} \right] + \ln \left[ \Omega^2 - \frac{1}{z^* z_1^* \Omega^{2n + 2m}} \right]
    - \ln \left[ 1 - z^* z_2^* \Omega^{2n + 2m + 4} \right] + \right. \\ &&+ \left.  \ln \left[ \Omega^2 + z^* z_2^* \Omega^{2n + 2m + 4} \right] + 
     \ln \left[ 1 - \frac{1}{z^* z_2^* \Omega^{2n + 2m}} \right] - \ln \left[ \Omega^2 - \frac{1}{z^* z_2^* \Omega^{2n + 2m}} \right]\right) = \\
     &&= -\frac{\pi \alpha_+^2 \left( \Omega^4 - 1 \right) \Omega^{4+4m+2n} e^{i \left( \phi + 2 \phi_1 \right)}}{\left( e^{ 2 i \phi_1 }  -\Omega^{2 + 4m} \right) \left( e^{ i \left( \phi + \phi_1 \right) } - \Omega^{2(m+n)} \right) \left( e^{ i \left( \phi + \phi_1 \right) } - \Omega^{2(2+m+n)} \right)}
\end{eqnarray}

\begin{eqnarray}
    &&\int_{\Omega}d^2\bR' T^o (n, z, w') \tilde{T}^e (m, z_1) \tilde{T}^e (m, z_2) = \Omega \left[ \frac{\mathcal{I}_{2} (Q_{m}^* (z_1), Q_{n}^* (z))}{Q_{m}^* (z_2) - Q_{m}^* (z_1)} + \frac{\mathcal{I}_{2} (Q_{m}^* (z_2), Q_{n}^* (z))}{Q_{m}^* (z_1) - Q_{m}^* (z_2)} \right] = \\ &&=
    \Omega \frac{\mathcal{I}_{2} (Q_{m}^* (z_1), Q_{n}^* (z)) - \mathcal{I}_{2} (Q_{m}^* (z_2), Q_{n}^* (z))}{Q_{m}^* (z_2) - Q_{m}^* (z_1)} = \\
    &&=  \frac{\pi \alpha_+^2 \Omega^3}{\frac{z_2^*+\Omega^{4m+2}{z_2}}{\Omega^{2m+1}} - \frac{z_1^*+\Omega^{4m+2}{z_1}}{\Omega^{2m+1}}} \left(
     \ln \left[ 1 - z z_1 \Omega^{2n + 2m + 4} \right] - \ln \left[ \Omega^2 - z z_1 \Omega^{2n + 2m + 4} \right] - \right. \\ 
    &&- \left.  \ln \left[ 1 - \frac{1}{z z_1 \Omega^{2n + 2m}} \right] + \ln \left[ \Omega^2 - \frac{1}{z z_1 \Omega^{2n + 2m}} \right]
    - \ln \left[ 1 - z z_2 \Omega^{2n + 2m + 4} \right] + \right. \\ &&+ \left.  \ln \left[ \Omega^2 + z z_2 \Omega^{2n + 2m + 4} \right] + 
     \ln \left[ 1 - \frac{1}{z z_2 \Omega^{2n + 2m}} \right] - \ln \left[ \Omega^2 - \frac{1}{z z_2 \Omega^{2n + 2m}} \right]\right) = \\
     &&= \frac{\pi \alpha_+^2 \left( \Omega^4 - 1 \right) \Omega^{4+4m+2n} e^{i \left( \phi + 2 \phi_1 \right)}}{\left( e^{ 2 i \phi_1 } \Omega^{2 + 4m}  - 1 \right) \left( e^{ i \left( \phi + \phi_1 \right) } \Omega^{2(m+n)} - 1 \right) \left( e^{ i \left( \phi + \phi_1 \right) } \Omega^{2(2+m+n)} - 1 \right)}
\end{eqnarray}

\begin{eqnarray}
    &&\int_{\Omega}d^2\bR' T^o (n, z, w') \tilde{T}^o (m, z_1) \tilde{T}^o (m, z_2) = \Omega \left[ \frac{\mathcal{I}_{2} (Q_{m} (z_1), Q_{n}^* (z))}{Q_{m} (z_2) - Q_{m} (z_1)} + \frac{\mathcal{I}_{2} (Q_{m} (z_2), Q_{n}^* (z))}{Q_{m} (z_1) - Q_{m} (z_2)} \right] = \\ &&=
    \Omega \frac{\mathcal{I}_{2} (Q_{m} (z_1), Q_{n}^* (z)) - \mathcal{I}_{2} (Q_{m} (z_2), Q_{n}^* (z))}{Q_{m} (z_2) - Q_{m} (z_1)} = \\
    &&=  \frac{\pi \alpha_+^2 \Omega^3}{\frac{z_2+\Omega^{4m+2}{z_2^*}}{\Omega^{2m+1}} - \frac{z_1+\Omega^{4m+2}{z_1^*}}{\Omega^{2m+1}}} \left(
     \ln \left[ 1 - z z_1^* \Omega^{2n + 2m + 4} \right] - \ln \left[ \Omega^2 - z z_1^* \Omega^{2n + 2m + 4} \right] - \right. \\ 
    &&- \left.  \ln \left[ 1 - \frac{1}{z z_1^* \Omega^{2n + 2m}} \right] + \ln \left[ \Omega^2 - \frac{1}{z z_1^* \Omega^{2n + 2m}} \right]
    - \ln \left[ 1 - z z_2^* \Omega^{2n + 2m + 4} \right] + \right. \\ &&+ \left.  \ln \left[ \Omega^2 + z z_2^* \Omega^{2n + 2m + 4} \right] + 
     \ln \left[ 1 - \frac{1}{z z_2^* \Omega^{2n + 2m}} \right] - \ln \left[ \Omega^2 - \frac{1}{z z_2^* \Omega^{2n + 2m}} \right]\right) = \\
     &&= -\frac{\pi \alpha_+^2 \left( \Omega^4 - 1 \right) \Omega^{4+4m+2n} e^{i \left( \phi + 2 \phi_1 \right)}}{\left( e^{ 2 i \phi_1 }  -\Omega^{2 + 4m} \right) \left( e^{ i \phi_1 } - e^{ i \phi } \Omega^{2(m+n)} \right) \left( e^{ i \phi_1 } - e^{ i \phi } \Omega^{2(2+m+n)} \right)}
\end{eqnarray}

\subsection{First basis function $\left(E^{\newtext{(1)}}_x-iE^{\newtext{(1)}}_y\right)^2$}

As in the case of vanishing anisotropy of the linear conductivity tensor we can introduce three distinct basis functions. In contrast to the vanishing linear conductivity anisotropy, \underline{all three basis functions depend on the angular size of the lead.}

Using the decomposition from above we can now rewrite the expression for the potential for the first basis function:
\begin{eqnarray}
&&\Phi^{\newtext{(2)}}_{ani}(\bR)= \frac{\alpha_+}{\bar{\sigma}}\int_{\Omega}d^2\bR' \left[\left(\frac{\partial}{\partial
Z'}G_{N}(\bR,\bR')\right)\left(\sigma^{(2)}_{-} e^{i\chi^{(2)}_{-}} + \Omega^2 \sigma^{(2)}_{+} e^{-i\chi^{(2)}_{+}} \right) \left(E^{\newtext{(1)}}_x-iE^{\newtext{(1)}}_y\right)^2+ \mathrm{c.c.} \right] = \nonumber \\
&&=\frac{\alpha_+ }{\bar{\sigma}}\int_{\Omega}d^2\bR' \left[ \left(\sigma^{(2)}_{-} e^{i\chi^{(2)}_{-}} + \Omega^2 \sigma^{(2)}_{+} e^{-i\chi^{(2)}_{+}} \right)\left(\frac{\partial}{\partial
Z'}G_{N}(\bR,\bR')\right)\left(E^{\newtext{(1)}}_x-iE^{\newtext{(1)}}_y\right)^2+ \mathrm{c.c.} \right] = \nonumber \\
&&=\frac{\alpha_+ }{\bar{\sigma}}\int_{\Omega}d^2\bR' \left[\frac{\left(\sigma^{(2)}_{-} e^{i\chi^{(2)}_{-}} + \Omega^2 \sigma^{(2)}_{+} e^{-i\chi^{(2)}_{+}} \right)}{2\pi \alpha_+} \left( \sum_{n=0,2,4,\ldots}^{\infty} T^e (n, z, w') +
\sum_{n=1,3,5\ldots}^{\infty} T^o (n, z, w') \right) \right. \nonumber \\ && \left. 
\left(\frac{-I}{\pi \bar{\sigma}\sqrt{1-\left(\frac{\Delta\sigma}{\bar{\sigma}}\right)^2}} 
\left( \sum_{n=0,2,4,\ldots}^{\infty} T^e (n, z_S) + \sum_{n=1,3,5,\ldots}^{\infty} T^o (n, z_S) 
 + \sum_{n=0,2,4,\ldots}^{\infty} \tilde{T}^e (n, z_S) + \sum_{n=1,3,5,\ldots}^{\infty} \tilde{T}^o (n, z_S) -(S\rightarrow D) \right)\right)^2+ \mathrm{c.c.} \right] \nonumber \\
 &&= \frac{I^2}{2\pi^3 \bar{\sigma}^3 \left( 1-\left(\frac{\Delta\sigma}{\bar{\sigma}}\right)^2 \right) }\int_{\Omega}d^2\bR' \left[ \left(\sigma^{(2)}_{-} e^{i\chi^{(2)}_{-}} + \Omega^2 \sigma^{(2)}_{+} e^{-i\chi^{(2)}_{+}} \right) \left( \sum_{n=0,2,4,\ldots}^{\infty} T^e (n, z, w') +
\sum_{n=1,3,5\ldots}^{\infty} T^o (n, z, w') \right) \right. \nonumber \\ && \left.
\left( \sum_{n=0,2,4,\ldots}^{\infty} T^e (n, z_S) + \sum_{n=1,3,5,\ldots}^{\infty} T^o (n, z_S) 
 + \sum_{n=0,2,4,\ldots}^{\infty} \tilde{T}^e (n, z_S) + \sum_{n=1,3,5,\ldots}^{\infty} \tilde{T}^o (n, z_S) -(S\rightarrow D) \right)^2+ \mathrm{c.c.} \right] = \nonumber \\
 &&= \frac{I^2}{2\pi^3 \bar{\sigma}^3 \left( 1-\left(\frac{\Delta\sigma}{\bar{\sigma}}\right)^2 \right) }\int_{\Omega}d^2\bR' \left[ \left(\sigma^{(2)}_{-} e^{i\chi^{(2)}_{-}} + \Omega^2 \sigma^{(2)}_{+} e^{-i\chi^{(2)}_{+}} \right) \left( \sum_{n=0,2,4,\ldots}^{\infty} T^e (n, z, w') +
\sum_{n=1,3,5\ldots}^{\infty} T^o (n, z, w') \right) \right. \nonumber \\ && \left.
\left( \sum_{n'=0,2,4,\ldots}^{\infty} T^e (n', z_S) + \sum_{n'=1,3,5,\ldots}^{\infty} T^o (n', z_S) 
 + \sum_{n'=0,2,4,\ldots}^{\infty} \tilde{T}^e (n', z_S) + \sum_{n'=1,3,5,\ldots}^{\infty} \tilde{T}^o (n', z_S) -(S\rightarrow D) \right)  \right. \nonumber \\ && \left.
\left( \sum_{m=0,2,4,\ldots}^{\infty} T^e (m, z_S) + \sum_{m=1,3,5,\ldots}^{\infty} T^o (m, z_S) 
 + \sum_{m=0,2,4,\ldots}^{\infty} \tilde{T}^e (m, z_S) + \sum_{m=1,3,5,\ldots}^{\infty} \tilde{T}^o (m, z_S) -(S\rightarrow D) \right)
 + \mathrm{c.c.} \right] \nonumber
\end{eqnarray}
This can be explicitly expanded to obtain the final expression for the first basis function:
\begin{eqnarray}
&&\Phi^{\newtext{(2)}}_{ani}(\bR)= \frac{I^2}{2\pi^3 \bar{\sigma}^3 \left( 1-\left(\frac{\Delta\sigma}{\bar{\sigma}}\right)^2 \right) }\int_{\Omega}d^2\bR' \left[ \left(\sigma^{(2)}_{-} e^{i\chi^{(2)}_{-}} + \Omega^2 \sigma^{(2)}_{+} e^{-i\chi^{(2)}_{+}} \right) \sum_{n,n',m}^{\infty} \left( T^e (n, z, w') + T^o (n, z, w') \right) \right.  \\ && \left.
\left( T^e (n', z_S) T^e (m, z_S) + T^e (n', z_S) T^o (m, z_S) + T^e (n', z_S) \tilde{T}^e (m, z_S) + T^e (n', z_S) \tilde{T}^o (m, z_S) + 
\right.  \right.  \\ &&+ \left. \left. 
T^o (n', z_S) T^e (m, z_S)  + T^o (n', z_S) T^o (m, z_S) + T^o (n', z_S) \tilde{T}^e (m, z_S) + T^o (n', z_S) \tilde{T}^o (m, z_S) +
\right.  \right.  \\ &&+ \left. \left. 
\tilde{T}^e (n', z_S) T^e (m, z_S) + \tilde{T}^e (n', z_S) T^o (m, z_S) + \tilde{T}^e (n', z_S) \tilde{T}^e (m, z_S) + \tilde{T}^e (n', z_S) \tilde{T}^o (m, z_S)
\right.  \right.  \\ &&+ \left. \left. 
\tilde{T}^o (n', z_S) T^e (m, z_S) + \tilde{T}^o (n', z_S) T^o (m, z_S) + \tilde{T}^o (n', z_S) \tilde{T}^e (m, z_S) + \tilde{T}^o (n', z_S) \tilde{T}^o (m, z_S) \right) + 
\right.   \\ &&- \left. 
(...)_{SD} - (...)_{DS} + (...)_{DD}
+ \mathrm{c.c.} \right],
\end{eqnarray}
where $()_{SD, DS, DD}$ indicate dependence on $z_S$ or $z_D$ and where integrals are expressed via base integrals.
Since the expression above contains all possible combinations of $TTT-$terms, the first basis function does depend on the size of the lead. The same is true for the other two basis functions which we present below.

\subsection{Second basis function $\left(E^{\newtext{(1)}}_x+iE^{\newtext{(1)}}_y\right)^2$}

We follow the same steps as in the previous subsection to get the expression for the potential for the second basis function:
\begin{eqnarray}
&&\Phi^{\newtext{(2)}}_{ani}(\bR)= 
\frac{\alpha_+}{\bar{\sigma}}\int_{\Omega}d^2\bR' \left[\left(\frac{\partial}{\partial
Z'}G_{N}(\bR,\bR')\right) 
\left(\sigma^{(2)}_{+} e^{i\chi^{(2)}_{+}} + \Omega^2 \sigma^{(2)}_{-} e^{-i\chi^{(2)}_{-}} \right) \left(E^{\newtext{(1)}}_x+iE^{\newtext{(1)}}_y\right)^2 
+ \mathrm{c.c.} \right] = \\
&&=\frac{\alpha_+  }{\bar{\sigma}}\int_{\Omega}d^2\bR' \left[\left(\sigma^{(2)}_{+} e^{i\chi^{(2)}_{+}} + \Omega^2 \sigma^{(2)}_{-} e^{-i\chi^{(2)}_{-}} \right) \left(\frac{\partial}{\partial
Z'}G_{N}(\bR,\bR')\right) 
\left(E^{\newtext{(1)}}_x+iE^{\newtext{(1)}}_y\right)^2 
+ \mathrm{c.c.} \right] = \\
 &&= \frac{I^2}{2\pi^3 \bar{\sigma}^3 \left( 1-\left(\frac{\Delta\sigma}{\bar{\sigma}}\right)^2 \right) }\int_{\Omega}d^2\bR' \left[ \left(\sigma^{(2)}_{+} e^{i\chi^{(2)}_{+}} + \Omega^2 \sigma^{(2)}_{-} e^{-i\chi^{(2)}_{-}} \right) \left( \sum_{n=0,2,4,\ldots}^{\infty} T^e (n, z, w') +
\sum_{n=1,3,5\ldots}^{\infty} T^o (n, z, w') \right) \right.  \\ && \left.
\left( \sum_{n'=0,2,4,\ldots}^{\infty} \frac{\tilde{T}^e (n', z_S)}{\Omega^2} + \sum_{n'=1,3,5,\ldots}^{\infty} \frac{\tilde{T}^o (n', z_S)}{\Omega^2} 
 + \sum_{n'=0,2,4,\ldots}^{\infty} \Omega^2 T^e (n', z_S) + \sum_{n'=1,3,5,\ldots}^{\infty} \Omega^2 T^o (n', z_S) -(S\rightarrow D) \right)  \right.  \\ && \left.
\left( \sum_{m=0,2,4,\ldots}^{\infty} \frac{\tilde{T}^e (m, z_S)}{\Omega^2} + \sum_{m=1,3,5,\ldots}^{\infty} \frac{\tilde{T}^o (m, z_S)}{\Omega^2} 
 + \sum_{m=0,2,4,\ldots}^{\infty} \Omega^2 T^e (m, z_S) + \sum_{m=1,3,5,\ldots}^{\infty} \Omega^2 T^o (m, z_S) -(S\rightarrow D) \right)
 + \mathrm{c.c.} \right] \nonumber
\end{eqnarray}
This can be explicitly expanded:
\begin{eqnarray}
&&\Phi^{\newtext{(2)}}_{ani}(\bR)= \frac{I^2}{2\pi^3 \bar{\sigma}^3 \left( 1-\left(\frac{\Delta\sigma}{\bar{\sigma}}\right)^2 \right) }\int_{\Omega}d^2\bR' \left[ \left(\sigma^{(2)}_{+} e^{i\chi^{(2)}_{+}} + \Omega^2 \sigma^{(2)}_{-} e^{-i\chi^{(2)}_{-}} \right) \sum_{n,n',m}^{\infty} \left( T^e (n, z, w') + T^o (n, z, w') \right) \right.  \\ && \left.
\left( \frac{\tilde{T}^e (n', z_S) \tilde{T}^e (m, z_S)}{\Omega^4} + \frac{\tilde{T}^e (n', z_S) \tilde{T}^o (m, z_S)}{\Omega^4} + \tilde{T}^e (n', z_S) T^e (m, z_S) + \tilde{T}^e (n', z_S) T^o (m, z_S) + 
\right.  \right.  \\ &&+ \left. \left. 
\frac{\tilde{T}^o (n', z_S) \tilde{T}^e (m, z_S)}{\Omega^4}  + \frac{\tilde{T}^o (n', z_S) \tilde{T}^o (m, z_S)}{\Omega^4} + \tilde{T}^o (n', z_S) T^e (m, z_S) + \tilde{T}^o (n', z_S) T^o (m, z_S) +
\right.  \right.  \\ &&+ \left. \left. 
T^e (n', z_S) \tilde{T}^e (m, z_S) + T^e (n', z_S) \tilde{T}^o (m, z_S) + \Omega^4 T^e (n', z_S) T^e (m, z_S) + \Omega^4 T^e (n', z_S) T^o (m, z_S)
\right.  \right.  \\ &&+ \left. \left. 
T^o (n', z_S) \tilde{T}^e (m, z_S) + T^o (n', z_S) \tilde{T}^o (m, z_S) + \Omega^4 T^o (n', z_S) T^e (m, z_S) + \Omega^4 T^o (n', z_S) T^o (m, z_S) \right) + 
\right.   \\ &&- \left. 
(...)_{SD} - (...)_{DS} + (...)_{DD}
+ \mathrm{c.c.} \right],
\end{eqnarray}
and all constituent integrals of three $T-$terms were previously expressed via base integrals.

\subsection{Third basis function 
$\left(E^{\newtext{(1)}}_x\right)^2+\left(E^{\newtext{(1)}}_y\right)^2$}

Finally, we proceed to the third basis function:
\begin{eqnarray}
&&\Phi^{\newtext{(2)}}_{ani}(\bR)= \frac{\alpha_+}{\bar{\sigma}}\int_{\Omega}d^2\bR' \left[\left(\frac{\partial}{\partial
Z'}G_{N}(\bR,\bR')\right)  \sigma^{(2)}_{0} \left( e^{i\chi^{(2)}_{0}} + \Omega^2  e^{-i\chi^{(2)}_{0}} \right) \left({E^{\newtext{(1)}}_x}^2+{E^{\newtext{(1)}}_y}^2\right)+ \mathrm{c.c.} \right] = \\
&&=\frac{\alpha_+ \sigma^{(2)}_{0}  }{\bar{\sigma}}\int_{\Omega}d^2\bR' \left[ \left( e^{i\chi^{(2)}_{0}} + \Omega^2  e^{-i\chi^{(2)}_{0}} \right) \left(\frac{\partial}{\partial
Z'}G_{N}(\bR,\bR')\right)  \left({E^{\newtext{(1)}}_x}^2+{E^{\newtext{(1)}}_y}^2\right)+ \mathrm{c.c.} \right] = \\
&&= \frac{\sigma^{(2)}_{0} I^2}{2\pi^3 \bar{\sigma}^3 \left( 1-\left(\frac{\Delta\sigma}{\bar{\sigma}}\right)^2 \right) }\int_{\Omega}d^2\bR' \left[ \left( e^{i\chi^{(2)}_{0}} + \Omega^2  e^{-i\chi^{(2)}_{0}} \right) \left( \sum_{n=0,2,4,\ldots}^{\infty} T^e (n, z, w') +
\sum_{n=1,3,5\ldots}^{\infty} T^o (n, z, w') \right) \right.  \\ && \left.
\left( \sum_{n'=0,2,4,\ldots}^{\infty} T^e (n', z_S) + \sum_{n'=1,3,5,\ldots}^{\infty} T^o (n', z_S) 
 + \sum_{n'=0,2,4,\ldots}^{\infty} \tilde{T}^e (n', z_S) + \sum_{n'=1,3,5,\ldots}^{\infty} \tilde{T}^o (n', z_S) -(S\rightarrow D) \right)  \right.  \\ && \left.
\left( \sum_{m=0,2,4,\ldots}^{\infty} \frac{\tilde{T}^e (m, z_S)}{\Omega^2} + \sum_{m=1,3,5,\ldots}^{\infty} \frac{\tilde{T}^o (m, z_S)}{\Omega^2} 
 + \sum_{m=0,2,4,\ldots}^{\infty} \Omega^2 T^e (m, z_S) + \sum_{m=1,3,5,\ldots}^{\infty} \Omega^2 T^o (m, z_S) -(S\rightarrow D) \right)
 + \mathrm{c.c.} \right], \nonumber
\end{eqnarray}
which can be explicitly expanded:
\begin{eqnarray}
&&\Phi^{\newtext{(2)}}_{ani}(\bR)= \frac{\sigma^{(2)}_{0} I^2}{2\pi^3 \bar{\sigma}^3 \left( 1-\left(\frac{\Delta\sigma}{\bar{\sigma}}\right)^2 \right) }\int_{\Omega}d^2\bR' \left[ \left( e^{i\chi^{(2)}_{0}} + \Omega^2  e^{-i\chi^{(2)}_{0}} \right) \sum_{n,n',m}^{\infty} \left( T^e (n, z, w') + T^o (n, z, w') \right) \right.  \\ && \left.
\left( \frac{T^e (n', z_S) \tilde{T}^e (m, z_S)}{\Omega^2} + \frac{T^e (n', z_S) \tilde{T}^o (m, z_S)}{\Omega^2} + \Omega^2 T^e (n', z_S) T^e (m, z_S) + \Omega^2 T^e (n', z_S) T^o (m, z_S) + 
\right.  \right.  \\ &&+ \left. \left. 
\frac{T^o (n', z_S) \tilde{T}^e (m, z_S)}{\Omega^2}  + \frac{T^o (n', z_S) \tilde{T}^o (m, z_S)}{\Omega^2} + \Omega^2 T^o (n', z_S) T^e (m, z_S) + \Omega^2 T^o (n', z_S) T^o (m, z_S) +
\right.  \right.  \\ &&+ \left. \left. 
\frac{\tilde{T}^e (n', z_S) \tilde{T}^e (m, z_S)}{\Omega^2} + \frac{\tilde{T}^e (n', z_S) \tilde{T}^o (m, z_S)}{\Omega^2} + \Omega^2 \tilde{T}^e (n', z_S) T^e (m, z_S) + \Omega^2 \tilde{T}^e (n', z_S) T^o (m, z_S)
\right.  \right.  \\ &&+ \left. \left. 
\frac{\tilde{T}^o (n', z_S) \tilde{T}^e (m, z_S)}{\Omega^2} + \frac{\tilde{T}^o (n', z_S) \tilde{T}^o (m, z_S)}{\Omega^2} + \Omega^2 \tilde{T}^o (n', z_S) T^e (m, z_S) + \Omega^2 \tilde{T}^o (n', z_S) T^o (m, z_S) \right) + 
\right.   \\ &&- \left. 
(...)_{SD} - (...)_{DS} + (...)_{DD}
+ \mathrm{c.c.} \right],
\end{eqnarray}
and all constituent integrals of three $T-$terms were expressed via base integrals previously.

\subsection{Rewriting the first two basis functions to have only $\sigma^{(2)}_{-}$- and $\sigma^{(2)}_{+}$-dependent terms}

The first two basis functions mix components of the second-order conductivity tensor $\sigma^{(2)}_{-}$ and $\sigma^{(2)}_{+}$. It might be more convenient to introduce functions that explicitly depend only on $\sigma^{(2)}_{-}$ and and $\sigma^{(2)}_{+}$, because in such a notation there is no mixing between terms corresponding to different components of the nonlinear conductivity tensor. The third basis function depends only on $\sigma^{(2)}_{0}$, therefore, it is already "self-generated".

We combine the components together to have expressions dependent only on $\sigma^{(2)}_{-}$ 
\begin{eqnarray}
&&\Phi^{\newtext{\left(2_{-}\right)}}_{ani}(\bR)= \frac{\sigma^{(2)}_{-} I^2}{2\pi^3 \bar{\sigma}^3 \left( 1-\left(\frac{\Delta\sigma}{\bar{\sigma}}\right)^2 \right) }\int_{\Omega}d^2\bR' \left[ \sum_{n,n',m}^{\infty} \left( T^e (n, z, w') + T^o (n, z, w') \right) \right. \nonumber \\  && \left.
\left( \left( e^{i\chi^{(2)}_{-}} + \Omega^2 e^{-i\chi^{(2)}_{-}} \right) \left( \tilde{T}^e (n', z_S) T^e (m, z_S) + \tilde{T}^e (n', z_S) T^o (m, z_S) + \tilde{T}^o (n', z_S) T^e (m, z_S) + \tilde{T}^o (n', z_S) T^o (m, z_S)   + \right.
\right.  \right. \nonumber \\ &&+ \left. \left.   \left.
T^e (n', z_S) \tilde{T}^e (m, z_S) + T^e (n', z_S) \tilde{T}^o (m, z_S)  + T^o (n', z_S) \tilde{T}^e (m, z_S) + T^o (n', z_S) \tilde{T}^o (m, z_S) \right) +
\right.  \right. \nonumber \\ &&+ \left. \left. 
\left( \left( e^{i\chi^{(2)}_{-}} + \frac{e^{-i\chi^{(2)}_{-}}}{\Omega^2} \right) \left( \tilde{T}^e (n', z_S) \tilde{T}^e (m, z_S) + \tilde{T}^e (n', z_S) \tilde{T}^o (m, z_S) + \tilde{T}^o (n', z_S) \tilde{T}^e (m, z_S) + \tilde{T}^o (n', z_S) \tilde{T}^o (m, z_S) \right) \right) 
\right.  \right. \nonumber \\ &&+ \left. \left. 
\left( e^{i\chi^{(2)}_{-}} + \Omega^6 e^{-i\chi^{(2)}_{-}} \right) \left( T^e (n', z_S) T^e (m, z_S) + T^e (n', z_S) T^o (m, z_S) + T^o (n', z_S) T^e (m, z_S) + T^o (n', z_S) T^o (m, z_S) \right) \right) + 
\right.  \nonumber \\ &&- \left. 
(...)_{SD} - (...)_{DS} + (...)_{DD}
+ \mathrm{c.c.} \right], \nonumber
\end{eqnarray}
and $\sigma^{(2)}_{+}$:
\begin{eqnarray}
&&\Phi^{\newtext{\left(2_{+}\right)}}_{ani}(\bR)= \frac{\sigma^{(2)}_{+} I^2}{2\pi^3 \bar{\sigma}^3 \left( 1-\left(\frac{\Delta\sigma}{\bar{\sigma}}\right)^2 \right) }\int_{\Omega}d^2\bR' \left[ \sum_{n,n',m}^{\infty} \left( T^e (n, z, w') + T^o (n, z, w') \right) \right. \nonumber  \\  && \left.
\left( \left( e^{i\chi^{(2)}_{+}} + \Omega^2 e^{-i\chi^{(2)}_{+}} \right) \left( \tilde{T}^e (n', z_S) T^e (m, z_S) + \tilde{T}^e (n', z_S) T^o (m, z_S) + \tilde{T}^o (n', z_S) T^e (m, z_S) + \tilde{T}^o (n', z_S) T^o (m, z_S)   + \right.
\right.  \right. \nonumber \\ &&+ \left. \left.   \left.
T^e (n', z_S) \tilde{T}^e (m, z_S) + T^e (n', z_S) \tilde{T}^o (m, z_S)  + T^o (n', z_S) \tilde{T}^e (m, z_S) + T^o (n', z_S) \tilde{T}^o (m, z_S) \right) +
\right.  \right. \nonumber \\ &&+ \left. \left. 
\left( \left( \frac{e^{i\chi^{(2)}_{+}}}{\Omega^4} + \Omega^2 e^{-i\chi^{(2)}_{+}} \right) \left( \tilde{T}^e (n', z_S) \tilde{T}^e (m, z_S) + \tilde{T}^e (n', z_S) \tilde{T}^o (m, z_S) + \tilde{T}^o (n', z_S) \tilde{T}^e (m, z_S) + \tilde{T}^o (n', z_S) \tilde{T}^o (m, z_S) \right) \right) 
\right.  \right. \nonumber \\ &&+ \left. \left. 
\left( \Omega^4 e^{i\chi^{(2)}_{+}} + \Omega^2 e^{-i\chi^{(2)}_{+}} \right) \left( T^e (n', z_S) T^e (m, z_S) + T^e (n', z_S) T^o (m, z_S) + T^o (n', z_S) T^e (m, z_S) + T^o (n', z_S) T^o (m, z_S) \right) \right) + 
\right.  \nonumber \\ &&- \left. 
(...)_{SD} - (...)_{DS} + (...)_{DD}
+ \mathrm{c.c.} \right], \nonumber
\end{eqnarray}
The expression for $\sigma^{(2)}_{0}$ is already "self-generated" in the original form; we provide it here for completeness:
\begin{eqnarray}
&&\Phi^{\newtext{(2_0)}}_{ani}(\bR)= \frac{\sigma^{(2)}_{0} I^2}{2\pi^3 \bar{\sigma}^3 \left( 1-\left(\frac{\Delta\sigma}{\bar{\sigma}}\right)^2 \right) }\int_{\Omega}d^2\bR' \left[ \left( e^{i\chi^{(2)}_{0}} + \Omega^2  e^{-i\chi^{(2)}_{0}} \right) \sum_{n,n',m}^{\infty} \left( T^e (n, z, w') + T^o (n, z, w') \right) \right. \nonumber \\ && \left.
\left( \frac{T^e (n', z_S) \tilde{T}^e (m, z_S)}{\Omega^2} + \frac{T^e (n', z_S) \tilde{T}^o (m, z_S)}{\Omega^2} + \Omega^2 T^e (n', z_S) T^e (m, z_S) + \Omega^2 T^e (n', z_S) T^o (m, z_S) + 
\right.  \right. \nonumber \\ &&+ \left. \left. 
\frac{T^o (n', z_S) \tilde{T}^e (m, z_S)}{\Omega^2}  + \frac{T^o (n', z_S) \tilde{T}^o (m, z_S)}{\Omega^2} + \Omega^2 T^o (n', z_S) T^e (m, z_S) + \Omega^2 T^o (n', z_S) T^o (m, z_S) +
\right.  \right. \nonumber \\ &&+ \left. \left. 
\frac{\tilde{T}^e (n', z_S) \tilde{T}^e (m, z_S)}{\Omega^2} + \frac{\tilde{T}^e (n', z_S) \tilde{T}^o (m, z_S)}{\Omega^2} + \Omega^2 \tilde{T}^e (n', z_S) T^e (m, z_S) + \Omega^2 \tilde{T}^e (n', z_S) T^o (m, z_S)
\right.  \right. \nonumber \\ &&+ \left. \left. 
\frac{\tilde{T}^o (n', z_S) \tilde{T}^e (m, z_S)}{\Omega^2} + \frac{\tilde{T}^o (n', z_S) \tilde{T}^o (m, z_S)}{\Omega^2} + \Omega^2 \tilde{T}^o (n', z_S) T^e (m, z_S) + \Omega^2 \tilde{T}^o (n', z_S) T^o (m, z_S) \right) + 
\right. \nonumber  \\ &&- \left. 
(...)_{SD} - (...)_{DS} + (...)_{DD}
+ \mathrm{c.c.} \right]. \nonumber
\end{eqnarray}

\subsection{A comment on the structure of the basis functions}

The structures of the basis functions presented in the previous subsection are exceptionally similar. It appears that the basis functions can be made identical for certain $\Omega$, however, this is true only for $\Omega=1$, so the functions are always distinct. To get identical basis functions for all three terms we need to require
\begin{eqnarray}
    \left( e^{i\chi^{(2)}_{-}} + \Omega^2 e^{-i\chi^{(2)}_{-}} \right) = \left( e^{i\chi^{(2)}_{+}} + \Omega^2 e^{-i\chi^{(2)}_{+}} \right), \\
    \left( e^{i\chi^{(2)}_{-}} + \frac{e^{-i\chi^{(2)}_{-}}}{\Omega^2} \right) = \left( \frac{e^{i\chi^{(2)}_{+}}}{\Omega^4} + \Omega^2 e^{-i\chi^{(2)}_{+}} \right), \\
    \left( e^{i\chi^{(2)}_{-}} + \Omega^6 e^{-i\chi^{(2)}_{-}} \right) = \left( \Omega^4 e^{i\chi^{(2)}_{+}} + \Omega^2 e^{-i\chi^{(2)}_{+}} \right).
\end{eqnarray}
For simplicity first consider $\chi^{(2)}_{-} = \chi^{(2)}_{+} = \chi^{(2)}_{0} = 0$. Then the first line is always trivially satisfied and we have
\begin{eqnarray}
    1 + \Omega^2 = 1 + \Omega^2 , \\
    1 + \frac{1}{\Omega^2} = \frac{1}{\Omega^4} + \Omega^2, \\
    1 + \Omega^6 = \Omega^4 + \Omega^2,
\end{eqnarray}
where the last two lines give exactly the same condition
\begin{equation}
    1 + \Omega^6 = \Omega^4 + \Omega^2,
\end{equation}
which is satisfied only for $\Omega=1$.

For the case of generic phases $\chi^{(2)}_{-}, \chi^{(2)}_{+}, \chi^{(2)}_{0}$ we have
\begin{eqnarray}
    \left( e^{i\chi^{(2)}_{-}} + \Omega^2 e^{-i\chi^{(2)}_{-}} \right) = \left( e^{i\chi^{(2)}_{+}} + \Omega^2 e^{-i\chi^{(2)}_{+}} \right), \\
    \left( \Omega^4 e^{i\chi^{(2)}_{-}} + \Omega^2 e^{-i\chi^{(2)}_{-}} \right) = \left( e^{i\chi^{(2)}_{+}} + \Omega^6 e^{-i\chi^{(2)}_{+}} \right), \\
    \left( e^{i\chi^{(2)}_{-}} + \Omega^6 e^{-i\chi^{(2)}_{-}} \right) = \left( \Omega^4 e^{i\chi^{(2)}_{+}} + \Omega^2 e^{-i\chi^{(2)}_{+}} \right).
\end{eqnarray}
If $\chi^{(2)}_{-} = \chi^{(2)}_{+}$, then the system can be satisfied only for $\Omega=1$.
For generic phases the first equation requires 
$$
\cos \chi^{(2)}_{-} = \cos \chi^{(2)}_{+}; \;\;\; \sin \chi^{(2)}_{-} = \sin \chi^{(2)}_{+},
$$
which means that the phases have to be identical. 

It turns out, though, that the shapes and magnitudes of basis functions look quite similar for $\Omega>0.5$ as is shown in Fig. \ref{FEM} by a direct calculation and comparison with finite element numerical solution. Therefore, for large anisotropy it might be not that easy to extract the tensor. However, we still can figure out what one in principle can extract from the data even if all three basis functions look almost the same (indistinguishable within the accuracy of experimental data). To do that consider the case with $\Omega \rightarrow 1$ and $\chi^{(2)}_{-} = \chi^{(2)}_{+} = \chi^{(2)}_{0} = 0$. Then all three basis functions are identical and their sum is governed by the prefactor $\sigma^{(2)}_{-} + \sigma^{(2)}_{+} + \sigma^{(2)}_{0}$, hence, we can only extract the value of this sum from the fitting procedure. We can relate the sum of these three coefficients to the components of the nonlinear conductivity tensor written in coordinate basis using the relations 
\begin{equation}
    \begin{gathered}
        \sigma^{(2)}_{-} = \frac{\mathfrak{c}_3^1 - \mathfrak{c}_1^2}{2}, \\
        \sigma^{(2)}_{+} = \frac{\mathfrak{c}_3^1 + \mathfrak{c}_1^2}{2}, \\
        \sigma^{(2)}_{0} = \mathfrak{c}_0^1.
    \end{gathered}
\end{equation}
with nonlinear tensor in the coordinate basis written as
$$
\tilde{\sigma}_{\alpha \mu \nu} = \mathfrak{c}_0^{\alpha} \tau_0 + \mathfrak{c}_1^{\alpha} \tau_1 + \mathfrak{c}_3^{\alpha} \tau_3.
$$
Then 
\begin{equation}
    \sigma^{(2)}_{-} + \sigma^{(2)}_{+} + \sigma^{(2)}_{0} = \frac{\mathfrak{c}_3^1 - \mathfrak{c}_1^2}{2} + \frac{\mathfrak{c}_3^1 + \mathfrak{c}_1^2}{2} + \mathfrak{c}_0^1 = \mathfrak{c}_3^1 + \mathfrak{c}_0^1,
\end{equation}
which means that \underline{we can only access $\tilde{\sigma}_{xxx}$ component of the nonlinear conductivity tensor} from experimentally measured data. Now let us generalize this result for arbitrary phases. In that case the equations read
\begin{eqnarray}
    \sigma^{(2)}_{-} e^{i \chi^{(2)}_{-}} + \sigma^{(2)}_{+} e^{i \chi^{(2)}_{+}} + \sigma^{(2)}_{0} e^{i \chi^{(2)}_{0}} = \mathfrak{c}_0^1 + \mathfrak{c}_3^1 + i \mathfrak{c}_0^2 + i \mathfrak{c}_3^2 \\
    \sigma^{(2)}_{-} e^{i \chi^{(2)}_{-}} + \sigma^{(2)}_{+} e^{i \chi^{(2)}_{+}} - \sigma^{(2)}_{0} e^{i \chi^{(2)}_{0}} = \mathfrak{c}_3^1 - \mathfrak{c}_0^1 + i \mathfrak{c}_3^2 - i \mathfrak{c}_0^2 \\ 
    \sigma^{(2)}_{-} e^{i \chi^{(2)}_{-}} - \sigma^{(2)}_{+} e^{i \chi^{(2)}_{+}} = -\mathfrak{c}_1^2 + i \mathfrak{c}_1^1,
\end{eqnarray}
while the prefactor in front of the sum of basis functions is
\begin{equation}
    \sigma^{(2)}_{-} \left( e^{i\chi^{(2)}_{-}} + e^{-i\chi^{(2)}_{-}} \right) + \sigma^{(2)}_{+} \left( e^{i\chi^{(2)}_{+}} + e^{-i\chi^{(2)}_{+}} \right) + \sigma^{(2)}_{0} \left( e^{i\chi^{(2)}_{0}} + e^{-i\chi^{(2)}_{0}} \right),
\end{equation}
which can be rewritten in coordinate notation as
\begin{eqnarray}
    &&\sigma^{(2)}_{-} \left( e^{i\chi^{(2)}_{-}} + e^{-i\chi^{(2)}_{-}} \right) + \sigma^{(2)}_{+} \left( e^{i\chi^{(2)}_{+}} + e^{-i\chi^{(2)}_{+}} \right) + \sigma^{(2)}_{0} \left( e^{i\chi^{(2)}_{0}} + e^{-i\chi^{(2)}_{0}} \right) = \left( \sigma^{(2)}_{-} e^{i \chi^{(2)}_{-}} + \sigma^{(2)}_{+} e^{i \chi^{(2)}_{+}} + \sigma^{(2)}_{0} e^{i \chi^{(2)}_{0}} \right) + \mathrm{c. c.} = \nonumber \\
    &&= \left( \mathfrak{c}_0^1 + \mathfrak{c}_3^1 + i \mathfrak{c}_0^2 + i \mathfrak{c}_3^2 \right) + \mathrm{c. c.} = 2 \left( \mathfrak{c}_0^1 + \mathfrak{c}_3^1 \right),
\end{eqnarray}
which is again the same component $\tilde{\sigma}_{xxx}$.

Our result shouldn't be surprising as
the limit of $\Omega=1$ is similar to the limit of a completely 1D system (an array of uncoupled wires): injected electrons don't scatter between the wires, therefore, if they have $x-$direction they cannot contribute to $y-$direction conductivity tensor because they are not deflected in the absence of scattering between the decoupled wires. Then this means that the basis functions become indistinguishable from each other and this prevents the tensor from getting extracted.

\section{Comparison of an analytical solution with the numerical solution via finite element method}

\begin{figure}[h!]
\includegraphics[width=0.95\linewidth]{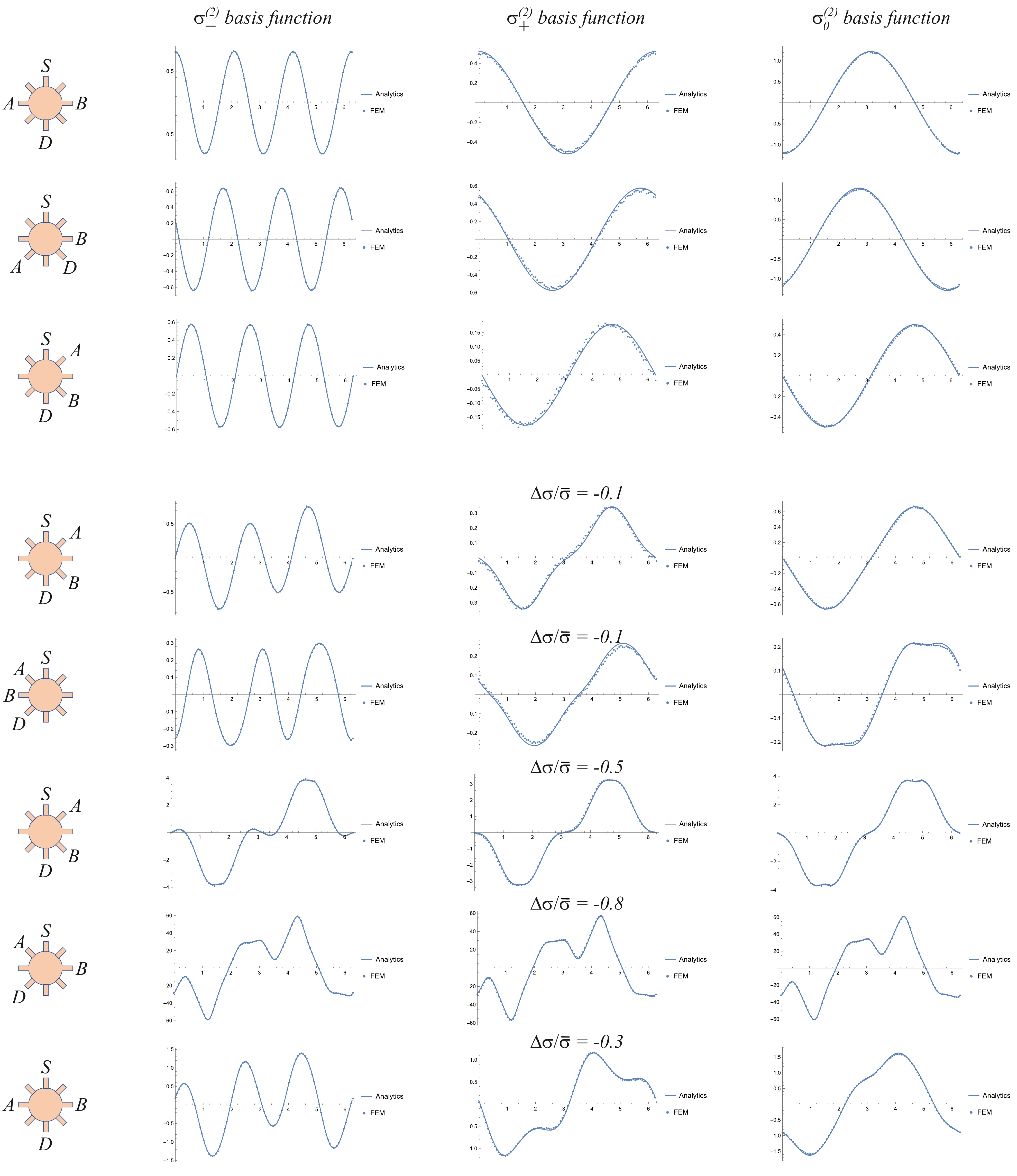}
\centering{}\caption{ \textbf{Comparison of analytical solution and numerical solution via Finite Element Method (FEM) for different measurement configurations, basis functions, and values of linear conductivity tensor anisotropy.} The horizontal axis is $\phi_R$ defined as in Fig. \ref{rot_sketch}. First three rows are plotted for zero linear anisotropy. For all rows except the last one $\chi^{(2)}_{-} = \chi^{(2)}_{+} = \chi^{(2)}_{0} =0$, for the last row $\chi^{(2)}_{-} = \chi^{(2)}_{+} = \chi^{(2)}_{0} =1$. For all calculations we set the overall prefactor that depends on $I$, $\sigma, \sigma^{(2)}$, and disk radius $a$ to be 1 and we used the angular width of the contact $\lambda = \frac{\pi}{30}$.  We see that the agreement between analytical and numerical results is perfect for every case. Note that for large anisotropy (0.8) the basis functions become hardly distinguishable. 
} 
\label{FEM}
\end{figure}

To check our analytical calculations we performed Finite Element Method (FEM) simulations of the nonlinear Ohm's law in the disk geometry using Neumann boundary conditions, fixed finite width of source and drain. To work in the quadratic in current regime we set the magnitude of the injected current to be small. The setup of the finite element analysis and the comparison with the analytical solution is shown in Fig. \ref{FEM}. We see that for all configurations the agreement between the two methods is nearly perfect. 

\section{Fitting protocol and error estimation}

Proper extraction of the nonlinear conductivity tensor in the disk geometry requires the knowledge of the linear conductivity tensor. Therefore, the first step is to analyze the linear transport data and extract the linear conductivity out of it. This procedure follows \cite{Oskar_solo,Zhang2022electronic}; we present it here for readers' convenience. 

As was shown in \cite{Oskar_solo}, in the case of a linear conductivity tensor being anisotropic with higher resistivity axis being along the $x$ coordinate axis and with nonzero Hall component $\sigma_H$, the linear potential is given by
\begin{eqnarray}
&&\Phi^{\newtext{(1)}}_{ani}(\bR)=\frac{I}{\pi} \frac{\sqrt{\sigma_{+} \sigma_{-}}}{\sigma_{+} \sigma_{-} + \sigma_H^2}
\left(\sum_{n=0,2,4,\ldots}^{\infty}
\ln
\frac{\left|1+\Omega^{4n+2} e^{-2 i \theta_D} -\Omega^{2n}\frac{Z}{\alpha_+} e^{- i \theta_D} \right|^2
}{\left|1+\Omega^{4n+2} e^{-2 i \theta_S} -\Omega^{2n}\frac{Z}{\alpha_+} e^{- i \theta_S} \right|^2
}+
\sum_{n=1,3,5,\ldots}^{\infty}
\ln
\frac{\left|1+\Omega^{4n+2} e^{2 i \theta_D} -\Omega^{2n}\frac{Z}{\alpha_+} e^{i \theta_D} \right|^2
}{\left|1+\Omega^{4n+2} e^{2 i \theta_S} -\Omega^{2n}\frac{Z}{\alpha_+} e^{i \theta_S} \right|^2
}\right)\nonumber\\ &&+ 
\frac{I}{\pi} \frac{\sigma_H}{\sigma_+ \sigma_- + \sigma_H^2} 
\left( \sum_{n=0,2,4,...}^{\infty} \left[  \arg \left( 1 + e^{-2 i \theta_D} \Omega^{2 + 4n} - e^{- i \theta_D} \frac{Z}{\alpha_+} \Omega^{2n} \right) - \arg \left( 1 + e^{-2 i \theta_S} \Omega^{2 + 4n} - e^{- i \theta_S} \frac{Z}{\alpha_+} \Omega^{2n} \right) \right] \right. \nonumber \\ &&+
\left. \sum_{n=1,3,5,...}^{\infty} \left[    \arg \left( 1 + e^{2 i \theta_D} \Omega^{2 + 4n} - e^{i \theta_D} \frac{Z}{\alpha_+} \Omega^{2n} \right) - \arg \left( 1 + e^{2 i \theta_S} \Omega^{2 + 4n} - e^{i \theta_S} \frac{Z}{\alpha_+} \Omega^{2n} \right) \right] \right),
\label{LinearFullEq}
\end{eqnarray}
where $\sigma_{+,-} = \mathfrak{k}_0 \pm \mathfrak{k}_3 = \bar{\sigma} \pm \Delta \sigma$, $\sigma_H = \mathfrak{k}_2$ -- is the linear Hall conductivity in the coordinate representation of the linear conductivity tensor, $Z = X+iY= x/\sqrt{1+\Delta \sigma/\bar{\sigma}} + i y /\sqrt{1-\Delta \sigma/\bar{\sigma}}$, where $x,y$ -- are coordinates of the point at which the linear potential is measured. Without loss of generality it is assumed that $\Delta \sigma/\bar{\sigma}<0$. Parameter $\alpha_+$ is defined as $\alpha_+ = \frac{a}{2} \left( 1/\sqrt{1+\Delta \sigma/\bar{\sigma}} + 1/\sqrt{1-\Delta \sigma/\bar{\sigma}} \right)$. Finally, we remind the reader that
$$
\Omega = \sqrt{\frac{\sqrt{1-\frac{\Delta \sigma}{\bar{\sigma}}}-\sqrt{1+\frac{\Delta \sigma}{\bar{\sigma}}}}{\sqrt{1-\frac{\Delta \sigma}{\bar{\sigma}}}+\sqrt{1+\frac{\Delta \sigma}{\bar{\sigma}}}}}.
$$

To fit the four-terminal linear transport measurement data, we modify Eq. \eqref{LinearFullEq} to take 4 inputs: $\phi_S,\phi_D,\phi_A,\phi_B$ and fit for 4 parameters $\alpha,\bar{\sigma}, \Delta\sigma,$ and $\sigma_H$. Each of $\phi_S,\phi_D,\phi_A,\phi_B$ is the angle of contacts used for that particular configuration, relative to the lab frame (note that the contacts are always on the circumference of the disk). Because the principle axis of the linear conductivity tensor does not necessarily align with the ``$x$ coordinate'' of the lab, an extra parameter $\alpha$, is introduced to compensate for the rotational mismatch between the principle axis and the $x$ coordinate through a coordinate transformation: $\phi_S,\phi_D,\phi_A,\phi_B$ into $\phi_S-\alpha,\phi_D-\alpha,\phi_A-\alpha,\phi_B-\alpha$. In other words, $\alpha$ defines the orientation of the principle axis in the lab frame. 
$\phi_S,\phi_D,\phi_A,\phi_B$, together with $\alpha$, ensure Eq. \eqref{LinearFullEq} can be applied to calculate the expected linear potential difference $V_{0}^{1\omega}$ of a given configuration and linear conductivity matrix. By varying $\phi_S,\phi_D,\phi_A,\phi_B$, we can create an array of expected values for different measurement configurations, denoted as $\vec{V}_0^{1w}$. Comparing this expected value against the measured voltage using the same set of configurations, we define the loss function:
\begin{eqnarray}
L=(\vec{V}_0^{1w}-\vec{V}_M^{1w}) \cdot (\vec{V}_0^{1w}-\vec{V}_M^{1w}),
\end{eqnarray}
where $\vec{V}_M^{1w}$ denotes the array of measured voltages. Minimizing this loss function is essentially a least square fit over the 4 fitting parameters $\alpha,\bar{\sigma}, \Delta\sigma,$ and $\sigma_H$, and hence the linear conductivity tensor is extracted (see Fig.~\ref{figSI_linear} for examples).\\ \ 

The above protocol describes the main logic of linear conductivity extraction for an ideal sample where contacts are all point-like. Below we discuss briefly the complexation from the finite width of the real sample contacts to the fitting and error estimation. Realistically, when performing the fitting, we adapt the generalized version of Eq. \eqref{LinearFullEq} (see  \cite{Oskar_solo}), which is essentially summing over a range of point-like source drain pairs ($\pm 9^\circ$ degree around the ideal source drain locations for our sample), with each pair carries a fraction of the total current. In this way, \eqref{LinearFullEq} is essentially updated to handle finite width source drain contacts. After the fit, we then handle the error from A-B contacts having finite width. We calculate the expected voltages of $V_{A+9^\circ,B+9^\circ},V_{A+9^\circ,B-9^\circ},V_{A-9^\circ,B+9^\circ},V_{A-9^\circ,B-9^\circ}$ and use the largest and smallest values among them to be the upper and lower bound of the error bar (gray shades in Fig.~\ref{figSI_linear} a,b,d,e,g,h). \\ \

\begin{figure*}
\includegraphics[width=0.5\linewidth]{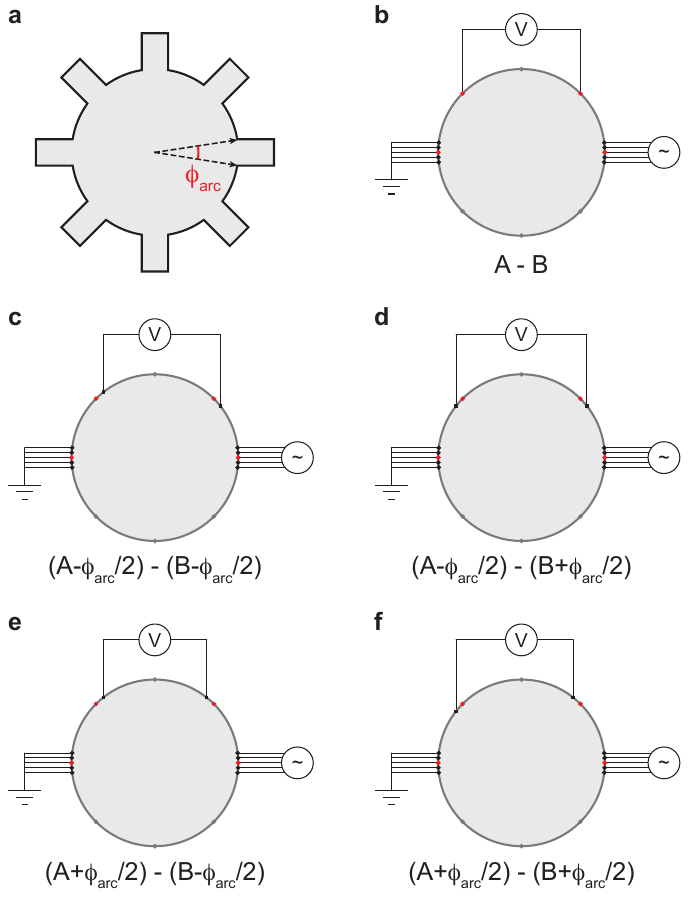}
\caption{ \newtext{{\bf{Estimation of error due to finite width of contacts} }
(a) A schematic of a disk-shaped sample showing finite contact width, described by a nonzero arc angle $\phi_\textrm{arc}$. (b) A schematic of handling current source and drain with finite width contact, using $V_\parallel$ at $\phi=0^\circ$ as an example. Black and red dots denote an approximation of finite width of current contacts with multiple sources and drains, spanning $\phi_\textrm{arc}$ around the ideal locations of the point-like contacts. (c-f) Schematics showing how error from finite voltage contact width is estimated. By placing point-like voltage contacts on the boundary of the finite width contacts, an upper and a lower bound can be estimated. Red dots denote the ideal locations of point-like contacts, black dots denote the extra effective point-like contacts due to finite width of the actual contacts. The procedure is the same as in \cite{Zhang2022electronic}. }
}
\label{fig_estimate}
\end{figure*}

\newtext{After the extraction, we estimate the error due to finite width of contacts. Suppose each of the sample contacts has an angular width (arc angle) of $\phi_\textrm{arc}$. Eqn. 54 in Ref.~\cite{Oskar_solo} generalizes the solution for linear potential in disk-shaped samples for the case of multiple current sources and drains. We take advantage of this expression to handle $\phi_\textrm{arc}$ of source and drain contacts by assigning multiple point-like sources and drains spanning $\phi_\textrm{arc}$ around locations of the ideal point-like source and drain on the rim (solution for linear potential has no divergences and to a good approximation is independent of a size of source/drain as long as they're not too large). We then estimate the uncertainty from finite size of contacts by comparing the four calculated voltages where we assume point-like voltage leads at ($A-\phi_\textrm{arc}/2$)-($B-\phi_\textrm{arc}/2$), ($A-\phi_\textrm{arc}/2$)-($B+\phi_\textrm{arc}/2$), ($A+\phi_\textrm{arc}/2$)-($B-\phi_\textrm{arc}/2$), and ($A+\phi_\textrm{arc}/2$)-($B+\phi_\textrm{arc}/2$), as shown in Fig.~\ref{fig_estimate}. We take the maximum of the four as the upper bound of the expectation and minimum of the four as the lower bound (gray shade in Fig.~\ref{full_linear} and Fig.~\ref{figSI_linear}). Such type of an analysis was previously used in \cite{Zhang2022electronic}. }

To fit the nonlinear transport data we devised the following fitting protocol. Every measurement configuration provides a particular value of the potential difference $V_{AB}$. The result of all measurements can be cast into the form of a vector $m$ of length $N$. There are two such vectors: one for linear signal, and one for nonlinear. One first analyzes the linear signal using the solution from \cite{Oskar_solo} in the way it was done in \cite{Zhang2022electronic}. The analysis of linear signal allows to determine what is the degree of linear tensor anisotropy (if any) and the direction of the anisotropy axis with respect to a chosen reference frame. 

The next step is to analyze the nonlinear data. In contrast to the case of linear signal the fit here is linear due to basis functions forming a linear combination.
The goal of fitting is to extract 6 independent components of nonlinear conductivity tensor from the data vector \newtext{$\vec{V}^{\textrm{NL-EXP}}$}. Mathematically, one can pose this problem as a system of linear equations:
\begin{equation}
    \newtext{\vec{V}^{\textrm{NL-EXP}}} = \hat{A} \cdot \vec{c},
\end{equation}
where \newtext{$\vec{V}^{\textrm{NL-EXP}}$} -- is the vector of experimental data containing $n$ data points, $\vec{c}$ - is the \newtext{unknown} vector of length 6 which depends on the components of nonlinear conductivity tensor \newtext{(to be determined via fitting)}, and $\hat{A}$ -- is the $n \times 6$ matrix which contains values of nonlinear potential for a given configuration. $i$-th column of matrix $\hat{A}$ contains 6 numbers; these numbers are the theoretical values of nonlinear electrostatic potential \newtext{difference $\Phi^{(2)} (A)-\Phi^{(2)} (B)$ between two measurement points A and B } for a given component of the nonlinear tensor in a given ($i$-th) measurement configuration \newtext{(i.e., for given source, drain, A, B locations and the angular size of source and drain)}. The vector $\vec{c}$ can be straightforwardly found:
\begin{equation}
    \vec{c} = \left( \hat{A}^{T} \hat{A} \right)^{-1} \hat{A}^{T} \cdot \newtext{\vec{V}^{\textrm{NL-EXP}}}.
    \label{fitting_eq}
\end{equation}
In the case of perfect data (no noise) this system is overcomplete as we use $n$ points to extract $6$ parameters. In the noisy case Eq. \eqref{fitting_eq} is an exact analog of the linear least squares fit. 

An explicit implementation of matrix $A$ depends on the results of a linear fit. For non-vanishing linear anisotropy one first needs to make sure that the experimental data is expressed in the reference frame aligned with principal axis ($x-$axis is along the higher resistance direction). After that one generates matrix $\hat{A}$ using the anisotropic basis functions with $\Delta \sigma / \bar{\sigma}$ extracted from the linear fit. In a simpler case of vanishing (small) linear anisotropy one doesn't need to adjust the reference frame of data vector. 

The variation of vector of fitted parameters $\vec{c}$ can be estimated using the standard formula for linear regression since Eq. \eqref{fitting_eq} is mathematically equivalent to linear least squares fit:
\begin{equation}
    \mathrm{var}(c_j) \simeq \frac{S}{n-6} \left[ \left( \hat{A}^{T} \hat{A} \right)^{-1} \right]_{jj},
\end{equation}
where $j=1..6$ labels fitted parameters, $n$ -- is the number of measurements being fitted, and $S$ -- is the residual sum of squares (RSS) given by
$$
S = \vec{e} \cdot \vec{e}
$$
with
$$
\vec{e} = \newtext{\vec{V}^{\textrm{NL-EXP}}} - \hat{A} \left( \hat{A}^{T} \hat{A} \right)^{-1} \hat{A}^{T} \cdot \newtext{\vec{V}^{\textrm{NL-EXP}}}.
$$
It is convenient to define a vector of parameters as 
\begin{equation}
    \vec{c} = \left( \sigma^{(2)}_{-} \cos \chi^{(2)}_{-}, \sigma^{(2)}_{-} \sin \chi^{(2)}_{-}, \sigma^{(2)}_{+} \cos \chi^{(2)}_{+}, \sigma^{(2)}_{+} \sin \chi^{(2)}_{+}, \sigma^{(2)}_{0} \cos \chi^{(2)}_{0}, \sigma^{(2)}_{0} \sin \chi^{(2)}_{0} \right).
\end{equation}
\newtext{Equivalently, we can write this as 
\begin{equation}
    \vec{c} = \left( \mathrm{Re} \Xi^{(2)}_{-}, \mathrm{Im} \Xi^{(2)}_{-}, \mathrm{Re} \Xi^{(2)}_{+}, \mathrm{Im} \Xi^{(2)}_{+}, \mathrm{Re} \Xi^{(2)}_{0}, \mathrm{Im} \Xi^{(2)}_{0} \right).
\end{equation} }
Then, the uncertainty of any function of these parameters can be estimated in the usual way:
\begin{equation}
    \delta F(\vec{c}) = \sqrt{ \sum_{j=1}^6 \left( \frac{\partial F}{\partial c_j} \delta c_j \right)^2},
\end{equation}
where $\delta c_j = \sqrt{\mathrm{var}(c_j)}$.

The protocol of nonlinear signal fitting doesn't incorporate the uncertainty of the extracted value of linear conductivity $\sigma$. In general, an identification of the linear conductivity uncertainty is not an easy task as it comes out as a result of a nonlinear regression fitting procedure. In our case, linear conductivity is isotropic to a good level of accuracy, therefore, to the leading order we need to estimate the uncertainty of the isotropic component of the linear conductivity tensor. To do so we consider parallel left and parallel right measurement configurations and calculate RSS for them. This then, for a given value of injected current and assuming perfectly isotropic linear conductivity in \eqref{LinearFullEq}, allows us to estimate the uncertainty of the linear conductivity magnitude extraction.

\newtext{One can take advantage of the fact that Eqs. \eqref{eq:phi21iso_different_limit}, \eqref{eq:phi22iso_different_limit}, \eqref{eq:phi23iso} give the same voltage drop between two contacts for certain measurement configurations. For example, two parallel configurations (see first two configurations shown in Fig. \ref{figSI_sample2}c,d) give the same expected voltage drop. Therefore, the data between these 2 configurations can be averaged and then fit as a data set for 1 configuration. We did this averaging for Sample 1 data from two parallel configurations; for Sample 2 we didn't perform such averaging  for any configurations. }

\section{Onsager relations for linear conductivity}
\label{sec:Linear_Onsager}

In this section we prove the Onsager relation for linear conductivity for a generic system with spin-orbit coupling and in the presence of a generic set of time-reversal symmetry breaking perturbations $( \Delta_i^e, \Delta_j^o )$, where $\Delta_i^e$ are even under time reversal and $\Delta_j^o$ are odd. These perturbations can be either external or internal, due to the mean fields which develop via a spontaneous symmetry breaking. For brevity we introduce 
$$B = ( \Delta_i^e, \Delta_j^o ), \; \; \; \tilde{B} = ( \Delta_i^e, -\Delta_j^o ).$$ 
The Hamiltonian of the system $H_{ ( \Delta_i^e, \Delta_j^o )}$ transforms under time reversal as 
\begin{eqnarray}
H_{ ( \Delta_i^e, \Delta_j^o )}&=&\mathcal{S}_y H^*_{ ( \Delta_i^e, -\Delta_j^o )}\mathcal{S}_y \;\;\;\Leftrightarrow\;\;
H^*_{ ( \Delta_i^e, \Delta_j^o )}=\mathcal{S}_y H_{ ( \Delta_i^e, -\Delta_j^o )}\mathcal{S}_y,
\end{eqnarray}
or in brief notation
\begin{eqnarray}
H_{ B}&=&\mathcal{S}_y H^*_{ \tilde{B} }\mathcal{S}_y \;\;\;\Leftrightarrow\;\;
H^*_{ B}=\mathcal{S}_y H_{ \tilde{B}}\mathcal{S}_y,
\label{TR_on_H}
\end{eqnarray}
where
$$
\mathcal{S}_y=\prod_{j'=1}^N s^y_{j'}\
$$
and $s^y_{j'}$ are Pauli matrices acting in spin space of $j'-$th particle. 

Within the standard first order perturbation theory an expectation value of a quantum-mechanical operator reads
\begin{eqnarray}
\langle \mathcal{O}_2\rangle (t)=-\frac{i}{\hbar}\int_{-\infty}^t dt' e^{st'}\text{Tr}\left(\hat{\rho}_B
\left[\mathcal{O}^{\left(H_B\right)}_2(t),\mathcal{O}^{\left(H_B\right)}_1(t')\right]\right),
\end{eqnarray}
where the factor $e^{st'}$ assures that the perturbation has been turned on slowly in the distant past, $\hat{\rho}_{ B}$ -- is the density matrix which, along with time-evolved operators, satisfies  
\begin{eqnarray}
&&\mathcal{O}^{\left(H_B\right)}_j(t)= e^{\frac{i}{\hbar}H_Bt} \mathcal{O}_j(t) e^{-\frac{i}{\hbar}H_Bt}\\
&&\hat{\rho}_B=\frac{e^{-\beta H_B}}{\text{Tr}\left(e^{-\beta H_B}\right)}.
\end{eqnarray}
Note, that the time dependence of $\mathcal{O}^{\left(H_{ B}\right)}_j(t)$ comes from both the quantum-mechanical evolution operator and from an explicit time dependence of the external perturbation $\mathcal{O}_j(t)$.
For the electrical current response, we perturb the system with an electric field ${\bf E}(t)$ and measure the electrical current, therefore
\begin{eqnarray}
\mathcal{O}_1(t)&=&eE_\nu(t)\sum_{j=1}^N r^\nu_{j}\\
\mathcal{O}_2&=&-eJ_\mu(\br),
\end{eqnarray}
where the electron charge is $-e$, $N$ -- is the number of charges in the system, $r_j^{\nu}$ -- are position operators for those charges, and $J_{\mu}(\br)$ -- is the number current density operator so that $\mathcal{O}_2$ is the charge current density operator.

Employing the identity
\begin{equation}
\begin{gathered}
\hat{\rho}_{ B}\int^\beta_0 d\lambda e^{\lambda H_{ B}}\left[\mathcal{O}^{\left(H_{ B}\right)}_1(t'),H_{ B}\right] e^{-\lambda H_{ B}}
=-\hat{\rho}_{ B} \int^\beta_0 d\lambda \frac{d}{d\lambda}
\left(e^{\lambda H_{ B}}\mathcal{O}^{\left(H_{ B} \right)}_1(t')e^{-\lambda H_{ B}}\right)
=\left[\hat{\rho}_{ B},\mathcal{O}^{\left(H_{ B}\right)}_1(t')\right] 
\end{gathered}
\end{equation}
and the cyclic property of the trace
\begin{eqnarray}
&&\text{Tr}\left(A[B,C]\right)=\text{Tr}\left(ABC-ACB\right)=\text{Tr}\left(CAB-ACB\right)=\text{Tr}\left([C,A]B\right)
\end{eqnarray}
we obtain
\begin{equation}
\begin{gathered}
\langle \mathcal{O}_2\rangle (t)=-\frac{i}{\hbar}\int_{-\infty}^t dt' e^{st'}\text{Tr}\left(\hat{\rho}_B
\left[\mathcal{O}^{\left(H_B\right)}_2(t),\mathcal{O}^{\left(H_B\right)}_1(t')\right]\right)\\
=
\frac{i}{\hbar}\int^\beta_0 d\lambda \int_{-\infty}^t dt' e^{st'}\text{Tr}\left(
\hat{\rho}_B e^{\lambda H_B}\left[\mathcal{O}^{\left(H_B\right)}_1(t'),H_B\right] e^{-\lambda H_B}\mathcal{O}^{\left(H_B\right)}_2(t)\right)\\
=
\frac{i}{\hbar}\int^\beta_0 d\lambda \int_{-\infty}^t dt' e^{st'}\text{Tr}\left(
e^{(\lambda-\beta) H_B}\left[\mathcal{O}^{\left(H_B\right)}_1(t'),H_B\right] e^{(\beta-\lambda) H_B}\hat{\rho}_B\mathcal{O}^{\left(H_B\right)}_2(t)\right)\\
=
\frac{i}{\hbar}\int^\beta_0 d\lambda \int_{-\infty}^t dt' e^{st'}\text{Tr}\left(\hat{\rho}_B e^{(\beta-\lambda) H_B}\mathcal{O}^{\left(H_B\right)}_2(t)
e^{(\lambda-\beta) H_B}\left[\mathcal{O}^{\left(H_B\right)}_1(t'),H_B\right]\right)\\
=
-\frac{i}{\hbar}\int_\beta^0 d\lambda' \int_{-\infty}^t dt' e^{st'}\text{Tr}\left(\hat{\rho}_Be^{\lambda' H_B}\mathcal{O}^{\left(H_B\right)}_2(t)
e^{-\lambda'H_B}\left[\mathcal{O}^{\left(H_B\right)}_1(t'),H_B\right]\right)\\
=
\frac{i}{\hbar}\int^\beta_0 d\lambda \int_{-\infty}^t dt' e^{st'}\text{Tr}\left(\hat{\rho}_B e^{\lambda H_B}\mathcal{O}^{\left(H_B\right)}_2(t)
e^{-\lambda H_B}\left[\mathcal{O}^{\left(H_B\right)}_1(t'),H_B\right]\right).
\end{gathered}
\end{equation}

The Heisenberg equations of motion allow us to express the commutator via the number density operator:
\begin{equation}
\begin{gathered}
\left[\mathcal{O}^{\left(H_B\right)}_1(t),H_B\right]= i\hbar eE_\nu(t)\sum_{j=1}^N \dot{r}^\nu_{j}(t) = \\ =
i\hbar eE_\nu(t)\int d^2\br \sum_{j=1}^N \dot{r}^\nu_{j}(t)\delta(\br-\br_j(t))=
i\hbar eE_\nu(t)\int d^2\br' J_\nu^{\left(H_B\right)}(\br',t).
\end{gathered}
\end{equation}

Using definitions of $\mathcal{O}_1, \mathcal{O}_2$ and the expression for the commutator after some algebra we obtain
\begin{eqnarray}
&&\langle -eJ_\mu(\br)\rangle (t)= 
e^2\int d^2\br'\int^\beta_0 d\lambda \int_{-\infty}^t dt' e^{st'}\text{Tr}\left(
\hat{\rho}_B e^{\lambda H_B} J_\nu^{\left(H_B \right)}(\br',t')
 e^{-\lambda H_B}
e^{\frac{i}{\hbar}H_B t} J_\mu(\br) e^{-\frac{i}{\hbar}H_B t}
\right)E_\nu(t')\\ 
&&=e^2\int d^2\br'\int^\beta_0 d\lambda \int_{-\infty}^t dt' e^{st'}\text{Tr}\left(
\hat{\rho}_B e^{\lambda H_B}
J_\nu(\br')
 e^{-\lambda H_B}
e^{\frac{i}{\hbar}H_B (t-t')} J_\mu(\br) e^{-\frac{i}{\hbar}H_B (t-t')}
\right)E_\nu(t') 
\label{eq:ordering1}.
\end{eqnarray}
Alternatively, we can rewrite the expression above as
\begin{eqnarray}
&&\langle -eJ_\mu(\br)\rangle (t)
= e^2\int d^2\br'\int^\beta_0 d\lambda \int_{-\infty}^t dt' e^{st'}\text{Tr}\left(
e^{(\lambda-\beta) H_B}
J_\nu(\br')
 e^{-(\lambda-\beta) H_B}\hat{\rho}_B
e^{\frac{i}{\hbar}H_B (t-t')} J_\mu(\br) e^{-\frac{i}{\hbar}H_B (t-t')}
\right)E_\nu(t')\nonumber\\
&&=
e^2\int d^2\br'\int^\beta_0 d\lambda' \int_{-\infty}^t dt' e^{st'}\text{Tr}\left(
e^{-\lambda' H_B}
J_\nu(\br')
 e^{\lambda' H_B}\hat{\rho}_B
e^{\frac{i}{\hbar}H_B (t-t')} J_\mu(\br) e^{-\frac{i}{\hbar}H_B (t-t')}
\right)E_\nu(t')\nonumber\\
&&=
e^2\int d^2\br'\int^\beta_0 d\lambda' \int_{-\infty}^t dt' e^{st'}\text{Tr}\left(
\hat{\rho}_B e^{\lambda' H_B} J_\mu(\br) e^{-\lambda' H_B}
e^{-\frac{i}{\hbar}H_B (t-t')}J_\nu(\br')
e^{\frac{i}{\hbar}H_B (t-t')}
\right)E_\nu(t').
\label{eq:ordering2}
\end{eqnarray}
In order to relate the expression (\ref{eq:ordering1}) to the expression (\ref{eq:ordering2}) we note that $\langle -eJ_\mu(\br)\rangle (t)$ is real since it is an expectation value of a Hermitian operator in a time evolved state (it has to be real order by order in perturbation theory). Therefore,  Eq. (\ref{eq:ordering1}) satisfies
\begin{eqnarray}
&&e^2\int d^2\br'\int^\beta_0 d\lambda \int_{-\infty}^t dt' e^{st'}\text{Tr}\left(
\hat{\rho}_B e^{\lambda H_B}
J_\nu(\br')
 e^{-\lambda H_B}
e^{\frac{i}{\hbar}H_B (t-t')} J_\mu(\br) e^{-\frac{i}{\hbar}H_B (t-t')}
\right)E_\nu(t')  \\
&=&e^2\int d^2\br'\int^\beta_0 d\lambda \int_{-\infty}^t dt' e^{st'}\text{Tr}\left(
\hat{\rho}^*_B e^{\lambda H^*_B}
J^*_\nu(\br')
 e^{-\lambda H^*_B}
e^{-\frac{i}{\hbar}H^*_B (t-t')} J^*_\mu(\br) e^{\frac{i}{\hbar}H^*_B (t-t')}
\right)E_\nu(t'). \label{starred_expr} 
\end{eqnarray}
Using properties of the Hamiltonian transformation under time reversal Eq. \eqref{TR_on_H} on Eq. \eqref{starred_expr} we then have
\begin{eqnarray}
&&e^2\int d^2\br'\int^\beta_0 d\lambda \int_{-\infty}^t dt' e^{st'}\text{Tr}\left(
\hat{\rho}_Be^{\lambda H_B}
J_\nu(\br')
 e^{-\lambda H_B}
e^{\frac{i}{\hbar}H_B(t-t')} J_\mu(\br) e^{-\frac{i}{\hbar}H_B(t-t')}
\right)E_\nu(t') \nonumber \\
&=&e^2\int d^2\br'\int^\beta_0 d\lambda \int_{-\infty}^t dt' e^{st'}\text{Tr}\left(
\hat{\rho}^*_B e^{\lambda H^*_B}
J^*_\nu(\br')
 e^{-\lambda H^*_B}
e^{-\frac{i}{\hbar}H^*_B (t-t')} J^*_\mu(\br) e^{\frac{i}{\hbar}H^*_B (t-t')}
\right)E_\nu(t') \nonumber \\
&=&e^2\int d^2\br'\int^\beta_0 d\lambda \int_{-\infty}^t dt' e^{st'}\text{Tr}\left(
\mathcal{S}_y\hat{\rho}_{ \tilde{B} }e^{\lambda H_{ \tilde{B} }}\mathcal{S}_y
J^*_\nu(\br')\mathcal{S}_y
 e^{-\lambda H_{ \tilde{B} }}
e^{-\frac{i}{\hbar}H_{ \tilde{B} }(t-t')} \mathcal{S}_y J^*_\mu(\br)\mathcal{S}_y e^{\frac{i}{\hbar}H_{ \tilde{B} }(t-t')}\mathcal{S}_y
\right)E_\nu(t')\nonumber\\
&=&e^2\int d^2\br'\int^\beta_0 d\lambda \int_{-\infty}^t dt' e^{st'}\text{Tr}\left(
\hat{\rho}_{ \tilde{B}}e^{\lambda H_{ \tilde{B}}}\mathcal{S}_y
J^*_\nu(\br')\mathcal{S}_y
 e^{-\lambda H_{ \tilde{B}}}
e^{-\frac{i}{\hbar}H_{ \tilde{B}}(t-t')} \mathcal{S}_y J^*_\mu(\br)\mathcal{S}_y e^{\frac{i}{\hbar}H_{ \tilde{B}}(t-t')}
\right)E_\nu(t')\nonumber\\
&=&e^2\int d^2\br'\int^\beta_0 d\lambda \int_{-\infty}^t dt' e^{st'}\text{Tr}\left(
\hat{\rho}_B e^{\lambda H_B} J_\mu(\br) e^{-\lambda H_B}
e^{-\frac{i}{\hbar}H_B(t-t')}J_\nu(\br')
e^{\frac{i}{\hbar}H_B(t-t')}
\right)E_\nu(t').
\end{eqnarray}
Integrating over all $\br$ to get the total current, using the fact that the electrical current operator is odd under the time reversal symmetry $\mathcal{S}_y J^*_\mu(\br)\mathcal{S}_y=-J_\mu(\br)$, and noting that the above equality holds for any $E_\nu(t')$, we find
\begin{eqnarray}
\sigma_{\mu\nu}(t-t',B)=\sigma_{\nu\mu}(t-t', \tilde{B}). \nonumber
\end{eqnarray}
Fourier transforming in time gives the Onsager relation 
\begin{eqnarray}
\sigma_{\mu\nu}(\omega,B)=\sigma_{\nu\mu}(\omega, \tilde{B})
\end{eqnarray}
or explicitly expressing via $( \Delta_i^e, \Delta_j^o )$:
\begin{eqnarray}
\sigma_{\mu\nu}(\omega, \Delta_i^e, \Delta_j^o)=\sigma_{\nu\mu}(\omega, \Delta_i^e, -\Delta_j^o).
\label{Onsager_lin_cond}
\end{eqnarray}

\section{Contributions to nonlinear conductivity tensor due to various mechanisms}

Nonlinear conductivity can be generated by various microscopic mechanisms. They can be split in two generic groups: intrinsic and extrinsic. Intrinsic contributions come from the systems itself either due to specific properties of the ground state, symmetries, topology and quantum geometry of the system. Extrinsic contributions are coming from effects of disorder. In this section we review various contributions to nonlinear conductivity tensor obtained in the relaxation time approximation. 

\subsection{Intrinsic contribution: second-order Drude contribution}

We employ Boltzmann kinetic equation in the relaxation time approximation to calculate components of the nonlinear conductivity tensor. Nonlinear conductivity in this approach is calculated via expanding the distribution function $n$ up to second order in electric field:
$$
f = f_0 + f_1 + f_2,
$$
where $f_0 = n_F$ -- is the Fermi function (equilibrium distribution function), $f_1 \sim E$, and $f_2 \sim E^2$. The validity of perturbative expression is controlled by the smallness of electric field $E$. In our case (the actual experimental setup) the field scales with injected current $I$ which is small, therefore, such an expansion is justified. We also assume that scattering time can be momentum dependent. 

Boltzmann kinetic equation in relaxation time approximation in the presence of electric field reads
\begin{equation}
    \dot{\mathbf{k}} \frac{\partial f}{\partial \mathbf{k}} = - \frac{f - f_0}{\tau \left(\mathbf{k} \right)},
\end{equation}
where $\hbar \dot{\mathbf{k}} = q \mathbf{E}$ with $q$ being the charge of carriers, and $\tau \left(\mathbf{k} \right)$ is the relaxation time. We now solve the equation perturbatively in powers of $E$. Adding linear in $E$ distribution function correction yields
\begin{equation}
\frac{q \mathbf{E}}{\hbar} \frac{\partial f_0}{\partial \mathbf{k}} = - \frac{f_0 + f_1 - f_0}{\tau \left(\mathbf{k} \right)} = - \frac{f_1}{\tau \left(\mathbf{k} \right)},
\end{equation}
hence 
\begin{equation}
    f_1 = - \frac{q \tau\left(\mathbf{k} \right) E_{\nu}}{\hbar} \frac{\partial f_0}{\partial k_{\nu}} = - \frac{q \tau\left(\mathbf{k} \right) E_{\nu}}{\hbar} \frac{\partial \epsilon}{\partial k_{\nu}} \frac{\partial f_0}{\partial \epsilon},  
\end{equation}
where $\epsilon = \epsilon (\mathbf{k})$ -- is the quasiparticle dispersion. 
Now we use the solution for $f_1$ to calculate the second-order correction to the distribution function by plugging it into the Boltzmann equation
\begin{equation}
    \frac{q \mathbf{E}}{\hbar} \frac{\partial f_1}{\partial \mathbf{k}} = - \frac{f_2 }{\tau \left(\mathbf{k} \right)},
\end{equation}
and obtaining
\begin{equation}
\begin{gathered}
    f_2 = -\frac{q \tau \left(\mathbf{k} \right) E_{\mu}}{\hbar} \frac{\partial f_1}{\partial k_{\mu}} = \frac{q^2 \tau \left(\mathbf{k} \right) E_{\mu} E_{\nu}}{\hbar^2} \frac{\partial }{\partial k_{\mu}} \left( \tau\left(\mathbf{k} \right)  \frac{\partial f_0}{\partial k_{\nu}} \right).
\end{gathered}
\end{equation}
Electric current is given by the relation between the carriers' velocity and the distribution function:
\begin{equation}
\begin{gathered}
    j_{\alpha} = q \int \frac{d^2 \mathbf{k}}{(2 \pi)^2} v_{\alpha} \left(f_1 + f_2 \right) = -\frac{q^2 E_{\nu} }{\hbar^2} \int \frac{d^2 \mathbf{k}}{(2 \pi)^2} \tau\left(\mathbf{k} \right) \left( \frac{\partial \epsilon}{\partial k_{\alpha}} \right)  \left( \frac{\partial \epsilon}{\partial k_{\nu}} \right) \left( \frac{\partial f_0}{\partial \epsilon} \right) + \\ + 
\frac{q^3 E_{\mu} E_{\nu}}{\hbar^3}  \int \frac{d^2 \mathbf{k}}{(2 \pi)^2}  \left( \frac{\partial \epsilon}{\partial k_{\alpha}} \right) \left[ \tau\left(\mathbf{k} \right) \left( \frac{\partial \tau\left(\mathbf{k} \right)}{\partial k_{\mu}} \right) \left( \frac{\partial f_0}{\partial k_{\nu}} \right) + \tau^2\left(\mathbf{k} \right) \left( \frac{\partial^2 f_0}{\partial k_{\mu} \partial k_{\nu}} \right) \right] .
\end{gathered}
\end{equation}
Then linear and nonlinear conductivity tensors are defined as $\sigma$ and $\tilde{\sigma}$ correspondingly:
\begin{equation}
    \begin{gathered}
        \sigma_{\alpha \nu} = -\frac{q^2 }{\hbar^2} \int \frac{d^2 \mathbf{k}}{(2 \pi)^2} \tau\left(\mathbf{k} \right) \left( \frac{\partial \epsilon}{\partial k_{\alpha}} \right)  \left( \frac{\partial \epsilon}{\partial k_{\nu}} \right) \left( \frac{\partial f_0}{\partial \epsilon} \right), \\
        \tilde{\sigma}_{\alpha \mu \nu} = -\frac{q^3}{\hbar^3} \int \frac{d^2 \mathbf{k}}{(2 \pi)^2}  \left[ \left( \frac{\partial^2 \epsilon}{\partial k_{\mu} \partial k_{\alpha}} \right) \tau^2\left(\mathbf{k} \right) + \left( \frac{\partial \epsilon}{\partial k_{\alpha}} \right) \tau\left(\mathbf{k} \right) \left( \frac{\partial \tau\left(\mathbf{k} \right)}{\partial k_{\mu}} \right) \right] \left( \frac{\partial \epsilon}{\partial k_{\nu}} \right) \left( \frac{\partial f_0}{\partial \epsilon} \right),
    \end{gathered}
    \label{Tensor_components_Boltzmann}
\end{equation}
where for $\tilde{\sigma}$ we applied integration by parts to get rid of the term that contains second derivative of $f_0$.
The expressions above correspond only to Drude contributions to linear and nonlinear conductivity tensors.

\subsection{Intrinsic contribution: Berry curvature dipole}

In 2D systems with Berry curvature and broken rotational symmetry there is another contribution to nonlinear conductivity. Such contribution comes from the Berry curvature dipole (BCD) and doesn't require time reversal (TR) symmetry breaking to be present. In DC limit, BCD-induced nonlinear conductivity tensor contains just the nonlinear Hall contribution given by \cite{Sodemann_Fu} 
\begin{equation}
    \mathcal{C}_a = \mathcal{D}_a = \frac{e^3 \tau}{2} \int \frac{d^2 \mathbf{k}}{(2 \pi)^2} f_0 \frac{\partial \Omega (\mathbf{k})}{\partial k_a},
\end{equation}
where $\Omega (\mathbf{k})$ -- is the Berry curvature and $f_0$ -- is the equilibrium distribution function, while $\mathcal{A}, \mathcal{B}$-vector contributions are 0. The largest symmetry for the system to exhibit BCD is a single mirror axis. Importantly, BCD is present for both TR-preserving and TR-breaking setups.

\subsection{Intrinsic contribution: quantum geometry}

Another intrinsic contribution to nonlinear conductivity tensor stems from nontrivial quantum geometry of the system which is imprinted in quantum geometric tensor. Such contribution to nonlinear conductivity requires time reversal (TR) symmetry to be broken and can contain both dissipative and non-dissipative contributions. There are several slightly different expressions presented in literature and we show them all here for completeness. 

In Ref. \cite{Gao2014PRLNHEmetric} the nonlinear Hall contribution stemming from quantum geometry is given by
\begin{equation}
    \tilde{\sigma}_{\alpha \mu \nu}^{NH} = \frac{e^3}{\hbar} \int \frac{d^2 \mathbf{k}}{(2 \pi)^2} \left( v_{\alpha} G_{\mu \nu} - v_{\mu} G_{\alpha \nu} \right) \frac{\partial f_0}{\partial \epsilon},
\end{equation}
where 
\begin{eqnarray}
G_{ij} = 2 \mathrm{Re} \sum_{n \neq 0} \frac{\left( V_i \right)_{0n} \left( V_j \right)_{0n}}{\left( \epsilon_0 - \epsilon_n \right)^3},
\label{G_tensor_Q_Niu}\\
\left( V_l \right)_{nm} = \left\langle n \right| v_l \left| m \right \rangle,
\end{eqnarray}
and $v_l = \frac{\partial \epsilon_0}{\partial k_l}$.

In Ref. \cite{AgarwalQG2023PRB} nonlinear conductivity tensor induced by quantum geometry is split into two parts. The non-dissipative part (dubbed "Purely Hall") is given by (using electron charge being $-e$)
\begin{equation}
    \tilde{\sigma}_{\alpha \mu \nu}^{PH}  = - \frac{e^3}{\hbar} \sum_{m,p} \int \frac{d^2 \mathbf{k}}{(2 \pi)^2} f_{0}^{(m)}  \left[ 2 \frac{\partial G_{\mu \nu}^{(mp)}}{\partial k_{\alpha}} - \left( \frac{\partial G_{\alpha \nu}^{(mp)}}{\partial k_{\mu}} + \frac{\partial G_{\alpha \mu}^{(mp)}}{\partial k_{\nu}} \right) \right]
\end{equation}
with $f_{0}^{(m)}$ being an equiligrium distribution function for band $m$, while dissipative part (dubbed "Purely Dissipative") reads
\begin{equation}
    \tilde{\sigma}_{\alpha \mu \nu}^{PD}  = \frac{e^3}{\hbar} \sum_{m,p} \int \frac{d^2 \mathbf{k}}{(2 \pi)^2} f_{0}^{(m)}  \left[ \frac{\partial G_{\mu \nu}^{(mp)}}{\partial k_{\alpha}} + \frac{\partial G_{\alpha \nu}^{(mp)}}{\partial k_{\mu}} + \frac{\partial G_{\alpha \mu}^{(mp)}}{\partial k_{\nu}} \right],
\end{equation}
so that the full quantum geometry-induced nonlinear conductivity tensor reads 
\begin{equation}
    \tilde{\sigma}_{\alpha \mu \nu}^{QG}  = - \frac{e^3}{\hbar} \sum_{m,p} \int \frac{d^2 \mathbf{k}}{(2 \pi)^2} f_{0}^{(m)}  \left[ \frac{\partial G_{\mu \nu}^{(mp)}}{\partial k_{\alpha}} - 2 \left( \frac{\partial G_{\alpha \nu}^{(mp)}}{\partial k_{\mu}} + \frac{\partial G_{\alpha \mu}^{(mp)}}{\partial k_{\nu}} \right) \right].
\end{equation}
Here
\begin{equation}
    G_{\alpha \mu}^{(mp)} = - \frac{\langle u_p (\mathbf{k}) | \partial_{\mathbf{k}_{\alpha}} u_m (\mathbf{k})  \rangle \langle u_m (\mathbf{k}) | \partial_{\mathbf{k}_{\mu}} u_p (\mathbf{k})  \rangle + \langle u_p (\mathbf{k}) | \partial_{\mathbf{k}_{\mu}} u_m (\mathbf{k})  \rangle \langle u_m (\mathbf{k}) | \partial_{\mathbf{k}_{\alpha}} u_p (\mathbf{k})  \rangle}{2 \left( \epsilon_m - \epsilon_p \right)},
\end{equation}
with $m,p$ being band indices. Note, that this expression is $\frac{1}{2}$ of the $G$ given in Eq. \eqref{G_tensor_Q_Niu} and that for a single band at the Fermi level the double sum is reduced to a single sum.

Finally, in Ref. \cite{KaplanPRL2024} the full quantum geometry-induced nonlinear conductivity tensor is given by 
\begin{equation}
    \tilde{\sigma}_{\alpha \mu \nu} = - \frac{e^3}{\hbar} \sum_{n} \int \frac{d^2 \mathbf{k}}{(2 \pi)^2} f_{0}^{(n)}  \left[ 2 \frac{\partial G_{\mu \nu}^{(n)}}{\partial k_{\alpha}} - \frac{1}{2} \left( \frac{\partial G_{\nu \alpha}^{(n)}}{\partial k_{\mu}} + \frac{\partial G_{\mu \alpha}^{(n)}}{\partial k_{\nu}} \right) \right],
\end{equation}
and $G$ follows the definition without $\frac{1}{2}$ factor
\begin{equation}
    G_{\alpha \mu}^{(n)} = - \sum_p \frac{\langle u_p (\mathbf{k}) | \partial_{\mathbf{k}_{\alpha}} u_n (\mathbf{k})  \rangle \langle u_n (\mathbf{k}) | \partial_{\mathbf{k}_{\mu}} u_p (\mathbf{k})  \rangle + \langle u_p (\mathbf{k}) | \partial_{\mathbf{k}_{\mu}} u_n (\mathbf{k})  \rangle \langle u_n (\mathbf{k}) | \partial_{\mathbf{k}_{\alpha}} u_p (\mathbf{k})  \rangle}{ \epsilon_n - \epsilon_p },
\end{equation}

\subsection{Extrinsic contribution: side jumps}

External contributions come from scattering of quasiparticles off impurities. For weak impurity scattering the symmetric part of the transition probability reads
\begin{equation}
    w^S_{\mathbf{p} \mathbf{p}'} = \frac{2 \pi}{\hbar} n_{imp} V_0^2 \left| \langle u (\mathbf{p}) |  u (\mathbf{p}')  \rangle \right|^2 \delta \left( \epsilon \left( \mathbf{p} \right) - \epsilon \left( \mathbf{p}' \right) \right),
\end{equation}
where $n_{imp}$ -- is the concentration of impurities and $V_0$ -- is the zero-momentum Fourier component of the disorder potential. It defines the scattering (relaxation) time 
\begin{equation}
    \frac{1}{\tau} = \int \frac{d^2 \mathbf{p}'}{(2 \pi)^2} w^S_{\mathbf{p} \mathbf{p}'} \left( 1 - \cos \theta_{\mathbf{p} \mathbf{p}'} \right). 
\end{equation}
Assuming momentum independence of $w^S_{\mathbf{p} \mathbf{p}'}$ one can obtain momentum-independent relaxation time.  
The coordinate shift associated with a side-jump --  a transverse displacement of the center of the scattering quasiparticle -- is given by 
\begin{equation}
    \delta \mathbf{r} \left( \mathbf{p}, \mathbf{p}' \right) = i \langle u (\mathbf{p}) | \partial_{\mathbf{p}} u (\mathbf{p})  \rangle - i \langle u (\mathbf{p}') | \partial_{\mathbf{p}'} u (\mathbf{p}')  \rangle - \left( \partial_{\mathbf{p}} + \partial_{\mathbf{p}'} \right) \mathrm{arg}  \langle u (\mathbf{p}) | u (\mathbf{p}')  \rangle.
\end{equation}
Such a coordinate shift results in an energy shift that modifies the scattering rate of quasiparticles. For a generic $2 \times 2$ Hamiltonian of the form $H (\mathbf{p}) = d_0 (\mathbf{p}) + \mathbf{d} (\mathbf{p}) \cdot \pmb{\sigma}$ the coordinate shift can be expressed as \cite{KoenigDzeroLevchenkoPesin}
\begin{equation}
    \left( \delta \mathbf{r} \left( \mathbf{p}, \mathbf{p}' \right) \right)_a = \pm \frac{\left( \frac{\partial \hat{d} (p)}{\partial p_a} + \frac{\partial \hat{d}(p')}{\partial p'_a}  \right) \cdot \left( \hat{d}(p) \times \hat{d}(p') \right)}{2 \left( 1 + \hat{d}(p) \cdot \hat{d}(p') \right)},
\end{equation}
where $\pm$ specifies conduction or valence band and $\hat{d} (p) = \mathbf{d} (\mathbf{p}) / \left| \mathbf{d} (\mathbf{p}) \right|$.
To describe the side jump contribution it is convenient to introduce the side-jump accumulation velocity \cite{KoenigDzeroLevchenkoPesin} 
\begin{equation}
    \mathbf{v}^{sj} \left( \mathbf{p} \right) = \int \frac{d^2 \mathbf{p}}{(2\pi)^2} w^S \left( \mathbf{p}, \mathbf{p}' \right) \delta \mathbf{r} \left( \mathbf{p}', \mathbf{p} \right) \delta \left( \epsilon \left( \mathbf{p} \right) - \epsilon \left( \mathbf{p}' \right) \right).
\end{equation}
The side jump contribution to nonlinear conductivity tensor for momentum-independent relaxation time then is given by 
\begin{equation}
    \tilde{\sigma}_{\alpha \mu \nu} = \frac{e^3 \tau^2}{\hbar} \int \frac{d^2 \mathbf{k}}{(2\pi)^2} \left[ - 2 v_{\nu} \frac{\partial v^{sj}_{\alpha} }{\partial k_{\mu}} + v_{\nu} \frac{\partial v^{sj}_{\mu} }{\partial k_{\alpha}} + v^{sj}_{\mu} \frac{\partial v_{\alpha} }{\partial k_{\nu}} + v^{sj}_{\nu} \frac{\partial v_{\alpha} }{\partial k_{\mu}} + v_{\mu} \frac{\partial v^{sj}_{\nu} }{\partial k_{\alpha}} \right] \frac{\partial f_0}{\partial \epsilon},
\end{equation}
where the first term corresponds to the accumulation part and the last four -- to the anomalous distribution.

\subsection{Extrinsic contribution: skew scattering}

External contributions come from scattering of quasiparticles off impurities. 
Skew scattering accounts for the antisymmetric part of the scattering rate, which appears due  to the internal chirality of charge carriers associated with, e.g., opposite signs of Berry curvature in systems with broken inversion symmetry. 
Antisymmetric transition probability appears in the third order expansion in impurity potential strength
\begin{equation}
    w^A_{\mathbf{p} \mathbf{p}'} = - \frac{(2 \pi)^2}{\hbar} n_{imp} V_0^3 \delta \left( \epsilon \left( \mathbf{p} \right) - \epsilon \left( \mathbf{p}' \right) \right) \int \frac{d^2 \mathbf{p}''}{(2 \pi)^2} \delta \left( \epsilon \left( \mathbf{p} \right) - \epsilon \left( \mathbf{p}'' \right) \right) \mathrm{Im} \left[ \langle u (\mathbf{p}) |  u (\mathbf{p}')  \rangle \langle u (\mathbf{p}') |  u (\mathbf{p}'') \langle u (\mathbf{p}'') |  u (\mathbf{p})  \rangle \rangle \right].
\end{equation}
For convenience one can introduce the "skew acceleration"
\begin{equation}
    \mathbf{A}^{sk}_{\mathbf{p}} = \int \frac{d^2 \mathbf{p}'}{(2 \pi)^2} w^A_{\mathbf{p} \mathbf{p}'} \left( \frac{\partial \epsilon (\mathbf{p})}{\partial \mathbf{p} } - \frac{\partial \epsilon (\mathbf{p}')}{\partial \mathbf{p}' } \right).
\end{equation}
Then the  skew scattering contribution to nonlinear conductivity tensor reads \cite{KoenigDzeroLevchenkoPesin} 
\begin{equation}
    \tilde{\sigma}_{\alpha \mu \nu} = \frac{e^3 \tau^3}{2 \hbar^2}  \int \frac{d^2 \mathbf{p}}{(2 \pi)^2} \left[ 2 \left( \frac{\partial A^{sk}_{\mu}}{ \partial k_{\alpha}} - \frac{\partial A^{sk}_{\alpha}}{ \partial k_{\mu}} \right) v_{\nu} - \left( A^{sk}_{\alpha} \frac{\partial v_{\mu}}{ \partial k_{\nu}} - A^{sk}_{\mu} \frac{\partial v_{\alpha}}{ \partial k_{\nu}} \right) \right] \frac{\partial f_0}{\partial \epsilon}.
\end{equation}

\section{Symmetries of the non-linear conductivity: nonlinear Boltzmann (Drude) contribution and the hierarchy between the $\mathcal{A}, \mathcal{B}, \mathcal{C}$}
\label{sec:Drude_symmetries}

Second-order Drude conductivity tensor to nonlinear conductivity tensor in the (momentum-independent) relaxation time approximation has some interesting symmetry properties. This allows us to establish a hierarchy of different terms constituting nonlinear conductivity tensor $\tilde{\sigma}$. In this section we show how these symmetry properties can be discovered.

For our purposes it is convenient to employ the version Eq. 12a of \cite{KaplanPRL2024}, which expresses the nonlinear Boltzmann (Drude) contribution to conductivity as
\begin{eqnarray}
\tilde{\sigma}_{abc}=-\frac{e^3\tau^2}{\hbar^3}\sum_{n}\int \frac{d^2 \mathbf{k}}{(2 \pi)^2} f(\eps_n(\bk,\Delta))\frac{\partial}{\partial k_a}\frac{\partial}{\partial k_b}\frac{\partial}{\partial k_c}\eps_n(\bk,\Delta).
\label{Binghai_Drude}
\end{eqnarray}
This expression is equivalent to the expression we derived earlier upon the integration by parts is applied to the integral. Let us now rewrite the expression for nonlinear current with conductivity given by Eq. \eqref{Binghai_Drude}. Let $k=k_x+ik_y$ and $\bar{k}=k_x-ik_y$. Introducing complex derivatives 
\begin{equation}
    \frac{\partial}{\partial k_x}+i\frac{\partial}{\partial k_y}=2\frac{\partial }{\partial \bar{k}}\; \; \; \text{and} \; \; \; \frac{\partial}{\partial k_x}-i\frac{\partial}{\partial k_y}=2\frac{\partial }{\partial k} \nonumber
\end{equation}
and after some algebraic manipulations we obtain an expression for nonlinear current in the complex representation

\begin{eqnarray}
j_x+ij_y&=&\left(-2\frac{e^3\tau^2}{\hbar^3}\sum_{n}\int \frac{d^2 \mathbf{k}}{(2 \pi)^2} f(\eps_n(\bk,\Delta))\frac{\partial}{\partial \bar{k}}\frac{\partial}{\partial \bar{k}}\frac{\partial}{\partial \bar{k}}\eps_n(\bk,\Delta)\right)(E_x-iE_y)^2\\
&+&\left(-2\frac{e^3\tau^2}{\hbar^3}\sum_{n}\int \frac{d^2 \mathbf{k}}{(2 \pi)^2} f(\eps_n(\bk,\Delta))\frac{\partial}{\partial \bar{k}}\frac{\partial}{\partial k}\frac{\partial}{\partial k}\eps_n(\bk,\Delta)\right)(E_x+iE_y)^2\\
&+&2\left(-2\frac{e^3\tau^2}{\hbar^3}\sum_{n}\int \frac{d^2 \mathbf{k}}{(2 \pi)^2} f(\eps_n(\bk,\Delta))\frac{\partial}{\partial \bar{k}}\frac{\partial}{\partial \bar{k}}\frac{\partial}{\partial k}\eps_n(\bk,\Delta)\right)\left(E^2_x+E^2_y\right),
\end{eqnarray}
where we collected together terms with different rotational symmetry properties. Explicit expressions for $\Xi^{(2)}_{-,+,0}$ then read
\begin{equation}
\begin{aligned}
&\Xi^{(2)}_{-}= -2\frac{e^3\tau^2}{\hbar^3}\sum_{n}\int \frac{d^2 \mathbf{k}}{(2 \pi)^2} f(\eps_n(\bk,\Delta))\frac{\partial}{\partial \bar{k}}\frac{\partial}{\partial \bar{k}}\frac{\partial}{\partial \bar{k}}\eps_n(\bk,\Delta)  \\
&\Xi^{(2)}_{+}= -2\frac{e^3\tau^2}{\hbar^3}\sum_{n}\int \frac{d^2 \mathbf{k}}{(2 \pi)^2} f(\eps_n(\bk,\Delta))\frac{\partial}{\partial \bar{k}}\frac{\partial}{\partial k}\frac{\partial}{\partial k}\eps_n(\bk,\Delta)  \\
&\Xi^{(2)}_{0}=  -4\frac{e^3\tau^2}{\hbar^3}\sum_{n}\int \frac{d^2 \mathbf{k}}{(2 \pi)^2} f(\eps_n(\bk,\Delta))\frac{\partial}{\partial \bar{k}}\frac{\partial}{\partial \bar{k}}\frac{\partial}{\partial k}\eps_n(\bk,\Delta).
\label{Drude_complex}
\end{aligned}
\end{equation}
Note, that the integrands for 1-fold terms $\Xi^{(2)}_{+,0}$ are complex conjugated of each other since electron dispersion $\eps_n(\bk,\Delta)$ is a real function. This indicates that the second order Drude contributions to $\Xi^{(2)}_{+,0}$ will experience a mirror axis as $\mathrm{arg} \Xi^{(2)}_{+} = - \mathrm{arg} \Xi^{(2)}_{0}$. The real physical mirror axis, however, should be the same for all three $\Xi^{(2)}$ coefficients, which is not guaranteed by Eq. \eqref{Drude_complex}. In addition to that, we notice that $\left | \Xi^{(2)}_{+} \right| = \frac{1}{2} \left | \Xi^{(2)}_{0} \right|$. Without loss of generality we can then rotate our reference frame to the mirror axis for which $\mathrm{arg} \Xi^{(2)}_{+} = \mathrm{arg} \Xi^{(2)}_{0} = 0$, i.e., $\Xi^{(2)}_{+,0} = \sigma^{(2)}_{+,0}$ is completely real. In such case we immediately obtain that 
\begin{equation}
\begin{gathered}
\mathcal{B} \cdot \mathcal{C} = 0, \\
\left| \mathcal{B} \right| = 3 \left| \mathcal{C} \right|,
\end{gathered}
\end{equation}
which closely resembles the structure of nonlinear conductivity tensor extracted from the experimental data provided the 3-fold contribution is at least an order of magnitude smaller. Note however, that one needs to ensure that boundary terms are properly incorporated while integrating by parts as these terms may provide a distinct contribution and the relations between $\left| \mathcal{B} \right|$ and $\left| \mathcal{C} \right|$ and between phases $\mathrm{arg} \Xi^{(2)}_{+}, \mathrm{arg} \Xi^{(2)}_{0}$ will become only approximate.

Eq. \eqref{Drude_complex} has also some important implications for time reversal and rotational symmetries. As an example consider a system with two valleys and one Fermi pocket per valley, which would be consistent with PIP$_2$ phase observed in Bernal Bilayer Graphene subject to an applied perpendicular displacement field. Consider now adding an order parameter $\Delta$ to such a setup and calculating nonlinear conductivity using Eq.  \eqref{Drude_complex}.
If the time reversal symmetry is preserved by the order parameter, the single particle energy at the two valleys satisfies
$\eps(\bk,\Delta)=\eps(-\bk,\Delta)$. In this case contributions from the two pockets cancel. However, if the order parameter is odd under time reversal symmetry, then the single particle energy at the two valleys satisfies
$\eps(\bk,\Delta)=\eps(-\bk,-\Delta)$. In this case contributions from the two pockets guarantee that the nonlinear conductivity \eqref{Drude_complex} is odd under $\Delta$.\\

This has interesting implications if, in addition to time reversal, the order parameter $\Delta$ also breaks rotational symmetry. If the order parameter transforms as an $E$ irrep (hence it will break rotational symmetry), we expect $\eps(\bk,\Delta)=\eps(C_3\bk,C_3\Delta)$.
We also have $C_3k=\omega k$ where $\omega=e^{2\pi i/3}$. Assuming that 
$C_3\left(\Delta_1+i\Delta_2\right)=\omega \left(\Delta_1+i\Delta_2\right)$, the 3-fold symmetric term
\begin{eqnarray}
\Xi^{(2)}_-(\Delta)&=&-2\frac{e^3\tau^2}{\hbar^3}\sum_{n}\int \frac{d^2 \mathbf{k}}{(2 \pi)^2} f(\eps_n(\bk,\Delta))\frac{\partial}{\partial \bar{k}}\frac{\partial}{\partial \bar{k}}\frac{\partial}{\partial \bar{k}}\eps_n(\bk,\Delta)
\end{eqnarray}
must satisfy
\begin{eqnarray}
\Xi^{(2)}_-(\Delta)&=&\Xi^{(2)}_-(C_3\Delta)
\end{eqnarray}
and because for TRS-odd order parameter it is odd in $\Delta$, its Taylor expansion must start with $\mathcal{O}(\Delta^3)$ terms.
In other words
\begin{eqnarray}
\Xi^{(2)}_-(\Delta)&=&\alpha_+ \left(\Delta_1+i\Delta_2\right)^3+\alpha_- \left(\Delta_1-i\Delta_2\right)^3+ \mathcal{O}(\Delta^5).
\label{Drude_3fold_nemat_dep}
\end{eqnarray}
On the other hand, the 1-fold terms
\begin{eqnarray}
\Xi_+(\Delta)&=&-2\frac{e^3\tau^2}{\hbar^3}\sum_{n}\int \frac{d^2 \mathbf{k}}{(2 \pi)^2} f(\eps_n(\bk,\Delta))\frac{\partial}{\partial \bar{k}}\frac{\partial}{\partial k}\frac{\partial}{\partial k}\eps_n(\bk,\Delta)\\
\Xi_0(\Delta)&=&-4\frac{e^3\tau^2}{\hbar^3}\sum_{n}\int \frac{d^2 \mathbf{k}}{(2 \pi)^2} f(\eps_n(\bk,\Delta))\frac{\partial}{\partial \bar{k}}\frac{\partial}{\partial \bar{k}}\frac{\partial}{\partial k}\eps_n(\bk,\Delta)
\end{eqnarray}
must satisfy
\begin{equation}
\begin{gathered}
\Xi_+(\Delta)=\omega^*\Xi_+(C_3\Delta)\\
\Xi_0(\Delta)=\omega\Xi_0(C_3\Delta).
\label{Drude_1fold_nemat_dep}
\end{gathered}
\end{equation}
Therefore,
\begin{eqnarray}
\Xi_+(\Delta)&=&\beta_+\left(\Delta_1-i\Delta_2\right)+\mathcal{O}(\Delta^3)\\
\Xi_0(\Delta)&=&\beta_0\left(\Delta_1+i\Delta_2\right)+\mathcal{O}(\Delta^3).
\end{eqnarray}
This allows us establish a hierarchy of magnitudes of nonlinear conductivity tensor components in the presence of rotational symmetry breaking for small magnitude of the order parameter $\Delta$.

\subsection{The hierarchy between $\mathcal{A}, \mathcal{B}, \mathcal{C}$}
\label{SI_sec_hierarchy}

Assuming a weak time reversal-breaking nematic order parameter (like charge $E_u$ or spin $E_g$ \cite{CvetkovicPRB2012}) we can qualitatively account for the observed hierarchy of nonlinear conductivity tensor components in both PIP$_2$ and Sym$_4$ phases of BBG. Here we discuss the hierarchy considering 2 large Fermi pockets, 1 pocket in each valley as an example. In the next section we provide a qualitative explanation for the observed phenomena within both PIP$_2$ and Sym$_4$ phases.   

Consider the following setup: there is 1 large Fermi pocket per valley and there is a  nematic order parameter that also breaks rotational symmetry and has small magnitude. For simplicity, consider this order parameter to be charge $E_u$, which breaks both rotational and time reversal symmetries. 
First of all, there is a natural explanation why the linear anisotropy is small in such setup. It is due to Onsager's theorem: with $C_3$ and time reversal symmetry breaking order parameter $\Delta$, the linear conductivity must be $\sigma_{\mu\nu}(\Delta)=\sigma_{\nu\mu}(-\Delta)$. This means that the anisotropic part of the conductivity tensor must be even in $\Delta$, and therefore its Taylor expansion starts with $\mathcal{O}(\Delta^2)$. The isotropic part is, of course, $\mathcal{O}(1)$. Therefore, if the rotational symmetry breaking order parameter also breaks TR symmetry, the anisotropy of linear conductivity tensor will be very small as long as the perturbation is not too strong. 

Now, for the nonlinear conductivity, if the main contribution comes from second-order  Boltzmann (Drude) contribution \eqref{Drude_complex}, the 3-fold term is odd in $\Delta$ and its Taylor expansion must start at $\mathcal{O}(\Delta^3)$ (again due to broken time reversal and $C_3$ rotational symmetry), while the 1-fold terms are also odd but start at $\mathcal{O}(\Delta)$. Therefore, it is possible to have a small linear anisotropy but very large nonlinear anisotropy if the $C_3$ breaking order parameter is odd under the time reversal symmetry and its magnitude is not too large.

\section{Beyond the relaxation time approximation: full $T$-matrix treatment of collision integral and skew scattering}

In this section we solve the Boltzmann kinetic equation with the full collision integral in the presence of an elastic impurity scattering. This allows us to solve the Boltzmann kinetic equation with the collision integral given by the full $T-$matrix. We outline the method below.

In the steady state the kinetic equation reads \cite{Sturman_1984}
\begin{equation}
    -\frac{e \mathbf{E}}{\hbar} \cdot \frac{\partial f (\mathbf{k})}{\partial \mathbf{k}} = - \int \frac{d^2 \mathbf{k}'}{(2 \pi)^2} W_{\mathbf{k}, \mathbf{k}'} \left( f (\mathbf{k}) - f (\mathbf{k}') \right),
\end{equation}
where the scattering matrix $W_{\mathbf{k}, \mathbf{k}'}$ is a real but not necessarily symmetric function of momenta $\mathbf{k}, \mathbf{k}'$. For elastic scattering 
\begin{equation}
    W_{\mathbf{k}, \mathbf{k}'} = \frac{2 \pi}{\hbar} \left| T_{\mathbf{k}, \mathbf{k}'}  \right|^2 \delta \left( \epsilon (\mathbf{k}) - \epsilon (\mathbf{k}') \right),
\end{equation}
where $T_{\mathbf{k}, \mathbf{k}'}$ -- is the scattering matrix or $T$-matrix. To obtain linear and nonlinear conductivities we expand the distribution function $f$ up to the second order in electric field. 

\subsection{First order in electric field}

To first order in electric field the kinetic equation takes the form
\begin{equation} 
    \begin{gathered}
        -\frac{e \mathbf{E}}{\hbar} \cdot \frac{\partial f_0 (\mathbf{k})}{\partial \mathbf{k}} = - \int \frac{d^2 \mathbf{k}'}{(2 \pi)^2} W_{\mathbf{k}, \mathbf{k}'} \left( f_1 (\mathbf{k}) - f_1 (\mathbf{k}') \right) \Rightarrow \\ 
        -\frac{e \mathbf{E}}{\hbar} \cdot \frac{\partial \epsilon (\mathbf{k})}{\partial \mathbf{k}} \frac{\partial f_0 (\epsilon)}{\partial \epsilon} \Biggr|_{\epsilon=\epsilon(\mathbf{k})} = - f_1 (\mathbf{k}) \left( \int \frac{d^2 \mathbf{k}'}{(2 \pi)^2} W_{\mathbf{k}, \mathbf{k}'} \right) + \int \frac{d^2 \mathbf{k}'}{(2 \pi)^2} W_{\mathbf{k}, \mathbf{k}'}   f_1 (\mathbf{k}'). 
    \end{gathered}
\end{equation}
For elastic scattering both $\mathbf{k}, \mathbf{k}'$ belong to the same fixed energy contour. The left hand side of the expression is peaked at the Fermi surface (FS) for small temperatures. Therefore, we seek a solution for $f_1 (\mathbf{k})$ which vanishes far away from the FS:
\begin{equation}
    f_1 (\mathbf{k}) = \Phi_1 (\mathbf{k}) \frac{\partial f_0 (\epsilon)}{\partial \epsilon} \Biggr|_{\epsilon=\epsilon(\mathbf{k})},
\end{equation}
where $\Phi_1 (\mathbf{k})$ is a slowly varying function of the coordinate perpendicular to the FS and should not be confused with basis functions for nonlinear potential from earlier sections. Using this parametrization of $f_1$ we obtain for the kinetic equation
\begin{eqnarray}
    -\frac{e \mathbf{E}}{\hbar} \cdot \frac{\partial \epsilon (\mathbf{k})}{\partial \mathbf{k}} \frac{\partial f_0 (\epsilon)}{\partial \epsilon} \Biggr|_{\epsilon=\epsilon(\mathbf{k})} &&= - \Phi_1 (\mathbf{k}) \frac{\partial f_0 (\epsilon)}{\partial \epsilon} \Biggr|_{\epsilon=\epsilon(\mathbf{k})} \left( \int \frac{d^2 \mathbf{k}'}{(2 \pi)^2} W_{\mathbf{k}, \mathbf{k}'} \right) + \int \frac{d^2 \mathbf{k}'}{(2 \pi)^2} W_{\mathbf{k}, \mathbf{k}'}   \Phi_1 (\mathbf{k}') \frac{\partial f_0 (\epsilon)}{\partial \epsilon} \Biggr|_{\epsilon=\epsilon(\mathbf{k'})}  \\ &&=
    - \Phi_1 (\mathbf{k}) \frac{\partial f_0 (\epsilon)}{\partial \epsilon} \Biggr|_{\epsilon=\epsilon(\mathbf{k})} \left( \int \frac{d^2 \mathbf{k}'}{(2 \pi)^2} W_{\mathbf{k}, \mathbf{k}'} \right) + \frac{\partial f_0 (\epsilon)}{\partial \epsilon} \Biggr|_{\epsilon=\epsilon(\mathbf{k})} \int \frac{d^2 \mathbf{k}'}{(2 \pi)^2} W_{\mathbf{k}, \mathbf{k}'}   \Phi_1 (\mathbf{k}'). 
\end{eqnarray}
Making use of expression for $W_{\mathbf{k}, \mathbf{k}'}$ and introducing velocity $\mathbf{v} ( \mathbf{k}) = \frac{1}{\hbar} \frac{\partial \epsilon (\mathbf{k})}{\partial \mathbf{k}}$
we obtain
\begin{eqnarray}
    e \mathbf{E} \cdot \mathbf{v} ( \mathbf{k}) = \Phi_1 (\mathbf{k})  \left( \int \frac{d^2 \mathbf{k}'}{(2 \pi)^2} \frac{2 \pi}{\hbar} \left| T_{\mathbf{k}, \mathbf{k}'}  \right|^2 \delta \left( \epsilon (\mathbf{k}) - \epsilon (\mathbf{k}') \right) \right) -  \int \frac{d^2 \mathbf{k}'}{(2 \pi)^2} \Phi_1 (\mathbf{k}') \frac{2 \pi}{\hbar} \left| T_{\mathbf{k}, \mathbf{k}'}  \right|^2 \delta \left( \epsilon (\mathbf{k}) - \epsilon (\mathbf{k}') \right).  
\end{eqnarray}
Since both momenta are located on the same surface of equal energy, we can rewrite the expression for $\Phi_1 (\mathbf{k})$ in its final form
\begin{eqnarray}
    e \hbar^2 \mathbf{E} \cdot \mathbf{v} ( \mathbf{k}_{\theta}) = \Phi_1 (\mathbf{k}_{\theta})  \left( \int \frac{d \theta'}{2 \pi} k_{\theta'} \frac{\left| T_{\mathbf{k}_{\theta}, \mathbf{k}_{\theta'}}  \right|^2}{\left| \mathbf{v} \left( \mathbf{k}_{\theta'} \right) \right|} \right) -  \int \frac{d \theta'}{2 \pi} k_{\theta'} \Phi_1 (\mathbf{k}_{\theta'})  \frac{\left| T_{\mathbf{k}_{\theta}, \mathbf{k}_{\theta'}}  \right|^2}{\left| \mathbf{v} \left( \mathbf{k}_{\theta'} \right) \right|},   
\end{eqnarray}
where angles $\theta, \theta'$ parameterize momenta $k_{\theta} = k (\theta), k_{\theta'}=k (\theta')$ on the equal energy contour.  

It is important to mention that $\Phi_1 (\mathbf{k}_{\theta'})  \propto e \mathbf{E} \cdot \mathbf{v} ( \mathbf{k}_{\theta})$ if $\left| T_{\mathbf{k}_{\theta}, \mathbf{k}_{\theta'}}  \right|^2$ is independent of $\theta, \theta'$. This is a consequence of the identity
\begin{equation}
    \int d \theta' k_{\theta'} \frac{\mathbf{v} \left( \mathbf{k}_{\theta'} \right)}{\left| \mathbf{v} \left( \mathbf{k}_{\theta'} \right) \right|} = \int d \theta' k_{\theta'} \hat{\mathbf{v}} \left( \mathbf{k}_{\theta'} \right) =0.
\end{equation}
To see this, note that $\hat{\mathbf{v}} \left( \mathbf{k}_{\theta'} \right)$ points radially out from the constant energy contour, therefore, $\hat{\mathbf{z}} \times \int d \theta' k_{\theta'} \hat{\mathbf{v}} \left( \mathbf{k}_{\theta'} \right)$ corresponds to the sum of infinitesimal vectors tangent to the constant energy contour (curve) as we go along this curve. The summation of those vectors head-to-tail must vanish as we complete the closed curve. Thus, if $\hat{\mathbf{z}} \times \int d \theta' k_{\theta'} \hat{\mathbf{v}} \left( \mathbf{k}_{\theta'} \right)$ vanishes, then so does $\int d \theta' k_{\theta'} \hat{\mathbf{v}} \left( \mathbf{k}_{\theta'} \right)$ as it is just a $90^{\circ}$ rotation of the vanishing vector.

\subsection{Second order in electric field}

We now proceed to the next order contribution. To the second order in electric field Boltzmann equation gives
\begin{equation} 
    \begin{gathered}
        -\frac{e \mathbf{E}}{\hbar} \cdot \frac{\partial f_1 (\mathbf{k})}{\partial \mathbf{k}} = - \int \frac{d^2 \mathbf{k}'}{(2 \pi)^2} W_{\mathbf{k}, \mathbf{k}'} \left( f_2 (\mathbf{k}) - f_2 (\mathbf{k}') \right) \Rightarrow \\ 
        -\frac{e \mathbf{E}}{\hbar} \cdot \frac{\partial}{\partial \mathbf{k}} \left( \Phi_1 (\mathbf{k}) \frac{\partial f_0 (\epsilon)}{\partial \epsilon} \Biggr|_{\epsilon=\epsilon(\mathbf{k})} \right) = - f_2 (\mathbf{k}) \left( \int \frac{d^2 \mathbf{k}'}{(2 \pi)^2} W_{\mathbf{k}, \mathbf{k}'} \right) + \int \frac{d^2 \mathbf{k}'}{(2 \pi)^2} W_{\mathbf{k}, \mathbf{k}'}   f_2 (\mathbf{k}'). 
    \end{gathered}
\end{equation}
At the second order in electric field we seek the solution of the form
\begin{equation}
    f_2 (\mathbf{k}) = \Phi_{2,1} (\mathbf{k}) \frac{\partial f_0 (\epsilon)}{\partial \epsilon} \Biggr|_{\epsilon=\epsilon(\mathbf{k})} + \Phi_{2,2} (\mathbf{k}) \frac{\partial^2 f_0 (\epsilon)}{\partial \epsilon^2} \Biggr|_{\epsilon=\epsilon(\mathbf{k})},
\end{equation}
where $\Phi_{2,1} (\mathbf{k})$ and $\Phi_{2,2} (\mathbf{k})$, like $\Phi_1 (\mathbf{k})$ in the linear case, are smooth functions of the component of $\mathbf{k}$ perpendicular to the constant energy contour. Substituting the parametrization of $f_2 (\mathbf{k})$ into the kinetic equation we obtain
\begin{eqnarray}
    &&-\frac{e \mathbf{E}}{\hbar} \cdot \frac{\partial}{\partial \mathbf{k}} \left( \Phi_1 (\mathbf{k}) \frac{\partial f_0 (\epsilon)}{\partial \epsilon} \Biggr|_{\epsilon=\epsilon(\mathbf{k})} \right) = - \left( \Phi_{2,1} (\mathbf{k}) \frac{\partial f_0 (\epsilon)}{\partial \epsilon} \Biggr|_{\epsilon=\epsilon(\mathbf{k})} + \Phi_{2,2} (\mathbf{k}) \frac{\partial^2 f_0 (\epsilon)}{\partial \epsilon^2} \Biggr|_{\epsilon=\epsilon(\mathbf{k})} \right) \left( \int \frac{d^2 \mathbf{k}'}{(2 \pi)^2} W_{\mathbf{k}, \mathbf{k}'} \right) \\ &&+ \int \frac{d^2 \mathbf{k}'}{(2 \pi)^2} W_{\mathbf{k}, \mathbf{k}'}   \left( \Phi_{2,1} (\mathbf{k}') \frac{\partial f_0 (\epsilon)}{\partial \epsilon} \Biggr|_{\epsilon=\epsilon(\mathbf{k}')} + \Phi_{2,2} (\mathbf{k}') \frac{\partial^2 f_0 (\epsilon)}{\partial \epsilon^2} \Biggr|_{\epsilon=\epsilon(\mathbf{k}')} \right) = \nonumber \\ &&= - \left( \Phi_{2,1} (\mathbf{k}) \frac{\partial f_0 (\epsilon)}{\partial \epsilon} \Biggr|_{\epsilon=\epsilon(\mathbf{k})} + \Phi_{2,2} (\mathbf{k}) \frac{\partial^2 f_0 (\epsilon)}{\partial \epsilon^2} \Biggr|_{\epsilon=\epsilon(\mathbf{k})} \right) \left( \int \frac{d^2 \mathbf{k}'}{(2 \pi)^2} W_{\mathbf{k}, \mathbf{k}'} \right) \\ &&+  \frac{\partial f_0 (\epsilon)}{\partial \epsilon} \Biggr|_{\epsilon=\epsilon(\mathbf{k})} \int \frac{d^2 \mathbf{k}'}{(2 \pi)^2} W_{\mathbf{k}, \mathbf{k}'}  \Phi_{2,1} (\mathbf{k}') \nonumber + \frac{\partial^2 f_0 (\epsilon)}{\partial \epsilon^2} \Biggr|_{\epsilon=\epsilon(\mathbf{k})} \int \frac{d^2 \mathbf{k}'}{(2 \pi)^2} W_{\mathbf{k}, \mathbf{k}'}    \Phi_{2,2} (\mathbf{k}'). 
\end{eqnarray}
We now make use of the expression for $W_{\mathbf{k}, \mathbf{k}'}$ and match the derivatives of the occupation functions on both sides of the expression to get 
\begin{equation}
\begin{gathered}
    -\frac{e \mathbf{E}}{\hbar} \cdot \frac{\partial \Phi_1 (\mathbf{k}) }{\partial \mathbf{k}} = -\Phi_{2,1} (\mathbf{k})  \left( \int \frac{d^2 \mathbf{k}'}{(2 \pi)^2} \frac{2 \pi}{\hbar} \left| T_{\mathbf{k}, \mathbf{k}'}  \right|^2 \delta \left( \epsilon (\mathbf{k}) - \epsilon (\mathbf{k}') \right) \right) +  \int \frac{d^2 \mathbf{k}'}{(2 \pi)^2} \Phi_{2,1} (\mathbf{k}') \frac{2 \pi}{\hbar} \left| T_{\mathbf{k}, \mathbf{k}'}  \right|^2 \delta \left( \epsilon (\mathbf{k}) - \epsilon (\mathbf{k}') \right), \\
    -e \mathbf{E} \cdot \mathbf{v} (\mathbf{k}) \Phi_1 (\mathbf{k})  = -\Phi_{2,2} (\mathbf{k})  \left( \int \frac{d^2 \mathbf{k}'}{(2 \pi)^2} \frac{2 \pi}{\hbar} \left| T_{\mathbf{k}, \mathbf{k}'}  \right|^2 \delta \left( \epsilon (\mathbf{k}) - \epsilon (\mathbf{k}') \right) \right) +  \int \frac{d^2 \mathbf{k}'}{(2 \pi)^2} \Phi_{2,2} (\mathbf{k}') \frac{2 \pi}{\hbar} \left| T_{\mathbf{k}, \mathbf{k}'}  \right|^2 \delta \left( \epsilon (\mathbf{k}) - \epsilon (\mathbf{k}') \right).
    \end{gathered} 
\end{equation}
These two equations have the same structure as the equation for $\Phi_1 (\mathbf{k})$ and can be cast into the same convenient form:
\begin{equation}
\begin{gathered}
    e \hbar \mathbf{E} \cdot \frac{\partial \Phi_1 ( \mathbf{k})}{\partial \mathbf{k}} \Biggr|_{\mathbf{k}=\mathbf{k}_{\theta}} = \Phi_{2,1} (\mathbf{k}_{\theta})  \left( \int \frac{d \theta'}{2 \pi} k_{\theta'} \frac{\left| T_{\mathbf{k}_{\theta}, \mathbf{k}_{\theta'}}  \right|^2}{\left| \mathbf{v} \left( \mathbf{k}_{\theta'} \right) \right|} \right) -  \int \frac{d \theta'}{2 \pi} k_{\theta'} \Phi_{2,1} (\mathbf{k}_{\theta'})  \frac{\left| T_{\mathbf{k}_{\theta}, \mathbf{k}_{\theta'}}  \right|^2}{\left| \mathbf{v} \left( \mathbf{k}_{\theta'} \right) \right|},   \\
    e \hbar^2 \mathbf{E} \cdot \mathbf{v} ( \mathbf{k}_{\theta}) \Phi_1 (\mathbf{k}_{\theta}) = \Phi_{2,2} (\mathbf{k}_{\theta})  \left( \int \frac{d \theta'}{2 \pi} k_{\theta'} \frac{\left| T_{\mathbf{k}_{\theta}, \mathbf{k}_{\theta'}}  \right|^2}{\left| \mathbf{v} \left( \mathbf{k}_{\theta'} \right) \right|} \right) -  \int \frac{d \theta'}{2 \pi} k_{\theta'} \Phi_{2,2} (\mathbf{k}_{\theta'})  \frac{\left| T_{\mathbf{k}_{\theta}, \mathbf{k}_{\theta'}}  \right|^2}{\left| \mathbf{v} \left( \mathbf{k}_{\theta'} \right) \right|}.   
\end{gathered}
\end{equation}
To calculate the right hand side of these expressions we need to specify the structure of the $T-$matrix. We do it below.

\subsection{Separating the kernel}

To actually solve for the distribution functions we make use of the collision integral kernel being separable:
\begin{equation}
    W_{\mathbf{k}, \mathbf{k}'} = \frac{2\pi}{\hbar} \delta \left( \epsilon (\mathbf{k}) - \epsilon (\mathbf{k}') \right) \sum_{j=0}^3 f_j (\mathbf{k}) g_j (\mathbf{k}') .
\end{equation}
At linear order in electric field the equation for $\Phi_1 (\mathbf{k})$ takes the form 
\begin{equation}
    e \hbar^2 \mathbf{E} \cdot \mathbf{v} ( \mathbf{k}) = \Phi_1 (\mathbf{k}) \sum_{j=0}^3 f_j (\mathbf{k}) \left( \int \frac{d^2 \mathbf{k}'}{2 \pi} g_j (\mathbf{k}') \delta \left( \epsilon (\mathbf{k}) - \epsilon (\mathbf{k}') \right) \right) - \sum_{j=0}^3 f_j (\mathbf{k}) \left( \int \frac{d^2 \mathbf{k}'}{2 \pi} \Phi_1 (\mathbf{k}') g_j (\mathbf{k}') \delta \left( \epsilon (\mathbf{k}) - \epsilon (\mathbf{k}') \right) \right).  
\end{equation}
Let us introduce auxiliary quantities
\begin{eqnarray}
    \Gamma (\mathbf{k}) = \sum_{j=0}^3 f_j (\mathbf{k}) \left( \int \frac{d^2 \mathbf{k}'}{2 \pi} g_j (\mathbf{k}') \delta \left( \epsilon (\mathbf{k}) - \epsilon (\mathbf{k}') \right) \right) = \sum_{j=0}^3 f_j (\mathbf{k}) \mathcal{G}_j (\epsilon (\mathbf{k})), \\
    \mathcal{C}_j^{(1)}(\omega) = \int \frac{d^2 \mathbf{k}'}{2 \pi} \Phi_1 (\mathbf{k}') g_j (\mathbf{k}') \delta \left( \omega - \epsilon (\mathbf{k}') \right),
\end{eqnarray}
that allow us to express $\Phi_1 (\mathbf{k})$ as
\begin{equation}
    \Phi_1 (\mathbf{k}) = \frac{e \hbar^2 \mathbf{E} \cdot \mathbf{v} ( \mathbf{k})}{\Gamma (\mathbf{k})} + \sum_{j=0}^3 \frac{f_j(\mathbf{k})}{\Gamma (\mathbf{k})} \mathcal{C}_j^{(1)} (\epsilon (\mathbf{k})).
\end{equation}
Analogous manipulations cab be performed at nonlinear order to the expressions with similar mathematical structure
\begin{eqnarray}
    \Phi_{2,1} (\mathbf{k}) = \frac{e \hbar \mathbf{E} }{\Gamma (\mathbf{k})} \cdot \frac{\partial \Phi_1 ( \mathbf{k})}{\partial \mathbf{k}} + \sum_{j=0}^3 \frac{f_j(\mathbf{k})}{\Gamma (\mathbf{k})} \mathcal{C}_j^{(2,1)} (\epsilon (\mathbf{k})), \\
    \Phi_{2,2} (\mathbf{k}) = \frac{e \hbar^2 \Phi_1 (\mathbf{k}) \mathbf{E} \cdot \mathbf{v} ( \mathbf{k})}{\Gamma (\mathbf{k})} + \sum_{j=0}^3 \frac{f_j(\mathbf{k})}{\Gamma (\mathbf{k})} \mathcal{C}_j^{(2,2)} (\epsilon (\mathbf{k})), \\
    \mathcal{C}_j^{(2,a)}(\omega) = \int \frac{d^2 \mathbf{k}'}{2 \pi} \Phi_{2,a} (\mathbf{k}') g_j (\mathbf{k}') \delta \left( \omega - \epsilon (\mathbf{k}') \right).
\end{eqnarray}
To calculate the nonlinear current and we will need to evaluate $\frac{\partial \Gamma ( \mathbf{k})}{\partial \mathbf{k}}$, $\frac{\partial \Phi_1 ( \mathbf{k})}{\partial \mathbf{k}}$ and $\frac{\partial \Phi_{2,2} ( \mathbf{k})}{\partial \mathbf{k}}$, which are given by
\begin{equation}
    \frac{\partial \Gamma ( \mathbf{k})}{\partial \mathbf{k}} = \sum_{j=0}^3 \frac{\partial f_j (\mathbf{k})}{\partial \mathbf{k}} \mathcal{G}_j (\epsilon (\mathbf{k})) + \sum_{j=0}^3 f_j (\mathbf{k}) \frac{\partial \epsilon (\mathbf{k})}{\partial \mathbf{k}} \frac{d\mathcal{G}_j (\omega)}{d \omega} \Biggr|_{\omega = \epsilon (\mathbf{k})},
\end{equation}
\begin{equation}
    \begin{gathered}
        \frac{\partial \Phi_1 ( \mathbf{k})}{\partial \mathbf{k}} = - \frac{e \hbar^2}{\Gamma^2 (\mathbf{k})} \frac{\partial \Gamma ( \mathbf{k})}{\partial \mathbf{k}} \left( \mathbf{E} \cdot \mathbf{v} (\mathbf{k}) \right) + \frac{e \hbar^2}{\Gamma (\mathbf{k})} \frac{\partial}{\partial \mathbf{k}} \left( \mathbf{E} \cdot \mathbf{v} (\mathbf{k}) \right) + \\ + \sum_{j=0}^3 \frac{\partial f_j (\mathbf{k})}{\partial \mathbf{k}} \frac{\mathcal{C}_j^{(1)} (\epsilon (\mathbf{k}))}{\Gamma (\mathbf{k})} - \sum_{j=0}^3 \frac{ f_j (\mathbf{k}) \mathcal{C}_j^{(1)} (\epsilon (\mathbf{k}))}{\Gamma^2 (\mathbf{k})} \frac{\partial \Gamma (\mathbf{k})}{\partial \mathbf{k} } + \sum_{j=0}^3 \frac{f_j (\mathbf{k})}{\Gamma (\mathbf{k})} \frac{\partial \mathcal{C}_j^{(1)} (\epsilon (\mathbf{k}))}{\partial \mathbf{k} },
    \end{gathered}
\end{equation}
\begin{equation}
    \begin{gathered}
        \frac{\partial \Phi_{2,2} ( \mathbf{k})}{\partial \mathbf{k}} = - \frac{e \hbar^2 \Phi_1 (\mathbf{k})}{\Gamma^2 (\mathbf{k})} \frac{\partial \Gamma ( \mathbf{k})}{\partial \mathbf{k}} \left( \mathbf{E} \cdot \mathbf{v} (\mathbf{k}) \right) + \frac{e \hbar^2 \Phi_1 (\mathbf{k})}{\Gamma (\mathbf{k})} \frac{\partial}{\partial \mathbf{k}} \left( \mathbf{E} \cdot \mathbf{v} (\mathbf{k}) \right) + \frac{e \hbar^2 \mathbf{E} \cdot \mathbf{v} (\mathbf{k})}{\Gamma (\mathbf{k})} \frac{\partial \Phi_1 ( \mathbf{k})}{\partial \mathbf{k}} + \\
        + \sum_{j=0}^3 \frac{\partial f_j (\mathbf{k})}{\partial \mathbf{k}} \frac{\mathcal{C}_j^{(2,2)} (\epsilon (\mathbf{k}))}{\Gamma (\mathbf{k})} - \sum_{j=0}^3 \frac{ f_j (\mathbf{k}) \mathcal{C}_j^{(2,2)} (\epsilon (\mathbf{k}))}{\Gamma^2 (\mathbf{k})} \frac{\partial \Gamma (\mathbf{k})}{\partial \mathbf{k} } + \sum_{j=0}^3 \frac{f_j (\mathbf{k})}{\Gamma (\mathbf{k})} \frac{\partial \mathcal{C}_j^{(2,2)} (\epsilon (\mathbf{k}))}{\partial \mathbf{k} }.
    \end{gathered}
\end{equation}

\subsection{Current density}

Current density is given by
\begin{equation}
    j_{\alpha} = - e \int \frac{d^2 \mathbf{k}}{(2 \pi)^2} v_{\alpha} (\mathbf{k}) f (\mathbf{k}),
\end{equation}
where, we remind, $\mathbf{v} ( \mathbf{k}) = \frac{1}{\hbar} \frac{\partial \epsilon (\mathbf{k})}{\partial \mathbf{k}}$.
At linear order in electric field we obtain
\begin{equation}
    j_{\alpha}^{(1)} = - e \int \frac{d^2 \mathbf{k}}{(2 \pi)^2} v_{\alpha} (\mathbf{k}) f_1 (\mathbf{k}) = - e \int \frac{d^2 \mathbf{k}}{(2 \pi)^2} v_{\alpha} (\mathbf{k}) \Phi_1 (\mathbf{k}) \frac{\partial f_0 (\epsilon)}{\partial \epsilon} \Biggr|_{\epsilon=\epsilon(\mathbf{k})}.
    \label{T-matr_lin_sigma}
\end{equation}
The first derivative of the distribution function pins the integration to the Fermi surface at low temperature. 

The second order contribution to current density reads
\begin{equation}
    \begin{gathered}
        j_{\alpha}^{(2)} = - e \int \frac{d^2 \mathbf{k}}{(2 \pi)^2} v_{\alpha} (\mathbf{k}) f_2 (\mathbf{k}) = - e \int \frac{d^2 \mathbf{k}}{(2 \pi)^2} v_{\alpha} (\mathbf{k}) \left( \Phi_{2,1} (\mathbf{k}) \frac{\partial f_0 (\epsilon)}{\partial \epsilon} \Biggr|_{\epsilon=\epsilon(\mathbf{k})} + \Phi_{2,2} (\mathbf{k}) \frac{\partial^2 f_0 (\epsilon)}{\partial \epsilon^2} \Biggr|_{\epsilon=\epsilon(\mathbf{k})} \right) \\ =  - e \int \frac{d^2 \mathbf{k}}{(2 \pi)^2} \left( v_{\alpha} (\mathbf{k}) \Phi_{2,1} (\mathbf{k}) - \frac{1}{\hbar} \frac{\partial \Phi_{2,2} (\mathbf{k})}{\partial k_{\alpha}} \right) \frac{\partial f_0 (\epsilon)}{\partial \epsilon} \Biggr|_{\epsilon=\epsilon(\mathbf{k})},
    \end{gathered}
    \label{T-matr_nonlin_sigma}
\end{equation}
where we made use of the identity
\begin{equation}
    v_{\alpha} (\mathbf{k}) \frac{\partial^2 f_0 (\epsilon)}{\partial \epsilon^2} \Biggr|_{\epsilon=\epsilon(\mathbf{k})} = \frac{1}{\hbar} \frac{\partial}{\partial k_{\alpha}} \frac{\partial f_0 (\epsilon)}{\partial \epsilon} \Biggr|_{\epsilon=\epsilon(\mathbf{k})}
\end{equation}
to get rid of the second derivative of the Fermi function. 
Eqs. \eqref{T-matr_lin_sigma}, \eqref{T-matr_nonlin_sigma} allow us to calculate both linear $\sigma_{\alpha \mu}$ and nonlinear $\tilde{\sigma}_{\alpha \mu \nu}$ conductivity tensors for a given scattering matrix.

\subsection{Scattering matrix}

The scattering $T$-matrix is given by the standard definition
\begin{equation}
    T_{\mathbf{k}, \mathbf{k}'} = \left \langle \mathbf{k} \right| V \left| \psi (\mathbf{k}') \right \rangle, 
\end{equation}
where $V$ - is the impurity potential and $\left| \psi (\mathbf{k}') \right \rangle$ is the eigenstate of the full Hamiltonian which satisfies the Lippman-Schwinger equation
\begin{equation}
    \left| \psi (\mathbf{k}') \right \rangle = \left| \mathbf{k}' \right \rangle + \left( \epsilon (\mathbf{k}') - H_0 + i0^{+} \right)^{-1} V \left| \psi (\mathbf{k}') \right \rangle.
\end{equation}
Using the Lippman-Schwinger expression the integral equation for $T$-matrix can be cast into the form 
\begin{equation}
    T (\omega) = V + V \left( \omega - H_{\mathbf{p}} + i0^{+}  \right)^{-1} T (\omega),
\end{equation}
where $\omega = \epsilon (\mathbf{k})$. We will be working in the leading order in the impurity concentration $n_{imp}$, which was estimated to be $\sim 6.6 \times 10^8 \text{cm}^{-2}$ \cite{Joucken2021}. Consider impurity potential given by
\begin{equation}
    V (\mathbf{r}) = \sum_{j=1}^{N_{imp}} U \left( \mathbf{r} - \mathbf{r}_j \right) = U_0 \sum_{j=1}^{N_{imp}} \delta \left( \mathbf{r} - \mathbf{r}_j \right)
\end{equation}
from  short-ranged random impurities. We also assume that impurities stimulate scattering only within each of the two valleys, i.e., there is inter-valley scattering is negligible. 
Assuming the disorder is self-averaging, we need to compute the disorder-averaged transition probability $\overline{\left| T_{\mathbf{k}, \mathbf{k}'}  \right|^2}$. Taking $T$-matrix to first order in impurity potential and averaging  over disorder yields
\begin{equation}
    \overline{\left| T_{\mathbf{k}, \mathbf{k}'}  \right|^2} \simeq \frac{1}{A^2} \overline{ \sum_{j,j'=1}^{N_{imp}} e^{-i \left( \mathbf{k} - \mathbf{k}' \right) \cdot \left( \mathbf{r}_j - \mathbf{r}'_j  \right) } \left| 
U_{\mathbf{k}, \mathbf{k}'} \right|^2 } = \frac{1}{A} \frac{N_{imp}}{A} \left| U_{\mathbf{k}, \mathbf{k}'} \right|^2 + \frac{N_{imp} \left( N_{imp} - 1 \right)}{A^2} \delta_{\mathbf{k}, \mathbf{k}'} \left| U_{\mathbf{k}, \mathbf{k}'} \right|^2, 
\label{disorder_average1}
\end{equation}
where 
$$
U_{\mathbf{k}, \mathbf{k}'} = \psi_{\mathbf{k}}^{\dagger} U \left( \mathbf{k} - \mathbf{k}' \right) \psi_{\mathbf{k}'} = \psi_{\mathbf{k}}^{\dagger} \left( \int d^2 \mathbf{r} e^{-i \left( \mathbf{k} - \mathbf{k}' \right) \cdot \mathbf{r}} U \left( \mathbf{r} \right) \right) \psi_{\mathbf{k}'}.
$$
The last term in Eq. \eqref{disorder_average1} can be absorbed into the redefinition of the Hamiltonian parameters, such as chemical potential. Importantly, one the the $A$ factors from the denominator is absorbed into the sum over momenta to provide the integral in the right hand side of the Boltzmann integral (the collision integral), therefore, it should be ignored when the right hand side is given by the integral (and not a sum over momenta).  

Performing similar calculations to higher orders in $U \left( \mathbf{r} \right)$ but resorting to the leading order in impurity concentration upon impurity averaging, which amounts to keeping only the terms with $\mathbf{r}_{j_1} = \mathbf{r}_{j_2} = \mathbf{r}_{j_3} = ...$, we obtain for the $T$-matrix 
\begin{equation}
    T_{\mathbf{k}, \mathbf{k}'} \simeq \frac{1}{A} \sum_{j=1}^{N_{imp}} e^{-i \left( \mathbf{k} - \mathbf{k}' \right) \cdot \mathbf{r}_j } \psi_{\mathbf{k}}^{\dagger} t_{\mathbf{k}, \mathbf{k}'} \psi_{\mathbf{k}'} + \mathcal{O} \left( \frac{n^2_{imp}}{A} \right),
\end{equation}
where
\begin{equation}
t_{\mathbf{k}, \mathbf{k}'} = U(\mathbf{k} - \mathbf{k}') + \int \frac{d^2 \mathbf{p}}{(2 \pi)^2} U(\mathbf{k} - \mathbf{p}) \left( \omega - H_{\mathbf{p}} + i0^{+}  \right)^{-1} t_{\mathbf{p}, \mathbf{k}'}.
\label{t_integral_equation}
\end{equation}
Therefore, the expression for $\left| T_{\mathbf{k}, \mathbf{k}'}  \right|^2$ reads
\begin{equation}
    \left| T_{\mathbf{k}, \mathbf{k}'}  \right|^2 \simeq \frac{n_{imp}}{A} \psi_{\mathbf{k}}^{\dagger} t_{\mathbf{k}, \mathbf{k}'} \psi_{\mathbf{k}'} \psi_{\mathbf{k}'}^{\dagger} t_{\mathbf{k}, \mathbf{k}'}^{\dagger} \psi_{\mathbf{k}}.
\end{equation}
For a $2 \times 2$ Hamiltonian one can employ the Pauli-matrix parametrization of the wave function
$$
\psi_{\mathbf{k}} \psi_{\mathbf{k}}^{\dagger} = \frac{1}{2} \left( 1_{2 \times 2} + \hat{\mathbf{n}}_{\mathbf{k}} \cdot \pmb{\sigma}  \right)
$$
and 
$$
t_{\mathbf{k}, \mathbf{k}'} = A_{\mathbf{k}, \mathbf{k}'} + \mathbf{B}_{\mathbf{k}, \mathbf{k}'} \cdot \pmb{\sigma} 
$$
we can express $\left| T_{\mathbf{k}, \mathbf{k}'}  \right|^2$ as 
\begin{equation}
    \begin{gathered}
        \left| T_{\mathbf{k}, \mathbf{k}'}  \right|^2 \simeq \frac{n_{imp}}{4A} \mathrm{Tr} \left[ \left( A_{\mathbf{k}, \mathbf{k}'} + \mathbf{B}_{\mathbf{k}, \mathbf{k}'} \cdot \pmb{\sigma}  \right) \left( 1_{2 \times 2} + \hat{\mathbf{n}}_{\mathbf{k}'} \cdot \pmb{\sigma}  \right) \left( A_{\mathbf{k}, \mathbf{k}'}^* + \mathbf{B}^*_{\mathbf{k}, \mathbf{k}'} \cdot \pmb{\sigma}  \right) \left( 1_{2 \times 2} + \hat{\mathbf{n}}_{\mathbf{k}} \cdot \pmb{\sigma}  \right) \right] \\
        = \frac{n_{imp}}{2A} \left[ \left| A_{\mathbf{k}, \mathbf{k}'}  \right|^2 \left( 1 + \hat{\mathbf{n}}_{\mathbf{k}} \cdot \hat{\mathbf{n}}_{\mathbf{k}'}  \right)  +  \left( A_{\mathbf{k}, \mathbf{k}'} \mathbf{B}^*_{\mathbf{k}, \mathbf{k}'} + A^*_{\mathbf{k}, \mathbf{k}'} \mathbf{B}_{\mathbf{k}, \mathbf{k}'} \right)  \cdot \left(  \hat{\mathbf{n}}_{\mathbf{k}} + \hat{\mathbf{n}}_{\mathbf{k}'}  \right) + \mathbf{B}_{\mathbf{k}, \mathbf{k}'} \cdot \mathbf{B}^*_{\mathbf{k}, \mathbf{k}'} \left( 1 - \hat{\mathbf{n}}_{\mathbf{k}} \cdot \hat{\mathbf{n}}_{\mathbf{k}'}  \right) + \right. \\
        \left. + \left( \mathbf{B}^*_{\mathbf{k}, \mathbf{k}'} \cdot \hat{\mathbf{n}}_{\mathbf{k}} \right) \left( \mathbf{B}_{\mathbf{k}, \mathbf{k}'} \cdot \hat{\mathbf{n}}_{\mathbf{k}'} \right) + \left( \mathbf{B}_{\mathbf{k}, \mathbf{k}'} \cdot \hat{\mathbf{n}}_{\mathbf{k}} \right) \left( \mathbf{B}^*_{\mathbf{k}, \mathbf{k}'} \cdot \hat{\mathbf{n}}_{\mathbf{k}'} \right) + \right. \\
        \left. + i \left( A_{\mathbf{k}, \mathbf{k}'} \mathbf{B}^*_{\mathbf{k}, \mathbf{k}'} - A^*_{\mathbf{k}, \mathbf{k}'} \mathbf{B}_{\mathbf{k}, \mathbf{k}'} \right) \cdot \left(  \hat{\mathbf{n}}_{\mathbf{k}} \times \hat{\mathbf{n}}_{\mathbf{k}'}  \right) + i \left( \mathbf{B}_{\mathbf{k}, \mathbf{k}'} \times \mathbf{B}^*_{\mathbf{k}, \mathbf{k}'} \right) \left(  \hat{\mathbf{n}}_{\mathbf{k}} -\hat{\mathbf{n}}_{\mathbf{k}'}  \right)
        \right], 
    \end{gathered}
\end{equation}
where $A_{\mathbf{k}, \mathbf{k}'}$ and $\mathbf{B}_{\mathbf{k}, \mathbf{k}'}$ are, in general, functions of momenta $\mathbf{k}, \mathbf{k}'$ and have to be defined by solving Eq. \eqref{t_integral_equation}.
For $A_{\mathbf{k}, \mathbf{k}'}, \mathbf{B}_{\mathbf{k}, \mathbf{k}'}$ independent of momenta, the last two terms lead to skew scattering as they are antisymmetric with respect to $\mathbf{k} \leftrightarrow \mathbf{k}'$; the rest contribute to the symmetric part of scattering probability.

\section{A phenomenological scenario which accounts for the structures of nonlinear conductivity tensor in PIP$_2$ and Sym$_4$ phases}

In this section we discuss a scenario which can account for the observed structures of $\sigma_{\alpha \mu}$ and $\tilde{\sigma}_{\alpha \mu \nu}$ in both PIP$_2$ and Sym$_4$ phases; \newtext{also see SI Section \ref{SI_sec_hierarchy} above}. However, the proposed nonlinear Drude scenario (as well as other known mechanisms of nonlinear transport) is incapable of explaining the magnitude of the effect, as we show below.

\subsection{A brief description of the scenario}

\begin{figure*}
\includegraphics[width=1\linewidth]{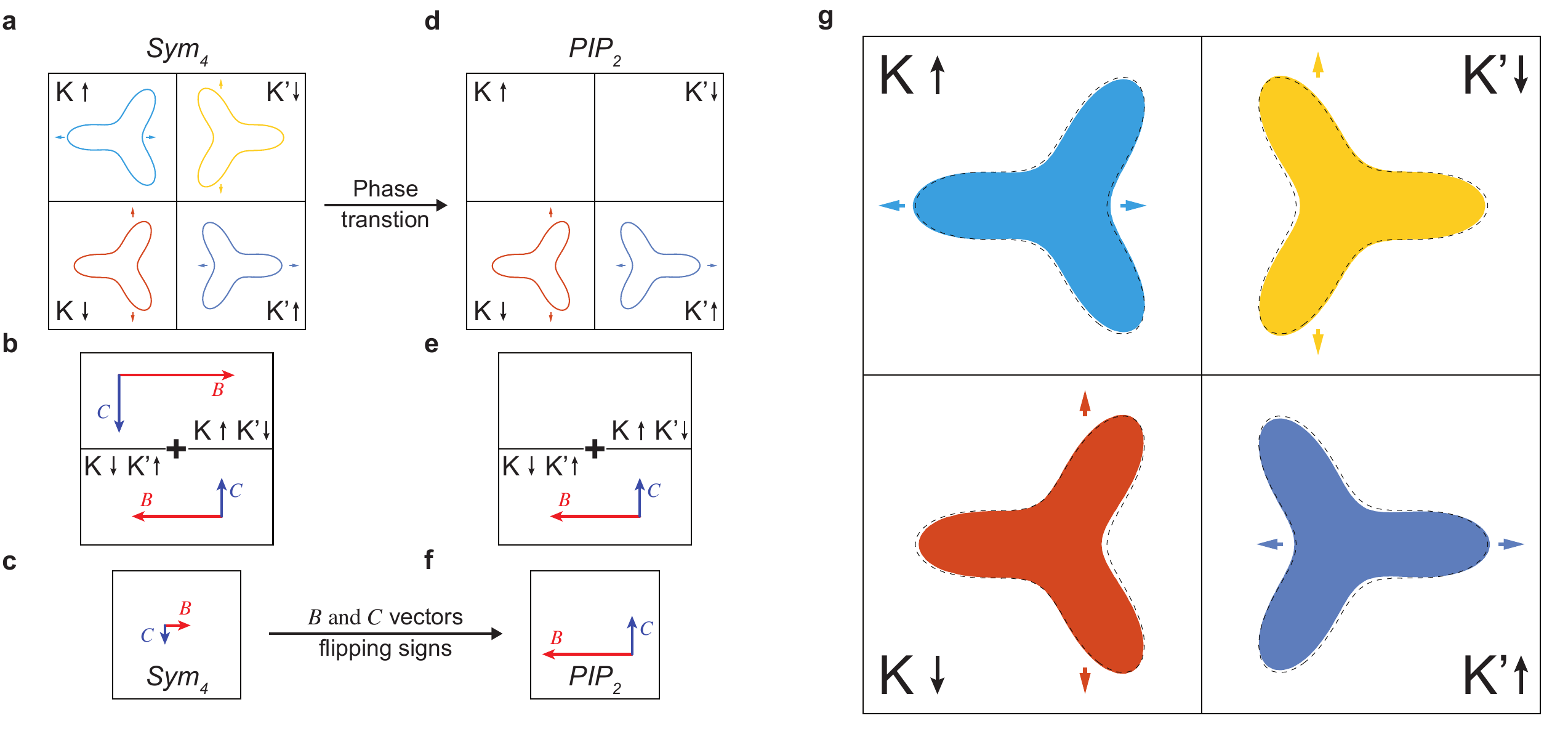}
\caption{{\bf{Fermi surface pockets and reorientation transition in nonlinear regime from nonlinear Drude conductivity}}
(a) Schematics of the Fermi surface pockets for different isospin flavors in the Sym$_4$ phase in the presence of the Kane-Mele SOC and $E_g$ spin-nematic order parameter. Arrows in each square describe the pocket distortion due to the nematic perturbation. The pocket size difference between the top and bottom rows is enlarged to be clear. (b) Hypothetical $\mathcal{B},\mathcal{C}$ contributions from each pairs of pockets for the Sym$_4$ phase. Due to opposite distortion between valleys, the top and bottom isospin pair give rise to  $\mathcal{B},\mathcal{C}$ contributions in opposite directions. Due to difference in pocket size, the top pair creates larger $\mathcal{B},\mathcal{C}$ contribution in magnitude than the bottom pair. Adding the top and bottom pair contribution together, (c) illustrates the total $\mathcal{B},\mathcal{C}$ for the Sym$_4$ phase. Note that scattering times in PIP$_2$ and Sym$_4$ phases are different as evident from linear conductivity measurements. (d) Schematics of the Fermi surface pockets for different isospin flavors for the PIP$_2$ phase, characterized by a valley antiferromagnetic order parameter. (e) Due to the sinking of the top pair in the PIP$_2$ phase, the only remaining contribution of $\mathcal{B},\mathcal{C}$ comes from the bottom isospin pair. Resulting a flip in sign of the total $\mathcal{B},\mathcal{C}$ as shown in (f). (g) A zoom in plot of panel (a) to show the distortion in each Fermi pocket. The black dashed contours represent the unperturbed Fermi surface contours.
}
\label{figSI_pockets}
\end{figure*}

One possible explanation for our observations can be given by calculating $\sigma_{\alpha \mu}$ and $\tilde{\sigma}_{\alpha\mu \nu}$ within Boltzmann theory in relaxation time approximation. 
Within this framework, a distortion in the Fermi surface (FS) naturally captures the linear and nonlinear conductivity tensors, along with the transition between PIP$_2$ and Sym$_4$ phases, given the presence of Kane-Mele-like spin-orbit coupling \cite{KaneMelePRL2005} $g_{KM}\sim 350$mK \cite{AndreaKITPSOC} and a weak  $E_g$ spin-nematic \cite{CvetkovicPRB2012} order parameter with magnitude $g_{E_g} \sim 5$K. First, we consider the Sym$_4$ phase that consists of four large Fermi surfaces with valley and spin flavors. According to quantum oscillation measurements (see Fig. \ref{figSI_QO} in SI) \cite{Andrea_BBG_Science2022}, the Luttinger volume of all 4 FSs in Sym$_4$ phase is almost identical, indicating that both $g_{E_g}$ and $g_{KM}$ are small. Small $g_{E_g}$ ensures that $\sigma_{\alpha \mu}$ is almost isotropic, which is consistent with our observations. Due to the unique order parameter, we divide 4 FSs into two pairs, where each pair is defined as two FSs with opposite spin and valley. The spin-nematic perturbation distorts two pairs of FSs differently, generating $\mathcal{B}$ and $\mathcal{C}$ with opposite signs, see Fig. \ref{figSI_pockets}.  The presence of $g_{KM}$ creates a small imbalance in the Luttinger volume between two pairs of FSs. Due to this imbalance, the average contribution from all FSs gives rise to a finite value in $\mathcal{B}$ and $\mathcal{C}$. At the same time, the second-order Drude contribution to $\mathcal{A}$ is vanishingly small due to the cancellation of contributions from the two valleys within each pair\newtext{; for underlying symmetry properties of nonlinear Drude contribution see SI Section \ref{sec:Drude_symmetries}.}  
Transition to PIP$_2$ eliminates the dominating FS pair in Sym$_4$, which naturally leads to a rotation in $\mathcal{B}$ and $\mathcal{C}$ (see Fig. \ref{figSI_pockets}). Notably, the spin-nematic order parameter breaks TR in addition to rotational symmetry for both Sym$_4$ and PIP$_2$ phases. Within this scenario one would need to postulate different scattering times in PIP$_2$ and in Sym$_4$ phases to match values of linear conductivity. However, the values of nonlinear conductivity are order of magnitudes off from the extracted from the data, as we show below. This indicates that nonlinear transport in Bernal Bilayer graphene might be governed by a new, unknown mechanism.

\subsection{Calculations for Bernal Bilayer Graphene in PIP$_2$ and Sym$_4$ phases}

We start with a $4 \times 4$ Hamiltonian \cite{JungMacD2014BBG} describing electronic states of Bernal Bilayer graphene near $K$ and $K'$ points in the Brillouin zone:
\begin{equation}
    H_0 = \begin{pmatrix}
    \newtext{\frac{D}{2}} & v_0 (\pm k_x - i k_y) & -v_4 (\pm k_x - i k_y) & -v_3 (\pm k_x + i k_y) \\
    v_0 (\pm k_x + i k_y) &  \newtext{\frac{D}{2}} & t_1 & -v_4 (\pm k_x - i k_y) \\
    -v_4 (\pm k_x + i k_y) & t_1 & \newtext{-\frac{D}{2}} & v_0 (\pm k_x - i k_y) \\
    -v_3 (\pm k_x - i k_y) & -v_4 (\pm k_x + i k_y) & v_0 (\pm k_x + i k_y) & \newtext{-\frac{D}{2}}
    \end{pmatrix},
    \label{4x4Hamiltonian}
\end{equation}
where $\pm$ corresponds to $K$ ($K'$) valley. 
The Hamiltonian \eqref{4x4Hamiltonian} is written in the basis $(A_t, B_t, A_b, B_b)$, where $A, B$ are sublattice sites, $t,b$ label top and bottom layers correspondingly, and $(t_0, t_1, t_3, t_4) = (-2.61,0.361,0.283,0.183)$eV, $v = -t_0 \sqrt{3} a/2 \hbar$, $v_3 = t_3 \sqrt{3} a/2 \hbar$, $v_4 = t_4 \sqrt{3} a/2 \hbar$. Additionally, $\newtext{D}$ -- is the magnitude of the out-of-plane displacement field and $a=2.46 \text{\AA}$ -- is the graphene lattice constant. \newtext{Finite displacement field is crucial for the observation of nonlinear conductivity as it breaks inversion symmetry and stimulates the effect, see Fig. \ref{figSI_Displacement} in the SI.} Incorporating the spin degree of freedom can be achieved by the standard means of a direct product with the identity matrix acting in the spin space. In our calculations we use the low-energy $2 \times 2$ Hamiltonian which is obtained by applying the Schrieffer-Wolff transformation to the $4 \times 4$ Hamiltonian. Such transformation requires reshuffling the indices to separate the high-energy sector of the Hamiltonian from the low-energy sector. After such reshuffling we work in the $(B_t, A_b, A_t, B_b)$ basis and the Hamiltonian reads
\begin{equation}
    H_0 = \begin{pmatrix}
    \newtext{\frac{D}{2}} & t_1 & v_0 (k_x + i k_y) & -v_4 (k_x - i k_y) \\
    t_1 &  \newtext{- \frac{D}{2}} & -v_4 (k_x + i k_y) & v_0 (k_x - i k_y) \\
    v_0 (k_x - i k_y) & -v_4 (k_x - i k_y) & \newtext{\frac{D}{2}} & -v_3 (k_x + i k_y)  \\
    -v_4 (k_x + i k_y) & v_0 (k_x + i k_y) & -v_3 (k_x - i k_y) & \newtext{-\frac{D}{2}}
    \end{pmatrix},
    \label{4x4Hamiltonian_block}
\end{equation}
where the top diagonal block corresponds to the high-energy sector (since $t_1$ is the highest energy scale present) and the bottom diagonal block corresponds to low-energy sector. To employ Schrieffer-Wolff transformation we consider the unperturbed Hamiltonian 
\begin{equation}
    H^{(0)} = \begin{pmatrix}
    \newtext{\frac{D}{2}} & t_1 & 0 & 0 \\
    t_1 &  \newtext{- \frac{D}{2}} & 0 & 0 \\
    0 & 0 & \newtext{\frac{D}{2}} & -v_3 (k_x + i k_y)  \\
    0 & 0 & -v_3 (k_x - i k_y) & \newtext{-\frac{D}{2}}
    \end{pmatrix},
    \label{4x4HamiltonianH0}
\end{equation}
and the perturbation
\begin{equation}
    \delta H = \begin{pmatrix}
    0 & 0 & v_0 (k_x + i k_y) & -v_4 (k_x - i k_y) \\
    0 & 0 & -v_4 (k_x + i k_y) & v_0 (k_x - i k_y) \\
    v_0 (k_x - i k_y) & -v_4 (k_x - i k_y) & 0 & 0  \\
    -v_4 (k_x + i k_y) & v_0 (k_x + i k_y) & 0 & 0
    \end{pmatrix}.
    \label{4x4HamiltonianHV}
\end{equation}
The low-energy Hamiltonian up to second order in perturbation is then obtained via the standard expression
\begin{equation}
    \langle \alpha | H' | \alpha' \rangle = \delta_{\alpha \alpha'} E_{\alpha} + \langle \alpha | \delta H | \alpha' \rangle + \frac{1}{2} \sum_{\beta \neq \alpha} \langle \alpha | \delta H | \beta \rangle \langle \beta | \delta H | \alpha' \rangle \left( \frac{1}{E_{\alpha} - E_{\beta}} + \frac{1}{E_{\alpha'} - E_{\beta'}} \right),
\end{equation}
where $E_{\alpha}$ are the eigenvalues of $H^{(0)}$.
An additional term describing Kane-Mele-like spin-orbit coupling \cite{KaneMelePRL2005} reads 
\begin{equation}
H_{SOC} = g_{KM} \sigma_3 \tau_3 s_3,
\end{equation}
where $\pmb{\tau}$, $\pmb{\sigma}$, and $\mathbf{s}$ -- are vectors of Pauli matrices acting in valley, sublattice, and spin spaces correspondingly and 
we use $g_{KM} = 0.03$meV $= 350$mK \cite{AndreaKITPSOC}. Therefore, our initial non-interacting Hamiltonian $H_0$ is given by
$$
H = H_0 + H_{SOC}. 
$$
While SOC is not necessary to describe nonlinear conductivity in the PIP$_2$ phase, it is crucial for the explanation of the signal in Sym$_4$ phase and for the "reorientation transition" -- the 180$^{\circ}$ rotation of $\mathcal{B}, \mathcal{C}$-vectors upon crossing the boundary between PIP$_2$ and Sym$_4$.

\begin{figure*}
\includegraphics[width=1\linewidth]{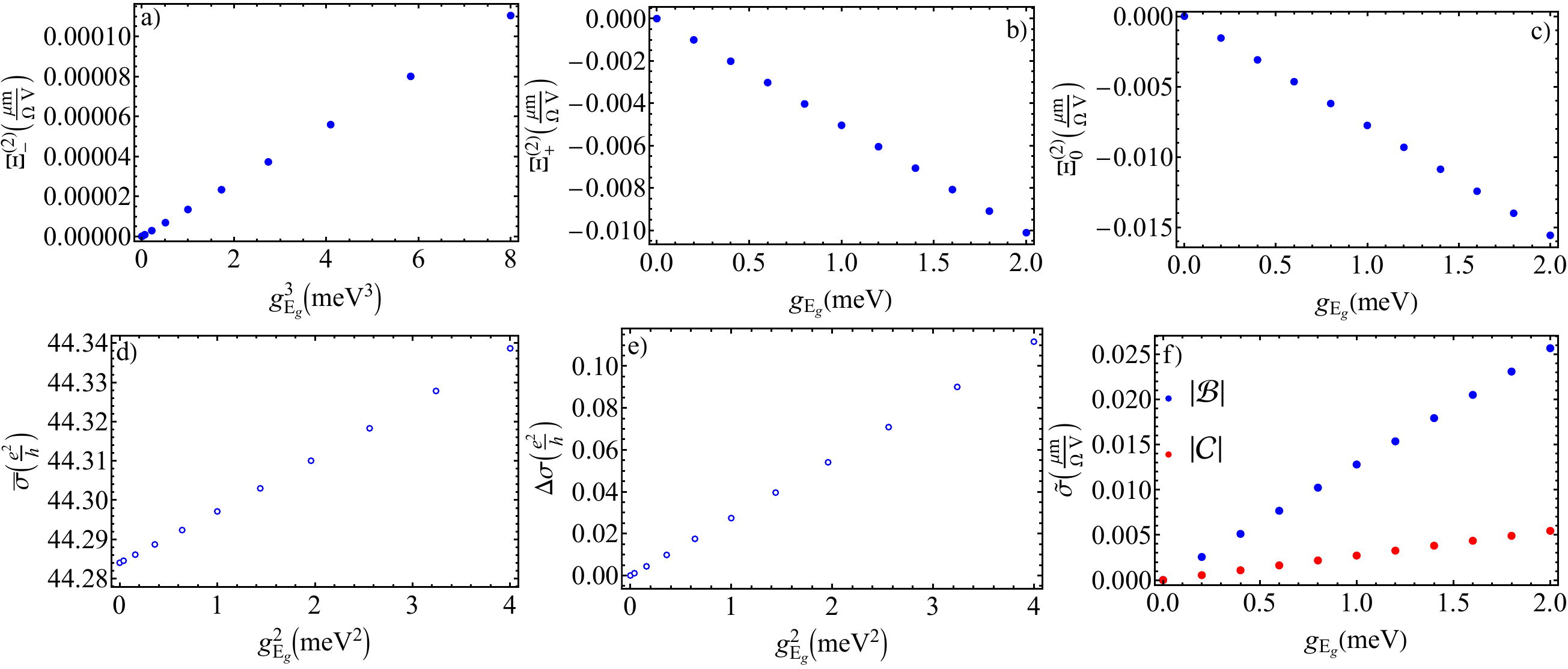}
\caption{{\bf{Dependence of Drude contribution to nonlinear conductivity on the magnitude of the spin-nematic order parameter $g_{E_g}$ in the PIP$_2$ phase}} Magnitude of $\Xi_{-}^{(2)}$ (a), $\Xi_{+}^{(2)}$ (b), and $\Xi_{0}^{(2)}$ (c) components of nonlinear conductivity tensor as a function of $g_{E_g}$ magnitude. THe functional dependencies on $g_{E_g}$ in all three panels satisfy the expected ones \eqref{Drude_3fold_nemat_dep}, \eqref{Drude_1fold_nemat_dep}. Dependence of linear conductivity isotropic part (d) and anisotropy (e) on $g_{E_g}$ magnitude. Panel (e) exhibits quadratic dependence of linear conductivity anisotropy on $g_{E_g}$, as expected from Onsager relations \eqref{Onsager_lin_cond}.  Panel (f) shows the calculated magnitudes of $\mathcal{B}$ and $\mathcal{C}$ vectors.  
In our calculations we used $\tau=3.33$ps, $\mu=-15.5$meV, $D=30$meV. }
\label{Drude_calc_PIP2}
\end{figure*}

Assuming a weak time reversal-breaking nematic order parameter (like charge $E_u$) we can qualitatively account for the observed hierarchy of nonlinear conductivity tensor components in both PIP$_2$ and Sym$_4$ phases of BBG. \newtext{Temperature dependence of nonlinear effect shown in Fig. \ref{figSI_temperature} sets the energy scale for the perturbation to be $\sim 10$K, i.e.,  $\sim 1$meV. }
To calculate the linear and nonlinear conductivity tensors in the PIP$_2$ phase we need to  postulate the nature of PIP$_2$ as well as the exact origin of the subleading order parameter that breaks rotational symmetry.
Specifically, we need to assume an order parameter that leads to the "removal" of 2 out of 4 large Fermi pockets, in accordance with quantum oscillations data. Our experimental data in the linear regime indicate that anomalous Hall conductivity in PIP$_2$ phase is vanishing, therefore, valley polarization can be eliminated from the pool of possible ordered states. Assuming no translation symmetry breaking, this leaves us with two potential order parameter candidates with zero transferred momentum $Q=0$ (a generalized Stoner instability): spin $A_{1g}$ (ferromagnetic, FM) and $A_{1u}$ (valley antiferromagnetic, AFM, a.k.a. staggered spin current \cite{CvetkovicPRB2012}). To be specific we assume valley AFM order in PIP$_2$ phase. We furthermore assume, in accordance with experimental data, that the magnitude of the PIP$_2$ order parameter is large and out of 4 initial Fermi pockets only 2 remain. This satisfies the Luttinger volume count as our data at PIP$_2$ is taken at half of the density where we took our measurements for Sym$_4$ phase. The Fermi pockets for the two phases are shown in Fig. \ref{figSI_pockets}.
To account for the broken rotational symmetry in the PIP$_2$ phase we add a perturbation that corresponds to the spin-nematic order parameter belonging to the $E_g$ irreducible representation. This order parameter manifestly breaks two symmetries: $C_3$ rotational symmetry and time-reversal symmetry.  It is given by \cite{CvetkovicPRB2012}
\begin{equation}
    H_{E_g} = g_{E_g} \tau_0 \otimes \pmb{\sigma} \otimes \mathbf{s}.
\end{equation} 
$H_{E_g}$ is added directly to the spinfull 4-band Hamiltonian $H_0 \otimes 1_{2 \times 2}$.
In our calculations we pick $\pmb{s}$ along the $z$-axis and $\pmb{\sigma}$ along the $x$-axis. The Fermi surfaces in the presence of a weak $E_g$ spin order parameter are shown in Fig. \ref{figSI_pockets}. We now use Eq. \eqref{Tensor_components_Boltzmann} with momentum-independent scattering time to calculate linear and nonlinear conductivity tensors in the PIP$_2$ phase and obtain that linear conductivity tensor is nearly isotropic with anomalous Hall contribution being nearly vanishing. For a weak $E_g$ spin nematic perturbation this is in accordance with Onsager relations from Sec. \ref{sec:Linear_Onsager}. The nonlinear conductivity tensor, on the other hand, has a very small 3-fold component $|\Xi_{-}^{(2)}|$ and 1-fold components satisfy $|\Xi_{+}^{(2)}| \sim \frac{|\Xi_{0}^{(2)}|}{2}$ which follows from properties of Drude contribution from Sec. \ref{sec:Drude_symmetries} and reproduces experimental ratios really well, see Fig. \ref{Drude_calc_PIP2}. However, the magnitudes of $\tilde{\sigma}_{\alpha \mu \nu}$ are significantly smaller than the extracted from the data in both phases.

\begin{figure*}
\includegraphics[width=1\linewidth]{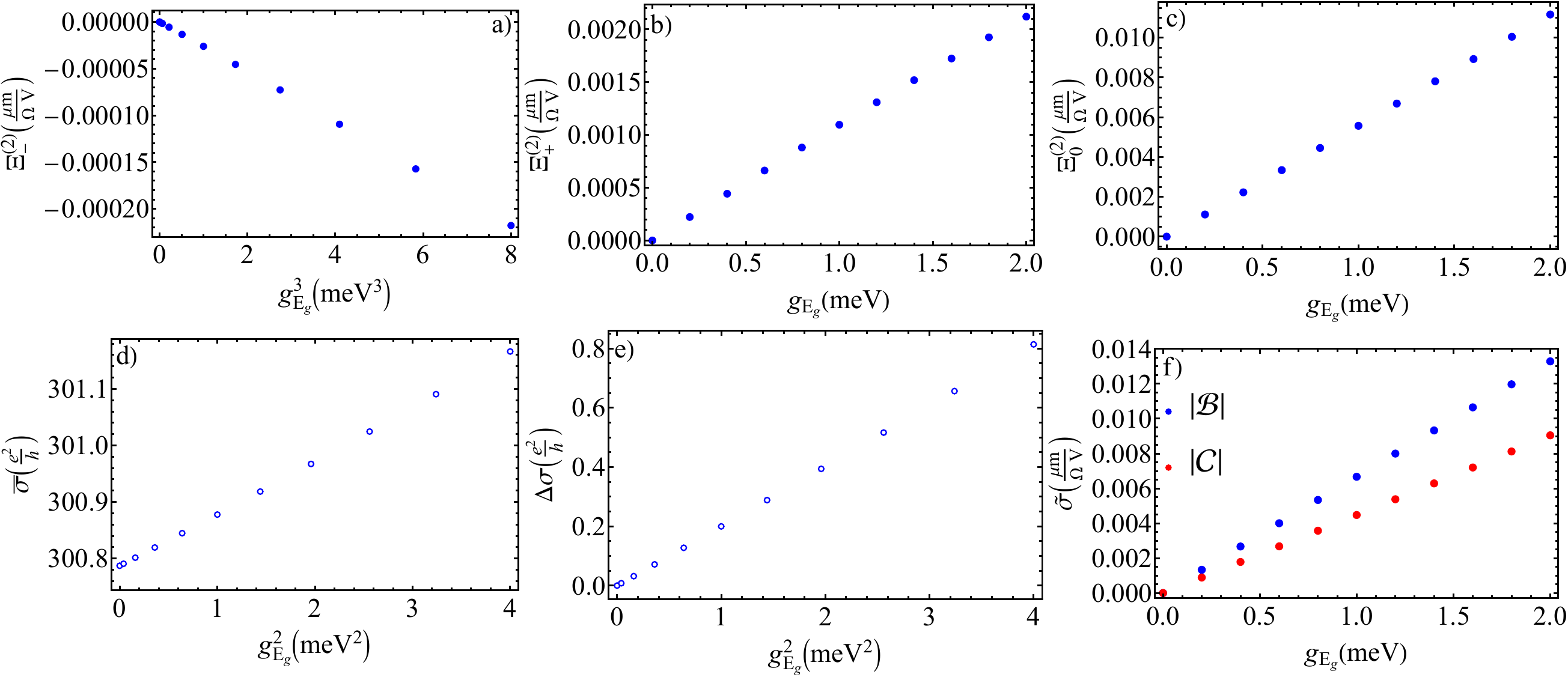}
\caption{{\bf{Dependence of Drude contribution to nonlinear conductivity on the magnitude of the spin-nematic order parameter $g_{E_g}$ in the Sym$_4$ phase}} Magnitude of $\Xi_{-}^{(2)}$ (a), $\Xi_{+}^{(2)}$ (b), and $\Xi_{0}^{(2)}$ (c) components of nonlinear conductivity tensor as a function of $g_{E_g}$ magnitude. Like for the PIP$_2$ phase, the functional dependencies on $g_{E_g}$ in all three panels satisfy the expected ones \eqref{Drude_3fold_nemat_dep}, \eqref{Drude_1fold_nemat_dep}. Notice that the signs of all three coefficients $\Xi_{-,+,0}^{(2)}$ are opposite to those in the PIP$_2$ two phase. Dependence of linear conductivity isotropic part (d) and anisotropy (e) on $g_{E_g}$ magnitude. Panel (e) exhibits quadratic dependence of linear conductivity anisotropy on $g_{E_g}$, as expected from Onsager relations \eqref{Onsager_lin_cond}.  Panel (f) shows the calculated magnitudes of $\mathcal{B}$ and $\mathcal{C}$ vectors.
In our calculations we used $\tau=11.5$ps, $\mu=-15.5$meV, $D=30$meV. }
\label{Drude_calc_Sym4}
\end{figure*}

The structure of $\tilde{\sigma}_{\alpha \mu \nu}$ in Sym$_4$ can be described by adding the two missing pockets back into the system. Their sizes are slightly larger than of the two pockets constituting the PIP$_2$ phase, however, this difference is small and wouldn't be distinguishable in quantum oscillations. Schematically, the transition between the PIP$_2$ and the Sym$_4$ phases and the corresponding change in the structure of  $\tilde{\sigma}_{\alpha \mu \nu}$ is shown in Fig. \ref{figSI_pockets}. We also show the calculated Drude contribution to nonlinear conductivity in Fig. \ref{Drude_calc_Sym4}. Comparing panels a), b), c) of Figs. \ref{Drude_calc_PIP2} and \ref{Drude_calc_Sym4} one notices that $\Xi^{(2)}$-coefficients in the PIP$_2$ phase have opposite signs compared to the Sym$_4$ phase. This is consistent with the observed $180^{\circ}$-reorientation of $\mathcal{B},\mathcal{C}$-vectors upon crossing the boundary between the two phases.

\begin{figure*}
\includegraphics[width=1\linewidth]{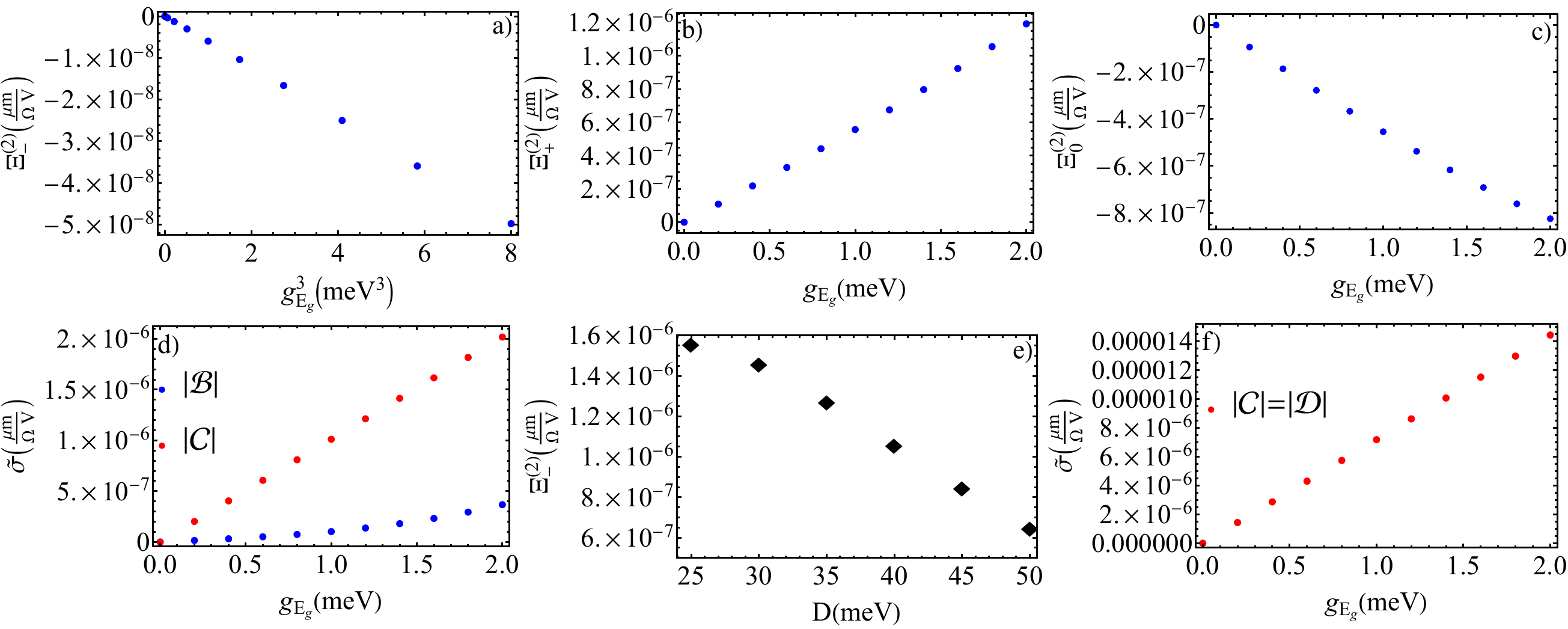}
\caption{{\bf{Dependence of Quantum geometric and Berry curvature dipole contributions to nonlinear conductivity on the magnitude of the spin-nematic order parameter $g_{E_g}$ in the PIP$_2$ phase}} Panels (a), (b), (c) show the dependence of quantum geometric contribution to $\Xi_{-,+,0}^{(2)}$ on the magnitude of $g_{E_g}$. Panel (d) shows the calculated magnitudes of $\mathcal{B}$ and $\mathcal{C}$ vectors. Panel (f) shows the dependence of Berry curvature dipole contribution on $g_{E_g}$. We observe that for the same parameters of the Hamiltonian and the scattering time the quantum geometric and Berry curvature dipole contributions are significantly smaller in magnitude than the nonlinear Drude contribution. Panel (e) shows the dependence of quantum geometric contribution on displacement field $D$ for a single pocket in one valley and in the absence of nematic order parameter. As one would naively expect from Eq. \eqref{G_tensor_Q_Niu}, the quantum geometric contribution becomes smaller in magnitude with an increase in displacement field. 
}
\label{QG_and_BCD_calc_PIP2}
\end{figure*}

In our calculations we find the other internal (BCD, quantum geometry)  contributions to be smaller in magnitude that those of nonlinear Drude, see Fig. \ref{QG_and_BCD_calc_PIP2} for PIP$_2$ phase. There is also an additional argument against the BCD scenario for nonlinear Hall effect in our setup, which we presented in the Main text. We repeat it here.  If the BCD alone is responsible for the nonlinear Hall effect, reversing $D$ is expected to influence the orientation of the $\mathcal{C}$-vector  ~\cite{Sodemann_Fu}. Alternatively, BCD-induced $\mathcal{C}$-vector could preserve its orientation while $\mathcal{A}$ and $\mathcal{B}$, given by quantum geometric or second-order Drude contributions, reverse sign under $D \rightarrow -D$, given that the order parameter associated with the nematic perturbation is linked to the direction of $D$. An example of this scenario is the charge $E_u$ order parameter. 
We examine these scenarios by extracting $\mathcal{A},\mathcal{B},\mathcal{C}$ vectors at $D= \pm 300$mV/nm. Fig. 4c of the Main text shows the angular dependence of nonlinear response taken at $D=-300$mV/nm and $n=-0.15\times 10^{12}$ cm$^{-2}$. The orientations of both the $\mathcal{C}$-vector and $\mathcal{B}$-vector remain the same as in Fig. 3d of the Main text. 
Therefore, the invariance of $\mathcal{B}$-vector and $\mathcal{C}$-vector, combined, indicates that nonlinear Hall effect is not induced by the BCD mechanism as the order parameter associated with the nematic perturbation appears invariant under a sign change in $D$. 

\begin{figure*}
\includegraphics[width=0.7\linewidth]{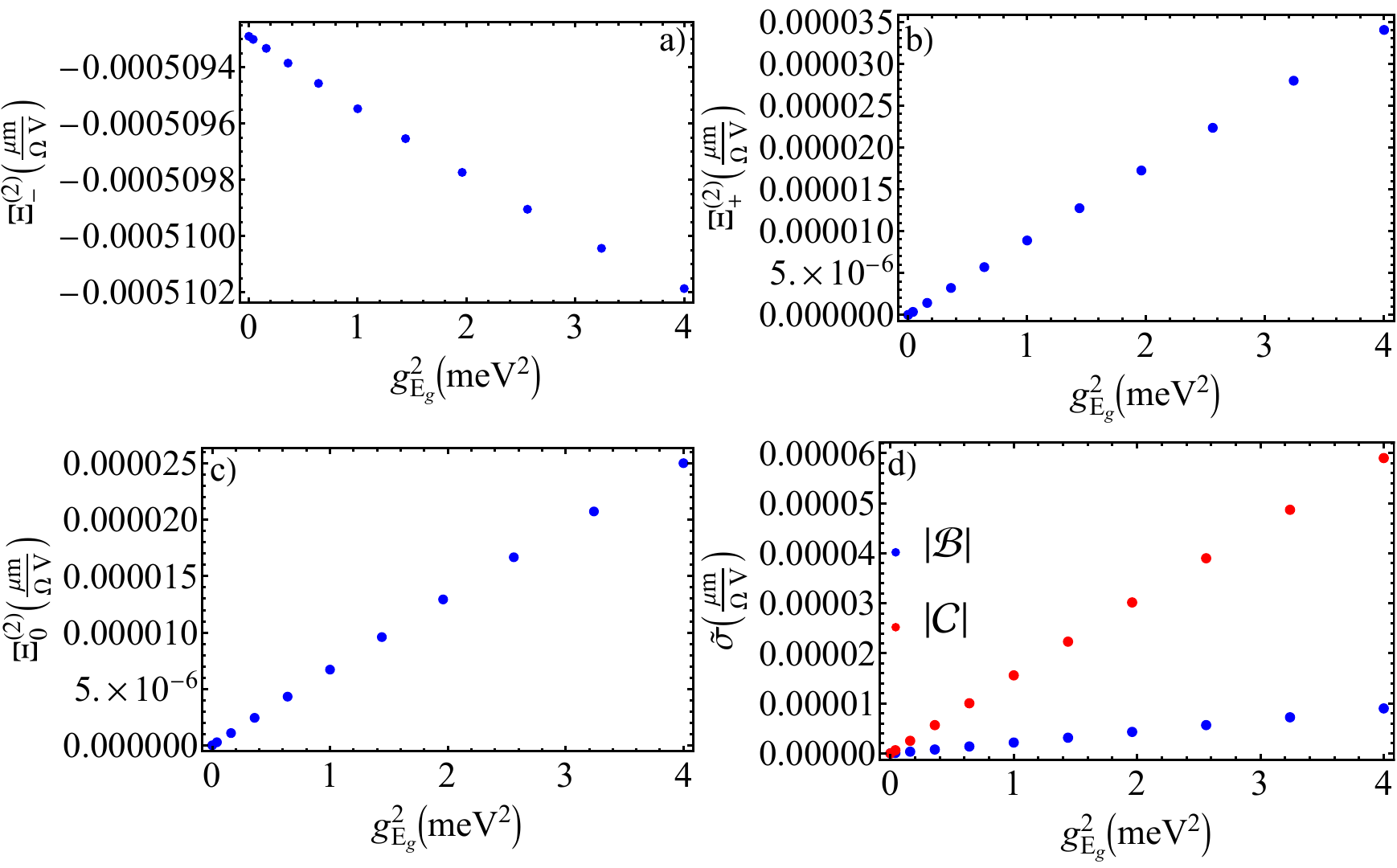}
\caption{{\bf{Dependence of side jump contribution to nonlinear conductivity on the magnitude of the spin-nematic order parameter $g_{E_g}$ in the PIP$_2$ phase}} Panels (a), (b), (c) show the dependence of side jump contribution to $\Xi_{-,+,0}^{(2)}$ on the magnitude of $g_{E_g}$. Note that the 3-fold contribution remains the largest of three even for $g_{E_g} = 2$meV, which clearly violates the experimentally observed hierarchy of $\mathcal{A}^3, \mathcal{B}, \mathcal{C}$. Panel (d) shows the dependence of Berry curvature dipole contribution on $g_{E_g}$. 
}
\label{Side_jump_calc_PIP2}
\end{figure*}

\begin{figure*}
\includegraphics[width=1.0\linewidth]{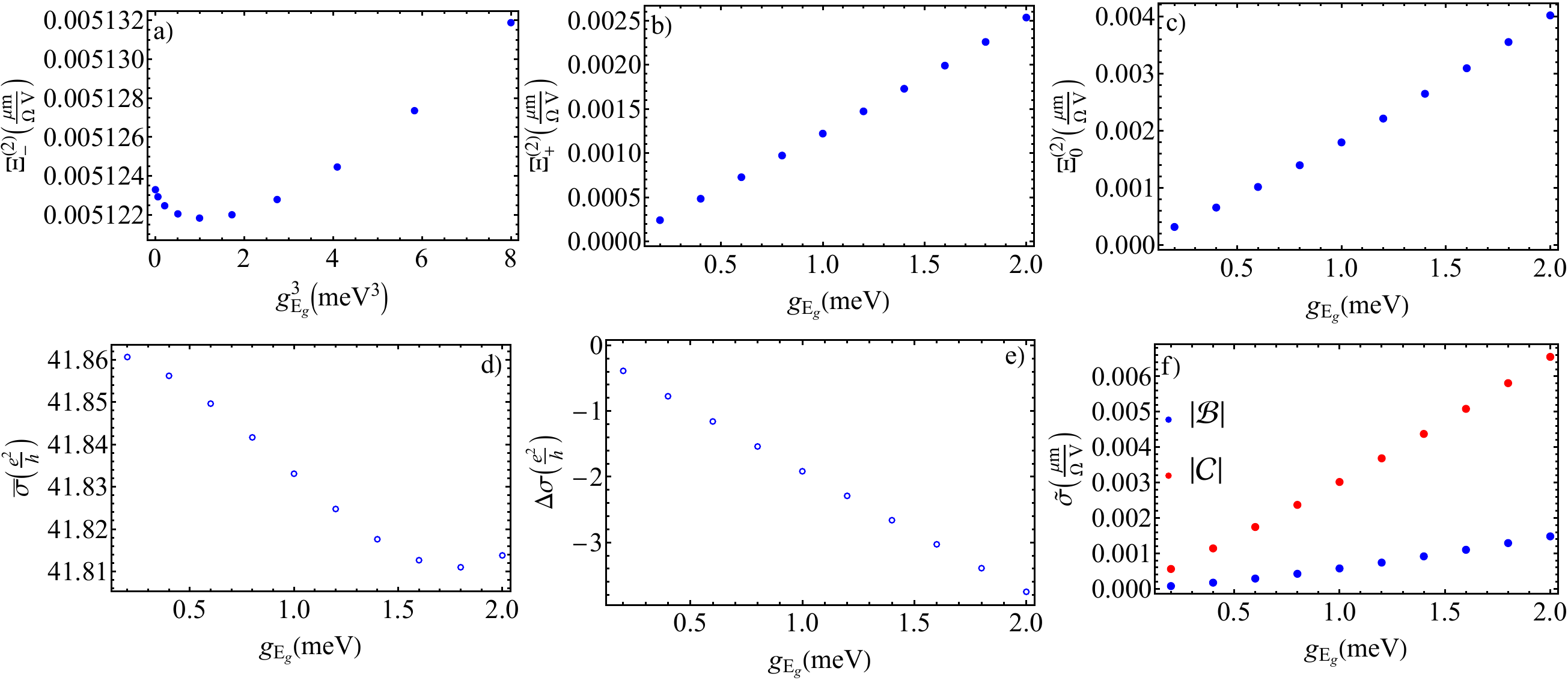}
\caption{{\bf{Dependence of skew scattering contribution to nonlinear conductivity on the magnitude of the charge-nematic order parameter $g_{E_g}$ in the PIP$_2$ phase}} To isolate the skew scattering contribution we calculate nonlinear conductivity for time-reversal preserving nematic perturbation, namely, for charge $E_g$ order parameter. Panels (a), (b), (c) show the dependence of side jump contribution to $\Xi_{-,+,0}^{(2)}$ on the magnitude of $g_{E_g}$. Note that like for the side jump case the 3-fold contribution remains the largest of three even for $g_{E_g} = 2$meV, which clearly violates the experimentally observed hierarchy of $\mathcal{A}^3, \mathcal{B}, \mathcal{C}$. Dependence of linear conductivity isotropic part (d) and anisotropy (e) on $g_{E_g}$ magnitude. Note that for TR-preserving perturbation linear conductivity anisotropy is higher than for TR-breaking perturbation.  Panel (f) shows the calculated magnitudes of $\mathcal{B}$ and $\mathcal{C}$ vectors. In our calculations we used $A_{\mathbf{k}, \mathbf{k}'} = U_0 A_{u.c.}$ and $\mathbf{B}_{\mathbf{k}, \mathbf{k}'}= \left( 0, 0, i A_{\mathbf{k}, \mathbf{k}'} \right)$ with $U_0 = 60$eV and $n_{imp} = 10^9 \text{cm}^{-2}$.  
}
\label{Skew_calc_PIP2}
\end{figure*}

Finally, we also calculate the side jump (Fig. \ref{Side_jump_calc_PIP2}) and skew scattering (Fig. \ref{Skew_calc_PIP2}) contributions to nonlinear conductivity for PIP$_2$ phase for a fixed value of linear conductivity $\simeq 40 \frac{e^2}{h}$ and observe that all known mechanisms lead to nonlinear conductivity magnitude being orders of magnitude smaller than what is observed in the data.

\end{document}